\documentclass[twocolumn]{aastex63}		         
\usepackage{lineno}
 
\usepackage{float}
\usepackage{comment}
\usepackage[]{natbib}
\usepackage{placeins}
\usepackage{graphicx}	
\usepackage{multirow}
\usepackage{amssymb}
\usepackage{amsmath}

\usepackage{hyperref}
\usepackage{booktabs}

\shorttitle{BH Candidates in Dwarf Galaxies Detected via [Fe X] Emission}

\newcounter{species} 
\def\ion#1#2{\hbox{\setcounter{species}{#2}#1\,{\scriptsize\Roman{species}}\relax}}

\newcommand{\fexl}{[\ion{Fe}{10}]$\lambda$6374}
\newcommand{\fex}{[\ion{Fe}{10}]}
\newcommand{\kms}{km~s$^{-1}$}

\def\lsim{\lower0.3em\hbox{$\,\buildrel <\over\sim\,$}}
\def\gsim{\lower0.3em\hbox{$\,\buildrel >\over\sim\,$}}

\usepackage[normalem]{ulem}
\def\crossout#1{}
\def\crossout#1{\sout{#1}}

\begin{document}
\title{A Sample of Massive Black Holes in Dwarf Galaxies Detected via \fex\ Coronal Line Emission: Active Galactic Nuclei and/or Tidal Disruption Events}

\correspondingauthor{Mallory Molina}
\email{mallory.molina@montana.edu}

\author[0000-0001-8440-3613]{Mallory Molina} 
\affil{eXtreme Gravity Institute, Department of Physics, Montana State University, Bozeman, MT 59717, USA}

\author[0000-0001-7158-614X]{Amy E. Reines}
\affil{eXtreme Gravity Institute, Department of Physics, Montana State University, Bozeman, MT 59717, USA}

\author{Lilikoi J. Latimer}
\affil{eXtreme Gravity Institute, Department of Physics, Montana State University, Bozeman, MT 59717, USA}

\author[0000-0003-4703-7276]{Vivienne Baldassare}
\affil{Department of Physics and Astronomy, Washington State University, Pullman, WA 99163, USA}

\author[0000-0002-4587-1905]{Sheyda Salehirad}
\affil{eXtreme Gravity Institute, Department of Physics, Montana State University, Bozeman, MT 59717, USA}

\begin{abstract}

The massive black hole (BH) population in dwarf galaxies ($M_{\rm BH} \lesssim 10^5~M_\odot$) can provide strong constraints on the origin of BH seeds.  However, traditional optical searches for active galactic nuclei (AGNs) only reliably detect high-accretion, relatively high-mass BHs in dwarf galaxies with low amounts of star formation, leaving a large portion of the overall BH population in dwarf galaxies relatively unexplored. Here, we present a sample of 81 dwarf galaxies ($M_\star \le 3 \times 10^9~M_\odot$) with detectable \fexl\ coronal line emission indicative of accretion onto massive BHs, only two of which were previously identified as optical AGNs. We analyze optical spectroscopy from the Sloan Digital Sky Survey and find \fexl\ luminosities in the range $L_{[Fe\textsc{x}]}\approx10^{36}$--$10^{39}$~erg~s$^{-1}$, with a median value of $1.6 \times 10^{38}$~erg~s$^{-1}$. The \fexl\ luminosities are generally much too high to be produced by stellar sources, including luminous Type IIn supernovae (SNe). Moreover, based on known SNe rates, we expect at most 8 Type IIn SNe in our sample. On the other hand, the \fexl\ luminosities are consistent with accretion onto massive BHs from AGNs or tidal disruption events (TDEs).  We find additional indicators of BH accretion in some cases using other emission line diagnostics, optical variability, X-ray and radio emission (or some combination of these). However, many of the galaxies in our sample only have evidence for a massive BH based on their \fexl\ luminosities.  This work highlights the power of coronal line emission to find BHs in dwarf galaxies missed by other selection techniques and to probe the BH population in bluer, lower mass dwarf galaxies.
\end{abstract}

\keywords{Active galaxies -- Active galactic nuclei -- Dwarf galaxies -- Low-luminosity active galactic nuclei -- Tidal disruption -- black holes}

\section{Introduction}\label{sec:intro}

Supermassive black holes (BHs) are known to live in the nuclei of almost all massive galaxies \citep[e.g.,][]{Kormendy1995,Kormendy2013}, but merger-driven galaxy growth has effectively erased their initial formation conditions \citep{Volonteri2010,Natarajan2014}. There are several proposed theories for the formation of the intial BH ``seeds'' in the early Universe, including remnants from Population III stars \citep{Bromm2011}, the direct collapse of gas clouds \citep{Loeb1994,Begelman2006,Lodato2006,Choi2015} and stellar collisions in dense star clusters \citep{Portegies2004,Devecchi2009,Davies2011,Lupi2014,Stone2017}. Each of these processes creates different size BH seeds; the remnants of Population III stars would create $M_{\rm BH}\sim100~M_\odot$ \citep{Bond1984,Heger2002}, while direct collapse and star cluster collisions would create $M_{\rm BH}\sim10^3$--$10^5~M_\odot$. 

While it is not feasible to observe the first BH seeds at high redshift, nearby dwarf galaxies can place strong constraints on their properties \citep[see][and references therein]{Reines2016,Greene2020}. Dwarf galaxies have relatively quiet merger histories compared to their more massive counterparts \citep{Bellovary2011}, and may experience strong supernova feedback that stunts BH growth \citep{Angles2017,Habouzit2017}. Therefore, if a dwarf galaxy does harbor a BH, it should have a mass relatively close to the initial seed mass (e.g., $M_{\rm BH}\lesssim10^5$).

Unfortunately, the smaller masses and lower luminosities of BHs in dwarf galaxies make them difficult to detect. In the optical regime, narrow-line ratios consistent with active galactic nuclei (AGN) photoionization as well as broad H$\alpha$ emission have been used to identify BHs in dwarf galaxies \citep[e.g.,][]{reines2013,Baldassare2016}. These studies incorporated the commonly used Baldwin, Phillips and Terlevich diagrams, \citep[BPT diagrams;][]{Baldwin1981} also referred to as the \citet[][VO87]{VO87} diagrams which use narrow-line ratios to differentiate between the spectral energy distributions of star forming regions and AGNs. However, these diagrams struggle with low-luminosity AGNs (LLAGNs), low ionization nuclear emission regions (LINERs) and contributions from shocks \citep[see][for a review]{Ho2008,Kewley2019}. Therefore, these studies often only detect the brightest, highest-accretion rate objects, leaving a large portion of the lower-mass, lower-accretion rate BHs relatively unexplored \citep{Greene2020}. Additionally, optical AGN indicators may be diluted by emission from the host galaxy, even in objects with low star-formation rates \citep[SFRs;][]{Moran2002,Groves2006,Stasinska2006,Cann2019}. Any combination of these factors can easily hide AGN activity in optical surveys. 

Infrared (IR) colors have also been used to identify AGNs in massive galaxies \citep{Jarrett2011,Mateos2012,Stern2012}, but is often confused with star formation in dwarf galaxies \citep[][]{Hainline2016,Condon2019,Latimer2021}. Hard X-ray emission is a clean diagnostic when searching for AGN activity at high luminosities, but can be confused with high-mass X-ray binaries at low luminosities and can also miss obscured AGNs \citep{hickox2018,Latimer2021}. Similarly, radio emission is not affected by reddening due to dust and is produced by almost all AGNs \citep[see][and references therein]{Ho2008}, but the expected radio emission from BHs in dwarf galaxies could be similar to that from \ion{H}{2} regions, supernova remnants or supernovae \citep[e.g.,][]{,Reines2008,Chomiuk2009,Johnson2009,aversa2011,Kepley2014,Varenius2019,reines2020}. Therefore, it is challenging to find low accretion-rate, low-mass BHs using these detection methods.

Recently, \cite{Molina2021} and \cite{Kimbro2021} confirmed the presence of low-accretion rate BHs in the dwarf galaxies SDSS J122011.26+302008.1 \citep[J1220+3020; ID 82 in][]{reines2020} and Mrk 709 S \citep{Reines2014}, respectively. Interestingly, both objects had clear detections of the coronal line \fexl. Due to its high ionization potential \citep[262.1~eV;][]{Oetken1977}, \fexl\ and other coronal lines are considered to be a reliable signature of AGN activity in galaxies \citep[e.g.,][]{penston1984,prieto2000,prieto2002,reunanen2003}. 
While coronal-line emission is often considered to be produced by gas photoionized by the hard AGN continuum \citep[e.g., ][]{Nussbaumer1970,Korista1989,Oliva1994,Pier1995}, \fexl\ emission can also be mechanically excited by radiative shock waves that are driven into the host galaxy by radio jets from the AGN \citep{wilson1999}. As both BHs studied in \cite{Molina2021} and \cite{Kimbro2021} had low accretion rates and strong radio detections, they conclude that the coronal-line emission is created mechanically by jet-driven winds. Given the presence of \fexl\ emission in two different objects, one of which was detected in the Sloan Digital Sky Survey (SDSS) 3\arcsec\ single-fiber spectrum, this discovery opens a new method for detecting low accretion-rate BHs in dwarf galaxies.

In this work we present the first systematic search for \fexl-selected massive BHs in dwarf galaxies, making use of SDSS 3\arcsec\ single-fiber spectra. We describe our sample selection in Section~\ref{sec:samp}. We then use various data sets to search for additional AGN markers in our sample. The analysis of the SDSS optical spectra, including several emission-line ratio diagnostics, are presented in Section~\ref{sec:sdss_spec}. We 
then search for AGN-like optical variability in Section~\ref{sec:lc_data}.
Finally, we use archival \textit{Chandra} data and radio surveys to search for X-ray and radio emission consistent with AGN activity in Sections~\ref{sec:xrayemis} and \ref{sec:radio_det}. 
We discuss our results in Section~\ref{sec:discussion} and summarize our findings in Section~\ref{sec:summary}. In this paper, we assume a $\Lambda\textrm{CDM}$ cosmology with $\Omega_{\rm m}=0.3$, $\Omega_{\Lambda}=0.7$ and H$_0=73$~km~s$^{-1}$~Mpc$^{-1}$.

\section{Dwarf Galaxies with \fexl\ Emission}\label{sec:samp}

\subsection{Sample Selection and Measurements}\label{sec:sampselect}

We selected our sample of dwarf galaxies from version \texttt{v1\_0\_1} of the NASA-Sloan Atlas \citep[NSA;][]{Blanton2011}. This updated version of the NSA was used to create the galaxy sample for the SDSS-IV Mapping Nearby Galaxies at Apache Point Observatory  \citep[MaNGA;][]{Bundy2015,Yan2016,Blanton2017} survey, and is publicly available via the SDSS Data Release 16 \citep[DR16;][]{Ahumada2020}. We note that any reference to NSA IDs in this paper use those from the \texttt{v1\_0\_1} catalog. In this work, we used the single-fiber SDSS spectra to identify our sample of dwarf galaxies with \fex\ emission. The spectra were taken with the 2.5-meter wide-field Sloan Foundation Telescope and the 640-fiber double spectrograph at Apache Point Observatory, New Mexico \citep{Gunn2006}. The fibers are 3\arcsec\ in diameter, cover a wavelength range of 3800--9200~\AA, and have an instrumental dispersion of 69~\kms\ per pixel. All spectra in DR16 were processed using the version \texttt{v5\_13\_0} of the \texttt{idlspec2d} pipeline \citep{Bolton2012,Dawson2013}, which performs both the data reduction and emission-line measurements. 

As we are interested in limiting our search to dwarf galaxies, we use the same mass constraint as \cite{reines2013}, i.e., $M_* \leq 3\times10^9~{\rm M}_\odot$, and require that each galaxy has a single-fiber SDSS spectrum. We identify 46,530 objects that meet these criteria. While we relied on the NSA to identify our parent sample of dwarf galaxies, we wrote our own software to search for and measure the \fexl\ line.

After we defined our parent sample of dwarf galaxies, we identified all galaxies with \fex\ emission. Previous work on \fex-emitting dwarf galaxies by \cite{Molina2021} and \cite{Kimbro2021} demonstrated that the \fex\ emission will be both relatively weak, even compared to the [\ion{O}{1}] doublet, and well-described by a single Gaussian. Additionally, \cite{Molina2021} found that the \fex\ emission in J1220+3020 was confined to the central 1\arcsec\ using Gemini IFU spectroscopy. We therefore expect weak \fexl\ emission in the much larger 3\arcsec\ aperture that may not exhibit the typical asymmetric profiles seen in AGNs in more massive galaxies \citep[e.g.,][]{Gelbord2009}. In order to account for both of these issues, we created a pipeline that fits both the [\ion{O}{1}] doublet and any potential, weak \fexl\ emission. Our processing pipeline utilizes the python packages \texttt{astropy} \citep{astropy2013,astropy2018}, \texttt{matplotlib} \citep{matplotlib} and \texttt{pyspeckit} \citep{pyspeckit}, which is a program that wraps around the python package \texttt{MPFIT} and is customized to fit astronomical spectra.

\begin{figure*}
    \centering
    \hspace{1.2cm}\includegraphics[width=0.35\textwidth]{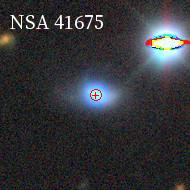}
    \hspace*{1.2cm}\includegraphics[width=0.48\textwidth]{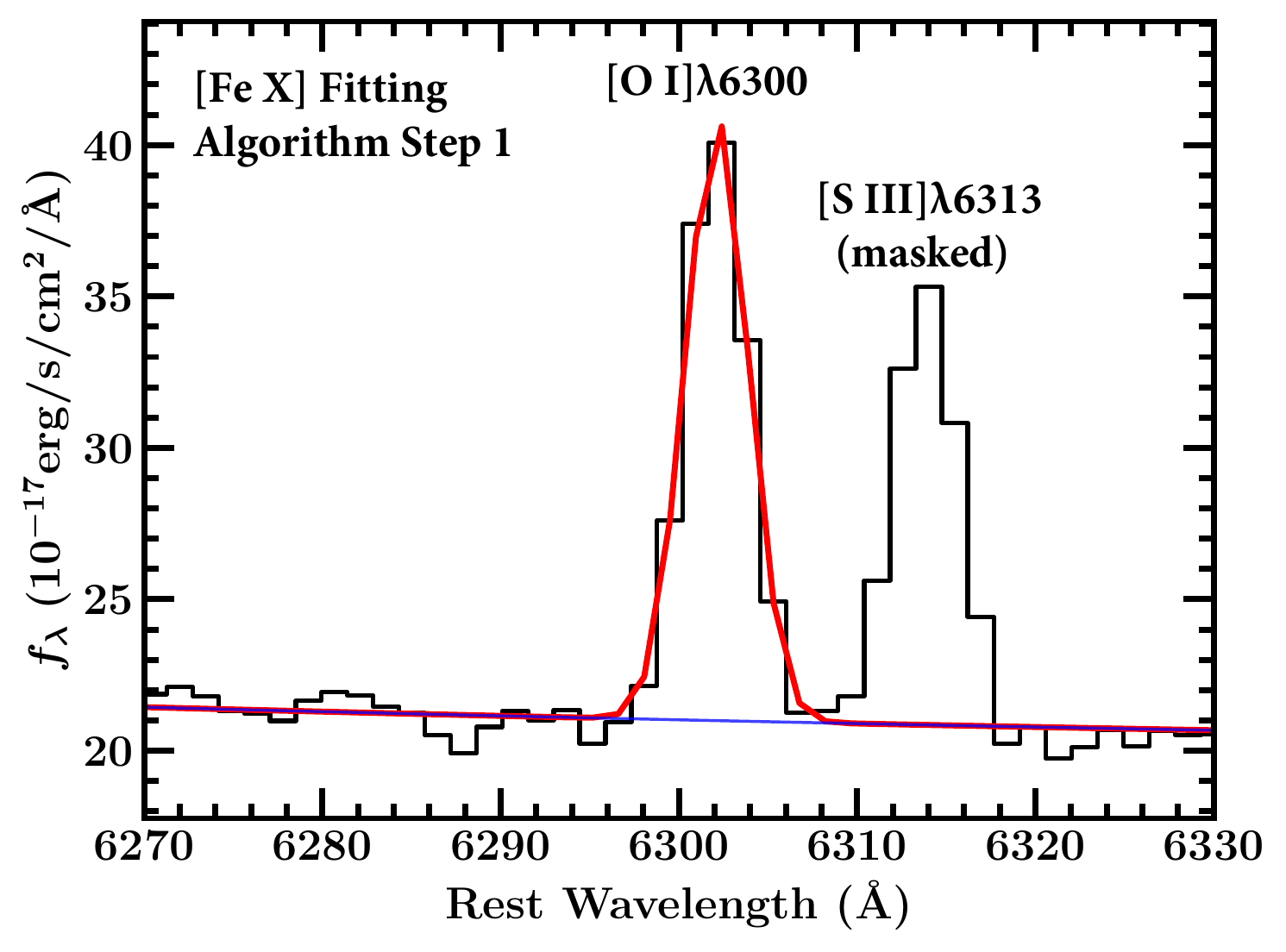}
    \includegraphics[width=0.48\textwidth]{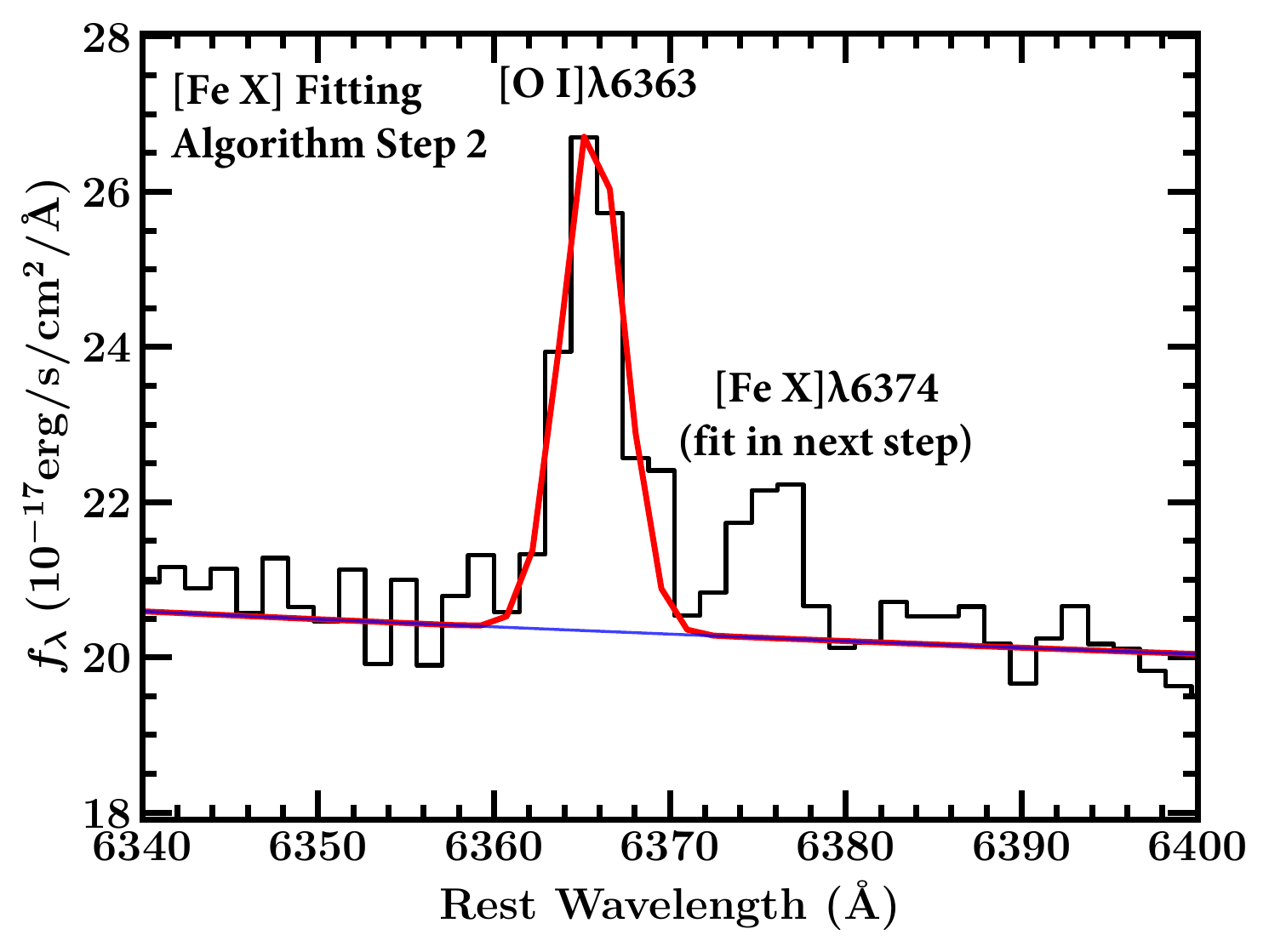}
     \includegraphics[width=0.48\textwidth]{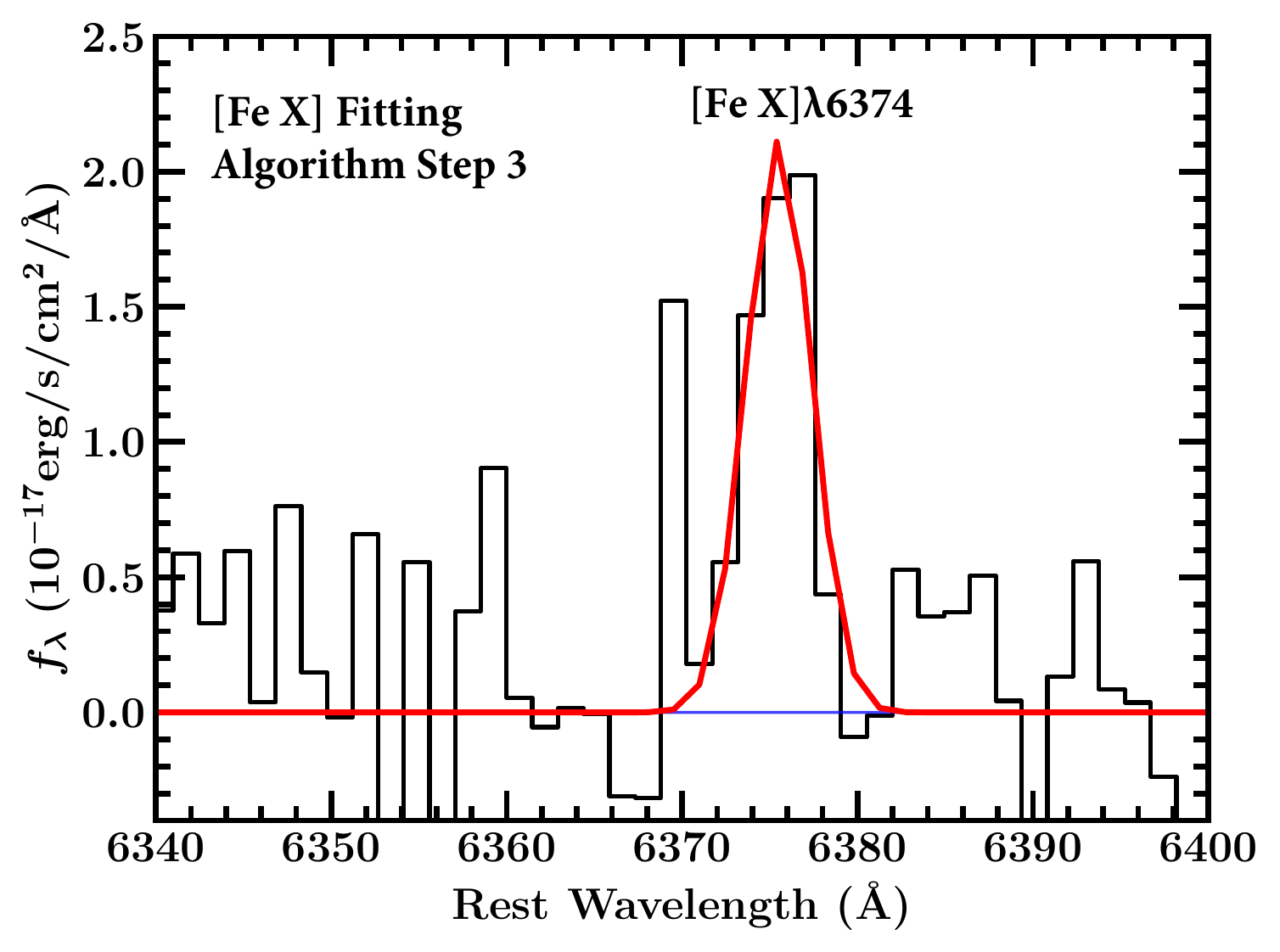}
    \caption{{{\it Top Left:} The DESI Legacy Imaging Survey SkyViewer $grz$ image of the \fex-emitting dwarf galaxy {NSA 41675}. The position of the galaxy nucleus as defined by the NSA is shown as the red cross, and the spatial extent of the SDSS 3\arcsec\ fiber is shown as the the red circle. {\it Top Right:} The fit of the [\ion{O}{3}]$\lambda$6300 line in NSA 41675, and first step in the \fex\ fitting algorithm. The data are in black, the model is shown as the solid red line, and the baseline is shown as the solid blue line. The [\ion{S}{3}]$\lambda$6313 line is detected and masked in this fit. {\it Bottom left: }The second step in the \fex\ fitting algorithm, with the data shown as a solid black line. In this step the [\ion{O}{3}]$\lambda$6363 emission-line model is precisely defined by the fitted [\ion{O}{3}]$\lambda$6300 line. The [\ion{O}{3}]$\lambda$6363 emission line and baseline fit (blue solid line) are added together to create the total model shown as the red line. The total model is subtracted to create the residual spectrum with the \fex\ line. {\it Bottom Right: }Fit to the \fexl\ emission line in the residual spectrum after subtracting the [\ion{O}{1}] emission and continuum. The data are in black, the model is over-plotted in red, {and the ``continuum'' baseline is shown in blue}. We show the galaxy and the fitted \fex\ emission line for all 81 dwarf galaxies in the Appendix.}}
    \label{fig:o16300}
\end{figure*}

We first required that the stronger line in the [\ion{O}{1}] doublet, [\ion{O}{1}]$\lambda$6300, was detected with a signal-to-noise (S/N) $\ge$ 3 in the automated SDSS spectroscopy pipeline, \texttt{idlspec2d}. This S/N check is important for two reasons: (1) strong [\ion{O}{1}] emission was seen in the two previous AGN candidates with \fexl\ emission and is generally associated with AGN activity \citep{kewley2006,reines2020,Molina2021,Kimbro2021}, and (2) a strong [\ion{O}{1}]$\lambda$6300 detection will ensure a clean subtraction of the weaker line in the doublet, [\ion{O}{1}]$\lambda$6363 which could be blended with the \fexl\ emission. 

If this criterion is met, we then fit the [\ion{O}{1}]$\lambda$6300 line using \texttt{pyspeckit}. We simultaneously fit a second-order polynomial to describe the {continuum around the [\ion{O}{1}]+\fex\ complex (spanning $\sim150$\AA)} and a single Gaussian to describe [\ion{O}{1}]$\lambda$6300. {We mask the emission line [\ion{S}{3}]$\lambda$6313 if present to ensure a good fit to the continuum. In all cases, the [\ion{O}{1}]$\lambda$6300 and [\ion{S}{3}]$\lambda$6313 emission lines were not blended.} We repeat the fitting process using a 2-Gaussian model, and select the 2-Gaussian model if it results in a reduced $\chi^2$ that is 10\% smaller than the 1-Gaussian model. {An example of an [\ion{O}{1}]$\lambda$6300 fit is shown in the top right panel of Figure~\ref{fig:o16300}.}

\begin{figure}
    \centering
    \includegraphics[width=0.48\textwidth]{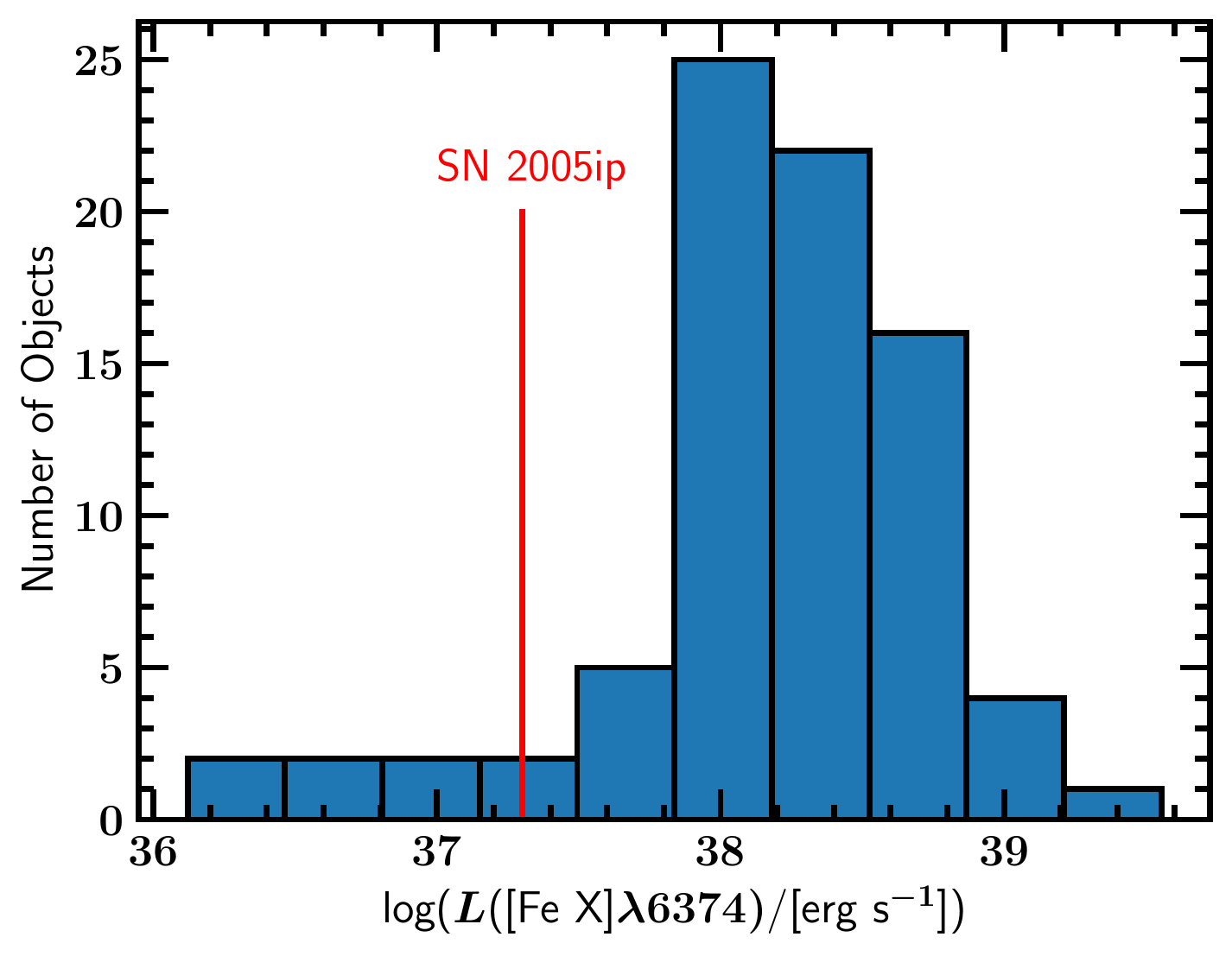}
    \caption{The distribution of \fexl\ luminosities in our sample. The red line denotes the \fex\ luminosity of the type IIn supernova SN 2005ip at its most luminous state at $\sim$100 days post-explosion. A majority of our objects fall well above that line, indicating that a single type IIn supernova cannot explain the observed emission. See Section~\ref{ssec:origin} for a more in-depth discussion.}
    \label{fig:fex_hist}
\end{figure}

We use the [\ion{O}{1}]$\lambda$6300 line as a template to perfectly describe the [\ion{O}{1}]$\lambda$6363 line. Specifically, we shift the model according to their rest wavelengths of 6365.536 and 6302.046, respectively, constrain the widths to be the same in velocity, and require a flux ratio of [\ion{O}{1}]$\lambda$6300/[\ion{O}{1}]$\lambda6363 = 3$. For the 2-Gaussian models, we also require that the relative position and height ratios are the same for the [\ion{O}{1}]$\lambda6363$ model. We then fit the {continuum around the [\ion{O}{1}]+\fex\ complex ($\sim150$\AA)} with a second-order polynomial and subtract both the continuum fit and the [\ion{O}{1}]$\lambda6363$ model. This process leaves a residual spectrum, with a potential \fexl\ line. {An example of an [\ion{O}{1}]$\lambda$6363 fit is shown in the bottom left panel of Figure~\ref{fig:o16300}.}

We then fit a single Gaussian to the residual spectrum to model the \fexl\ emission line and measure the root mean square (rms) of the two continuum windows on either side of \fex, which are each approximately $50$\AA\ wide. This measurement represents the $\sigma$, or noise in the residual spectrum. We accept the fit if the full-width at half maximum (FWHM) of the Gaussian is greater than or equal to the instrumental resolution at the observed wavelength of the \fexl\ line, the peak of the Gaussian is at least 2$\sigma$ above the noise in the residual spectrum and the integrated flux has a ${\rm S/N} \ge2$. Using these criteria, the pipeline identified 213 galaxies with detected \fex\ emission. {In all cases, a single Gaussian best describes the observed \fex\ emission, and the residuals are consistent with the surrounding noise.}

The automated pipeline described above was purposely inclusive as \fex\ is predicted to be a weak line. We therefore expected some of the flagged objects to be ``false positives'', such that the ``detected'' \fex\ emission is actually noise. We visually inspected all of the fits and removed a total of 130 objects that appeared to be false positives. 70 out of the 130 rejected \fex\ Gaussian fits were driven by one or two pixels, while the remaining 60 were very broad fits to noise. After that cut, we then removed 2 additional galaxies from the sample that were directly behind larger, foreground galaxies. Our final sample consists of 81 objects. Figure~\ref{fig:o16300} shows an example of an accepted \fexl\ fit, along with the image of the host galaxy from the Dark Energy Spectroscopic Instrument (DESI) Legacy Imaging Survey \citep{Dey2019}. The rest of the \fex\ emission line and galaxy pairs are shown in Appendix~\ref{app:prop}. {We note that some of the \fex\ profiles do appear somewhat asymmetric similar to that seen in previous coronal-line studies \citep[e.g.,][]{Gelbord2009}, but given their low S/N we do not attempt to characterize their profiles.} The 81 dwarf galaxies in our final sample have a range of \fexl\ luminosities given by $L_{[Fe\textsc{x}]}\approx10^{36}$--$10^{39}$~erg~s$^{-1}$, and a median value of $1.6 \times 10^{38}$~erg~s$^{-1}$. We show the distribution of the \fex\ luminosities in Figure~\ref{fig:fex_hist}.  

We find that all of the SDSS spectra are within 3\arcsec\ of the galaxy nucleus as defined by the NSA, and only 4 spectra have centroids farther than 1\farcs5 away from the galactic center. We list the relative distance between the spectroscopic fiber and the galaxy center and the basic properties of the host galaxy in Table~\ref{table:galprop}. The \fex\ line measurements are listed in Table~\ref{table:fex}. Notes on individual objects, such as their inclusion in previous studies, are also given in the Appendix.

\movetabledown=0.5in
\startlongtable
\begin{deluxetable*}{lccccccccccc}
  \tablecaption{{Sample of 81 \fexl-emitting Dwarf Galaxies} \label{table:galprop}}
\tabletypesize{\footnotesize}
\setlength{\tabcolsep}{4pt}
\renewcommand{\arraystretch}{1.}
\tablewidth{2pt}
\tablehead{ 
\vspace{-2mm}\\
\multicolumn{10}{c}{Galaxy Properties}  &\multicolumn{2}{c}{Previous Studies}\vspace{-2mm}\\
 \multicolumn{10}{c}{\hrulefill} &  \multicolumn{2}{c}{\hrulefill}\\
{NSAID} & {RA$_{\rm Gal}$} & {DEC$_{\rm Gal}$} & {RA$_{\rm Spec}$} & {DEC$_{\rm Spec}$} & {$\Delta$Spec}& {$z$} & {$\log(M_*/M_\odot)$} & {SFR}& {12+log([O/H])} & {Reference} & {Class.}\\
(1) & (2) & (3) & (4) & (5) & (6) & (7) & (8) & (9) & (10) & (11) & (12)}
\startdata
{50}     & {146.00780} & {-0.64226} & {146.00779} & {-0.64227} & {0.05} & {0.00478} & {7.8} & {0.17} & {7.7} & {Reines+20} & {SF}\\
{6055}   & {173.70484} & {-1.06324} & {173.70481} & {-1.06331} & {0.27} & {0.04641} & {9.5} & {0.40} & {8.5} & {BGG20} & {AGN}\\
{10870}  & {199.48543} & {1.11862}  & {199.48541} & {1.11860}  & {0.1} & {0.03106}  & {9.4} & {0.40} & {8.5} & {} & {}\\
{15288}  & {225.98677} & {-0.70008} & {225.98675} & {-0.70007} & {0.08} & {0.14134} & {9.4} & {6.13} & {8.2} & {} & {}\\
{21569}  & {199.19674} & {-3.04933} & {199.19659} & {-3.04942} & {0.63} & {0.01876} & {9.1} & {0.10} & {8.3} & {} & {}\\
{30143}  & {343.15207} & {-0.55478} & {343.15162} & {-0.55493} & {1.71} & {0.0543}  & {9.1} & {0.95} & {8.3} & {} & {}\\
41675  & 8.07754   & 15.00393 &   8.07747 & 15.00394 & 0.25 & 0.01787 & 8.2 & 0.51 & 8.1 &  & \\
{46653}  & {118.01749} & {42.75672} & {118.01755} & {42.75673} & {0.16} & {0.04113} & {9.3} & {0.38} & {8.5} & {} & {}\\
{50800}  & {131.36501} & {53.14802} & {131.36507} & {53.14805} & {0.17} & {0.03107} & {8.5} & {1.59} & {8.1} & {} & {}\\
{61382}  & {177.82335} & {67.31233} & {177.82341} & {67.31234} & {0.09} & {0.04599} & {9.4} & {0.45} & {8.3} & {} & {}\\
 61601 & 175.77729  & 68.12157 & 175.77723 & 68.12160 & 0.13 & 0.04892 & 8.9 & 1.88 & 8.2 & &\\   
63560 & 211.73123 & 65.91572 & 211.73115 &  65.91572 & 0.12 & 0.05995 & 9.4 & 3.15 & 8.4  &        &    \\             
95985 & 234.26741 & 55.26406 & 234.26740 &  55.26406 & 0.02 & 0.00225 & 7.5 & 0.03 & 8.1  & BWA20     & AGN   \\             
102990 & 316.78814 & -7.42285 & 316.78811 &  -7.42287 & 0.13 & 0.02830 & 8.8 & 0.17 & 8.4  &        &    \\             
107232 &   9.14935 &-10.05103 &   9.14942 & -10.05099 & 0.29 & 0.02080 & 9.5 & 0.38 & 8.6  &        &    \\             
108471 &  12.84411 &-11.14529 &  12.84402 & -11.14531 & 0.33 & 0.02085 & 9.0 & 0.20 & 8.4  &        &    \\             
115843 &  53.95511 & -0.65364 &  53.95512 &  -0.65364 & 0.04 & 0.03067 & 8.9 & 0.17 & 8.4  &        &    \\             
131809 & 140.10780 & 50.82773 & 140.10764 &  50.82762 & 0.54 & 0.03444 & 9.2 & 0.24 & 8.5  & BGG20     & AGN   \\             
140290 &  47.31564 & -0.94906 &  47.31566 &  -0.94902 & 0.16 & 0.05718 & 8.8 & 0.59 & 8.2  &        &    \\             
140756 &  49.07568 &  1.00611 &  49.07570 &   1.00607 & 0.16 & 0.03311 & 8.7 & 0.31 & 8.3  &        &    \\             
141378 &  52.61371 & -0.85374 &  52.61371 &  -0.85377 & 0.11 & 0.05243 & 9.3 & 0.57 & 8.5  &        &    \\             
141686 &  47.26619 &  0.64633 &  47.26618 &   0.64630 & 0.11 & 0.03009 & 8.1 & 0.21 & 7.9  &        &    \\             
144266 & 253.94519 & 36.67987 & 253.94519 &  36.67990 & 0.11 & 0.06234 & 9.0 & 1.21 & 8.2  &        &    \\             
162936 & 126.20782 & 38.58743 & 126.20768 &  38.58737 & 0.45 & 0.03041 & 8.8 & 0.16 & 8.5  &        &    \\             
164449 & 137.39846 & 45.95532 & 137.39844 &  45.95529 & 0.12 & 0.02733 & 8.8 & 0.13 & 8.5  &        &    \\             
174326 & 127.60643 & 33.05816 & 127.60641 &  33.05816 & 0.06 & 0.02086 & 9.1 & 0.19 & 8.5  &        &    \\             
201943 & 314.23197 & -0.40761 & 314.23196 &  -0.40758 & 0.11 & 0.02946 & 8.7 &   & 8.3  &        &    \\             
213780 &  53.97134 & -0.66295 &  53.97135 &  -0.66295 & 0.04 & 0.03016 & 8.8 & 0.21 & 8.4  &        &    \\             
220716 & 229.68660 & 51.20768 & 229.68655 &  51.20773 & 0.21 & 0.01418 & 8.8 & 0.05 & 8.3  &        &    \\             
237974 & 180.71961 & 10.24299 & 180.71954 &  10.24300 & 0.25 & 0.06167 & 9.4 & 0.87 & 8.4  &        &    \\             
256515 & 185.24312 & 56.35518 & 185.24313 &  56.35515 & 0.11 & 0.03461 & 9.5 & 0.38 & 8.6  &        &    \\             
256802 & 185.92845 & 58.24618 & 185.92837 &  58.24615 & 0.19 & 0.01437 & 9.4 & 0.14 & 8.4  & Reines+13 & AGN   \\             
258576 & 201.97411 & 55.64514 & 201.97408 &  55.64516 & 0.09 & 0.03845 & 8.9 & 0.38 & 8.4  &        &    \\             
270622 & 162.36197 & 42.24777 & 162.36176 &  42.24761 & 0.80 & 0.02410 & 8.2 & 0.06 & 8.4  &        &    \\             
275961 & 207.85572 & 40.21324 & 207.85573 &  40.21327 & 0.11 & 0.00823 & 9.4 & 0.25 & 8.6  & Reines+13 & Compos\\ite          
282978 & 230.02922 & 37.99824 & 230.02922 &  37.99821 & 0.11 & 0.06432 & 9.5 & 0.75 & 8.5  &        &    \\             
300542 & 198.27999 & 46.09894 & 198.27964 &  46.09844 & 2.00 & 0.02965 & 9.3 & 0.12 & 8.4  & Reines+13 & Sne   \\             
305229 &  19.43879 & -1.17564 &  19.43877 &  -1.17558 & 0.23 & 0.10881 & 8.9 & 1.87 & 8.1  &        &    \\             
321197 & 160.50153 & 12.33491 & 160.50154 &  12.33493 & 0.08 & 0.00257 & 8.1 & 0.01 & 8.3  &        &    \\             
328894 & 180.55578 &  6.28306 & 180.55576 &   6.28311 & 0.19 & 0.03634 & 8.7 & 0.34 & 8.5  &        &    \\             
331005 & 189.78945 &  8.14025 & 189.78975 &   8.13952 & 2.84 & 0.04948 & 8.9 & 0.44 & 8.5  &        &    \\             
331627 & 213.86196 & 36.40828 & 213.86185 &  36.40825 & 0.34 & 0.02817 & 9.4 & 0.52 & 8.5  &        &    \\             
345375 & 191.88671 & 13.61817 & 191.88671 &  13.61815 & 0.07 & 0.08469 & 9.2 & 1.16 & 8.5  &        &    \\             
366027 & 133.68095 &  8.94791 & 133.68087 &   8.94795 & 0.32 & 0.02962 & 9.1 & 0.32 & 8.1  &        &    \\             
393631 & 214.09711 & 33.12361 & 214.09713 &  33.12359 & 0.09 & 0.08324 & 9.4 & 2.88 & 8.5  &        &    \\             
401495 & 123.37373 & 54.74468 & 123.37383 &  54.74482 & 0.55 & 0.03234 & 8.7 & 0.27 & 8.4  &        &    \\             
402252 & 125.55366 & 58.02834 & 125.55370 &  58.02836 & 0.10 & 0.01645 & 8.3 & 0.06 & 8.2  &        &    \\             
409020 & 137.05685 & 26.70089 & 137.05690 &  26.70085 & 0.22 & 0.07682 & 9.2 & 3.02 & 8.2  &        &    \\             
418928 & 169.03563 & 30.37582 & 169.03562 &  30.37583 & 0.05 & 0.04164 & 9.4 & 0.25 & 8.6  &        &    \\             
421523 & 192.17151 & 38.57556 & 192.17129 &  38.57556 & 0.62 & 0.04121 & 9.4 & 0.27 & 8.6  &        &    \\             
422297 & 185.79998 & 37.99107 & 185.79997 &  37.99108 & 0.05 & 0.00752 & 7.8 & 0.01 & 8.2  &        &    \\             
426303 & 162.33047 & 37.87624 & 162.33031 &  37.87614 & 0.58 & 0.02526 & 9.2 & 0.18 & 8.4  &        &    \\             
426824 & 176.12647 & 32.56818 & 176.12646 &  32.56815 & 0.11 & 0.03072 & 8.7 & 0.30 & 8.3  &        &    \\             
427201 & 199.01634 & 29.38181 & 199.01632 &  29.38169 & 0.44 & 0.03779 & 9.1 & 4.92 & 8.3  & Reines+13 & SF+BHa\\
437653 & 198.02611 & 39.15767 & 198.02636 &  39.15771 & 0.71 & 0.02147 & 9.2 & 0.16 & 8.5  &        &    \\             
442120 & 204.73782 & 36.59586 & 204.73785 &  36.59584 & 0.11 & 0.01940 & 8.8 & 0.09 & 8.3  &        &    \\             
442212 & 203.50222 & 36.70000 & 203.50220 &  36.70002 & 0.09 & 0.05988 & 8.6 & 1.43 & 8.0  &        &    \\             
442427 & 201.58173 & 36.86264 & 201.58171 &  36.86266 & 0.09 & 0.05588 & 9.4 & 0.91 & 8.5  &        &    \\             
452253 & 216.75820 & 28.25949 & 216.75819 &  28.25950 & 0.05 & 0.04027 & 9.3 & 0.27 & 8.5  &        &    \\             
452914 & 216.15493 & 26.34413 & 216.15507 &  26.34419 & 0.50 & 0.03539 & 9.2 & 0.12 & 8.6  &        &    \\             
457940 & 224.50743 & 26.22497 & 224.50738 &  26.22500 & 0.19 & 0.03552 & 9.2 & 0.38 & 8.5  &        &    \\             
458829 & 161.33505 &  9.39698 & 161.33508 &   9.39697 & 0.11 & 0.05487 & 9.1 & 7.54 & 8.2  &        &    \\             
472272 & 245.46907 & 15.31555 & 245.46905 &  15.31555 & 0.07 & 0.03434 & 8.7 & 2.92 & 8.3  &        &    \\             
472333 & 246.57067 & 16.32011 & 246.57065 &  16.32012 & 0.08 & 0.01295 & 9.1 & 0.07 & 8.5  &        &    \\             
473078 & 167.42962 & 25.63336 & 167.42962 &  25.63335 & 0.04 & 0.04121 & 9.4 & 0.30 & 8.6  &        &    \\             
491412 & 139.91023 & 19.59117 & 139.91025 &  19.59119 & 0.10 & 0.05696 & 9.1 & 2.10 & 8.2  &        &    \\             
501341 & 157.95713 & 23.41739 & 157.95712 &  23.41741 & 0.08 & 0.00408 & 7.8 & 0.03 & 8.1  &        &    \\             
524413 & 178.57025 & 24.97274 & 178.57023 &  24.97274 & 0.07 & 0.02866 & 9.2 & 0.20 & 8.5  &        &    \\             
525522 & 237.71971 & 15.73100 & 237.71970 &  15.73099 & 0.05 & 0.03506 & 9.2 & 0.25 & 8.4  &        &    \\             
532026 & 138.83771 & 12.43752 & 138.83769 &  12.43755 & 0.13 & 0.03330 & 9.1 & 0.47 & 8.4  &        &    \\             
533731 & 147.32514 & 16.87894 & 147.32513 &  16.87896 & 0.08 & 0.05199 & 9.7 & 7.18 & 8.4  & Kimbro+21 & AGN   \\             
536931 & 157.45538 & 16.18098 & 157.45535 &  16.18091 & 0.27 & 0.01089 & 9.0 & 0.18 & 8.3  &        &    \\             
541112 & 202.54870 & 16.65336 & 202.54873 &  16.65332 & 0.18 & 0.03016 & 9.0 & 0.27 & 8.3  &        &    \\             
553772 & 208.69741 & 15.50116 & 208.69740 &  15.50120 & 0.15 & 0.02404 & 9.2 & 0.20 & 8.5  &        &    \\             
557918 & 206.56320 & 17.91621 & 206.56323 &  17.91624 & 0.15 & 0.02571 & 9.3 & 0.25 & 8.4  &        &    \\             
571629 & 231.46649 & 17.70089 & 231.46631 &  17.70061 & 1.18 & 0.02922 & 9.3 & 0.18 & 8.5  &        &    \\             
577169 & 129.78605 & 33.30856 & 129.78601 &  33.30857 & 0.13 & 0.05438 & 9.1 & 1.23 & 8.3  &        &    \\             
625882 & 166.28381 & 44.74638 & 166.28383 &  44.74646 & 0.29 & 0.02154 & 9.3 & 5.11 & 8.2  &        &    \\             
632355 & 225.98191 &  0.43211 & 225.98190 &   0.43214 & 0.11 & 0.00529 & 7.8 & 0.04 & 8.0  &        &    \\             
656364 & 163.93653 & 46.37194 & 163.93590 &  46.37180 & 1.64 & 0.05065 & 8.5 & 0.74 & 8.4  &        &    \\             
670127 & 191.12724 & 40.73691 & 191.12728 &  40.73690 & 0.11 & 0.01790 & 9.3 & 1.29 & 8.4  &        &    \\ 
\enddata
\tablecomments{Column (1): The NSA IDs from version \texttt{v1\_0\_1} of the NSA Catalog. Columns (2)--(5): The RA and DEC values in degrees of the galaxy center (denoted with Gal) and the spectroscopic SDSS fiber (denoted with Spec). Column (6): The offset between the spectroscopic fiber and the galaxy center in arcseconds. Columns (7) -- (8): The redshift and stellar mass (in units of solar masses) from the NSA, assuming $h=0.73$. Column (9): The total star formation rate in units of $M_\odot$~yr$^{-1}$ calculated from the {\it GALEX} FUV and WISE W4 measurements as described in Section~\ref{ssec:sprop}. Column (10): Metallicity calculated via \cite{pettini2004}, assuming the [\ion{N}{2}] and H$\alpha$ flux measurements that are measured directly from the SDSS 3\arcsec\ spectra in this work. The emission-line flux measurements are presented in Table~\ref{table:eline}. Column (11): The inclusion of this object in previous studies; Reines+13 refers to \cite{reines2013}, BGG20 refers to \cite{Baldassare2020}, BWA20 refers to \cite{Birchall2020}, Reines+20 refers to \cite{reines2020} and Kimbro+21 refers to \cite{Kimbro2021}. Column (12) provides the designation of the object from previous studies.}
\end{deluxetable*}

\begin{deluxetable*}{lccccc}
\label{table:fex}
\tablecaption{\fexl\ Measurements}
\setlength{\tabcolsep}{7pt}
\tablehead{
\colhead{NSAID} & {$f$(\fex)} & log($L_{\rm [Fe X]}$) & $\sigma_{\rm peak}$ & $S/N_{\rm line}$ &  $L$(\fex)$_{\rm gal}/L$(\fex)$_{\rm SN2005ip}$\\
(1) & (2) & (3) & (4) & (5) & (6)}
\startdata
50 &    $11.7\pm3.8$ & $36.7\pm0.5$ & 3.1 & 3.1 & 0.3\\
6055 &  $6.1\pm3.0 $&  $38.4\pm0.3$ & 2.1 & 2.0 & 12.2\\
10870 & $10.2\pm 4.6$& $38.3\pm0.4$ & 2.3 & 2.2 & 9.2\\
15288 & $8.9 \pm4.3 $& $39.6\pm0.3$ & 3.2 & 2.1 & 165.4\\
21569 & $8.6 \pm4.3 $& $37.8\pm0.3$ & 2.9 & 2.0 & 2.8\\
30143 & $11.5\pm 4.4$& $38.8\pm0.4$ & 2.2 & 2.6 & 31.6\\
41675 & $9.8 \pm4.0 $& $37.8\pm0.4$ & 3.9 & 2.5 & 2.9\\
46653 & $4.0 \pm2.0 $& $38.1\pm0.3$ & 3.6 & 2.0 & 6.3\\
50800 & $8.2 \pm3.8 $& $38.2\pm0.3$ & 2.3 & 2.2 & 7.4\\
61382 & $4.2 \pm1.9 $& $38.3\pm0.4$ & 4.2 & 2.2 & 8.3\\
\enddata
\tablecomments{The entirety of Table~\ref{table:fex} is published in the electronic edition of {\it The Astrophysical Journal}. We show a portion here to give information on its form and content. Column (1): NSA identification number. Column (2): The measured \fexl\ flux in units of $10^{-17}$~erg~s$^{-1}$~cm$^{-2}$. Column (3): The measured \fexl\ luminosity in units of erg~s$^{-1}$. Column (4): The strength of the line peak in units of the residual spectrum noise, $\sigma$. Column (5): The S/N of the integrated line flux. Column (6): The ratio of the measured \fexl\ luminosity from column (3) to the peak \fexl\ lumionsity in SN2005ip, $\sim2\times10^{37}$~erg~s$^{-1}$.}
\end{deluxetable*}

\subsection{Properties of the Host Galaxies}\label{ssec:sprop}

\begin{figure}
    \centering
    \includegraphics[width=0.48\textwidth]{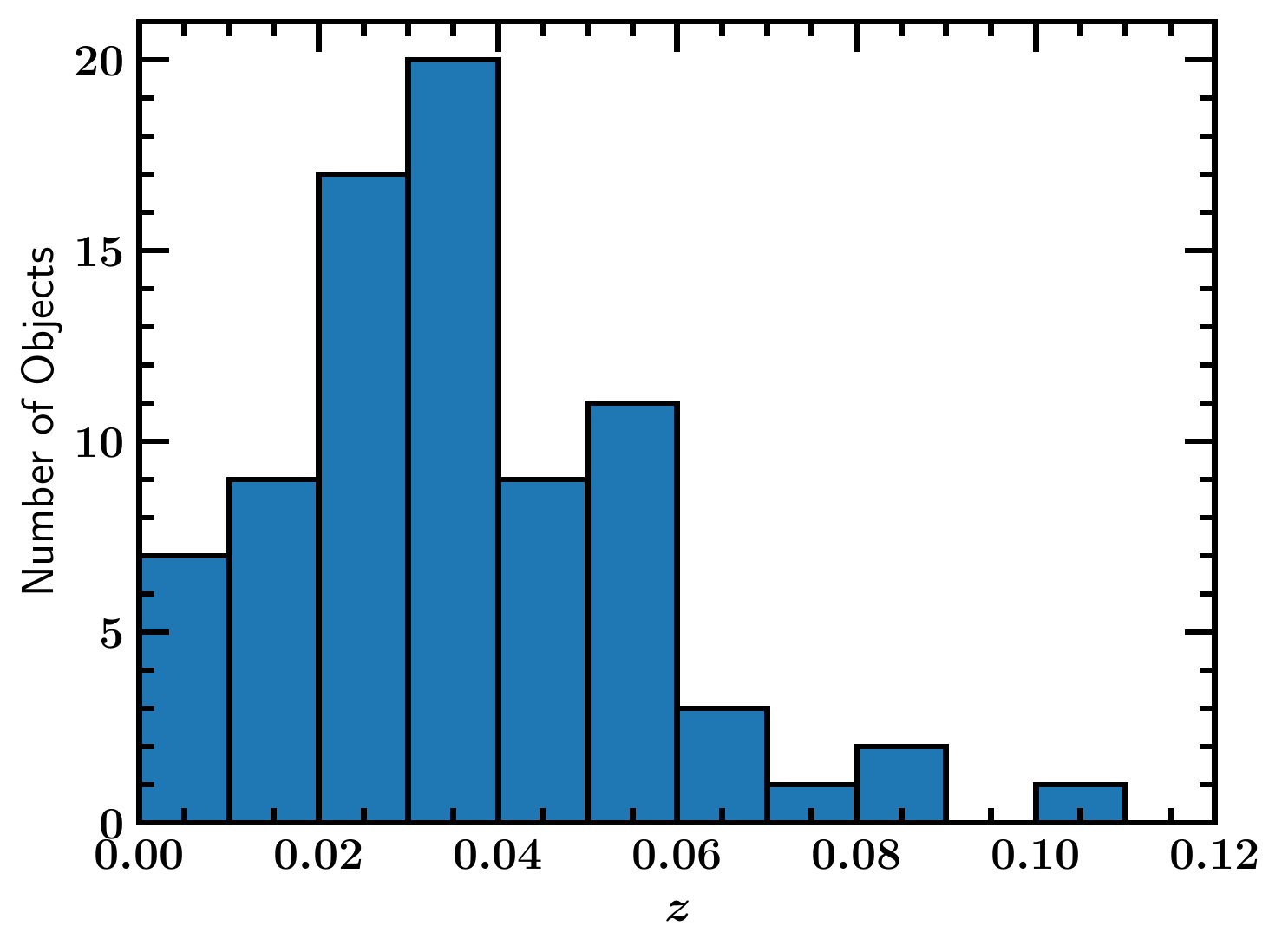}
    \caption{{The distribution of the host galaxy redshifts in our sample. Only two objects have $z > 0.1$, and 76\% have $z< 0.05$.}}
    \label{fig:zhist}
\end{figure}

The \fexl-emitting dwarf galaxies span a mass range of $7.5\leq \log(M_*/M_\odot) \leq 9.5$, with a median mass of $\log(M_*/M_\odot)=9.1$. All reported masses are from the NSA except for Mrk 709S (NSA 533731), which is taken from \cite{Reines2014}. The stellar mass for Mrk 709S from \cite{Reines2014} is estimated to be $\log(M_*/M_\odot)=9.4$, which is derived using photometry that separates the northern and southern galaxies in the pair. The NSA automated pipeline value is $\log(M_*/M_\odot)=9.7$ which is likely higher due to additional light from the other components in the system. {The galaxies span a redshift range of $0.00225\leq z\leq 0.14134$, with a median $z=0.03107$. All of our objects are relatively nearby; only two objects have $z > 0.1$, while 76\% of the galaxies have $z < 0.05$. The distribution of the galaxy redshifts in our sample is shown in Figure~\ref{fig:zhist}.} 

\begin{figure}
    \centering
    \includegraphics[width=0.48\textwidth]{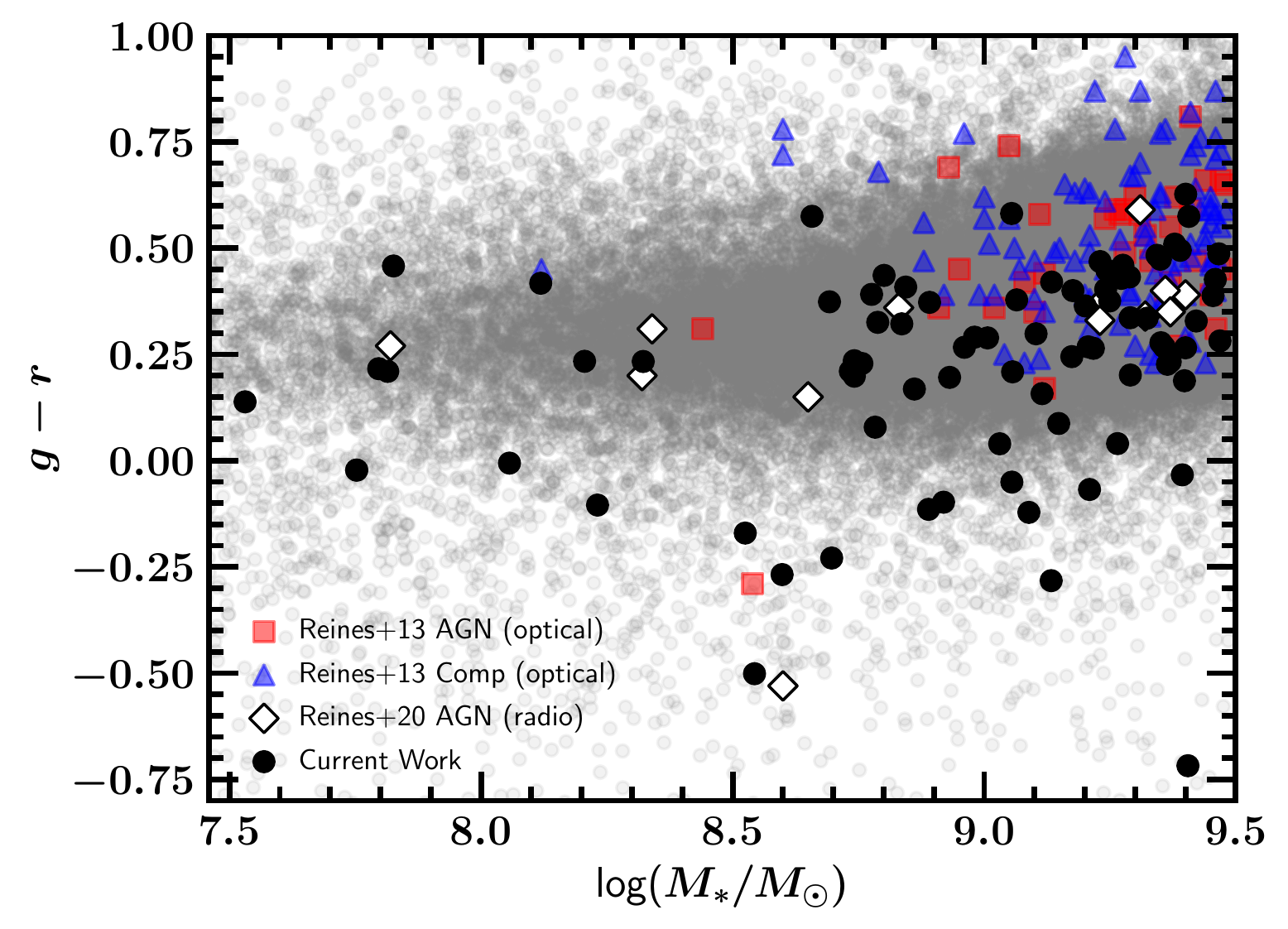}
    \caption{The $g-r$ vs.~stellar mass plot. The \fexl-emitting AGN candidates from this work are shown in black circles, the optically-selected AGN and composite objects from \cite{reines2013} are shown as red squares and blue triangles, respectively and the radio-selected AGNs from \cite{reines2020} are shown as unfilled white diamonds. These four samples are plotted over the distribution of dwarf galaxies from the NSA. The \cite{reines2013} optically-selected sample tends to have higher-mass, redder host galaxies, {while the \cite{reines2020} radio-selected sample and the \fex-selected galaxies in this work tend to trace a lower-mass, bluer population.}}
    \label{fig:agn_comp}
\end{figure}

The galaxies in our sample vary in size, with $r$-band half-light radii ranging from $\sim$0.7\arcsec--16.2\arcsec\ (0.1--16.3~kpc) with a median value of 3.6\arcsec\ (2.1~kpc). While 11 objects in our sample were classified as spiral galaxies by Galaxy Zoo 1 \citep{Lintott2011}, a majority appear irregular in morphology. {One interesting feature of this sample is how blue, and thus vigorously star-forming, a majority of the galaxies are, as clearly seen in the Legacy Survey images in the Appendix. In fact, the median $g-r$ color is 0.37, and spans the range $-0.42\leq g-r \leq 0.68$. We show the $g-r$ vs.~$M_*$ diagram in Figure~\ref{fig:agn_comp}.}

We quantified the global star formation rates (SFRs) of the dwarf galaxies via their far-ultraviolet (FUV; 1528 \AA) and mid-infrared (MIR; 25 $\mu$m) luminosities \citep{kennicutt2012,hao11} using the following equations:

\begin{equation}
\begin{split}
    \log({\rm SFR}/[M_\odot~{\rm yr}^{-1}]) = \log(L({\rm FUV})_{\rm corr}-43.35\\
    L({\rm FUV})_{\rm corr} = L({\rm FUV})_{\rm obs} + 3.89L(25~\mu{\rm m}).
    \end{split}
    \end{equation}

We use the elliptical Petrosian \textit{Galaxy Evolution Explorer (GALEX)} measurements reported in the NSA for our FUV measurements, and rely on the W4 (22~$\mu$m) luminosities from the All Wide-field Infrared Survey Explorer (ALLWISE) catalog. As the ratio between 22 $\mu$m and 25 $\mu$m luminosities is expected to be of order unity \citep{jarrett13}, and all but one of the objects are detected in {\it WISE}, we use the W4 observations in place of 25~$\mu$m measurements from the \textit{Infrared Astronomical Satellite (IRAS)}.

\begin{figure}
    \centering
    \includegraphics[width=0.48\textwidth]{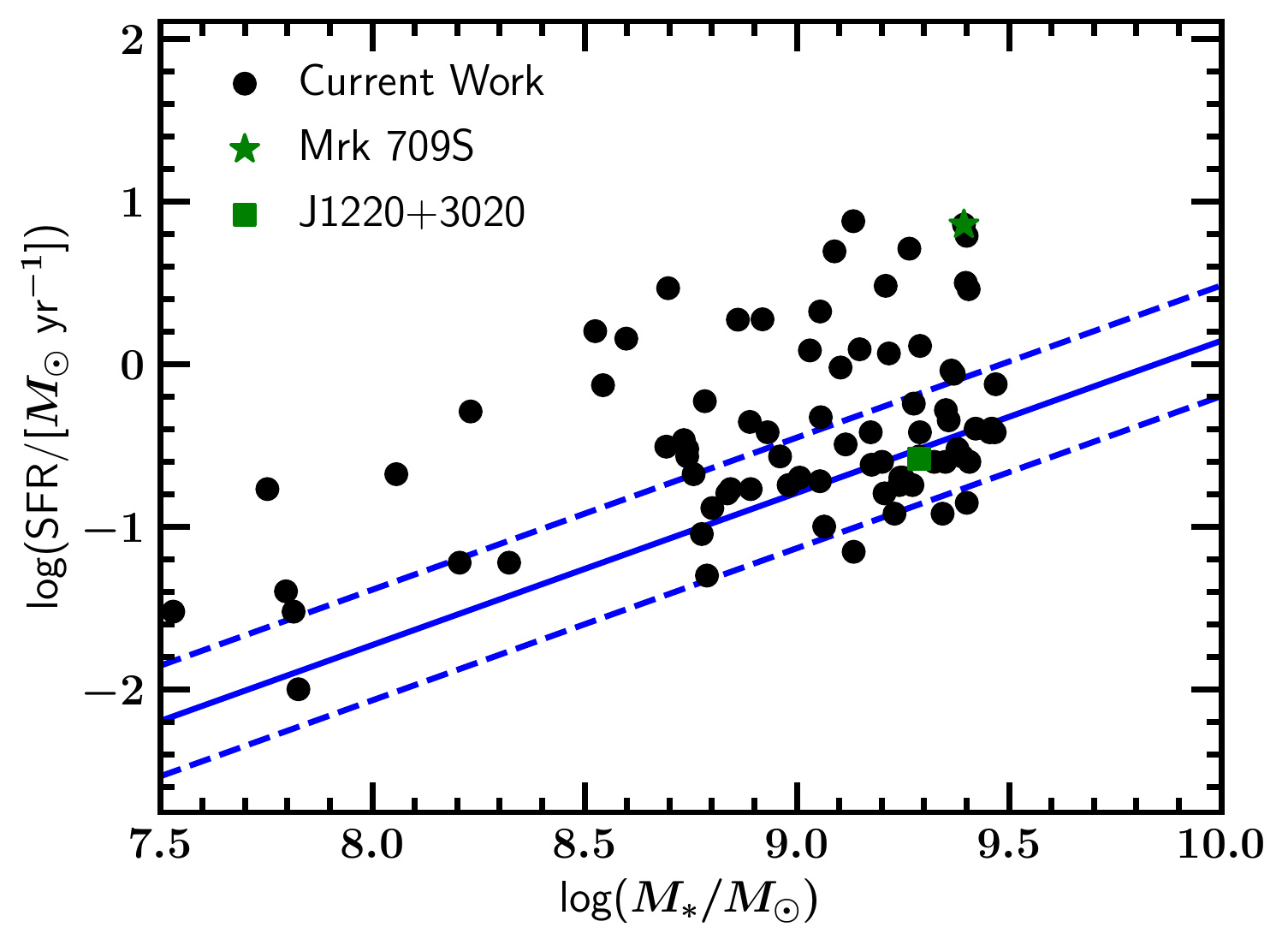}
    \caption{SFR vs.~stellar mass for the 81 dwarf galaxies in our sample. We calculated the SFR using {\it GALEX} and {\it WISE} data as described in Section~\ref{ssec:sprop}. The star forming main sequence from~\cite{McGaugh2017} is shown as a blue solid line, and the 1-$\sigma$ intrinsic scatter in the fit is shown by the two blue dashed lines. A total of 37 objects lie above the 1-$\sigma$ scatter of the relation, indicating the presence of vigorous star formation.}
    \label{fig:mstar_sfr}
\end{figure}

\begin{figure}
    \centering
    \includegraphics[width=0.48\textwidth]{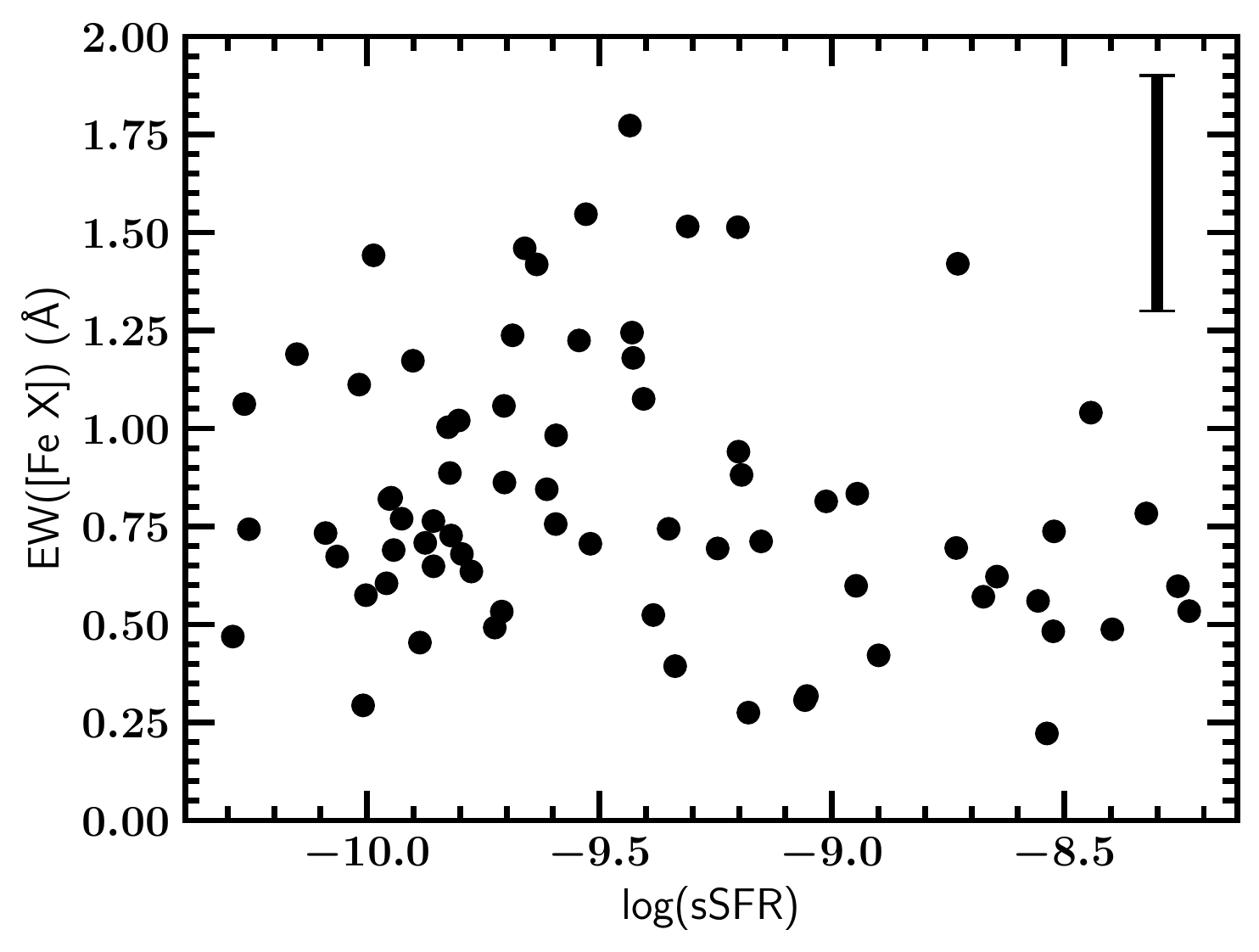}
    \caption{The \fexl\ EW vs.~mass-specific SFR for the \fexl-emitting dwarf galaxy sample. {We show the characteristic uncertainty for \fex\ EW in the top right.} We note that there is no clear trend, indicating that the \fex\ emission is likely not related to stellar processes associated with young stellar populations. {Similarly, we see no trend with the \fex\ luminosity and the mass-specific SFR.} See Section~\ref{ssec:origin} for a more in-depth discussion.}
    \label{fig:ewfex}
\end{figure}

The SFR vs.~stellar mass plot is presented in Figure~\ref{fig:mstar_sfr}, with Mrk 709S and J1220+3020, the objects studied in \cite{Kimbro2021} and \cite{Molina2021}, denoted by the green star and square respectively. Out of the two, Mrk 709S is included in our sample, while J1220+3020 does not have detectable \fex\ emission in its SDSS spectrum. We over-plot the star forming main sequence using dwarf galaxies as calculated by \cite{McGaugh2017}. We note that 37 of our galaxies lie above the 1-$\sigma$ intrinsic scatter in the relation, indicating particularly high SFRs. Indeed, we find that our galaxies have a median mass-specific SFR (${\rm sSFR} = {\rm SFR}/{\rm M}_*$) of $\log({\rm sSFR})\sim-9.6$, with the highest log$({\rm sSFR})=-8.2$. We note that there is no correlation between the \fexl\ equivalent width (EW) and the sSFR, as seen in Figure~\ref{fig:ewfex}. A similar result is seen when comparing the \fex\ EWs to both the $g-r$ and $g-i$ colors. {There is also no observed trend between the \fex\ luminosities and the sSFR, $g-r$ or $g-i$ colors.} Therefore, it is unlikely that the \fexl\ emission is driven by stellar processes associated with young stellar populations. We will revisit this discussion in more depth in Section~\ref{ssec:origin}.

We also estimate the metallicities of our galaxies via our log([\ion{N}{2}]/H$\alpha$) ratio measurements, described in Section~\ref{sec:sdss_spec}, using the relation from \cite{pettini2004}. The metallicities range from $12 + \log([{\rm O}/{\rm H}])=7.8\mbox{--}8.6$, and are all below the solar metallicity of 8.7 \citep{Allende2001}. {We note that the \cite{pettini2004} metallicity relation is calibrated for star-forming regions, and thus does not account for AGN activity. However, all but two objects in our sample have [\ion{N}{2}]/H$\alpha$ ratios consistent with star formation as shown in Figure~\ref{fig:bpt}. Furthermore, the optical metallicity indicators that include contributions from AGNs in \cite{storchi1998} are calibrated for metallicites $12 + \log([{\rm O}/{\rm H}])\geq 8.4$. 77/81 of the measured metallicities in our sample using the [\ion{N}{2}]/H$\alpha$ relation in Section 3 of \cite{storchi1998} are less than 8.4, making that measurement unreliable. Therefore, we adopt the \cite{pettini2004} metallicity models to calculate the host galaxy metallicity.} The basic properties of the galaxies including their SFRs and metallicities and any overlap with previous studies are given in Table~\ref{table:galprop}. 

\section{Additional Spectral Analysis}\label{sec:sdss_spec}

In addition to searching for \fexl\ in the SDSS spectra of dwarf galaxies as described above, we are also interested in searching for AGN activity using other diagnostic emission lines.
We created custom codes in python to subtract the stellar continuum, separate the broad and narrow components in H$\alpha$ and H$\beta$ and measure the fluxes of various emission lines. Our codes used the same packages described in Section~\ref{sec:samp} as well as the Penalized PiXel-Fitting \citep[\texttt{pPXF}; ][]{Cappellari2017} code to fit the stellar continuum. While \texttt{pPXF} can fit both the emission lines and the stellar continua simultaneously, we wrote our own emission-line fitting software that follows the methodology described in \cite{reines2013}.

\subsection{Stellar Continuum Subtraction}
Most of the galaxies in our sample are quite blue, indicating a significant amount of star formation. The stellar continuum, which usually dominates the continuum in the SDSS spectra, can contain stellar absorption lines which can severely affect our hydrogen Balmer emission-line measurements. This is especially true for potential weak broad H$\alpha$ and H$\beta$ emission. We therefore remove the stellar continuum using the \texttt{pPXF} fitting code. As our primary goal is to measure the emission lines, we sought good fits to the stellar continua but did not fully explore the entire parameter space. 

We used the \citet[][BC03]{Bruzual2003} template spectra, which were used in \cite{Tremonti2004}. They were originally selected to describe the SDSS-DR1 galaxies in the $D_n(4000)$--H$\delta_A$ plane. The template spectra include instantaneous burst single stellar population (SSP) models at ten different ages (0.005, 0.025, 0.10, 0.29, 0.64, 0.90, 1.4, 2.5, 5, and 11~Gyr) and three different metallicities ($Z= 0.008, 0.02, 0.05$). For the 81 galaxies in our sample, we modeled the spectra with \texttt{pPXF} using single-metallicity SSP models and a low-order multiplicative polynomial to account for reddening due to extinction. We accepted the fit for the template family with the lowest reduced $\chi^2$, which for all cases was the sub-solar $Z=0.008$ models. One object, NSA 50, did not show any stellar absorption lines and \texttt{pPXF} did not fit any stellar templates to the galaxy continuum. For this object, we employed a third-order polynomial to remove the continuum emission in the SDSS spectrum. {We show an example of a stellar continuum fit in the top panel of Figure~\ref{fig:eline_fits}.}

\subsection{Emission Line Measurements}\label{ssec:elmeas}
We fitted the emission lines in the continuum-subtracted SDSS spectra following the same methodology as \cite{reines2013}. We include a linear fit to the continuum in all of the emission-line fitting processes described below to account for uncertainty associated with the stellar continuum fit. 

We first fit the [\ion{S}{2}]$\lambda\lambda$6716,6731 doublet with a single Gaussian for each line, assuming the same width in velocity, and constraining their separation according to their laboratory wavelengths. We then perform a second fit on [\ion{S}{2}]$\lambda\lambda$6716,6731, but using two Gaussians per line. In this case, we constrain the relative heights, widths and positions of the two-component model to be the same for the two lines. We accept the 2-Gaussian model if the reduced $\chi^2$ is at least 10\% lower than that of the single Gaussian model. 

We then use the [\ion{S}{2}] model as a template for the [\ion{N}{2}]$\lambda\lambda$6548,6583 doublet and the narrow component of the H$\alpha$ line {as all three are known to have well-matched profiles \citep[e.g.,][]{filippenko1988,filippenko1989,Ho1997,Greene2004,reines2013}. Specifically, we require} that the widths are the same in velocity for the single-Gaussian model, and the relative velocity widths, position and height ratios are constrained to be the same in the 2-Gaussian model. If the [\ion{S}{2}] model is a single Gaussian, we allow the H$\alpha$ narrow line width to increase by as much as {25\%. This constraint accounts for the fact that H$\alpha$ is emitted physically closer to the power source than [\ion{S}{2}] \citep{Ho2008} and thus likely has a broader profile, and has been shown to accurately describe the narrow H$\alpha$ emission seen in dwarf galaxies by \cite{reines2013}.} For the [\ion{N}{2}] doublet, we further constrain the relative flux to be [\ion{N}{2}]$\lambda$6583/[\ion{N}{2}]$\lambda 6548=3$. This fit is performed again, this time with the inclusion of a broad H$\alpha$ component. We accept the broad+narrow H$\alpha$ fit if the reduced $\chi^2$ is at least 20\% smaller than that of the narrow-only H$\alpha$ fit, and the broad H$\alpha$ component has a FWHM of at least 500~km~s$^{-1}$. We accept the [\ion{O}{1}]$\lambda\lambda$6300,6363 and \fexl\ measurement as described in Section \ref{sec:sampselect} as there are no strong stellar absorption lines in that spectral region. We then fit the H$\beta$ emission line both with and without a broad component, using the same constraints used for the H$\alpha$ line. We accept the broad+narrow component fit if the reduced $\chi^2$ is 20\% lower than the narrow-only fit.

\begin{figure*}
    \centering
    \includegraphics[width=\textwidth]{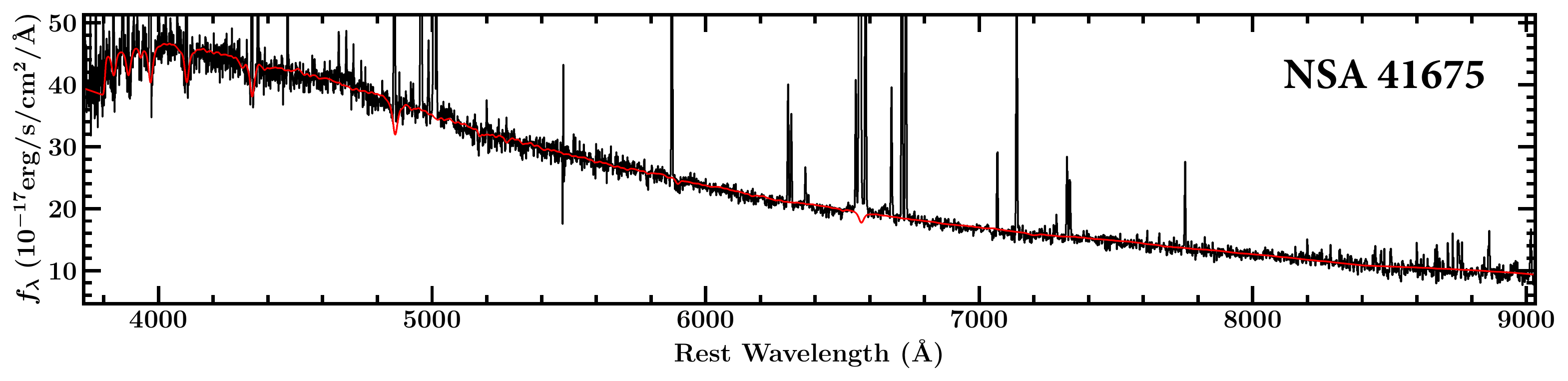}
    \includegraphics[width=0.32\textwidth]{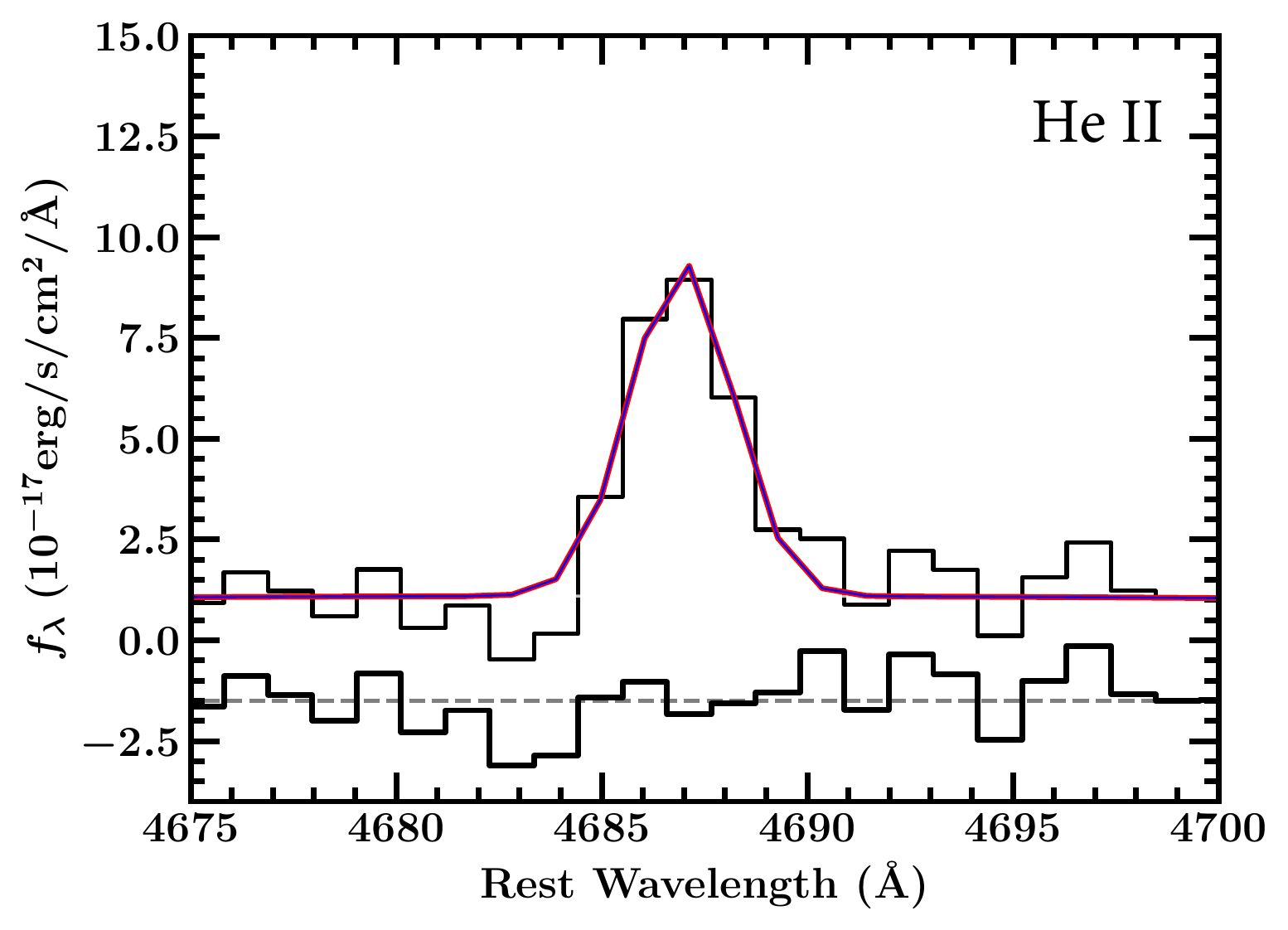}
    \includegraphics[width=0.32\textwidth]{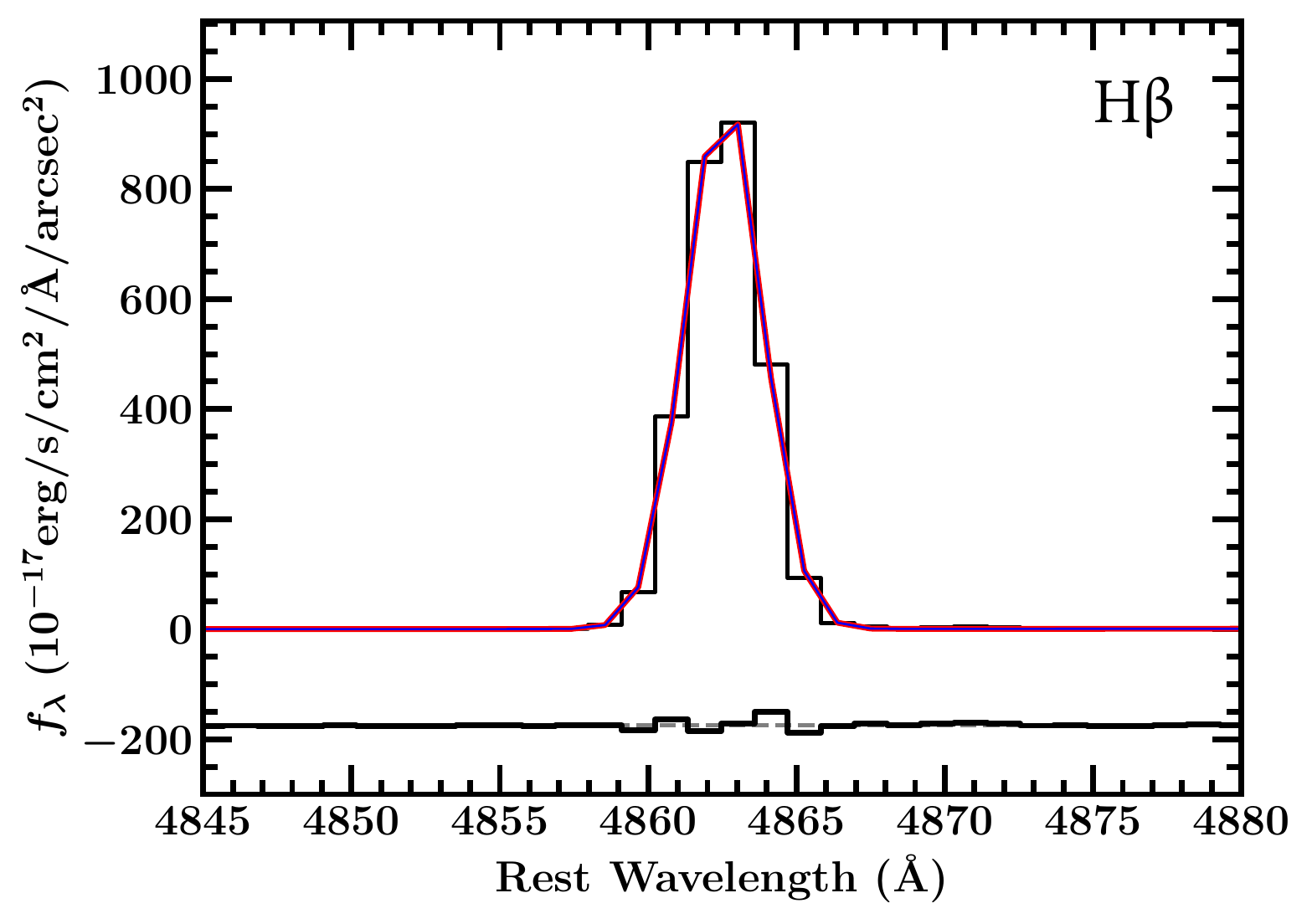}
    \includegraphics[width=0.32\textwidth]{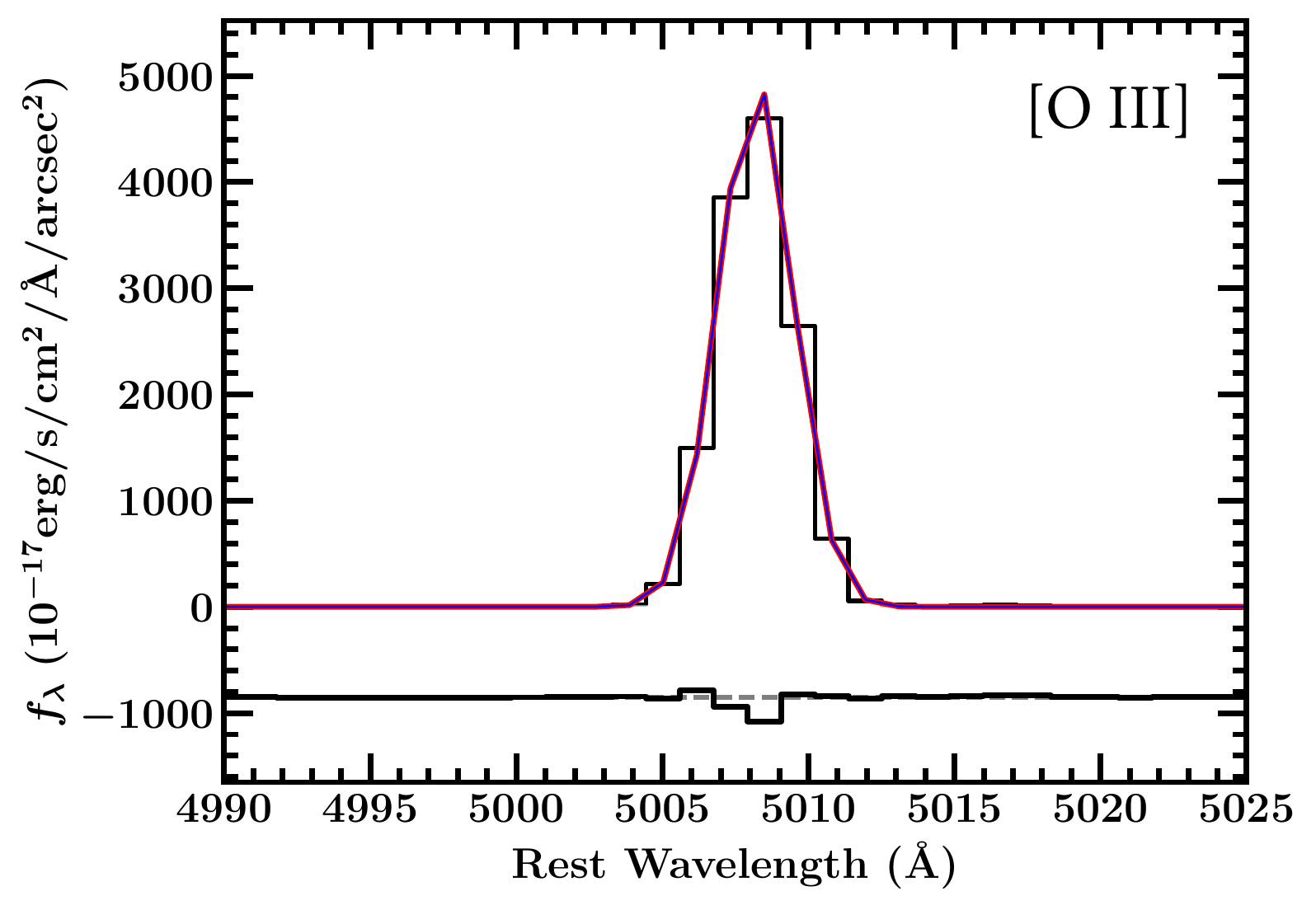}
    \includegraphics[width=0.32\textwidth]{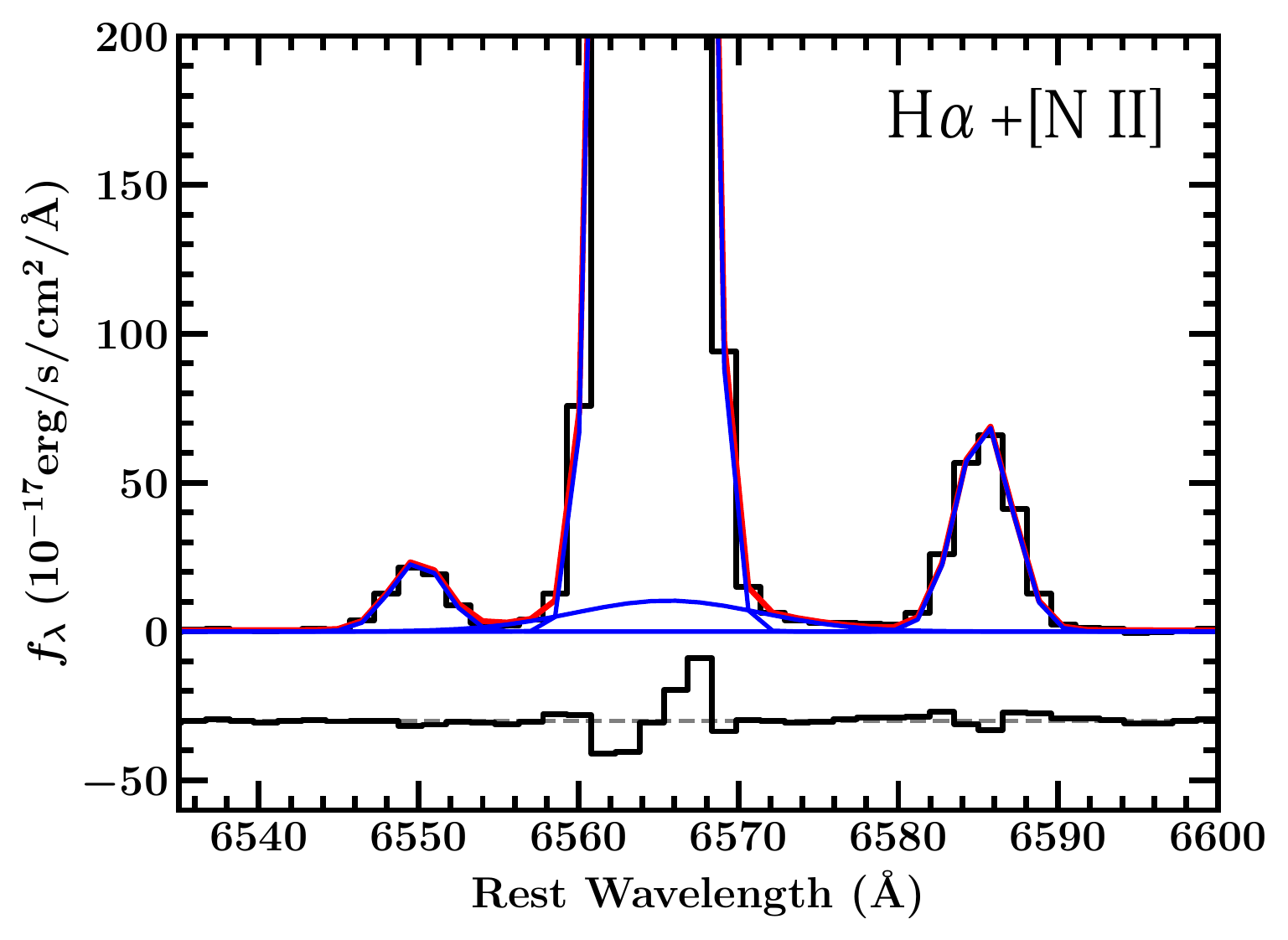}
    \includegraphics[width=0.32\textwidth]{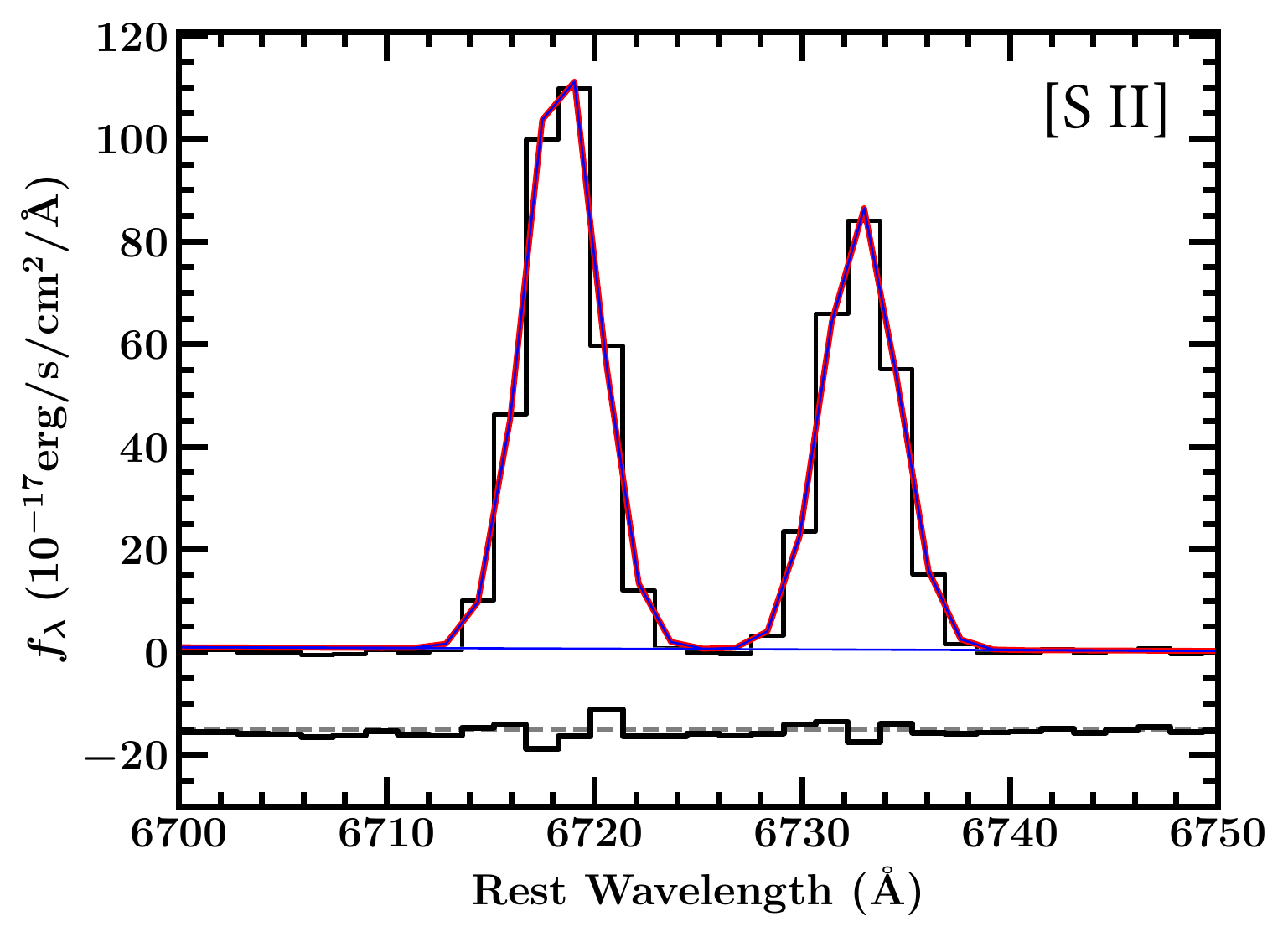}
    \includegraphics[width=0.32\textwidth]{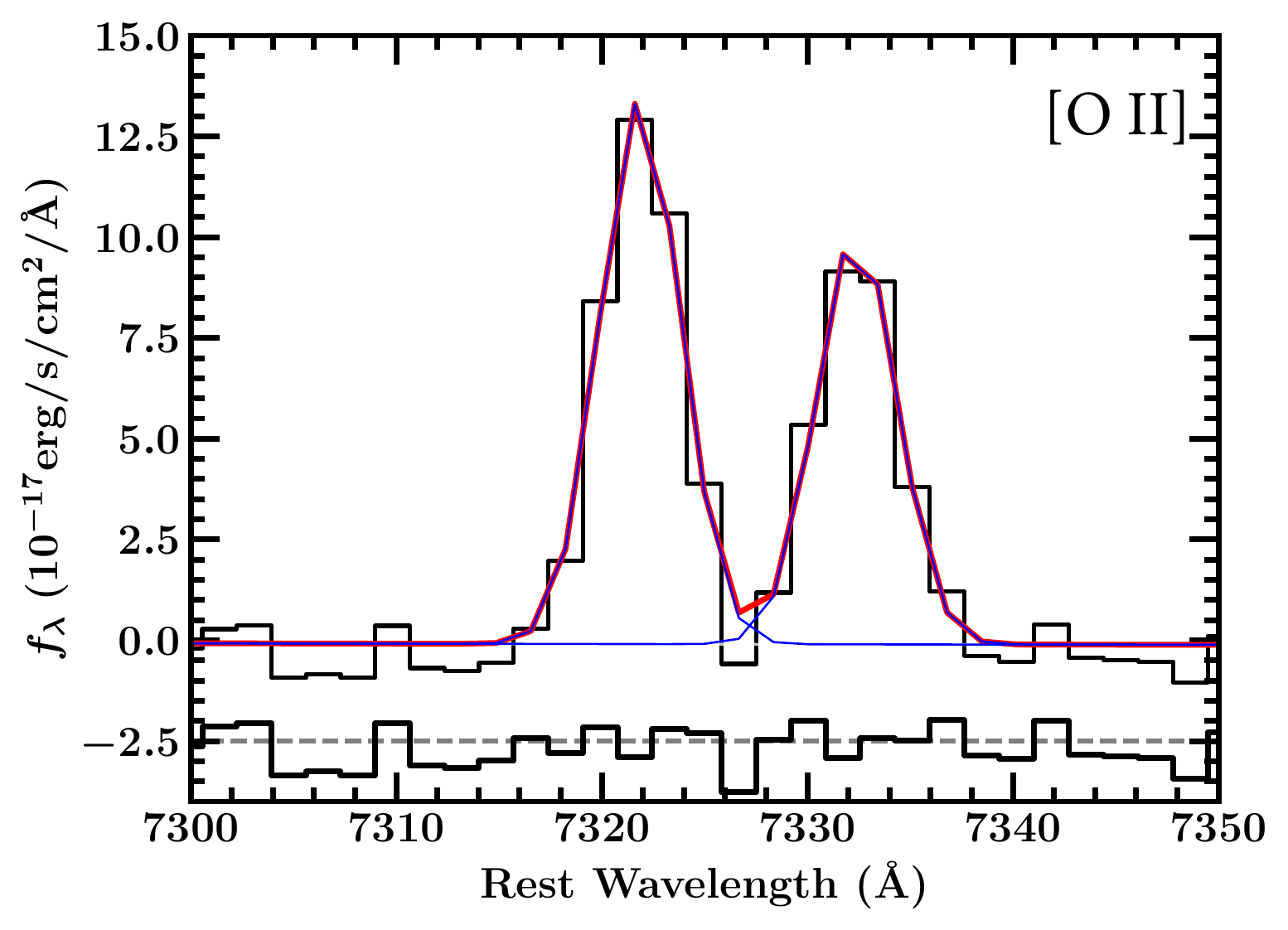}
    \caption{{{\it Top:} The redshift-corrected spectrum of NSA 41675. The data are in black and the stellar continuum fit is over-plotted in red. {\it Middle and Bottom:} The emission-line fits for NSA 41675 on the continuum-subtracted spectrum. We do not show the [\ion{O}{1}]$\lambda\lambda$6300,6363 and \fexl\ fits as they are presented in Figure~\ref{fig:o16300}. The data are shown in black, the total model is over-plotted in red and the individual components are shown in blue. The residuals are shown below each fit. 
    See Section~\ref{sssec:bha} for details.}}
        \label{fig:eline_fits}
\end{figure*}

Finally, we perform fits on the [\ion{O}{3}]$\lambda\lambda$4959,5007 doublets, as well as the \ion{He}{2}$\lambda$4686, [\ion{Fe}{7}]$\lambda$6087, and [\ion{O}{2}]$\lambda\lambda7320,7330$ emission lines. We fit the [\ion{O}{3}] emission with one- and two-Gaussian models, and accept the two-Gaussian model if the reduced $\chi^2$ is improved by at least 20\%. We placed no constraints on the [\ion{Fe}{7}]$\lambda$6087 and \ion{He}{2}$\lambda$4686 emission lines apart from only using a single-Gaussian fit. We constrained the widths of the [\ion{O}{2}]$\lambda\lambda7320,7330$ doublet to be the same, but did not put any constraints on the flux ratio.  
The broad- and narrow-line measurements for all lines except \fexl\ are given in {Table~\ref{table:eline}}, and the \fexl\ measurements are given in Table \ref{table:fex}. {We show example fits for all the measured emission lines in Figure~\ref{fig:eline_fits}.}

\startlongtable
\begin{deluxetable*}{lcccccccccccc}
  \tablecaption{{Emission Line Fluxes}\label{table:eline}}
\tabletypesize{\footnotesize}
\setlength{\tabcolsep}{2pt}
\renewcommand{\arraystretch}{1.}
\tablewidth{0pt}
\tablehead{
\colhead{NSAID} & \colhead{\ion{He}{2}$\lambda4686$} & \colhead{H$\beta$(n)} & \colhead{H$\beta$(b)}&  \colhead{[\ion{O}{3}]$\lambda5007$} & \colhead{[\ion{O}{1}]$\lambda6300$} & \colhead{H$\alpha$(n)} & \colhead{H$\alpha$(b)} & \colhead{[\ion{N}{2}]$\lambda6583$} & \colhead{[\ion{S}{2}]$\lambda6716$} & 
\colhead{[\ion{S}{2}]$\lambda6731$} & \colhead{[\ion{O}{2}]$\lambda7320$} & \colhead{[\ion{O}{2}]$\lambda7330$}
}
\startdata
50 & $77\pm6$ & $6740\pm40$ & $\dots$ & $38100\pm300$ & $188\pm6$ & $29400\pm200$& $\dots$ & $282\pm4$ & $630\pm6$ & $499\pm5$ & $149\pm5$ & $121\pm5$\\
6055   & $\dots$  & $82\pm5$ & $\dots$   & $80\pm20$ & $14\pm3$ & $289\pm4$ & $\dots$ & $61\pm2$ & $75\pm3$ & $55\pm3$ & $\dots$ & $\dots$\\
10870  & $\dots$  & $37\pm5$  & $\dots$  & $35\pm5$ & $6\pm2$ & $116\pm4$ & $\dots$ & $28\pm2$ & $37\pm4$ & $20\pm3$ & $\dots$ & $\dots$\\
15288  & $\dots$  & $464\pm6$  & $\dots$ & $2110\pm50$ & $37\pm3$ & $1560\pm20$ & $\dots$ & $80\pm2$ & $123\pm4$ & $97\pm3$ & $12\pm5$ & $13\pm4$\\
21569  & $\dots$  & $294\pm8$ & $\dots$   & $680\pm10$ & $32\pm4$ & $1152\pm9$ & $\dots$ & $120\pm3$ & $174\pm5$ & $124\pm4$ & $11\pm3$ & $9\pm3$\\
30143  & $\dots$  & $95\pm3$  & $\dots$   & $244\pm7$ & $10\pm2$ & $295\pm7$ & $\dots$  & $25\pm2$ & $49\pm4$ & $34\pm3$ & $\dots$ & $\dots$\\
41675  & $26\pm7$ & $3140\pm20$ & $\dots$   & $15900\pm100$ & $89\pm5$ & $9300\pm100$ & $150\pm30$ & $305\pm4$ & $521\pm8$ & $383\pm7$ & $66\pm4$ & $50\pm3$\\
46653  & $\dots$  & $57\pm3$ & $\dots$    & $72\pm3$ & $7\pm1$ & $196\pm4$ & $\dots$  & $41\pm1$ & $49\pm2$ & $37\pm2$ & $\dots$ & $\dots$\\
50800  & $\dots$  & $1960\pm20$ & $\dots$ & $11400\pm300$ & $100\pm5$ & $6760\pm60$  & $100\pm10$ & $239\pm3$ & $394\pm7$ & $276\pm5$ & $56\pm3$ & $45\pm3$\\
61382  & $\dots$  & $55\pm3$ & $\dots$   & $130\pm40$ & $10\pm3$ & $172\pm4$ & $\dots$ & $16\pm1$ & $33\pm2$ & $20\pm2$ & $\dots$ & $\dots$\\
\enddata
\tablecomments{The entirety of Table~\ref{table:eline} is published in the electronic edition of {\it The Astrophysical Journal}. We show a portion here to give information on its form and content. All fluxes are presented in units of $10^{-17}$~erg~s$^{-1}$~cm$^{-2}$, and are not corrected for reddening. While the weaker lines in the [\ion{O}{3}] and [\ion{N}{2}] doublets were simultaneously fit with the stronger doublet line, we do not list their fluxes as they are weaker by a fixed factor of 3. The [\ion{O}{1}] doublet is fit as described in Section~\ref{sec:sampselect}. H$\alpha$ and H$\beta$ fluxes are split into the narrow (n) and broad (b) components, where applicable.}
\end{deluxetable*}

\subsection{AGN Indicators}
{The 81 \fex-emitting dwarf galaxies presented in this work are considered strong AGN candidates given their high \fex\ luminosities ($L_{\rm [Fe X]}\gtrsim 10^{36}$~erg~s$^{-1}$). In order to search for additional evidence of AGN activity, we employed several supplemental AGN detection techniques using our emission-line measurements. We note that the vigorous star formation seen in the host galaxies may dilute or effectively hide the AGN signal in more traditional emission-line diagnostic indicators, similar to the two previously detected AGN with coronal-line emission, J1220+3020 and Mrk 709S \citep[see][for details]{Molina2021,Kimbro2021}. We describe each detection technique in detail below.}

\subsubsection{\ion{He}{2} and [\ion{Fe}{7}]$\lambda$6087} 

The emission line \ion{He}{2}~$\lambda4686$ has a relatively high ionization potential (54.4 eV), and thus needs a hard ionization field to be produced. Therefore, luminous \ion{He}{2} detections {can be} a useful indicator of AGN activity.
The log(\ion{He}{2}/H$\beta)$ ratio has been used to search for AGN activity in dwarf galaxies by \cite{Sartori2015}, given that \ion{He}{2} is less affected by star formation than [\ion{O}{3}]$\lambda5007$. In our sample, 5 objects show \ion{He}{2} emission with a $\rm{S/N} >3$.
We adopt a stricter criterion than \cite{Sartori2015} for AGN selection, setting a limit of log(\ion{He}{2}/H$\beta) > -1$ as neither X-ray Binaries (XRBs) nor Wolf-Rayet (WR) stars can reproduce that line ratio \citep{Shirazi2012,Schaerer2019}.  We find that 4 out of the 5 objects {(NSA IDs 50, 41675, 256802 and 275961)} {meet this criterion. We therefore consider these objects as very 
strong AGN candidates given their \fex\ and \ion{He}{2} emission.} Meanwhile, the only object to exhibit [\ion{Fe}{7}]$\lambda$6087 emission is NSAID 256802, which was previously identified as an AGN in \citet[][their ID 20]{reines2013}.

\subsubsection{Broad H$\alpha$}\label{sssec:bha}
{Virialized gas that orbits close to a BH will exhibit broad-line emission due to the bulk motion of the gas. This emission will be most clearly seen in the recombination hydrogen Balmer lines, particularly H$\alpha$ as it is less affected by dust attenuation than H$\beta$. {However, not all AGNs exhibit broad-line emission \citep[see][ and references therein]{Ho2008}, and we therefore do not exclude objects without broad-line emission from our AGN candidate sample.} We detected strong broad H$\alpha$ emission {($L[{\rm H}\alpha]>10^{40}$~erg~s$^{-1}$)} in 2/81 objects. Both galaxies, NSA 256802 and 427201, were included in \citet[][their ID 20 and ID N]{reines2013}. NSA 256802 was determined to be a broad-line AGN by \cite{reines2013}. While NSA 427201 was classified as star-forming with broad H$\alpha$ emission in \cite{reines2013}, we note that it could be a TDE (See Section~\ref{ssec:origin}). Weak broad-line emission {($L[{\rm H}\alpha]<10^{40}$~erg~s$^{-1}$)} was detected in an additional 9 BPT star-forming objects, but since weak broad H$\alpha$ can be produced by stellar processes \citep[e.g.,][]{Baldassare2016}, the origin of the broad H$\alpha$ is unclear.}

\begin{figure*}[t]
    \centering
    \includegraphics[width=\textwidth]{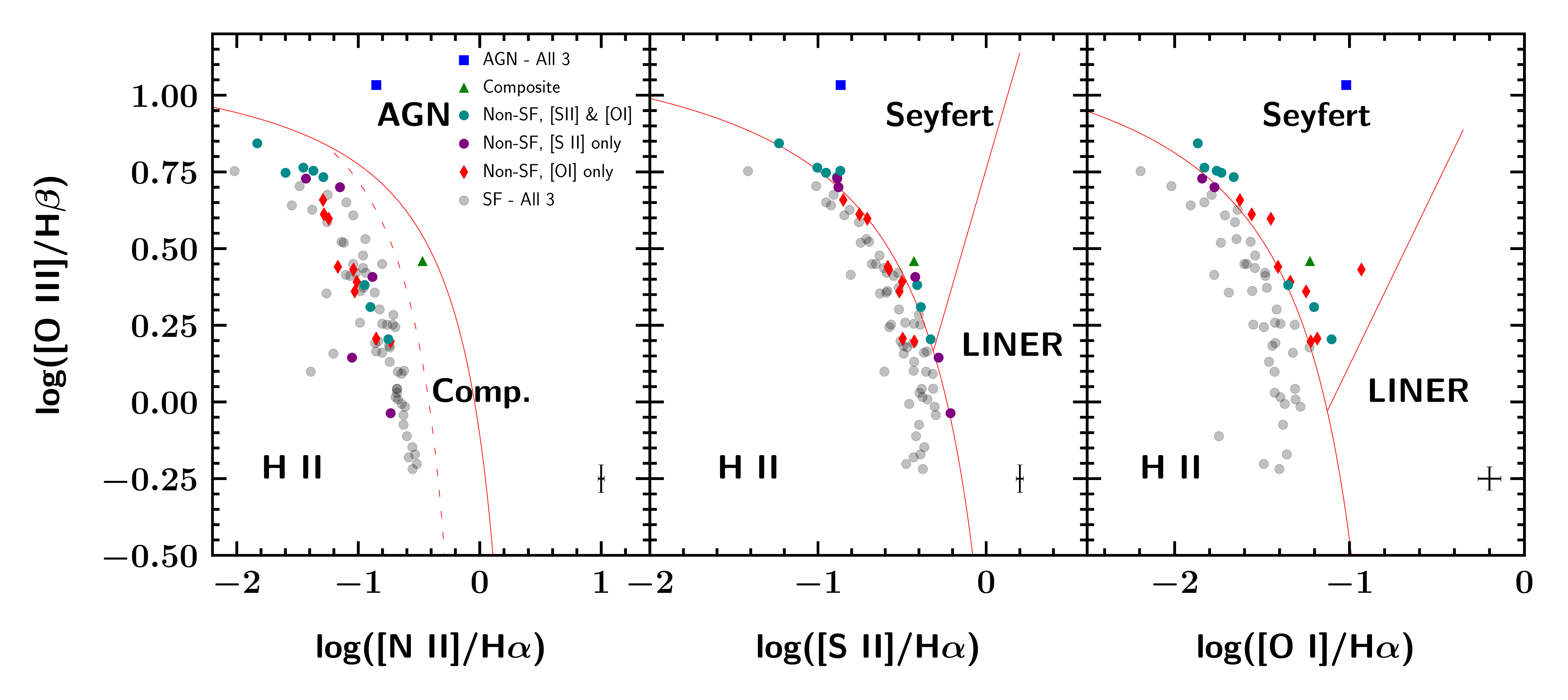}
    \caption{The BPT or VO87 diagrams for the 81 objects in our sample. {Given their high \fex\ luminosities, all of these objects are considered strong AGN candidates whose AGN signal may be hidden in traditional diagnostics due to the intense host-galaxy star formation.} The theoretical extreme starburst and Seyfert lines from \cite{kewley2006} are shown as solid red lines, while the empirical composite line from \cite{Kauffmann2003} is shown as the red dashed line. We show characteristic uncertainties for each line ratio in the bottom right corner of each panel. One object is AGN-like in all three diagrams (shown by the blue square), and one object falls in the composite region (green triangle). Both of these objects are also in the \cite{reines2013} survey (IDs 20 and 105, respectively). Of the remaining objects, 10\% are non-star-forming in both the [\ion{S}{2}]/H$\alpha$ and [\ion{O}{1}]/H$\alpha$ diagrams (shown as teal circles), 6\% are non-star-forming in only the [\ion{S}{2}]/H$\alpha$ diagram (shown as purple circles), and 11\% are non-star-forming in only the [\ion{O}{1}]/H$\alpha$ diagram (red diamonds).} 
    \label{fig:bpt}
\end{figure*}

\subsubsection{Narrow Line Diagnostic Diagrams} \label{ssec:bbpt}
Diagnostic emission-line ratio diagrams
can help separate galaxies dominated by AGN activity from star-forming galaxies. 
We created the narrow-line diagnostic diagrams shown in Figure~\ref{fig:bpt}, which are presented in \citet[][VO87]{VO87} and are based on the BPT diagrams \citep{Baldwin1981}. These diagrams take [\ion{O}{3}]/H$\beta$ versus [\ion{N}{2}]/H$\alpha$, [\ion{S}{2}]/H$\alpha$, and [\ion{O}{1}]/H$\alpha$. We include both the empirical composite line from \cite{Kauffmann2003}, and the theoretical extreme starburst lines and Seyfert-LINER lines from \cite{kewley2006}. The AGN and Composite objects denoted by the blue square and green triangle, respectively, are from \citet[][their IDs 20 and 105]{reines2013}. 

Approximately 30\% of our sample appears Seyfert- or LINER-like in at least one of the diagnostic diagrams, where 1 object is Seyfert-like in all three diagrams, 10\% appear non-star-forming in both the [\ion{S}{2}]/H$\alpha$ and [\ion{O}{1}]/H$\alpha$ diagrams, 6\% are non-star-forming in only the [\ion{S}{2}]/H$\alpha$ diagram, and 11\% are non-star-forming in only the [\ion{O}{1}]/H$\alpha$ diagram. Given the known metallicity dependence of [\ion{N}{2}]/H$\alpha$ ratio, the sensitivity of [\ion{S}{2}]/H$\alpha$ to shocks, and the inability of standard stellar processes to produce enhanced [\ion{O}{1}] emission \citep[e.g.,][]{kewley2006,Kewley2019,Molina2018}, we consider the objects that show Seyfert or LINER-like [\ion{O}{1}] emission to be the strongest AGN candidates based on the VO87 diagrams. {We do note that the emission-line ratios from the 3\arcsec\ SDSS single-fiber spectra for both Mrk 709S and J1220+3020 appear star-forming in all three VO87 diagrams, even though both have confirmed AGN activity and detected \fex\ emission \citep{reines2020,Molina2021,Kimbro2021}. Therefore, we still consider all of these objects as AGN candidates based on their \fex\ emission.}

\begin{figure}
    \centering
    \includegraphics[width=0.5\textwidth]{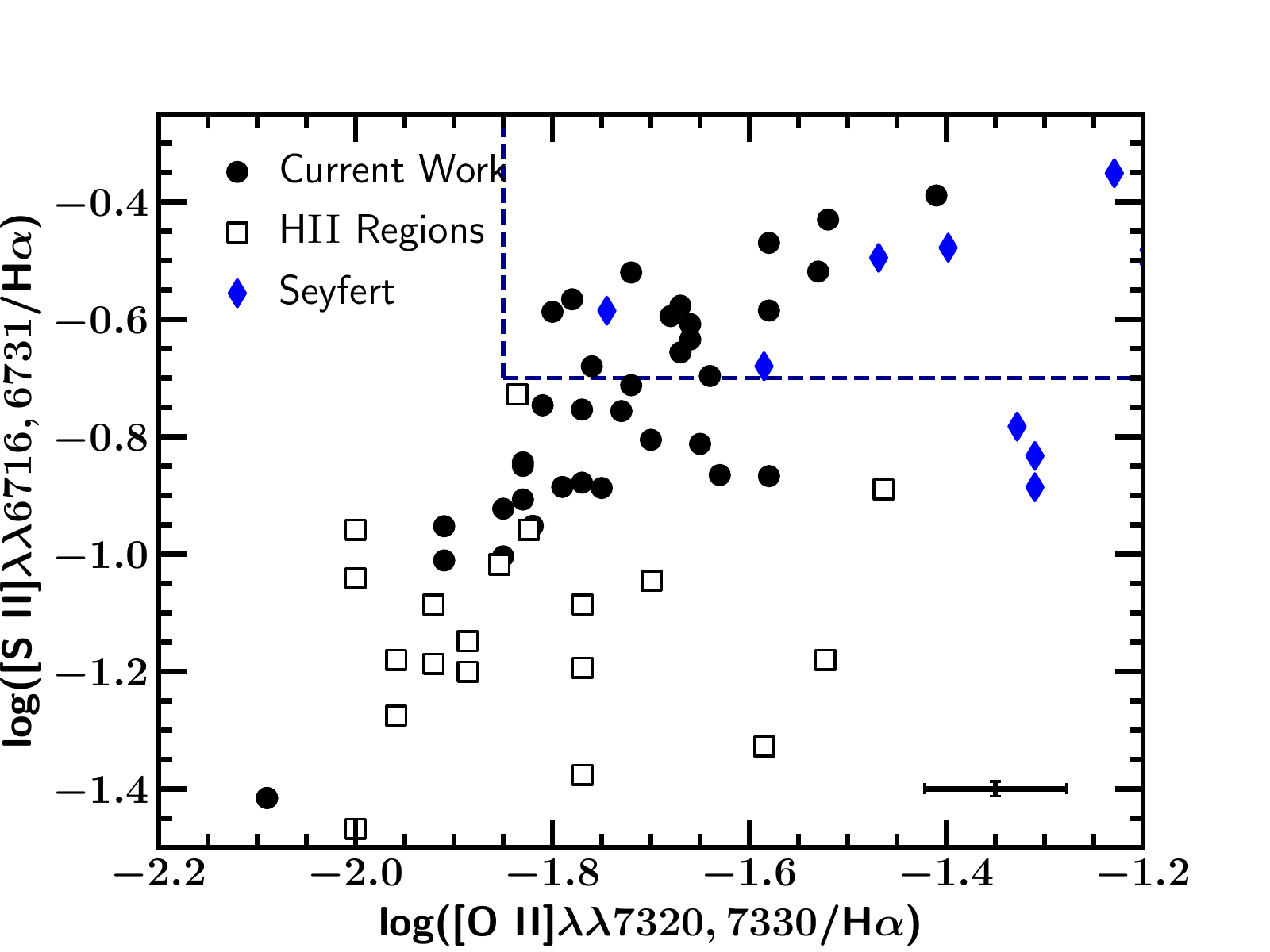}
    \caption{The log([\ion{S}{2}]$\lambda\lambda6716,6731$/H$\alpha$) vs.~log([\ion{O}{2}]$\lambda\lambda7320,7330$/H$\alpha$) diagram for our sample. Here we only include the 36 objects that have [\ion{O}{2}]$\lambda\lambda7320,7330$ detections. The data from this work are shown as black filled circles, and the star forming or \ion{H}{2} regions and Seyfert galaxies from \cite{osterbrock1992} are shown as unfilled squares and blue diamonds, respectively. The characteristic error bars for the line ratios in our sample is shown in the bottom left. The dashed box indicates the region we consider AGN-like given the clustering of the Seyfert and \ion{H}{2} region data. 16 out of the 36 objects with [\ion{O}{2}]$\lambda\lambda7320,7330$ detections are consistent with AGN activity. {We note that given their high \fex\ luminosities, all of these objects are considered strong AGN candidates whose AGN signal may be hidden in traditional diagnostics due to the intense host-galaxy star formation.}}
    \label{fig:s2o2}
\end{figure}

In addition to the typical VO87 diagrams, there are near-infrared (NIR) diagnostic diagrams that can be used to distinguish AGN from stellar activity. The [\ion{S}{2}]$\lambda\lambda9069,9531$/H$\alpha$ vs.~[\ion{S}{2}]$\lambda\lambda6716,6731$/H$\alpha$, [\ion{S}{2}]$\lambda\lambda9069,9531$/H$\alpha$ vs.~[\ion{O}{2}]$\lambda\lambda7320,7330$/H$\alpha$ and [\ion{S}{2}]$\lambda\lambda6716,6731$/H$\alpha$ vs.~[\ion{O}{2}]$\lambda\lambda7320,7330$/H$\alpha$ from \cite{osterbrock1992} were recently used by \cite{bohn2021} to determine the presence of AGN activity in dwarf galaxies. Given the wavelength range of the SDSS spectra, we utilize only one of the three diagrams, [\ion{S}{2}]$\lambda\lambda6716,6731$/H$\alpha$ vs.~[\ion{O}{2}]$\lambda\lambda7320,7330$/H$\alpha$ to search for AGN activity in our sample. We show this diagram in Figure~\ref{fig:s2o2}.

These diagrams are strictly empirical, and so we employ cuts in our data to identify objects that are clustered with the Seyfert-like objects in the diagram, denoted by the dark-blue dashed line in Figure~\ref{fig:s2o2}. {We employ a purity cut, such that all \ion{H}{2} regions are avoided in our AGN selection region. We thus }consider the 16 objects with log([\ion{S}{2}]$\lambda\lambda6716,6731$/H$\alpha) >-0.7$ and log([\ion{O}{2}]$\lambda\lambda7320,7330$/H$\alpha)>-1.95$ to be strong AGN candidates based on this NIR diagnostic diagram {and their strong \fex\ emission}. {We present the results of both the VO87 and NIR diagnostic diagrams in Table~\ref{table:nldiagram}.}

\begin{deluxetable}{rcccc}
  \tablecaption{{Results of Narrow-Line Diagnostic Diagrams}}
\tabletypesize{\footnotesize}
\setlength{\tabcolsep}{1pt}
\tablewidth{0pt}
\tablehead{ 
\colhead{} & \multicolumn{3}{c}{VO87 Diagrams} & \multicolumn{1}{c}{NIR Diagram}\\
\cmidrule(lr){2-4} \cmidrule(lr){5-5}
{NSA ID} & {[\ion{N}{2}]/H$\alpha$} & {[\ion{S}{2}]/H$\alpha$} & {[\ion{O}{1}]/H$\alpha$} & {[\ion{O}{2}]/H$\alpha$}\\
 (1) & (2) & (3) & (4) & (5)}
\startdata
50 & SF & Sy & Sy  & SF\\
15288 & SF & SF & Sy  & SF\\
21569 & SF & SF & SF  & AGN\\
50800 & SF & Sy & Sy & SF\\
61382 & SF & SF & Sy & \dots\\
63560 & SF & SF & SF & AGN\\
95985 & SF & Sy & Sy  & SF\\
102990 & SF & SF & SF & AGN\\
140290 & SF & Sy & Sy & \dots\\
140756 & SF & Sy & SF & \dots\\
141686 & SF & Sy & Sy & \dots\\
144266 & SF & SF & Sy  & AGN\\
162936 & SF & LINER & SF & \dots\\
164449 & SF & Sy & SF & \dots\\
174326 & SF & SF & SF & AGN\\
213780 & SF & SF & SF & AGN\\
220716 & SF & SF & Sy & \dots\\
237974 & SF & SF & Sy & \dots\\
256802 & AGN & Sy & Sy  & SF\\
270622 & SF & Sy & Sy  & \dots\\
275961 & Comp. & Sy & Sy & AGN\\
300542 & SF & Sy & Sy & AGN\\
305229 & SF & Sy & Sy & SF\\
328894 & SF & Sy & Sy & \dots\\
345375 & SF & SF & Sy & AGN\\
393631 & SF & SF & SF & AGN\\
401495 & SF & Sy & SF & \dots\\
426824 & SF & SF & Sy & \dots\\
442120 & SF & SF & SF & AGN\\
442212 & SF & Sy & Sy & SF\\
442427 & SF & LINER & SF & \dots\\
457940 & SF & Sy & Sy & \dots\\
458829 & SF & Sy & Sy & SF\\
491412 & SF & SF & Sy & SF\\
501341 & SF & Sy & SF & SF\\
524413 & SF & SF & Sy & \dots\\
532026 & SF & SF & SF & AGN\\
553772 & SF & SF & SF & AGN\\
577169 & SF & SF & SF & AGN\\
656364 & SF & SF & SF & AGN\\
670127 & SF & SF & SF & AGN\\
\enddata
\tablecomments{We only present objects that are Seyfert (Sy), Composite (Comp.), LINER or AGN-like in at least one diagram in this table. {The remaining 40 objects were either star-forming in all three VO87 diagrams, and are either consistent with star formation or do not have detected [\ion{O}{2}] emission needed for the NIR diagram.} Column (1): NSA IDs. Columns (2-4): Classifications from the [\ion{O}{3}]/H$\beta$ vs. [\ion{N}{2}]/H$\alpha$, [\ion{S}{2}]/H$\alpha$, and [\ion{O}{1}]/H$\alpha$ diagrams, as shown in Figure~\ref{fig:bpt}. Column (5): NIR [\ion{S}{2}]/H$\alpha$ vs. [\ion{O}{2}]/H$\alpha$ diagram as shown in Figure~\ref{fig:s2o2}.}
\label{table:nldiagram}
\end{deluxetable}

\section{Optical Variability}\label{sec:lc_data}

The variability {observed in some} AGN light curves is well-described by a damped random walk (DRW) model \citep[e.g.,][]{Kelly2009,MacLeod2010}, which is distinct from the variability expected from supernovae and supernova remnants (i.e., a clear decline in brightness after the initial peak post-explosion). While studying optical variability is a well-known technique used to detect AGN activity in high-mass galaxies, recent work by \cite{Baldassare2018,Baldassare2020} has demonstrated that it {can be effective in identifying optically-variable AGN in the} dwarf galaxy regime. 

In order to search for AGN-like variability in our sample, we created and analyzed light curves of the physical region covered by the SDSS spectroscopic fiber using data from the Palomar Transient Factory \citep[PTF;][]{Rau2009,Law2009}. The PTF is a wide-field optical survey designed to systematically explore the optical variable and transient sky. Data collection for the PTF began in 2009, which is $\sim9$ years after a majority of the SDSS single-fiber spectra were taken. Conversely, most of the optical decline seen in supernovae occur within the first 3--5 years \citep[e.g.][]{Filippenko1997,smith2009,Smith2017}. Therefore, while we can search for variability consistent with a DRW model and thus indicative of AGN activity, we may not be able to attribute signatures of SNe-like variability to the observed emission seen in the SDSS spectra.

\subsection{Data Analysis and Results}
We followed the methodology first presented in \cite{Baldassare2020}, which we briefly describe here. We rely on the $R$-band images from PTF as they have significantly better sky coverage, long baselines and larger number of observations than the $g$-band data, and limit our data collection to PTF images with seeing better than 3\arcsec. These cuts result in approximately 70 observations per object. We then create a template of the host galaxy emission using the Difference Imaging and Analysis Pipeline 2 (DIAPL2), which is a modified version of the Difference Imaging Analysis software \citep{Wozniak2000}. This template is then modified to match the seeing and background levels of each individual exposure. The template is then subtracted from each exposure creating a difference image.

We then construct light curves by performing aperture photometry on the template and difference images, such that the flux value for each data point is the template plus the difference image measurements. We use an aperture of 3\arcsec\ centered on the position of the SDSS fiber as listed in the NSAv1. After our light curves are constructed, we search for DRW-like variability using the software QSO\_fit \citep{Butler2011}. Objects are then classified as candidate variable AGN if they have $\sigma_{\rm var}>2$, $\sigma_{\rm QSO}> 2$ and $\sigma_{\rm QSO}\gtrsim\sigma_{\rm notQSO}$. Here, $\sigma_{\rm var}$ gives the significance that the object is variable; $\sigma_{\rm QSO}$ is the significance that the $\chi^2$ for the DRW model is better than that for non-AGN-like variability and $\sigma_{\rm notQSO}$ gives the significance that the source variability is better described by random variability. Out of the 50 total objects with adequate PTF data, the only galaxy to show AGN-like variability was NSA 131809 (NSAv0 26828), which was also detected as variable AGN candidate in \cite{Baldassare2020}. We also found supernova-like variability in NSA 331627, however this SNe event {occurred about 7 years} after the SDSS spectrum was taken and thus it cannot explain the observed \fex\ emission. {We note that the previously identified AGN NSA 256802 \citep[ID 20 from][]{reines2013} did not exhibit variability in the PTF data, demonstrating that not all AGN may be optically variable. We therefore do not exclude any objects as AGN candidates based on this detection technique.}

\section{X-ray Detections}\label{sec:xrayemis}
\begin{deluxetable}{clccc}
\tabletypesize{\footnotesize}
\tablecaption{\textit{Chandra} Observations}
\tablewidth{0pt}
\tablehead{
\colhead{NSAID}  & \colhead{Date observed}  & \colhead{Obs ID} & \colhead{Exp.\ time (ks)} & \colhead{$N_{\rm background}$} }
\startdata
256802      & 2010 Jul 25 & 11464 & 1.8   & 0.0034 \\
533731      & 2013 Jan 13 & 13929 & 20.8 & 0.0075 \\
472272      & 2013 Jun 02 & 14911 & 15.8 & 0.0062 \\
625882      & 2018 Jun 06 & 19441 & 29.6 & 0.1417 \\
50          & 2017 Mar 11 & 19463 & 14.9 & 0.1199 \\
275961      & 2018 Aug 17 & 20773 & 15.0 & 0.1751 
\enddata
\tablecomments{$N_{\rm background}$ is the number of expected 2-10 keV $N(>S)$ background sources within $3r_{50}$, using \citet{moretti03}. }
\label{tab:cxo}
\end{deluxetable}
\begin{deluxetable*}{lcccccccccc}
\tabletypesize{\footnotesize}
\tablecaption{X-ray Sources}
\tablewidth{0pt}
\setlength{\tabcolsep}{4pt}
\tablehead{
\colhead{NSAID} & \colhead{R.A.} & \colhead{Decl.}& {Offset} & \multicolumn{2}{c}{Net Counts} & \multicolumn{2}{c}{Flux ($10^{-15}$ erg s$^{-1}$ cm$^{-2}$)} & \multicolumn{2}{c}{Luminosity (log(erg s$^{-1}$))}\\
\cmidrule(l){5-6} \cmidrule(l){7-8} \cmidrule(l){9-10}
\colhead{ } & \colhead{(deg)} & \colhead{(deg)} & \colhead{(arcsec)} & \colhead{0.5-2 keV} & \colhead{2-7 keV} & \colhead{0.5-2 keV} & \colhead{2-10 keV} & \colhead{0.5-2 keV} & \colhead{2-10 keV} \\
\colhead{(1)} & \colhead{(2)} & \colhead{(3)} & \colhead{(4)} & \colhead{(5)} & \colhead{(6)} & \colhead{(7)} & \colhead{(8)} & \colhead{(9)} & \colhead{(10)}}
\startdata
256802$-$X1   & 185.928452 & 58.246088 & 0.3  & 296.50$ \pm 29.78$       & 124.76$ \pm 19.94$      & 719.25  & 1475.86 & 41.6    & 41.8  \\
533731$-$X1	 & 147.325040 & 16.878966 & 0.3 & 23.31$ \pm 8.92 $        & 8.55$ \pm 5.26$         & 5.73    & 8.88    & 40.5    & 40.7  \\
472272$-$X1	 & 245.468974 & 15.315684 & 0.5  & 9.27$ \pm 5.32$          & 3.08$^{+4.11}_{-2.46}$  & 3.20    & 4.31    & 39.9    & 40.0 \\
625882$-$X1   & 166.283858 & 44.748226 & 6.4  & 10.26$^{+7.47}_{-4.91}$  & 7.25$ \pm 4.92$         & 2.91    & 5.52    & 39.4    & 39.7 \\
625882$-$X2   & 166.284416 & 44.747104 & 2.8 & 17.40$ \pm 7.79$         & 15.01$^{+8.72}_{-6.12}$ & 4.94    & 11.41   & 39.7    & 40.0 \\
625882$-$X3\tablenotemark{a}   & 166.284419 & 44.747115 & 2.8  & 9.00$ \pm 5.29$          & \nodata                 & 2.56    & $<$1.67 & 39.4    & $<$39.2 \\
50$-$X1	 & 146.008247 & $-0.643140$ & 3.5 & 95.64$\pm 17.31$         & 67.44$ \pm 14.89$       & 49.44   & 100.45  & 39.5    & 39.8 \\
50$-$X2 	 & 146.007968 & $-0.642340$ & 0.7 & 5.17$^{+4.87}_{-3.22}$   & 12.22$^{+8.15}_{-5.47}$ & 2.68    & 18.23   & 38.2   & {39.0}\\
275961$-$X1	 & 207.855629 & 40.213289 & 0.3  & 10.54$^{+7.41}_{-4.87}$  & 36.25$ \pm 11.1$        & 6.09    & 55.24   & 38.9    & 39.9\\
\enddata
\tablecomments{Column 1: Identification number in the NSA. 
Column 2: right ascension of hard X-ray source (when present). 
Column 3: declination of hard X-ray source (when present).
Column 4: distance between hard X-ray source and center of SDSS spectroscopic Fiber.
Columns 5-6: net counts after applying a 90\% aperture correction.  Error bars represent 90\% confidence intervals.
Columns 7-8: fluxes corrected for Galactic absorption.
Columns 9-10: log luminosities corrected for Galactic absorption; calculated using a photon index of $\Gamma = 1.8$.
}\vspace{-2mm}
\tablenotetext{a}{No corresponding hard band X-ray source was detected. The reported flux and luminosity are upper limits.}
\label{tab:xray}
\end{deluxetable*}

{Hard X-ray photons are emitted by the hot corona surrounding the accretion disk \citep[e.g.,][]{Liang1984,Haardt1991}.} If the observed hard X-ray luminosity is much larger than that expected from X-ray binaries, this can signify the presence of an AGN in a galaxy.  However, AGNs with low luminosities (e.g., due to low accretion rates) may not exhibit enhanced X-ray emission.
To search for AGN activity at X-ray wavelengths, we cross-referenced our sample with the {\it Chandra} data archive\footnote{\url{https://cxc.harvard.edu/cda/}} using a search radius of 5\arcsec. Out of the 81 galaxies in our sample, 6 have been previously observed.  The \textit{Chandra} observations were taken between 2010 Jul 25 and 2018 Aug 17, with exposure times ranging between 1.8 and 29.6 ks, as summarized in Table~\ref{tab:cxo}. While some of these objects have been presented in various references \citep[e.g.,][]{Dong2012,Reines2014,He2019,Kimbro2021}, we downloaded the data and performed a consistent analysis for this work. We followed the general methodology laid out in \citet{Latimer2021}, which we summarize below.

\subsection{Data Reduction and Measurements} \label{sec:xraydatareduct}

We used \texttt{CIAO} v4.13 \citep{fruscione06} to reduce each observation, applying calibration files (CALDB 4.9.5) to create new level 2 event files. We then aligned the \textit{Chandra} astrometry to the SDSS frame. We created a list of X-ray point sources on the S3 chip using \texttt{wavdetect} on the image filtered from 0.5-7 keV, excluding any wavdetect sources falling within 3$r_{50}$ of the target galaxy. We then used the \texttt{CIAO} function \texttt{wcs\_match} to match the remaining X-ray sources to optical point sources in the SDSS DR12 catalog with $i$-band magnitudes $<22$. This resulted in updated astrometry for four of the six galaxies, with astrometric shifts ranging between $\pm$0.12-1.70 pixels ($0\farcs06$-$0\farcs83$).

Next we searched for X-ray point sources that could indicate the presence of AGNs in the galaxies. First, we checked our images for background flares and remove time intervals where the background rate was $>3\sigma$. We then re-ran \texttt{wavedetect} on the corrected images of the S3 chip, filtered from 0.5-7 keV. 
We also set a significance threshold of $10^{-6}$, corresponding to approximately one false source detection over the entire S3 chip. Further analysis is restricted to \texttt{wavdetect} sources that lie within 3$r_{50}$ of the target galaxies.

We extracted source counts using circular apertures corresponding to the 90\% enclosed energy fraction at 4.5 keV, or ${\sim}2 \arcsec$ for our X-ray sources. Background counts were estimated using source-centered circular annuli having an inner radius equal to the source aperture radius and an outer radius of $12~\times$ the inner radius. We considered a source to be detected if the source counts are above the background counts in the source aperture to within a 95\% confidence level, following the Bayesian methods in \cite{kraft91} for Poisson counting statistics in the presence of a background. All detected sources pass this test.

We calculated net counts by subtracting the background counts in the source aperture from the source counts and applying a 90\% aperture correction. We detected a total of {nine} X-ray point sources across six galaxies, and report their properties in Table \ref{tab:xray}. {An additional X-ray source was detected near NSAID 275961; we exclude this from analysis because it was $\sim24\arcsec$ offset from the SDSS spectroscopic fiber, which is 3 times larger than the $r$-band half-light radius and thus not within the galaxy.} Unabsorbed hard (2-10 keV) and soft (0.5-2 keV) X-ray fluxes are calculated using the \texttt{CIAO} function \texttt{srcflux}. We use Galactic column densities from \cite{dickey90} maps and a power-law spectral model with photon index $\Gamma=1.8$, which is typical for low-luminosity AGN \citep{Ho2008,ho09} and ultraluminous X-ray sources at these energies \citep{swartz08}. The 2-10 keV luminosities have a range of log$(L_{\rm 2-10 keV}/{\rm erg~s}^{-1}) = 39.0$--$41.8$. We use the estimated minimum fluxes to provide upper limits on X-ray source luminosities for sources that are not detected in both bands (e.g. source X3 in NSAID 625882). Table \ref{tab:xray} summarizes the unabsorbed fluxes and corresponding luminosities. Any potential absorption intrinsic to the sources is not accounted for, so these values should be taken as lower limits.

\begin{figure}[t]
\centering
\includegraphics[width=0.48\textwidth]{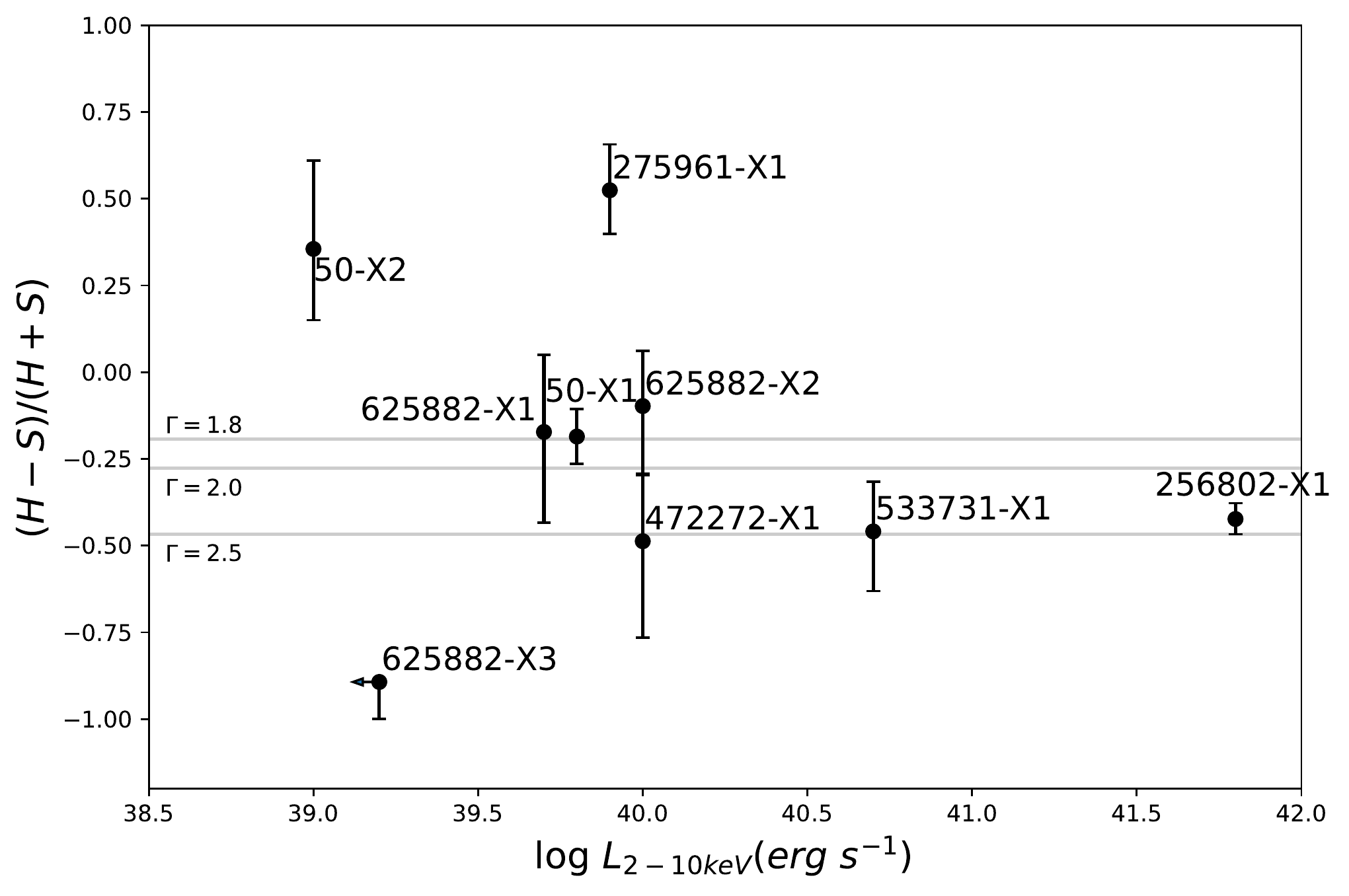}
\caption{ Hardness ratio vs.\ log 2-10 keV X-ray luminosity for our 9 X-ray sources that were detected in at least one of the two bands. The hardness ratio was calculated using BEHR (see Section \ref{sec:xraydatareduct}). The error bars are the 68\% confidence intervals.  Hardness ratios for unabsorbed power laws with $\Gamma=1.8$, 2.0, and 2.5 are shown as grey solid lines.
}
\label{fig:xrayhrratio}
\end{figure}

We estimated hardness ratios using the Bayesian Estimation of Hardness Ratios code \citep[BEHR;][]{park06}, as it is useful in the Poisson regime of low counts and additionally works even if only one of the hard or soft bands has a source detection. Here we define hardness ratio as $(H-S)/(H+S)$, where $H$ and $S$ are the number of detected counts in the hard (2-7 keV) and soft (0.5-2 keV) X-ray bands, respectively. We have {nine} X-ray sources with detections in at least one of the two bands, with the resulting hardness ratios displayed in Figure \ref{fig:xrayhrratio}. We also use the Portable, Interactive Multi-Mission Simulator (PIMMS)\footnote{\url{https://heasarc.gsfc.nasa.gov/cgi-bin/Tools/w3pimms/w3pimms.pl}} to estimate the hardness ratios for unabsorbed power laws with $\Gamma = 1.8$, 2.0, and 2.5. We plot these in Figure \ref{fig:xrayhrratio} as well.

We estimated the 95\% positional uncertainties for each source using Equation 5 from \cite{hong05}, an empirical formula involving the \texttt{wavdetect} counts and off-axis position of the sources. Error bars on net counts represent 90\% confidence intervals. For source counts $<10$, we calculate errors using the formalism of \cite{kraft91}, which takes background counts into account. For source counts $\geq 10$ we use confidence intervals from \cite{gehrels86}, which assume negligible background counts.

Finally, we find how many background X-ray sources we would expect to fall within 3$r_{50}$ of the target galaxy. We use Equation 2 from \cite{moretti03}, which takes in an input flux and returns the expected number of background sources we would expect to see with that flux or higher (per square degree). For our input fluxes we use our minimum flux sensitivity, which we estimate by assuming a source with 2 counts. The resulting 2-10 keV (0.5-2 keV) minimum fluxes range from $S_{min} \sim 1.7\textrm{-}26.2$ ($0.5\textrm{-}5.1$) in units of $10^{-15}$ erg s$^{-1}$ cm$^{-2}$ for the six \textit{Chandra} observations. Using these fluxes results in the expected number of hard band (soft band) background sources within 3$r_{50}$ of our target galaxies ranging between ${\sim}$0.003-0.175 (0.006-0.121).

\subsection{Contribution from X-ray Binaries}

We compared the measured X-ray emission to that expected from XRBs in order to search for enhanced X-ray emission that would likely indicate the presence of an AGN. The expected galaxy-wide X-ray luminosity from XRBs scales with SFR for high-mass XRBs \citep{grimm03} and with stellar mass for low-mass XRBs \citep{gilfanov04}. Using the relation from \cite{gilfanov04} to estimate the expected low-mass XRB luminosities from stellar masses, we find that our observed X-ray luminosities are higher than the expected low-mass XRB luminosities by at least {${\sim}1.3$} dex. We conclude that low-mass XRBs are unlikely to meaningfully contribute to the X-ray luminosities of our observed sources, and thus limit our focus to high-mass XRBs {(HMXRBs)}.

We generally follow the same method as outlined in \cite{Latimer2021}. To predict the expected X-ray luminosity from HMXRBs, we use the model from \cite{lehmer21}, which relates the gas-phase metallicity, SFR, and expected XRB X-ray luminosity. Note that the X-ray luminosities used in this model are in the 0.5-8 keV band; to match, we re-analyze our \textit{Chandra} data in this band, following the same methodology presented in Section \ref{sec:xraydatareduct}. Also, while the scatter in the \cite{lehmer21} relation is metallicity-dependant, for metallicities $> 8.0$ the scatter is more consistent and we find a 1$\sigma$ scatter of ${\sim}0.2$ dex. This applies to all of our X-ray detected galaxies with the exception of NSAID 50. 
We use the SFRs and metallicities described in Section~\ref{ssec:sprop}, which are summarized in Table~\ref{table:galprop}. 

\begin{figure}[t!]
\centering
\includegraphics[width=0.48\textwidth]{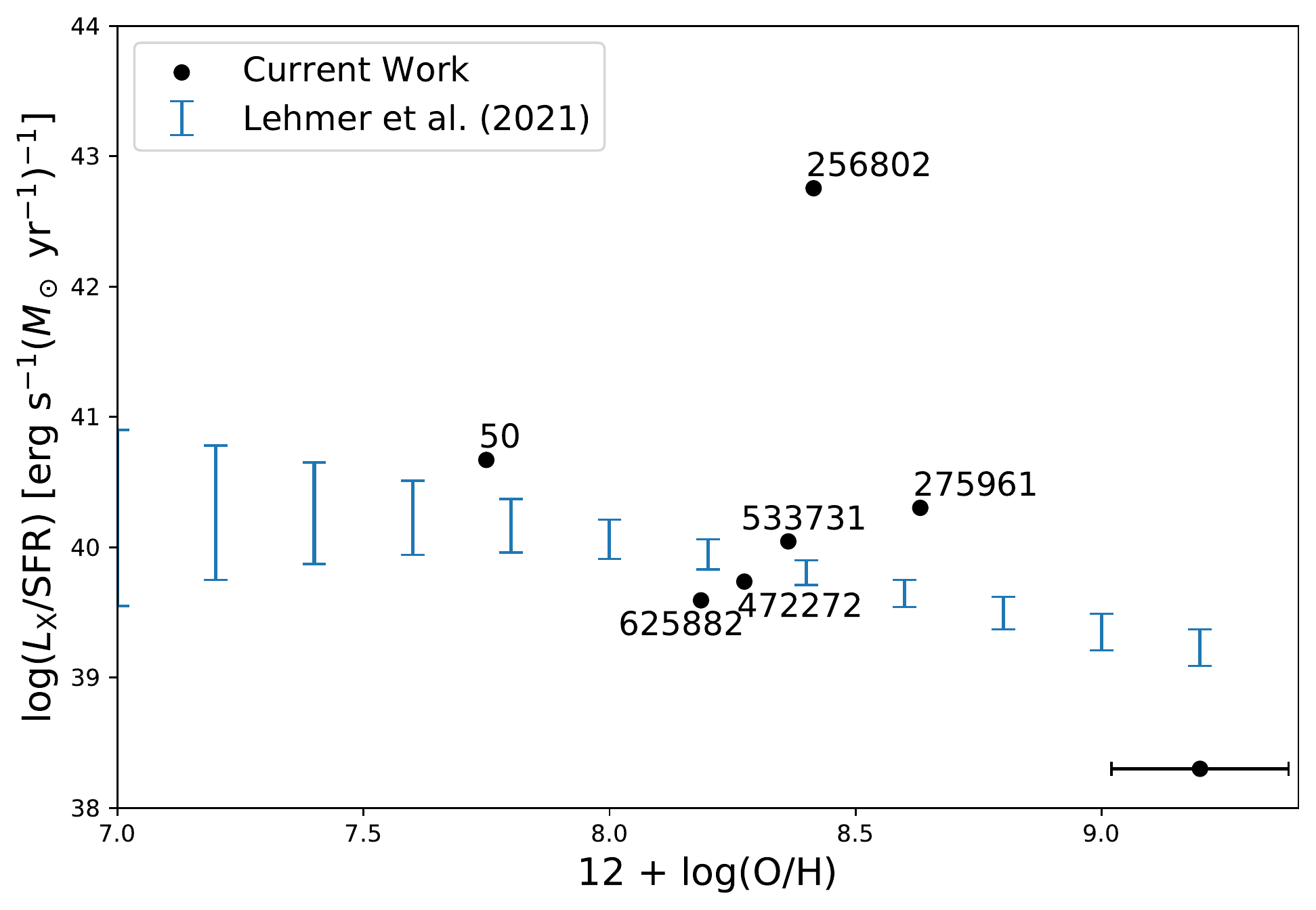}
\caption{ Log ratio of X-ray luminosity over host galaxy SFR vs.\ metallicity. The values and error bars in blue are the expected X-ray luminosity from XRBs from \cite{lehmer21}. The ${\sim} 0.18$ uncertainty in the metallicities is shown in black in the lower right-hand corner. {While only two objects (NSA 256802 and 275961) are X-ray-detected AGNs, we still consider all of the objects in our sample as AGN candidates due to their strong \fex\ emission.}}
\label{fig:expxrbalt}
\end{figure}

A comparison between the expected cumulative XRB luminosity for each galaxy and our observed galaxy-wide X-ray luminosities can be seen in Figure \ref{fig:expxrbalt}. Two galaxies have observed values significantly ($>3\sigma$) above expectations for XRBs: NSAIDS 256802 and 275961, which are ${\sim}15\sigma$ and {${\sim}3.5\sigma$ higher than the expected values from XRBs}, respectively. We therefore consider these two objects X-ray-detected AGNs. In addition to the archival {\it Chandra} data analyzed here, we note that one other galaxy in our sample, NSA 95985 (SDSS J153704.18+551550.5), was determined to be an X-ray-selected AGN candidate by \cite{Birchall2020}, who utilized the 3XMM DR7 catalog \citep{Rosen2016} to search for AGN activity.

\section{Radio Detections}\label{sec:radio_det}

\subsection{Catalog Data}

\begin{figure*}[t]
    \centering
    \includegraphics[width=0.19\textwidth]{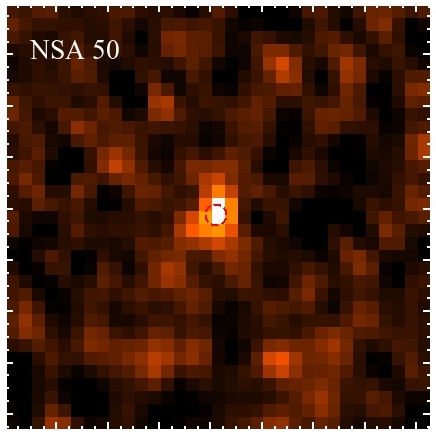}
    \includegraphics[width=0.19\textwidth]{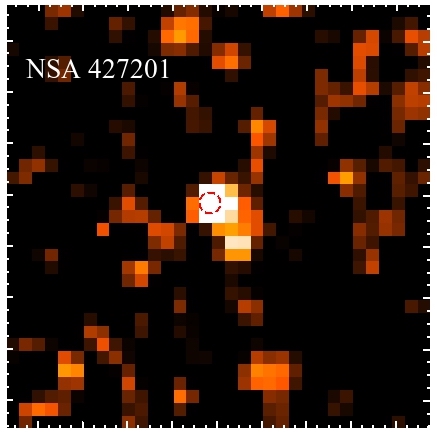}
    \includegraphics[width=0.19\textwidth]{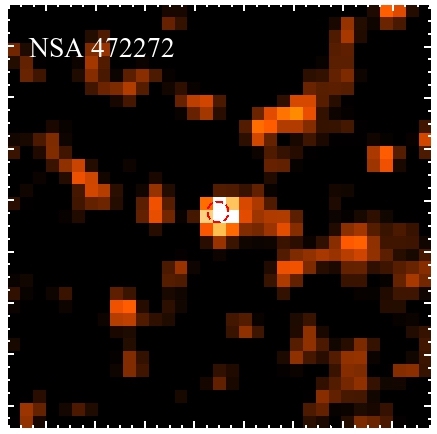}
    \includegraphics[width=0.19\textwidth]{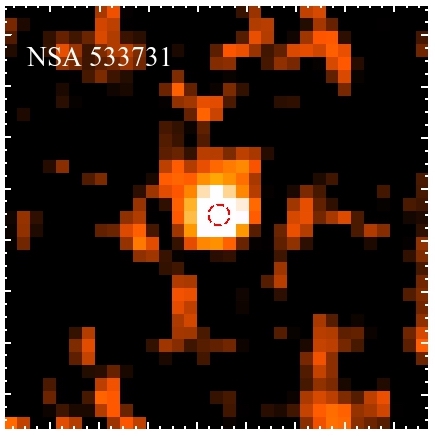}
    \includegraphics[width=0.19\textwidth]{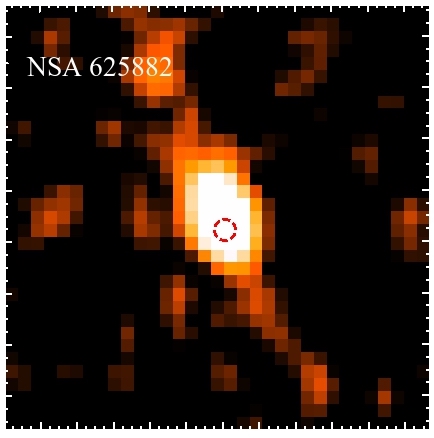}
    \caption{The FIRST 1.4~GHz observations for the 5 detected galaxies in our sample. The red dotted circle shows the position and spatial extent of the SDSS 3\arcsec\ fiber. Most of the objects have centrally-concentrated radio emission, except for NSA 625882 which clearly extended. {All but NSA 50 have radio emission strongly indicating AGN activity. However, all of the objects in our sample are still considered AGN candidates due to their strong \fex\ emission.}}
    \label{fig:radio}
\end{figure*}

Almost all AGNs, including LLAGNs produce radio emission, which is not obscured by dust attenuation \citep[see][and references therein]{Ho2008}. Additionally, bright radio emission consistent with an accreting BH was seen in both Mrk 709S \citep{Reines2014,Kimbro2021} and SDSS J122011.26+30200 \citep[ID82;][]{reines2020,Molina2021}, both of which also presented strong \fexl\ emission in their optical spectra. Therefore, we cross-referenced our sample of \fex-detected dwarf galaxies with both the Very Large Array (VLA) Faint Images of the Radio Sky at Twenty-centimeters (FIRST) survey \citep{Becker1995} and the {first epoch of the} Very Large Array Sky Survey \citep[VLASS; ][]{Lacy2020} using a maximum separation of 5\arcsec. {The single-epoch VLASS data comprises} 3~GHz observations with a spatial resolution of 2\farcs5, and a sensitivity threshold of $\approx0.12$~mJy, while the FIRST survey includes 1.4~GHz observations with a sensitivity of 0.15~mJy and a spatial resolution of 5\arcsec. Five total objects have FIRST detections, and three of those are also detected in VLASS. The FIRST and VLASS Quick Look Catalog \citep{Gordon2020} detections and integrated fluxes for our sample are given in Table~\ref{table:radio}, and the FIRST radio detections are shown in Figure~\ref{fig:radio}. We note that two objects in our sample, NSA 427201 and 50, were also in the \cite{reines2020} VLA sample as their IDs 98 and 38, respectively.

\begin{deluxetable*}{lcccccccc}
\label{table:radio}
\tablecaption{Radio Detections}
\setlength{\tabcolsep}{10pt}
\tablehead{
\colhead{NSAID} & \multicolumn{2}{c}{FIRST\tablenotemark{a}} & \multicolumn{6}{c}{VLASS\tablenotemark{b}}\vspace{-2mm}\\
\cmidrule(l){2-3} \cmidrule(l){4-9}
{} & $S_{1.4~\textrm{GHz}}$ & Offset & $S_{3~\textrm{GHz}}$ & log($L_{3~\textrm{GHz}}$) & Major Axis & Minor Axis & P.A. & Offset}
\startdata
472272 & 1.07 & 0.9 & \dots & \dots & \dots & \dots & \dots & \dots\\
50\tablenotemark{c} & 1.30 & 0.4 & \dots & \dots & \dots & \dots & \dots & \dots\\
427201\tablenotemark{c} & 1.59 & 2.1 & $0.7\pm0.2$ & $21.30$ & $3.0\pm0.6$ & $2.4\pm0.4$ & $166\pm38$ & 0.8\\
533731 & 3.14 & 0.8 & $1.7\pm0.5$ & 21.95 & $4\pm1$ & $2.9\pm0.5$ & $53\pm27$ & 0.4\\
625882 & 11.64 & 2.5 & $5.9\pm0.7$ & 21.74 & $5.8\pm0.6$ & $4.8\pm0.4$ & $174\pm24$ & 0.3\\
\enddata
\vspace{-1mm}
\tablenotetext{a}{The FIRST integrated 1.4~GHz flux densities are reported in units of mJy, and the offset is given in reference to the position of the SDSS spectroscopic fiber in units of arcseconds.}\vspace{-1mm}
\tablenotetext{b}{The VLASS Quick Look Catalog integrated 3~GHz flux densities are reported in units of mJy, while the luminosities are in W~Hz$^{-1}$. The major and minor axes of the resolved components and offset between the radio detection and the SDSS spectroscopic fiber are in arcseconds. The position angle (P.A.) is given in degrees east of north.}\vspace{-1mm}
\tablenotetext{c}{NSA 427201 and 50 were also in the \cite{reines2020} sample as IDs 98 and 38, respectively.}
\end{deluxetable*}

\subsection{Contribution from HII Regions, SNRs and SNe}\label{ssec:radio_res}

In order to test for radio AGN activity, we employed the same criteria as \cite{reines2020}. We relied on the FIRST 1.4~GHz data in our analysis as all 5 objects are detected in FIRST. {We first compared the estimated SFR from the radio source to the global SFR of the galaxy calculated via {\it GALEX}+WISE data, assuming that the radio emission is thermal.} We used equation 2 from \cite{condon1992} to calculate the production rate of Lyman continuum photons, $Q_{\textrm{Lyc}}$: 
\begin{equation}\label{eqn:qlyc}
\begin{split}
    \left(\frac{Q_{\textrm{Lyc}}}{s^{-1}}\right)\gtrsim 6.3\times10^{52}\left(\frac{T_e}{10^4~\textrm{K}}\right)^{-0.45}\left(\frac{\nu}{\textrm{GHz}}\right)^{0.1}\\
     \times\left(\frac{L_{\nu,\textrm{thermal}}}{10^{20}~\textrm{W~Hz}^{-1}}\right).
    \end{split}
\end{equation}

Here, we assume $T_e=10^4$~K and $\nu=1.4$~GHz to match the FIRST observations. We estimate $Q_{\textrm{Lyc}}\sim10^{52\mbox{--} 54}$~photons~s$^{-1}$ for all {5 radio sources}. We then estimate the instantaneous SFR for the {radio source, assuming it is purely thermal,} using equation  2 from \cite{kennicutt1998}: 
\begin{equation}\label{eqn:h2therm}
    \textrm{SFR} (M_\odot~yr^{-1}) = 1.08\times10^{-53} Q_{\textrm{Lyc}}(s^{-1}).
\end{equation}
 We find that the estimated global SFRs from Section~\ref{ssec:sprop} are lower than the estimates from the radio detections by a factor of $\sim2.5\mbox{--}15$, implying that thermal \ion{H}{2} regions {cannot fully explain the observed} radio emission.

We then compared our FIRST detections to that expected from individual supernovae or supernova remnants. In order to estimate the radio luminosity of the brightest supernova or supernova remnant in a galaxy, we use the following equation from \cite{Chomiuk2009}:
\begin{equation}\label{eqn:snei}
    L_{1.4}^{\textrm{max}}=(95^{+31}_{-23})SFR^{0.98\pm0.12}.
\end{equation}

Assuming the estimated global {{\it GALEX}+WISE} SFRs from Section~\ref{ssec:sprop}, we estimate the highest luminosity of an individual SRN/SNe to be $L_{1.4~\textrm{GHz}}\approx0.1\mbox{--}6.6\times10^{26}$~erg~s$^{-1}$, which is $\sim30\mbox{--}260$ times smaller than the measured 1.4~GHz luminosity. We therefore rule out individual SNRs/SNe as the source of the observed radio emission. 

Given the large radio emission region, if the origin is stellar it may be due to a population of SNRs/SNe. Following the procedure described in \cite{reines2020}, we calculate the total expected 1.4~GHz luminosity from all SNRs/SNe using the equation 
\begin{equation}\label{eqn:sne_pop}
    L_{\textrm{total}}=\int n(L)\,L\,dL
\end{equation}
from \cite{Chomiuk2009} where 
\begin{equation}\label{eqn:sne_popnl}
    n(L) = \frac{dN}{dL} = 92 \times SFR \times L^{-2.07}.
\end{equation}
We take the integral from 0.1 to $10^4$ in order to probe the entire range of luminosities presented in \cite{Chomiuk2009} and the SFRs from Table~\ref{table:galprop}, resulting in predicted values of $L_{1.4~\textrm{GHz}}=0.1\mbox{--}6\times10^{27}$~erg~s$^{-1}$. This is lower than the FIRST radio detections in all 5 objects by a factor of $4\mbox{--}27$. We therefore conclude that the 1.4~GHz radio emission cannot be explained solely by population of SNRs/SNe {in a majority of our objects.} 

\begin{deluxetable*}{lccccccccc}
\label{table:rad_res}
\tablecaption{Comparison of Observed Radio Emission to Estimated Values from Stellar Processes}
\setlength{\tabcolsep}{7pt}
\tablehead{
\colhead{NSAID} & {SFR$_{\rm rad}$} &{SFR$_{\rm rad}$/SFR$_{\rm FUV+IR}$} & $L^{\rm max}_{\nu}$ & $L_{\rm obs}/L^{\rm max}_{\nu}$ & $L_{\rm total}$ & $L_{\rm obs}/L_{\rm total}$ & $L_{\rm HII+SNe}$ & $L_{\rm obs}/L_{\rm HII+SNe}$ & Class.\\
(1) & (2) & (3) & (4) & (5) & (6) & (7) & (8) & (9) & 10}
\startdata
\multicolumn{10}{c}{FIRST Data (1.4~GHz)}\vspace*{1mm}\\
\hline
472272 & 17.9 & 6.1 & 2.7 & 93.7 & 25.0 & 10.2 & 6.5 & 3.9 & AGN\\
50 & 0.4 & 2.5 & 0.2 & 35.9 & 1.5 & 4.1 & 0.4 & 1.6 & SF\\
427201 & 32.2 & 6.6 & 4.5 & 101.3 & 42.0 & 10.9 & 10.9& 4.2 & AGN\\
533731 & 120.6 & 16.8 & 6.6 & 261.3 & 61.4 & 27.9 & 15.9 & 10.8 & AGN\\
625882 & 76.7 & 15.0 & 4.7 & 232.0 & 43.7 & 25.0 & 11.3 & 9.6 & AGN\\
\hline
\multicolumn{10}{c}{VLASS Data (3~GHz)}\vspace*{1mm}\\
\hline
427201 & 14.2 & 2.9 & 3.1 & 65.3 & 28.7 & 7.0 & 10.9& 2.2 & AGN\\
533731 & 65.3 & 9.1 & 4.5 & 207.2 & 41.9 & 22.1 & 15.9 & 6.8 & AGN\\
625882 & 38.9 & 7.6 & 3.2 & 172.2 & 29.8 & 18.5 & 11.3 & 5.7 & AGN\\
\enddata
\tablecomments{Column (1): NSAID. Column (2): The estimated SFR in units of $M_\odot$~yr$^{-1}$ using the observed radio emission in either FIRST or VLASS, assuming Equations~\ref{eqn:qlyc} and \ref{eqn:h2therm}. Column (3): Comparison of the observed SFR given in Table~\ref{table:galprop} to estimated SFR from Column 2. Column (4): Estimated maximum 1.4 or 3~GHz luminosity, in units of $10^{26}$~erg~s$^{-1}$~Hz$^{-1}$ for an individual SNe or SNR assuming equation~\ref{eqn:snei}. Column (5): Comparison of observed luminosity (FIRST or VLASS) to estimated luminosity from Column 4. Column (6): Total expected luminosity (at either 1.4 or 3~GHz and in units of $10^{26}$~erg~s$^{-1}$~Hz$^{-1}$) for a population of SNe/SNRs, assuming equation~\ref{eqn:sne_pop} and \ref{eqn:sne_popnl}. Column (7): Comparison of observed luminosity (FIRST or VLASS) to estimated luminosity from Column 6. Column (8): Summation of expected the luminosity (at either 1.4 or 3~GHz) in units of $10^{20}$~W~Hz$^{-1}$ from thermal \ion{H}{2} regions and populations of SNe, following Equations~\ref{eqn:qlyc}, \ref{eqn:h2therm}, \ref{eqn:sne_pop} and \ref{eqn:sne_popnl}. Column (9): Comparison of observed luminosity (FIRST or VLASS) to the estimated luminosity from Column 8. Column (10): Classification of observed radio luminosity--``AGN'' indicates that the radio emission is too luminous to be consistent with star formation, while ``SF'' indicates the radio emission is consistent with star formation.}
\end{deluxetable*}

We finally considered a combination of thermal \ion{H}{2} emission and population of SNRs/SNe. We summed the measurements from the SNR/SNe population {assuming the global {\it GALEX}+WISE SFRs as} described above with the expected thermal emission from \ion{H}{2} regions using the global SFRs given in Table~\ref{table:galprop}. The combination of these two values results in a predicted $L_{1.4~{\rm GHz}}=0.04\mbox{--}1.59\times10^{21}$~W~Hz$^{-1}$. All objects but NSA 50 have a measured 1.4~GHz luminosity that is a factor of $\gtrsim 4$ higher than the predicted value. We therefore conclude that the 1.4~GHz radio emission in all objects except for NSA 50 is likely the result of AGN activity. We also note that the high-resolution VLA data of NSA 50 studied in \citet[][their ID 38]{reines2020} was also consistent with star formation. Therefore, we cannot rule out star formation as the origin of the radio emission in this galaxy.

We performed the same analysis with the VLASS 3~GHz data, scaling the relations from luminosity densities at 1.4~GHz to 3~GHz assuming a spectral index of $\alpha=-0.5$ \citep{Chomiuk2009}. We find that the VLASS emission is also consistent with AGN activity for all three {detected} galaxies. {We present the comparison of both the VLASS and FIRST detections to predicted stellar processes in Table~\ref{table:rad_res}.}

\section{Discussion}\label{sec:discussion}

\subsection{Origin of the \fex\ Emission}\label{ssec:origin}

While we have set out to identify active massive BHs in dwarf galaxies via the detection of \fexl, here we consider whether other potential sources of coronal line emission can account for the observed \fexl\ in our sample. As a reminder, the galaxies in our sample have $L_{[Fe\textsc{x}]}\approx10^{36}$--$10^{39}$~erg~s$^{-1}$, with a median value of $\sim10^{38}$~erg~s$^{-1}$. 

\vspace{.2cm}

\noindent {\it Novae:}

{We do not consider novae to be a viable explanation for the observed \fex\ emission. Novae occur when a white dwarf accretes a sufficient amount of material from its late-type main-sequence stellar companion that a thermonuclear runaway is triggered resulting in an explosive ejection \citep{Bode2008}. The `He/N' class, or novae with prominent He and N lines, are known to create \fexl\ emission, and represent $\sim15$\% of the total nova population in the Milky Way \citep{Williams2012}, and about $\sim50$\% in the Large Magellanic Cloud \citep[LMC;][]{Shafter2013}. The five `He/N' novae with the strongest detections of \fex\ compared to the H$\alpha$+[\ion{N}{2}] emission are V3666 Oph, V723 Cas \citep{Rudy2021}, V3890 Sgr \citep{Williams1991,Williams1994} V1974 Cyg \citep{moromartin2001,vanlandigham2005} and V574 Pup \citep{Lynch2007,Walter2012}. Out of these 5, only V3666 and V3890 Sgr are known to have the maximum \fex\ brightness occur within the first year of the nova's maximum brightness \citep{Rudy2021}.}

{If we assume the H$\alpha$+[\ion{N}{2}] luminosities as a function of time from maximum brightness presented in Figure 3 of \cite{Tappert2020}, then $\sim 1,000$ and $\sim 10,000$ `He/N' novae are needed to explain an $L_{\rm [FeX]}\approx 10^{36}$~erg~s$^{-1}$, which corresponds to the minimum \fex\ luminosity in our sample. Even if we adopt the Milky Way nova rate of $\sim25$~yr$^{-1}$, which is much larger than that seen in dwarf galaxies \citep[e.g.,][]{Shafter2014} and assume that the `He/N' nova are 50\% of the total nova population, we would only expect $\sim13$ `He/N' novae within 1 year. This number is significantly smaller than the $\sim1,000$ needed to create the minimum observed \fex\ emission in our sample. We therefore conclude that novae are not responsible for the \fex\ emission observed in our dwarf galaxies.
}

\vspace{.2cm}

\noindent {\it {Wolf-Rayet Stars, Planetary Nebulae and HMXRBs:}} 

We also do not consider Wolf-Rayet stars, planetary nebulae or {high-mass XRBs} to be viable explanations for the observed \fexl\ emission in our sample. {While both Wolf-Rayet stars and planetary nebulae can produce coronal-line emission, they do not present \fexl\ emission \citep[][]{schaerer1999,pottasch2009,Zhang2012}, and no \fex\ emission has been reported in Wolf-Rayet galaxies \citep[e.g.,][]{Karthick2014,Fernandes2004}. Meanwhile, the Be star-neutron star X-ray binary, which is the largest sub-group of high-mass XRBs, does not produce any observable optical coronal emission lines \citep[e.g.,][]{Coe2021}.} 

\vspace{.2cm}

\noindent {\it {\ion{He}{2} Star-Forming Galaxies and Starburst Winds:}}

{Star-forming galaxies with \ion{He}{2} emission can produce optical [\ion{Fe}{3}] lines \citep[e.g.,][]{Kehrig2018}, but previous studies do not report any observed \fexl\ emission \citep{Shirazi2012,Kehrig2018}. Starburst-driven winds and superbubbles are also not a likely explanation of the \fex\ emission. While these strong outflows can produce strong UV [\ion{O}{6}]$\lambda\lambda$1032,1038 coronal emission lines \citep[e.g.,][]{Heckman2001}, most of the optical Fe emission will come from supernovae and not the out-flowing wind itself \citep{Silich2001,Danehkar2021}. Specifically in younger stellar populations ($t_{\rm age}<20$--30~Myr), a majority of the Fe emission comes from type II supernovae, which we address below.}

\vspace{.2cm}

\noindent {\it Type IIn Supernovae:}

We consider supernovae (SNe) as a possible origin for the observed \fex\ emission, however this does not appear to be a plausible explanation for the majority of our \fex-emitting dwarf galaxies. Few SNe create detectable \fexl\ emission, and the \fexl\ emission will fade on timescales of a few years post-explosion and have typical luminosities $L_{\rm [FeX]}< 10^{33}$~erg~s$^{-1}$ \citep{Dopita1997,Komossa2008,Komossa2009,smith2009}. Therefore, thousands of typical SNe would be needed to explain the observed \fex\ emission in our galaxies. This scenario seems unlikely, especially given the lack of an observed correlation between the \fex\ EW and mass-specific SFR (Figure~\ref{fig:ewfex}). {Additionally, we see no clear trend between the \fexl\ luminosity and the mass-specific SFR, $g-i$ or $g-r$ colors, which we might expect if the emission is predominately driven by supernovae.}

One prominent exception to this rule is SN2005ip, which is a Type IIn SNe with a rich and luminous coronal-line forest rarely seen in SNe, even among other Type IIn SNe, and is more reminiscent of AGN spectra \citep{smith2009,Smith2017}. After studying SN2005ip for the first $\sim3$ years after its discovery, \cite{smith2009} conclude that the luminous coronal-line emission is {due to SN2005ip's SNe wind properties. The wind density was just low enough to create the Type IIn spectrum, but not so dense that the interaction with the circumstellar medium became optically thick. The clumpy nature of the wind provides dense material, through which intense shock waves and X-rays can escape to excite the surrounding material. Both of these physical effects are thought to drive the incredibly strong coronal-line emission seen in SNe 2005ip.} 

We compared the observed \fexl\ emission in our sample to the peak \fexl\ luminosity of SN2005ip, $\sim2\times10^{37}$~erg~s$^{-1}$, that occurred within 100 days after its discovery \citep{smith2009}. A majority of the galaxies in our sample have \fexl\ luminosities above $\sim10^{38}$~erg~s$^{-1}$, well above that expected from SNe like SN2005ip, as shown in Figure~\ref{fig:fex_hist}. We list the \fex\ luminosities of the galaxies in our sample and its comparison to that of SN2005ip in Table~\ref{table:fex}. Indeed, only 9/81 objects have \fex\ luminosities less than twice that seen in SN2005ip; the remaining 72 galaxies would require $>2-160$~Type IIn supernovae like SN2005ip that had exploded within a 100-day span before the SDSS spectroscopic observation to explain the observed emission.

In order to determine if large populations of Type IIn SNe could create the observed \fex\ emission in our galaxies, we estimated the expected number Type IIn SNe in our parent sample. We first estimated the Type II supernova rate using the relation between the mass-weighted Type II supernova rate ($R_{II,M}$) and the mass-specific SFR via the information in Figure 7 from \cite{Graur2015} and the masses and mass-specific SFRs from our \fexl\ sample. Our galaxies have a median Type II supernova rate of $\sim0.002$~yr$^{-1}$, and span a range of $\sim0.0002\mbox{--}0.008$~yr$^{-1}$. Type IIn SNe are rare, and only account for $\sim9$\%  of the entire Type II SNe population \citep{Smith2014}. We therefore adopt the median Type II SNe rate (0.002~yr$^{-1}$) and weight it by 0.09 to estimate the Type IIn SNe rate (0.00018~yr$^{-1}$). We then multiply this rate by the number of galaxies in our parent sample (46,530), and conclude that we should expect a rate of $\sim8.3$ Type IIn SNe per year in our parent sample.

An accurate conversion from this SNe rate to the total expected number of SNe in our parent sample is time-dependent and not straightforward to calculate \citep[see Appendix A in][]{Leaman2011}. However, as the \fex\ luminosity remained relatively high within the first year for the Type IIn SNe 2005ip \citep{smith2009}, as long as any comparable SNe events occurred before the spectral observations of our galaxies, we should detect them in the SDSS spectra. Therefore, we expect at most $\sim$8 Type IIn SNe in our parent sample of 46,530 galaxies. Since SNe similar to SN2005ip are rare, even among Type IIn SNe \citep{smith2009,Smith2017}, we conclude that this estimate of the number of Type IIn SNe in our parent sample is likely highly overestimated.  We also note that out of the $\sim740,000$ SDSS galaxies studied in \cite{Graur2015}, only 16 Type II SNe were detected. Given that only $\sim$9\% of Type II SNe are Type IIn, we expect that the \cite{Graur2015} sample should contain only 1--2 Type IIn SNe.  

We also note that the typical spectral features expected within the first $\sim3$ years post explosion in Type IIn supernovae, including a forest of Fe coronal lines, and broad \ion{He}{1} \citep[on the order of $\sim1000$~km~s$^{-1}$;][]{smith2009} were not observed in any of our SDSS spectra. Furthermore, none of our objects show the characteristic 3-component H$\alpha$ emission expected in the first few years post-explosion in Type IIn SNe \citep[][]{stathakis1991,turatto1993,smith2009}. The two most well-known examples of Type IIn SNe show nearly identical H$\alpha$ profiles, with a broad component of width $\pm16,000$~km~s$^{-1}$, a second, intermediate-width component at $\pm11,000$~km~s$^{-1}$ and a narrow-line component at $\pm100$--$200$~km~s$^{-1}$. This suggests that if we did catch a type IIn supernova, it was likely not observed in the first $\sim3$--5 years, which is the only time that the \fexl\ could be as strong as that seen in AGN activity \citep{smith2009,Smith2017}. 

Given the high \fex\ line luminosities, the lack of other Type IIn spectral features and the low number of expected Type IIn SNe in our sample, we conclude that SNe are not responsible for the \fex\ emission in the vast majority of our dwarf galaxies.  The origin of the \fexl\ emission is most likely related to accretion onto massive BHs.  Both AGNs and TDEs are plausible explanations for the \fex\ emission. 

\vspace{.2cm}

\noindent {\it Tidal Disruption Events:}

{A tidal disruption event, or TDE, occurs when a star gets sufficiently close to a massive BH such that it gets ripped apart by tidal forces. A certain class of TDEs called extreme coronal line emitters (ECLEs) are known to produce coronal-line emission with $L_{\rm [FeX]} \sim 10^{38-40}$~erg~s$^{-1}$ \citep{Komossa2008,Wang2011,Wang2012}, consistent with the \fex\ luminosities in our sample. 
Additionally, the BHs associated with ECLEs have typical masses of $M_{\rm BH}\sim10^{5\mbox{--}6}$~M$_\odot$, which are also consistent with the BH masses expected for our sample of dwarf galaxies. We note, however, that none of our objects overlap with the known ECLEs \citep{Komossa2008,Wang2011,Wang2012}, or with the open TDE catalog\footnote{\url{https://tde.space/}}.}

A majority of known ECLEs exhibit strong coronal-line Fe forests, including \fexl\ that is comparable in strength to [\ion{O}{3}]$\lambda$5007 \citep{Wang2012}. In contrast, typical coronal-line strengths in Seyfert galaxies are only a few percent of [\ion{O}{3}] \citep{Murayama1998,Nagao2000}. ECLE coronal-line emission is transient, and fades on a timescale of 4--5 years \citep{Yang2013}. In addition to the coronal-line emission, 5/7 ECLEs studied in \cite{Wang2012} have complex, blue- and red-shifted broad-line \ion{He}{2} and H$\alpha$ emission lines, and are mainly found in galaxies with large, low surface-brightness disks \citep[see Figure 9 in][]{Wang2012}. ECLEs also tend to fall in the star-forming and/or composite regions of the [\ion{N}{2}]/H$\alpha$ and [\ion{S}{2}]/H$\alpha$ diagrams, and the Seyfert/LINER regions of the [\ion{O}{1}]/H$\alpha$ diagram, which is similar to the objects in our sample.

\begin{figure}
    \centering
    \includegraphics[width=0.48\textwidth]{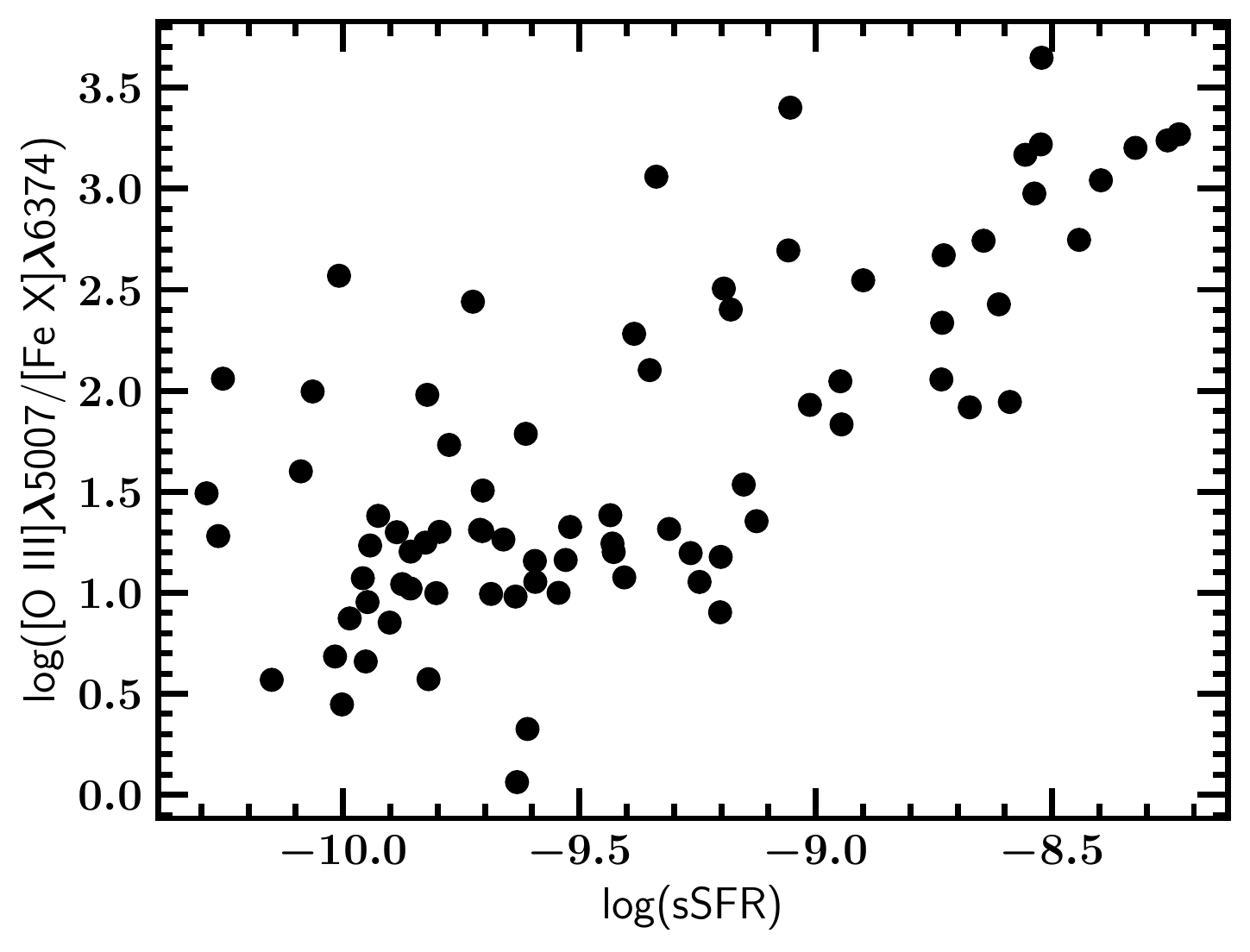}
    \caption{The reddening-corrected [\ion{O}{3}]$\lambda$5007/\fexl\ line ratio vs.~the sSFR for the \fexl-emitting galaxies. The ratio increases with sSFR, which could indicate the star formation is hiding the signature TDE-like line ratio.}
    \label{fig:o3fex}
\end{figure}

Most of our objects do not have the characteristic [\ion{O}{3}]/\fex\ ratios seen in ECLEs. However, the high levels of star formation in our objects compared to those in \cite{Wang2012} could be diluting the TDE signal in a similar fashion to traditional optical AGN indicators \citep[e.g.,][]{Moran2002,Cann2019}. In fact, we note that the [\ion{O}{3}]/\fex\ ratio is directly correlated to the mass-specific SFR as shown in Figure~\ref{fig:o3fex}. Therefore, some of these \fex-emitting objects could be ECLEs that are diluted by star formation and thus are only identified via their strong \fex\ emission.

The galaxy NSA 427201 is a particularly strong TDE candidate. This object was observed to have star-forming like narrow-line ratios in the [\ion{N}{2}]/H$\alpha$, [\ion{S}{2}]/H$\alpha$ and [\ion{O}{1}]/H$\alpha$ diagrams, with broad-line H$\alpha$ emission in \cite{reines2013}. Optical spectroscopic follow-up was performed by \cite{Baldassare2016}, who found no broad H$\alpha$ emission, and concluded that the observed emission was likely due to transient activity. In this work, we find that the observed \fexl\ emission is $19$ times higher than that seen in SN2005ip, which is similar to the ECLE studied in \cite{Wang2011}. Additionally, the broad H$\alpha$ emission is observed to fade by at least a factor of 10, if not completely disappear $\sim8$ years after a TDE \citep{Yang2013}. This is consistent with the non-detection of broad emission in the optical follow-up of \cite{Baldassare2016} that occurred 11 years after the SDSS spectroscopic observation. While the [\ion{O}{3}]/\fex\ ratio is not of order unity, as seen in most ECLEs, the sSFR of NSA 427201 is one of the highest in the sample at -8.4 and it has a ${\rm SFR}=4.92$~M$_\odot$~yr$^{-1}$. Therefore, it is very possible that the extreme level of star formation in this dwarf galaxy could have diluted the other traditional TDE indicators. 

\vspace{.2cm}

\noindent {\it AGNs:}

Finally, we consider the highly energetic processes associated with AGN activity as the most likely origin of the observed \fex\ emission in a majority of our objects. Coronal-line detection is often considered a reliable signature of AGN activity \citep[e.g.,][]{penston1984,prieto2000,prieto2002,reunanen2003,satyapal2008,goulding2009,cerquiera2021}, and is typically associated with either gas photoionized by the hard AGN continuum \citep[e.g.,][]{Nussbaumer1970,Korista1989,Oliva1994,Pier1995,Negus2021}, or radiative shock waves driven into the host galaxy by radio jets from the AGN \citep[e.g.,][]{wilson1999,Molina2021}. 

Coronal-line emission is frequently seen, but not necessarily expected in AGN spectra. Out of the 47 quasar and Seyfert-like AGN in massive galaxies studied by \cite{riffel2006}, 67\% of the objects had only one NIR coronal-line detected, and the remaining 33\% exhibited no coronal-line emission. These results were later confirmed with a much larger sample by \cite{Lamperti2017}. While optical Fe coronal-line emission tends to be observed more frequently \citep{cerquiera2021}, they are much weaker, usually 1--10\% of that seen in [\ion{O}{3}]$\lambda$5007 \citep{Murayama1998,Nagao2000}. {Additionally, coronal-line emission is often blue-shifted relative to the narrow lines when produced via photoionization \citep[e.g.,][]{Gelbord2009}. Approximately 50\% of our sample show such blue-shifted \fex\ emission, which could indicate that the \fex\ emission is the result of photoionization and not radiative, out-flowing winds.}

Based on the high \fex\ luminosities and additional indicators of AGN activity, we assume that a majority of \fex-emitting galaxies in this work host massive BHs. Assuming the M$_{\rm BH}$--M$_{\rm tot}$ relationship of \cite{reinesvolonteri2015} and the stellar mass values from the NSA, we expect the BH candidate masses to be within the range 10$^{4\mbox{--}6}$~M$_\odot$, with a median value of $10^{5}$~M$_\odot$. Given this result, we consider the structure of the central engine and its effect on the AGN indicators employed in this work below.

\subsection{Biases in AGN Selection Methods and the Effect on the \fex-selected Sample of Dwarf Galaxies}

The AGN selection techniques explored in this work focus on distinct regions of the electromagnetic spectrum that probe emission from different parts of an AGN. Most of these techniques assume that the accretion disk is well-explained by the radiatively efficient, geometrically thin, optically thick model first presented by \cite{Shakura1976}. In this model, the disk will produce {optical and UV} photons that will inverse Compton scatter away from the disk to produce hard X-ray emission in the hot, tenuous coronal region \citep{Liang1984,Haardt1991} which photoionizes the observed \fexl\ and other coronal-line emission \citep[e.g.,][]{Nussbaumer1970,Korista1989,Oliva1994,Pier1995}. Gas within the broad line region (BLR; 10--100 light days in Seyfert 1 galaxies) will have high Doppler motions, creating broad-line emission. Meanwhile the narrow line region (NLR) is produced by gas farther away from the central BH (100--300~pc) creating narrower emission-line widths \citep{Netzer2013}. The emission in the narrow-line region will also be photoionized by the central engine, creating the characteristic AGN-like emission seen in the VO87 (BPT-style), NIR and \ion{He}{2}/H$\beta$ diagrams \citep{VO87,Baldwin1981,kewley2006,osterbrock1992,Sartori2015}. Additionally, AGN jets produce synchrotron emission in the radio that is not affected by dust attenuation and emitted at some level by almost all AGN \citep{Ho2008}. Finally, almost all emission from AGNs is observed to vary on characteristic timescales related to the processes from which the emission is created \citep[e.g.,][]{Ulrich1997}. 

Conversely, radiatively inefficient accretion flows \citep[RIAFs; ][]{Narayan1995,Blandford1999} have very different accretion-disk structures. The accretion rate in RIAFs is so low ($L_{\rm bol}/L_{\rm Edd}\lesssim10^{-2}$) that at a certain transition radius, $R_t$, the collisional cooling timescale equals the accretion timescale. Outside of $R_t$ the accretion disk is geometrically thin and optical thick. However inside $R_t$ the disk becomes geometrically thick and optical thin. In this regime, the electrons are cooled by bremsstrahlung, synchrotron and Compton up-scattering while the ions remain at the virial temperature. This change in accretion-disk structure could remove some of the more traditional AGN-like narrow-line ratios and broad H$\alpha$ emission \citep{Ho2008,Trump2011} and are predicted to have strong radio outflows \citep{Meier2001,Fender2000}. These radio outflows {can create radiative winds that} collisionally excite the observed \fexl\ emission seen in RIAFs \citep[e.g.,][]{wilson1999}. This would result in a predominantly wind-driven energy budget, which is expected as the accretion rate decreases \citep{Melendez2008,diamondstanic2009}.

The difference between these two accretion flows typically seen in AGNs will affect the viability of the various AGN detection methods explored here. For example, the optical spectroscopic tests such as the narrow-line diagnostic diagrams, \ion{He}{2}/H$\beta$ line ratio and broad H$\alpha$ emission can easily miss AGNs with RIAFs \citep{Ho2008,Trump2011} or radiatively efficient AGNs that reside within host galaxies with even modest star formation rates \cite[e.g.,][]{Moran2002}. While X-ray and radio detections are relatively cleaner sources of AGN activity, the signal could be comparable to that of the ambient star formation in the host galaxy which can also effectively hide the AGN signal \citep[e.g.,][]{reines2020,Latimer2021}. Finally, AGN variability is rarely seen in AGNs with RIAFs \citep{Trump2009}, which could lead to a systematic bias towards radiatively efficient central engines. These effects could explain why many of our \fex-selected AGN candidates do not show other traditional AGN indicators.

The two most recent discoveries of AGN-related \fexl\ emission in dwarf galaxies \citep[e.g.,][]{Molina2021,Kimbro2021} used multi-wavelength data to identify a RIAF as the most likely BH accretion structure. We therefore anticipate that some fraction of our objects are also likely AGNs with RIAFs, which are biased against in most of the AGN selection techniques used in this work. Given RIAFs are predicted to have strong radio outflows, searching for radio emission is clearly the preferred AGN detection technique. While only a few objects in our sample have strong radio detections in FIRST, many AGNs even in the low-redshift universe, do not produce enough radio continuum to be detected by the FIRST survey \citep{Condon2019}. We also included data from {the first epoch of} VLASS, which has significantly better angular resolution but a sensitivity threshold similar to that of FIRST (0.12~mJy vs.~0.15~mJy). {Thus dedicated radio and X-ray follow-up could help determine if weakly accreting BHs with RIAFs are present in these objects.}

Meanwhile, some of the objects in this sample, such as NSA 256802, clearly harbor radiatively efficient AGNs. Therefore the corona is likely photoionizing the observed \fexl\ emission in these galaxies. We do note that NSA 256802 is particularly unique in this sample as it is the only object that is AGN-like using the BPT-like narrow-line diagrams. However, optical AGN signatures can easily be hidden by a modest amount of star formation in the host galaxy \citep{Moran2002}. The galaxies in our sample are very blue and have substantial mass-specific SFRs, indicating a high level of star formation which can easily dilute the more traditional narrow-line diagnostic indicators \citep{Cann2019,Trump2015}. Therefore, it is also possible that a significant fraction of these galaxies harbor thin-disk AGNs. Spectroscopy with the {\it James Webb Space Telescope} ({\it JWST}) would be useful for identifying near-IR coronal lines that can be used to identify AGN photoionization in radiatively efficient BHs \citep[e.g.,][]{Satyapal2015,Cann2018}.

\subsection{Comparison to Optical and Radio-Selected Samples of AGNs in Dwarf Galaxies}

We first compare our objects to the AGN and composite samples from \citet{reines2013} that were identified using the [\ion{O}{3}]/H$\beta$ vs.~[\ion{N}{2}]/H$\alpha$ diagram. As the \cite{reines2013} objects are luminous enough to be detected in the 3\arcsec\ SDSS aperture, and exhibit narrow-line ratios consistent with AGN activity, they are likely BHs with radiatively efficient flows.  Indeed, the broad-line AGNs and composites in \citet{reines2013} have {Eddington ratios spanning $\sim0.1$--50\%, comparable to high-redshift, massive broad-line quasars \citep{Baldassare2017}.} Therefore, they represent the most energetic, most luminous end of the dwarf galaxy AGN population. 

In the work presented here, we used the updated NSA v1\_0\_1 catalog, which has a 100\% overlap with the v0\_1\_2 catalog used by \citet{reines2013}. {Interestingly, while 136 objects were detected as either AGN or composite galaxies by Reines et al., only two of those objects are included in our \fex\ sample. When we compare the properties of our sample to the \cite{reines2013} AGN and composite samples as shown in Figure~\ref{fig:agn_comp}, we find that a majority our sample covers the lower-mass, bluer region of the $g-r$ vs.~$M_*$ diagram.} In fact, K--S two sample tests comparing the $g-r$ colors and mass between the \fexl-selected AGN candidate sample and the AGN and composite samples from \citet{reines2013} show that they are statistically distinct at the 99\% confidence level. 

The statistical difference seen in the \citet{reines2013} AGNs/composites and the \fexl-selected AGNs presented here could be a result of various phenomena including: (1) AGNs with \fexl\ emission in dwarf galaxies are biased towards a RIAF-like engines, (2) the \fexl\ emission line is significantly harder to produce via stellar emission than [\ion{N}{2}], allowing us to detect AGNs in bluer galaxies with more star formation, and (3) a combination of (1) and (2). 

We also compared our sample to the radio-selected AGN candidates in \cite{reines2020}. Similar to our \fex-selected sample, many of the \cite{reines2020} objects did not have obvious AGN-like emission in their SDSS spectra, even when the radio source overlapped with the SDSS fiber (although many had enhanced [\ion{O}{1}]/H$\alpha$). {We find our sample has a similar distribution in color and mass to the \cite{reines2020} AGN candidates, as seen in Figure~\ref{fig:agn_comp}.} 

At first glance, once could conclude that the similarity between the two samples may indicate the presence of radio activity in the \fexl-selected sample. However, it may just be that both of these samples access part of the dwarf galaxy population that is not accessible using traditional optical diagnostics with SDSS spectroscopy {as desmonstrated by Figure~\ref{fig:agn_comp}.} Thus \fex\ surveys provide a pathway for finding BHs in a dwarf-galaxy population that is notoriously difficult to explore with other detection methods.

\section{Summary and Conclusions}\label{sec:summary}
In this work, we present the first systematic search for \fexl-selected massive BHs in dwarf galaxies.  Our main results are as follows: 

\begin{enumerate}
    \item We identified \fexl\ emission in the SDSS single-fiber spectroscopy of 81 dwarf galaxies.  The \fexl\ luminosities are in the range {$10^{36\mbox{--}39}$~erg~s$^{-1}$, with a median of $1.6\times10^{38}$~erg~s$^{-1}$}, and indicate massive BHs are likely present in the majority of galaxies in our sample. AGNs and/or TDEs are the most likely origin of the \fexl\ emission.
    
    \item The \fexl\ luminosities are sufficiently high to exclude stellar origins of the emission (e.g., Wolf-Rayet stars, planetary nebula, supernovae) in most of our sample galaxies. There are a couple of known luminous Type IIn SNe with $L_{[\rm FeX]} \lesssim 2\times10^{37}$~erg~s$^{-1}$; however, a majority of our objects have \fex\ luminosities well above that value. Furthermore, we expect at most 8 Type IIn SNe in our sample, indicating that the \fex\ emission in most of these objects is driven by massive BH activity.
    
     \item \fexl\ emission can be produced by AGN photoionization or shocks from AGN jets. Given the results of previous studies of BHs in dwarf galaxies with strong \fexl\ emission \citep{Molina2021,Kimbro2021}, we expect some of the AGNs in this work will also be weakly-accreting BHs, potentially with RIAF structures. Follow-up observations, particularly at X-ray and radio wavelengths, would help us assess the nature of the accretion flows around the BHs in these dwarf galaxies.
    
    \item We searched for additional AGN indicators in our sample of \fex-detected dwarf galaxies using optical spectroscopy and optical variability, as well as existing X-ray and radio observations. While 51\% of the objects have multiple AGN signatures, many galaxies in our sample are only selected as active BHs via their high \fexl\ luminosity.

    \item The objects in our sample also have \fex\ luminosities and estimated BH masses consistent with that seen in some TDEs. While our galaxies do not show other traditional TDE indicators (i.e., [\ion{O}{3}$]/[\ion{Fe}{10}]\sim1$), it is possible that these signatures are being diluted by high levels of star formation. One galaxy in our sample, NSA 427201, is a particularly strong TDE candidate, demonstrating that \fex\ could potentially trace a population of TDEs in star-forming dwarfs.
    
    \item Overall, the galaxies in our sample exhibit bluer colors and lower stellar masses than {the traditional BPT-selected AGN \citep{reines2013} in dwarf galaxies}, indicating that \fexl\ searches can identify BH activity in galaxies with higher levels of ongoing star formation.

\end{enumerate}
 
 We conclude that \fexl\ is a powerful new tool to search for the elusive low-mass AGN, and potentially TDE, populations in dwarf galaxies. Furthermore, \fex\ searches can be used to construct samples of candidate NIR coronal-line emitting galaxies to observe with {\it JWST} \citep[e.g.,][]{Cann2018}. Similar to radio searches, \fexl\ searches can detect BH activity in lower-mass galaxies with significant star formation that are difficult to find in traditional optical surveys. We also demonstrate that this work can be completed using single-epoch optical surveys, which allows for efficient searches for BH activity in dwarf galaxies. Therefore, BHs detected via \fexl\ emission can help constrain the low-mass end of the BH population in dwarf galaxies and shed light on the origin of BH seeds. \citep[e.g.,][]{Reines2016, Greene2020}.


\acknowledgements
We thank the anonymous referee for their helpful comments
that improved this work. AER acknowledges support for this work provided by Montana State University and NASA through EPSCoR grant number 80NSSC20M0231.

This research uses services or data provided by the Astro Data Lab at NSF's National Optical-Infrared Astronomy Research Laboratory. NOIRLab is operated by the Association of Universities for Research in Astronomy (AURA), Inc. under a cooperative agreement with the National Science Foundation.

Funding for the Sloan Digital Sky 
Survey IV has been provided by the 
Alfred P. Sloan Foundation, the U.S. 
Department of Energy Office of 
Science, and the Participating 
Institutions. 

SDSS-IV acknowledges support and 
resources from the Center for High 
Performance Computing  at the 
University of Utah. The SDSS 
website is www.sdss.org.

SDSS-IV is managed by the 
Astrophysical Research Consortium 
for the Participating Institutions 
of the SDSS Collaboration including 
the Brazilian Participation Group, 
the Carnegie Institution for Science, 
Carnegie Mellon University, Center for 
Astrophysics | Harvard \& 
Smithsonian, the Chilean Participation 
Group, the French Participation Group, 
Instituto de Astrof\'isica de 
Canarias, The Johns Hopkins 
University, Kavli Institute for the 
Physics and Mathematics of the 
Universe (IPMU) / University of 
Tokyo, the Korean Participation Group, 
Lawrence Berkeley National Laboratory, 
Leibniz Institut f\"ur Astrophysik 
Potsdam (AIP),  Max-Planck-Institut 
f\"ur Astronomie (MPIA Heidelberg), 
Max-Planck-Institut f\"ur 
Astrophysik (MPA Garching), 
Max-Planck-Institut f\"ur 
Extraterrestrische Physik (MPE), 
National Astronomical Observatories of 
China, New Mexico State University, 
New York University, University of 
Notre Dame, Observat\'ario 
Nacional / MCTI, The Ohio State 
University, Pennsylvania State 
University, Shanghai 
Astronomical Observatory, United 
Kingdom Participation Group, 
Universidad Nacional Aut\'onoma 
de M\'exico, University of Arizona, 
University of Colorado Boulder, 
University of Oxford, University of 
Portsmouth, University of Utah, 
University of Virginia, University 
of Washington, University of 
Wisconsin, Vanderbilt University, 
and Yale University.

The Legacy Surveys consist of three individual and complementary projects: the Dark Energy Camera Legacy Survey (DECaLS; Proposal ID 2014B-0404; PIs: David Schlegel and Arjun Dey), the Beijing-Arizona Sky Survey (BASS; NOAO Prop. ID 2015A-0801; PIs: Zhou Xu and Xiaohui Fan), and the Mayall z-band Legacy Survey (MzLS; Prop. ID 2016A-0453; PI: Arjun Dey). DECaLS, BASS and MzLS together include data obtained, respectively, at the Blanco telescope, Cerro Tololo Inter-American Observatory, NSF’s NOIRLab; the Bok telescope, Steward Observatory, University of Arizona; and the Mayall telescope, Kitt Peak National Observatory, NOIRLab. The Legacy Surveys project is honored to be permitted to conduct astronomical research on Iolkam Du’ag (Kitt Peak), a mountain with particular significance to the Tohono O’odham Nation.

NOIRLab is operated by the Association of Universities for Research in Astronomy (AURA) under a cooperative agreement with the National Science Foundation.

This project used data obtained with the Dark Energy Camera (DECam), which was constructed by the Dark Energy Survey (DES) collaboration. Funding for the DES Projects has been provided by the U.S. Department of Energy, the U.S. National Science Foundation, the Ministry of Science and Education of Spain, the Science and Technology Facilities Council of the United Kingdom, the Higher Education Funding Council for England, the National Center for Supercomputing Applications at the University of Illinois at Urbana-Champaign, the Kavli Institute of Cosmological Physics at the University of Chicago, Center for Cosmology and Astro-Particle Physics at the Ohio State University, the Mitchell Institute for Fundamental Physics and Astronomy at Texas A\&M University, Financiadora de Estudos e Projetos, Fundacao Carlos Chagas Filho de Amparo, Financiadora de Estudos e Projetos, Fundacao Carlos Chagas Filho de Amparo a Pesquisa do Estado do Rio de Janeiro, Conselho Nacional de Desenvolvimento Cientifico e Tecnologico and the Ministerio da Ciencia, Tecnologia e Inovacao, the Deutsche Forschungsgemeinschaft and the Collaborating Institutions in the Dark Energy Survey. The Collaborating Institutions are Argonne National Laboratory, the University of California at Santa Cruz, the University of Cambridge, Centro de Investigaciones Energeticas, Medioambientales y Tecnologicas-Madrid, the University of Chicago, University College London, the DES-Brazil Consortium, the University of Edinburgh, the Eidgenossische Technische Hochschule (ETH) Zurich, Fermi National Accelerator Laboratory, the University of Illinois at Urbana-Champaign, the Institut de Ciencies de l’Espai (IEEC/CSIC), the Institut de Fisica d’Altes Energies, Lawrence Berkeley National Laboratory, the Ludwig Maximilians Universitat Munchen and the associated Excellence Cluster Universe, the University of Michigan, NSF’s NOIRLab, the University of Nottingham, the Ohio State University, the University of Pennsylvania, the University of Portsmouth, SLAC National Accelerator Laboratory, Stanford University, the University of Sussex, and Texas A\&M University.

BASS is a key project of the Telescope Access Program (TAP), which has been funded by the National Astronomical Observatories of China, the Chinese Academy of Sciences (the Strategic Priority Research Program “The Emergence of Cosmological Structures” Grant No. XDB09000000), and the Special Fund for Astronomy from the Ministry of Finance. The BASS is also supported by the External Cooperation Program of Chinese Academy of Sciences (Grant No. 114A11KYSB20160057), and Chinese National Natural Science Foundation (Grant No. 11433005).

The Legacy Survey team makes use of data products from the Near-Earth Object Wide-field Infrared Survey Explorer (NEOWISE), which is a project of the Jet Propulsion Laboratory/California Institute of Technology. NEOWISE is funded by the National Aeronautics and Space Administration.

The Legacy Surveys imaging of the DESI footprint is supported by the Director, Office of Science, Office of High Energy Physics of the U.S. Department of Energy under Contract No. DE-AC02-05CH1123, by the National Energy Research Scientific Computing Center, a DOE Office of Science User Facility under the same contract; and by the U.S. National Science Foundation, Division of Astronomical Sciences under Contract No. AST-0950945 to NOAO.

This research has made use of data obtained from the Chandra Data Archive and the Chandra Source Catalog, and software provided by the Chandra X-ray Center (CXC) in the application packages CIAO and Sherpa.

 This research made use of Astropy,\footnote{http://www.astropy.org} a community-developed core Python package for Astronomy \citep{astropy2013, astropy2018}.

\software{
Astropy \citep{astropy2013,astropy2018},
\texttt{CIAO} v4.13 \citep{fruscione06},
Matplotlib \citep{matplotlib},
pyspeckit \citep{pyspeckit}}


\appendix

\section{Galaxy Images and \fexl\ Spectra}\label{app:prop}

\begin{figure*}[h]
    \centering
    \includegraphics[width=0.13\textwidth]{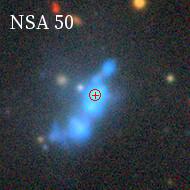}
    \hspace{-3mm}
    \includegraphics[width=0.19\textwidth]{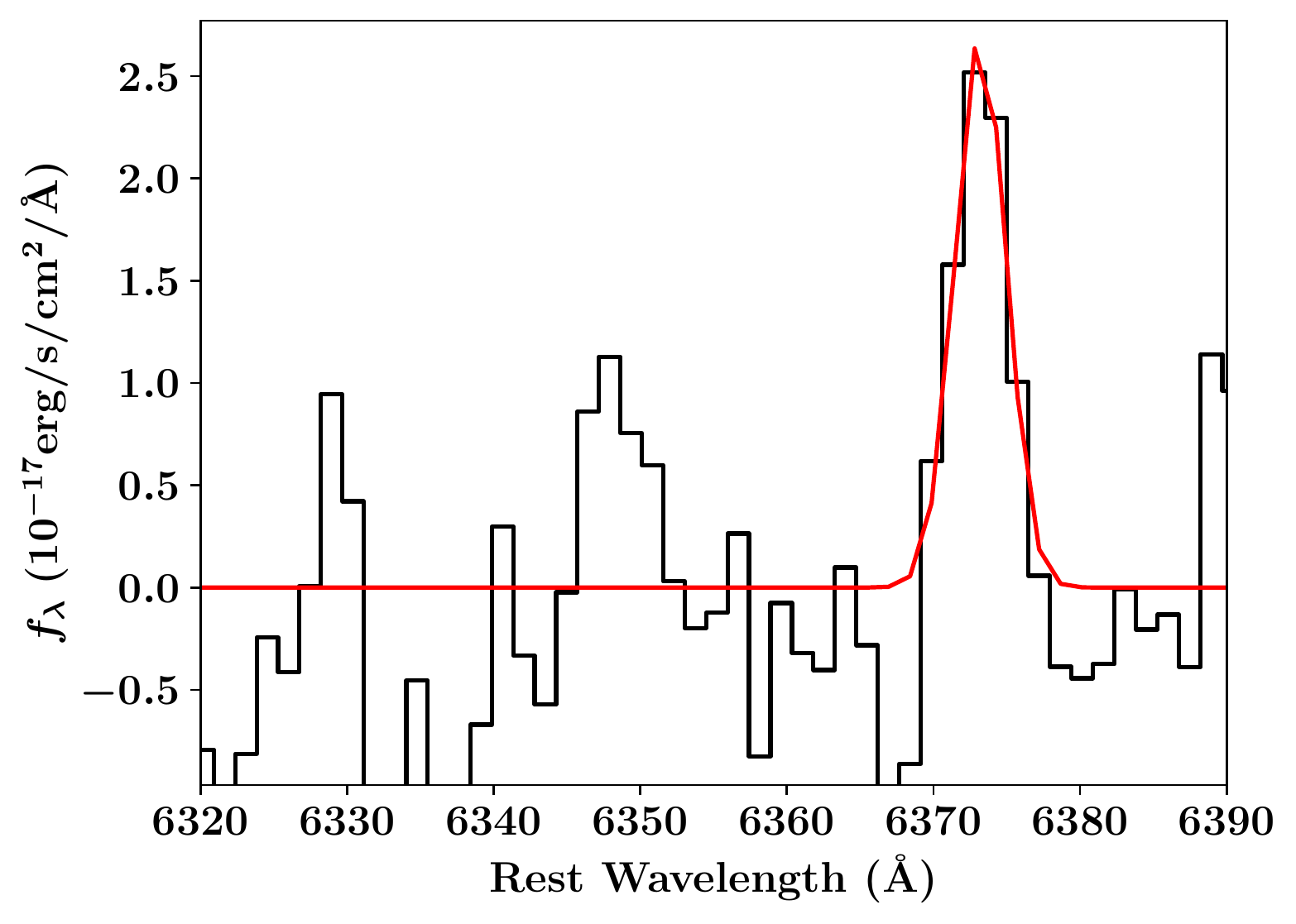}
    \hspace{1.5mm}
    \includegraphics[width=0.13\textwidth]{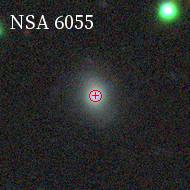}
        \hspace{-3mm}
    \includegraphics[width=0.19\textwidth]{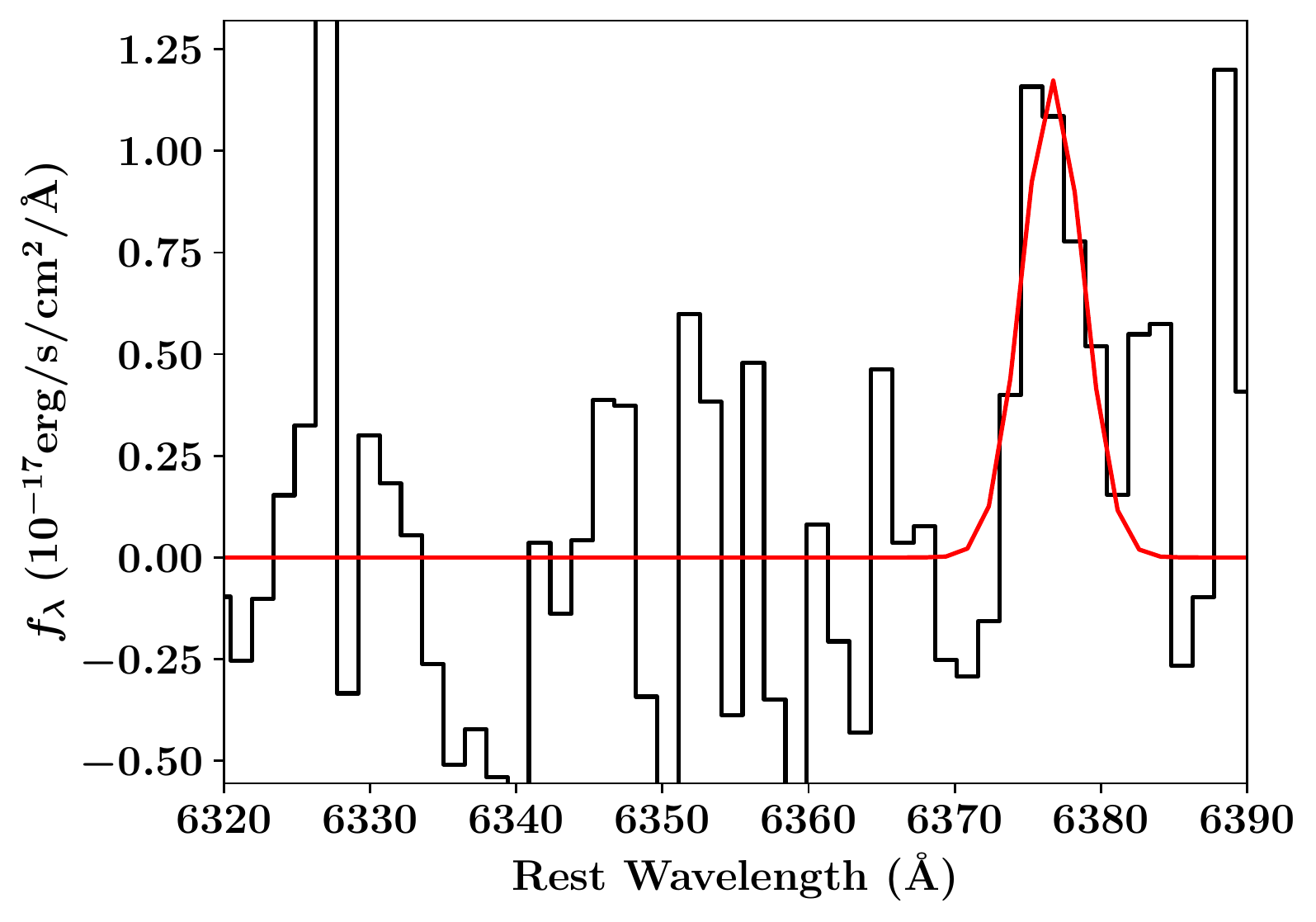}  
        \hspace{1.5mm}
    \includegraphics[width=0.13\textwidth]{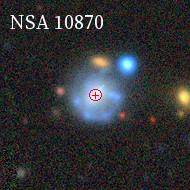}
        \hspace{-3mm}
    \includegraphics[width=0.19\textwidth]{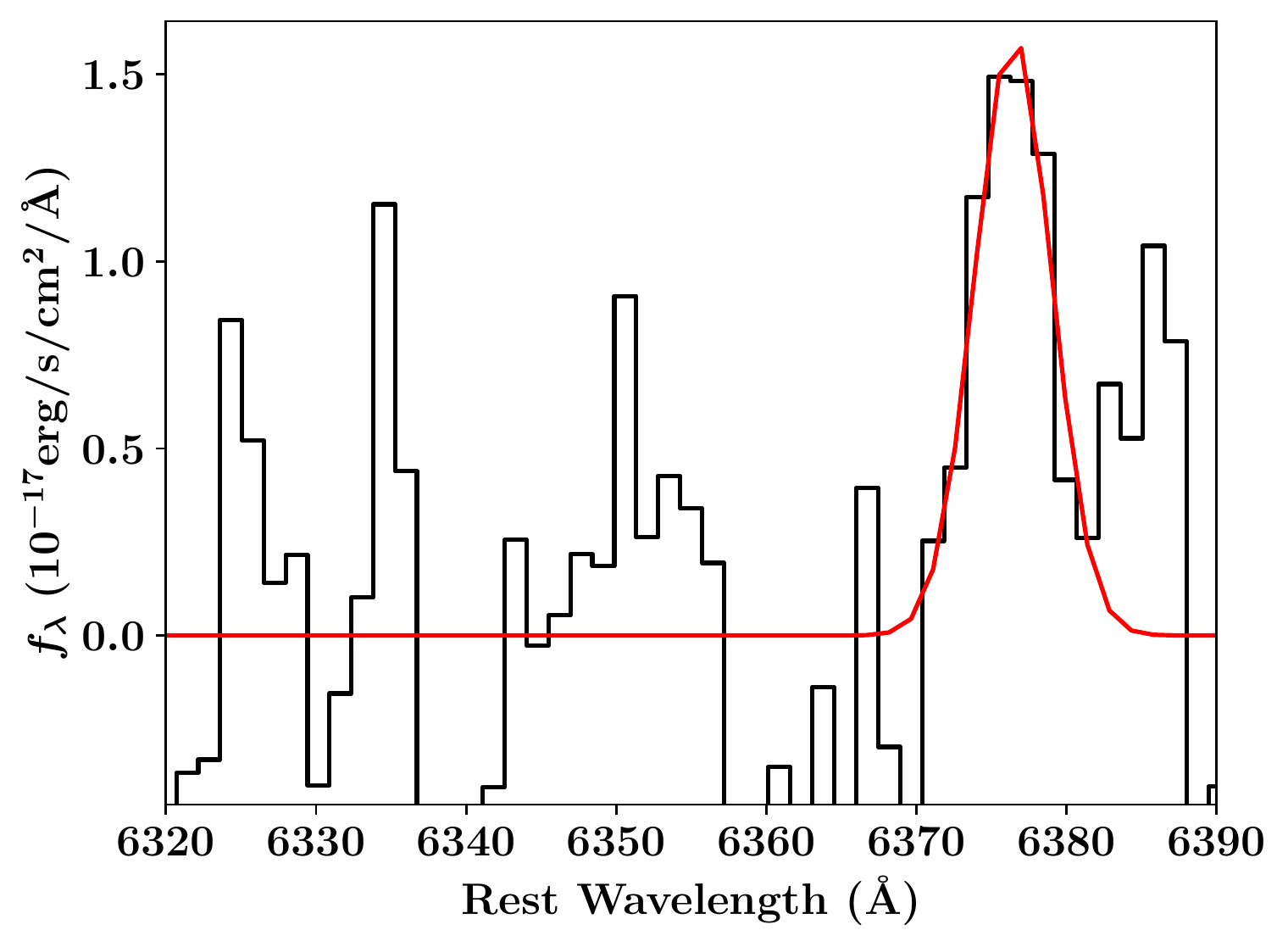} 
        \includegraphics[width=0.13\textwidth]{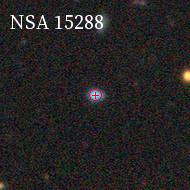}
    \hspace{-3mm}
    \includegraphics[width=0.19\textwidth]{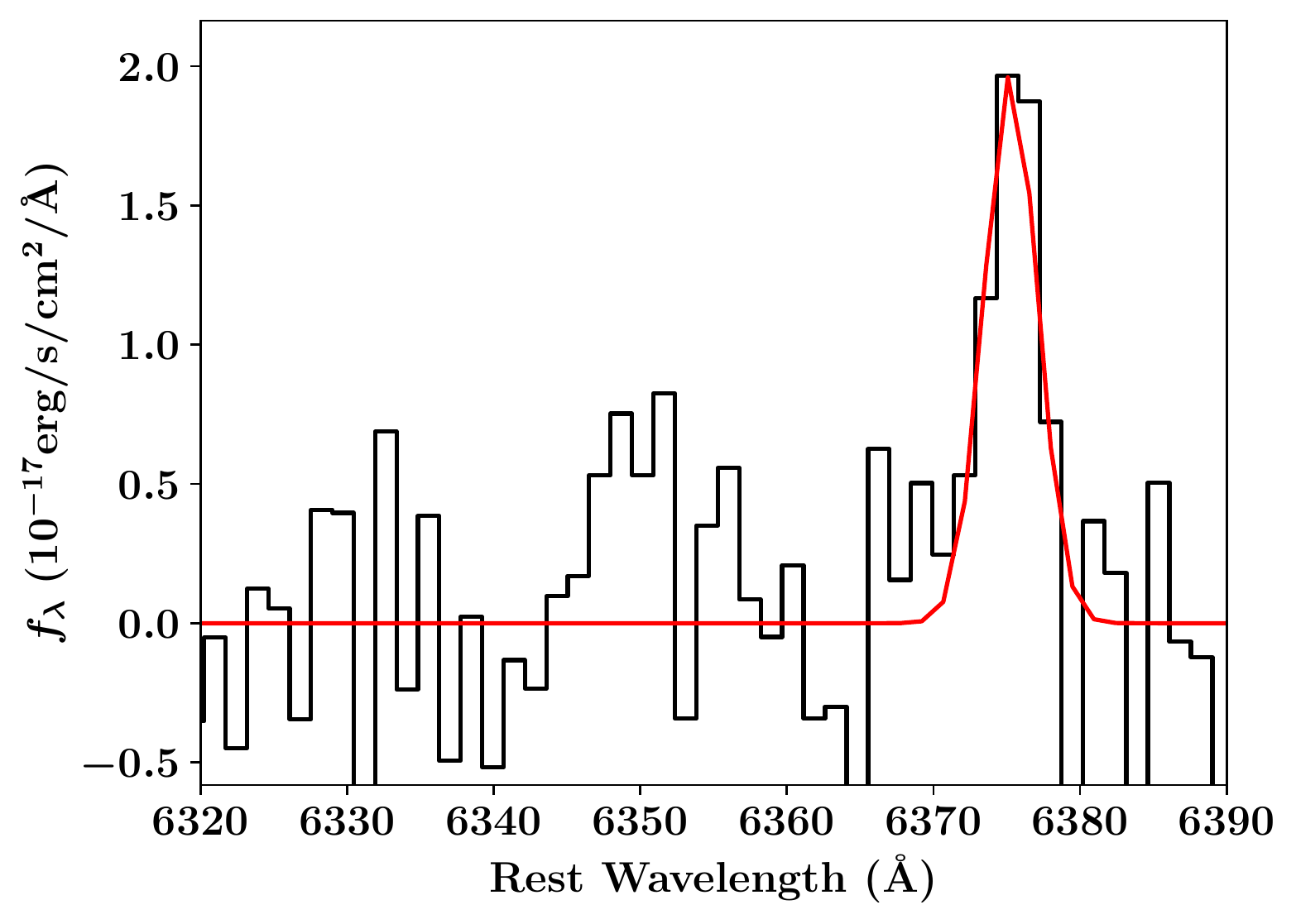}
    \hspace{1.5mm}
    \includegraphics[width=0.13\textwidth]{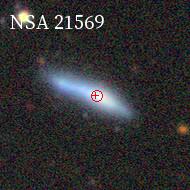}
        \hspace{-3mm}
    \includegraphics[width=0.19\textwidth]{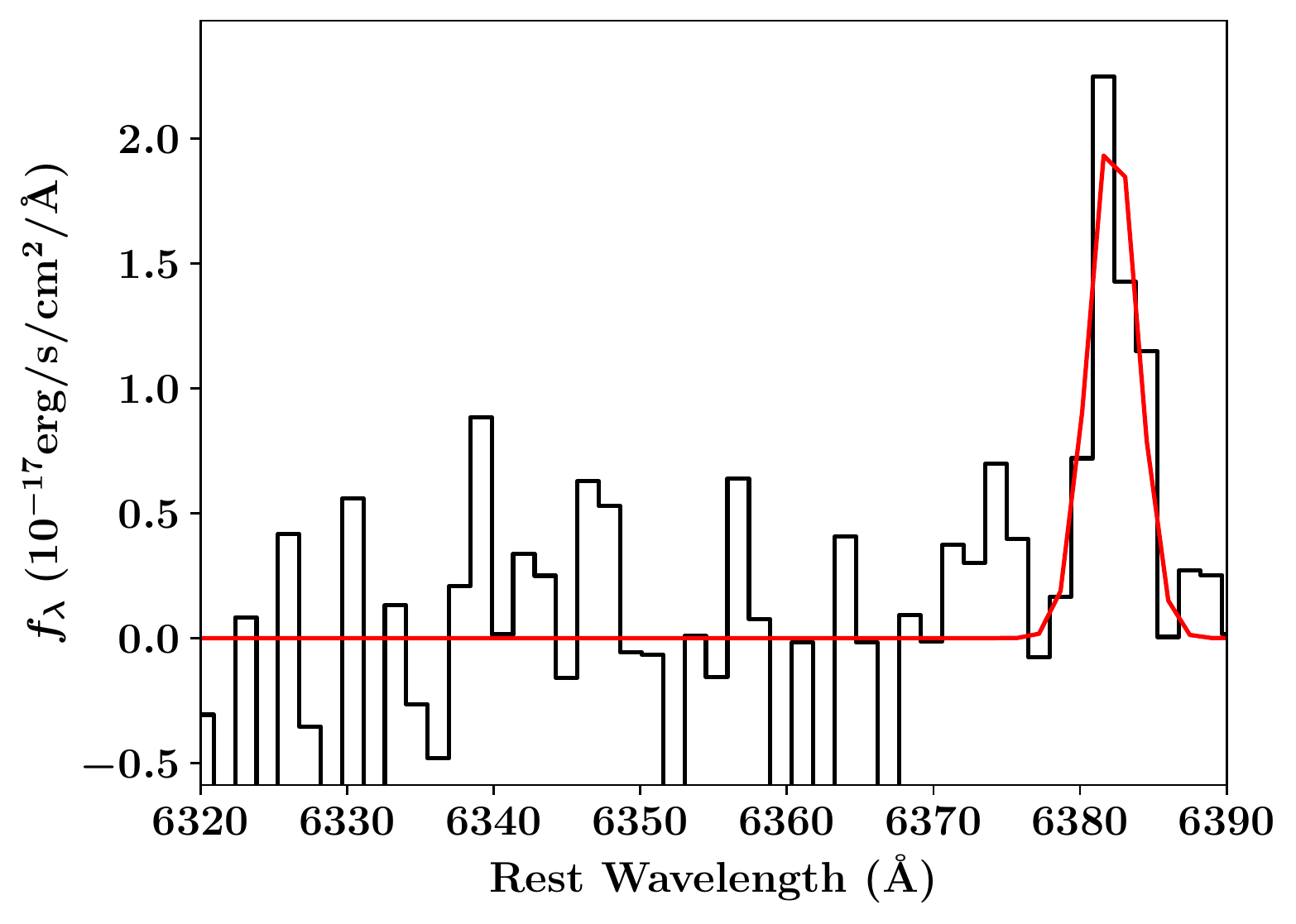}  
        \hspace{1.5mm}
    \includegraphics[width=0.13\textwidth]{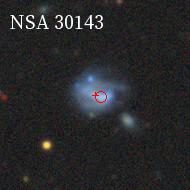}
        \hspace{-3mm}
    \includegraphics[width=0.19\textwidth]{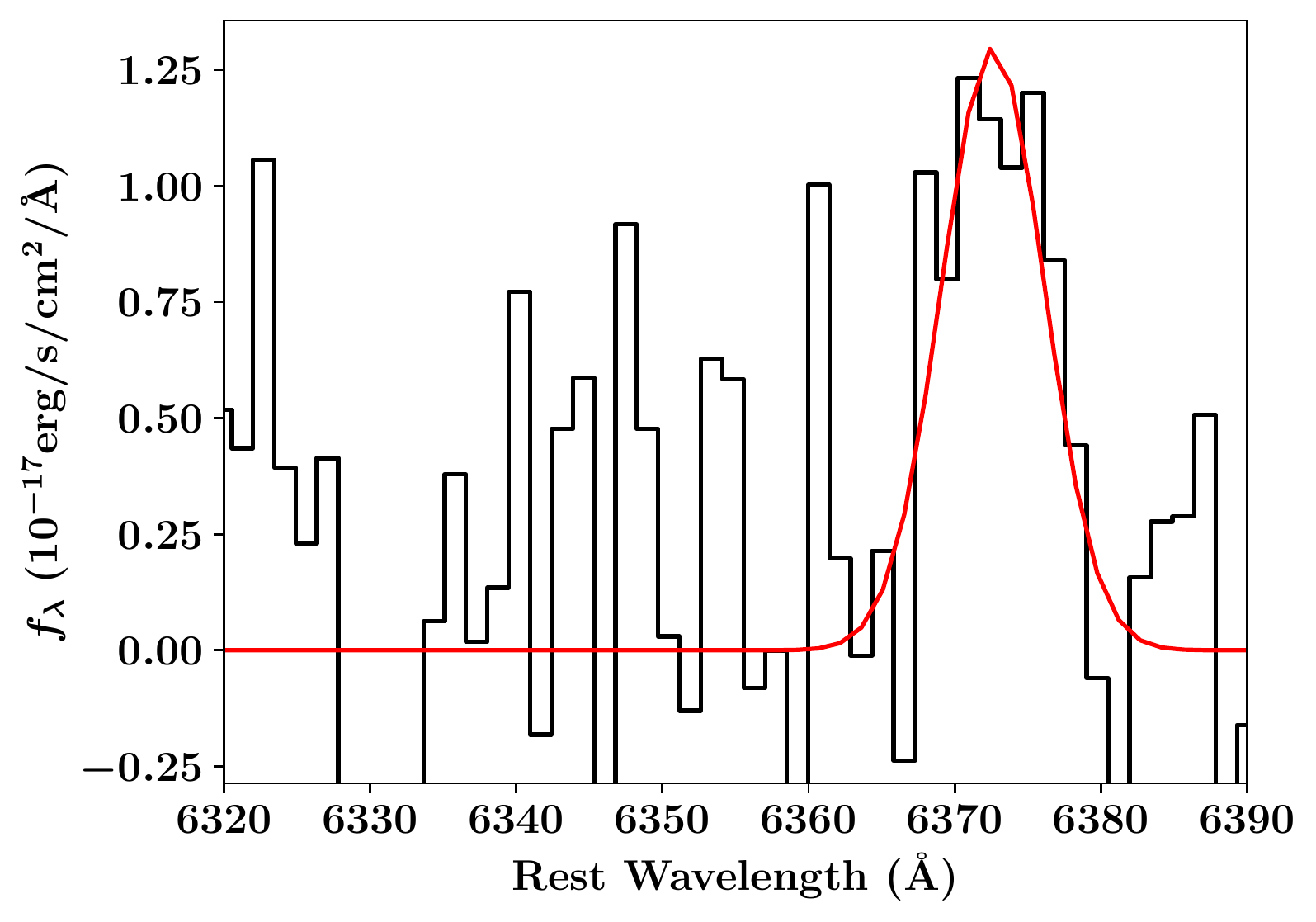}  
    \includegraphics[width=0.13\textwidth]{nsa41675_marked.jpg}
    \hspace{-3mm}
    \includegraphics[width=0.19\textwidth]{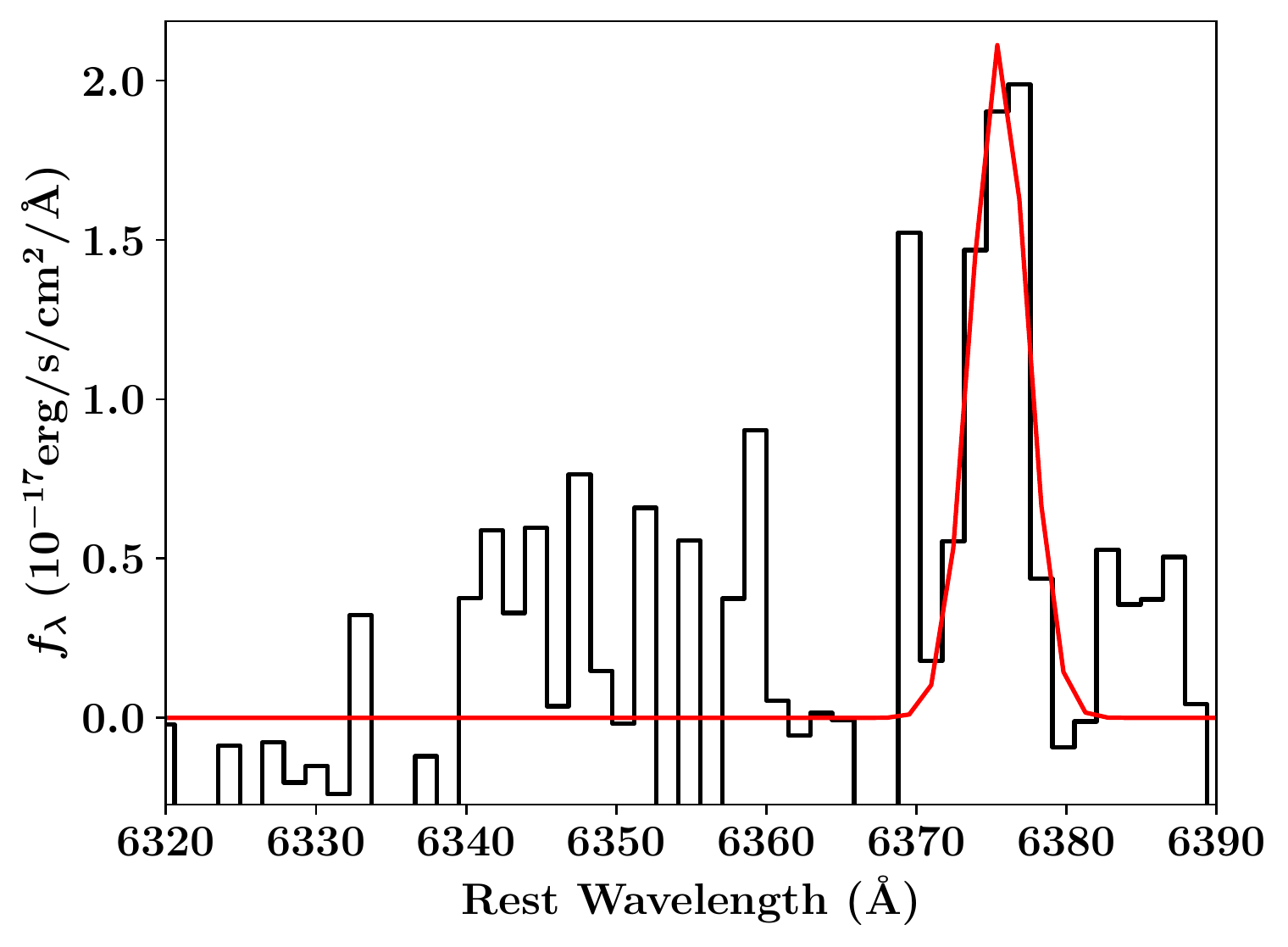}
    \hspace{1.5mm}
    \includegraphics[width=0.13\textwidth]{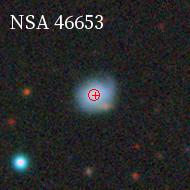}
        \hspace{-3mm}
    \includegraphics[width=0.19\textwidth]{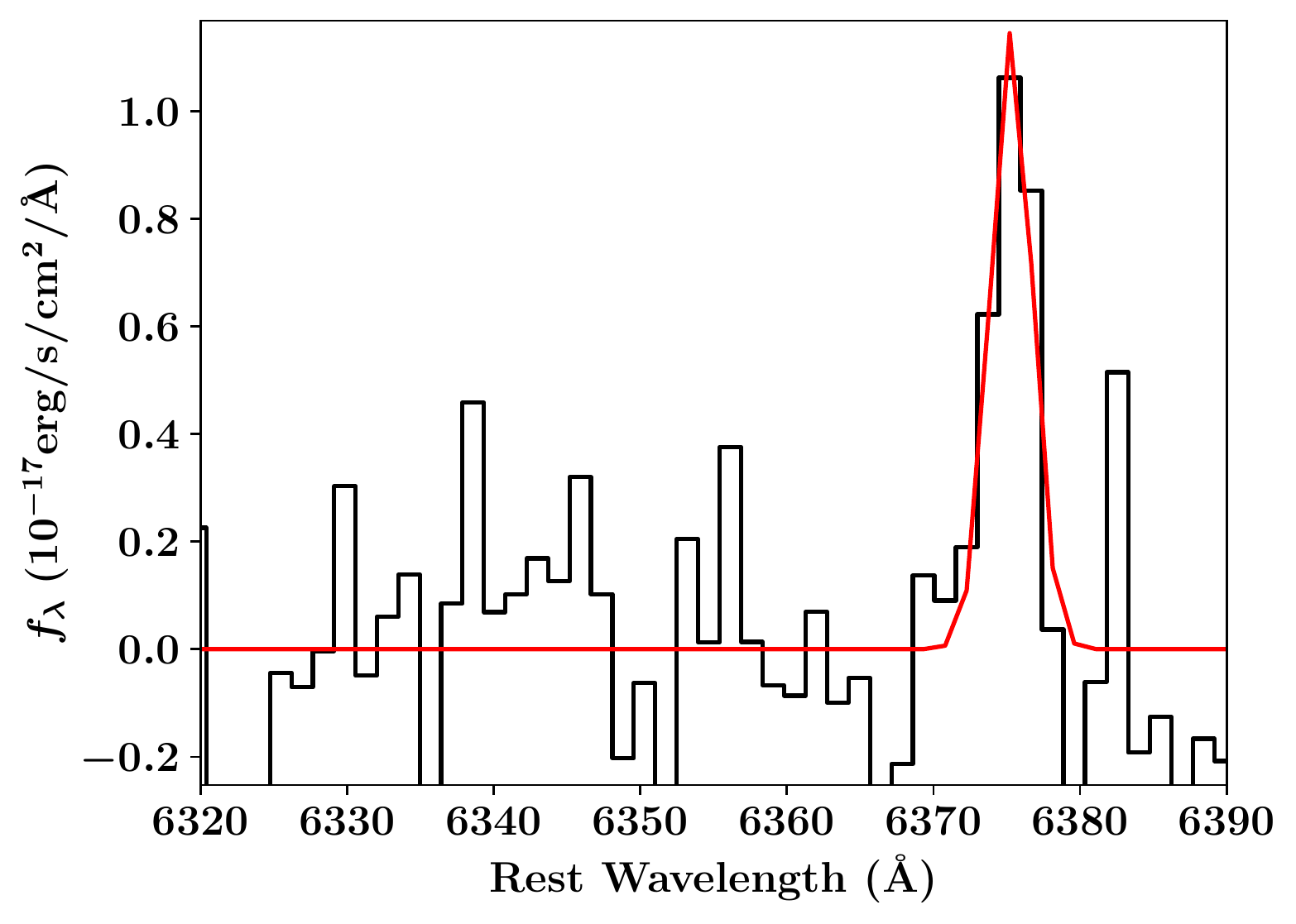}  
        \hspace{1.5mm}
    \includegraphics[width=0.13\textwidth]{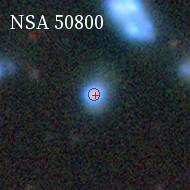}
        \hspace{-3mm}
    \includegraphics[width=0.19\textwidth]{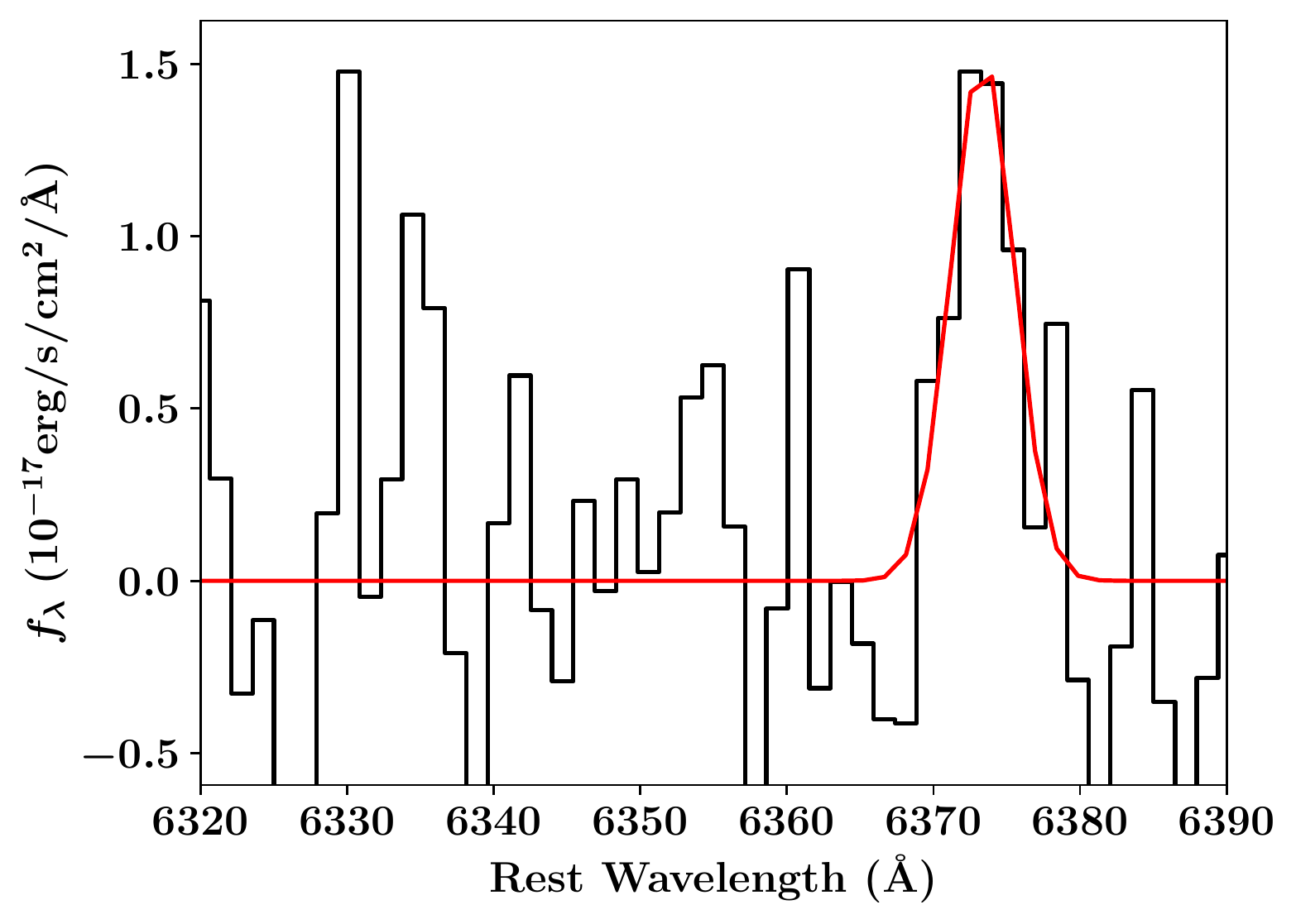} 
        \includegraphics[width=0.13\textwidth]{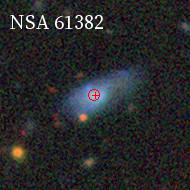}
    \hspace{-3mm}
    \includegraphics[width=0.19\textwidth]{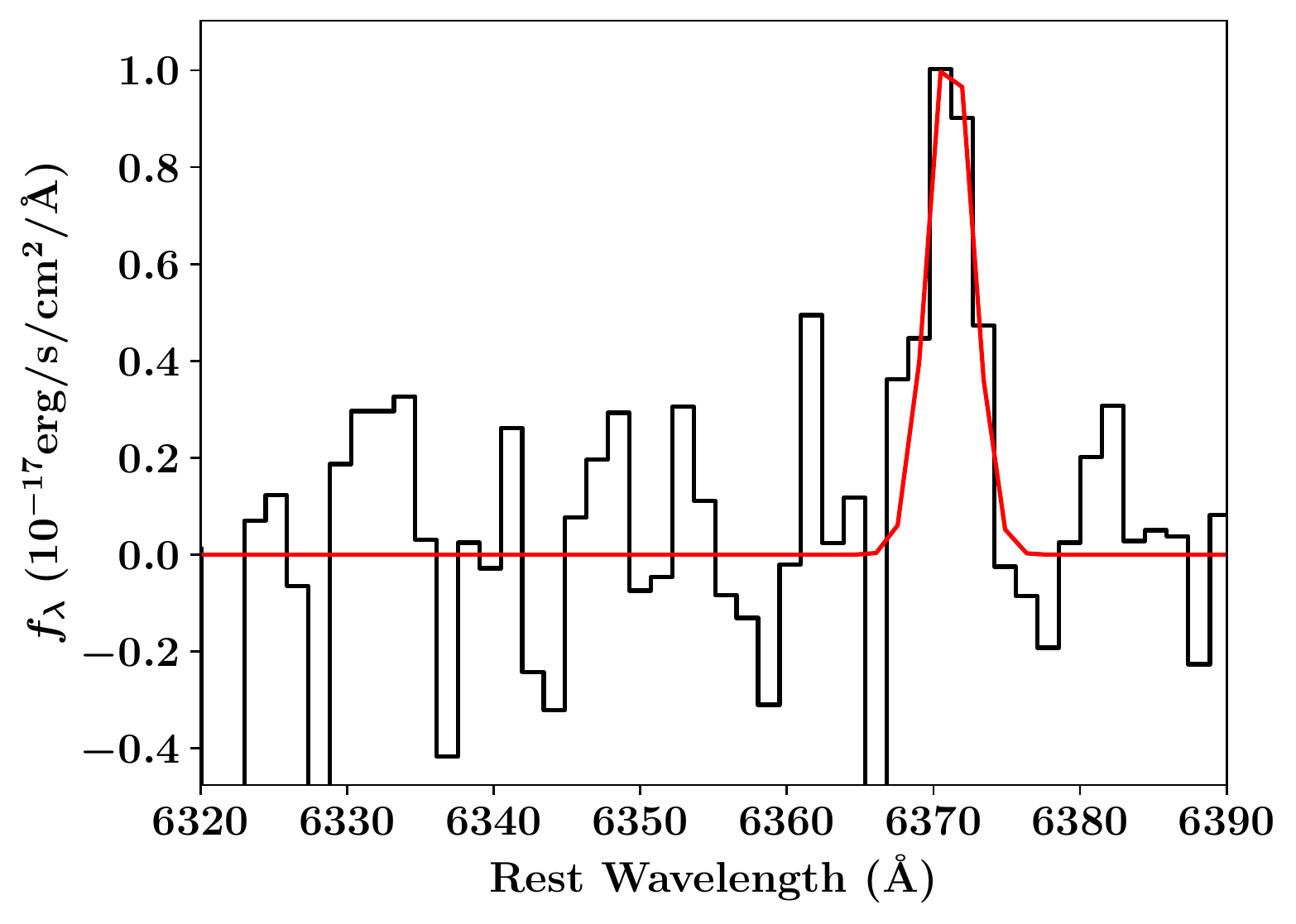}
    \hspace{1.5mm}
    \includegraphics[width=0.13\textwidth]{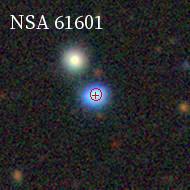}
        \hspace{-3mm}
    \includegraphics[width=0.19\textwidth]{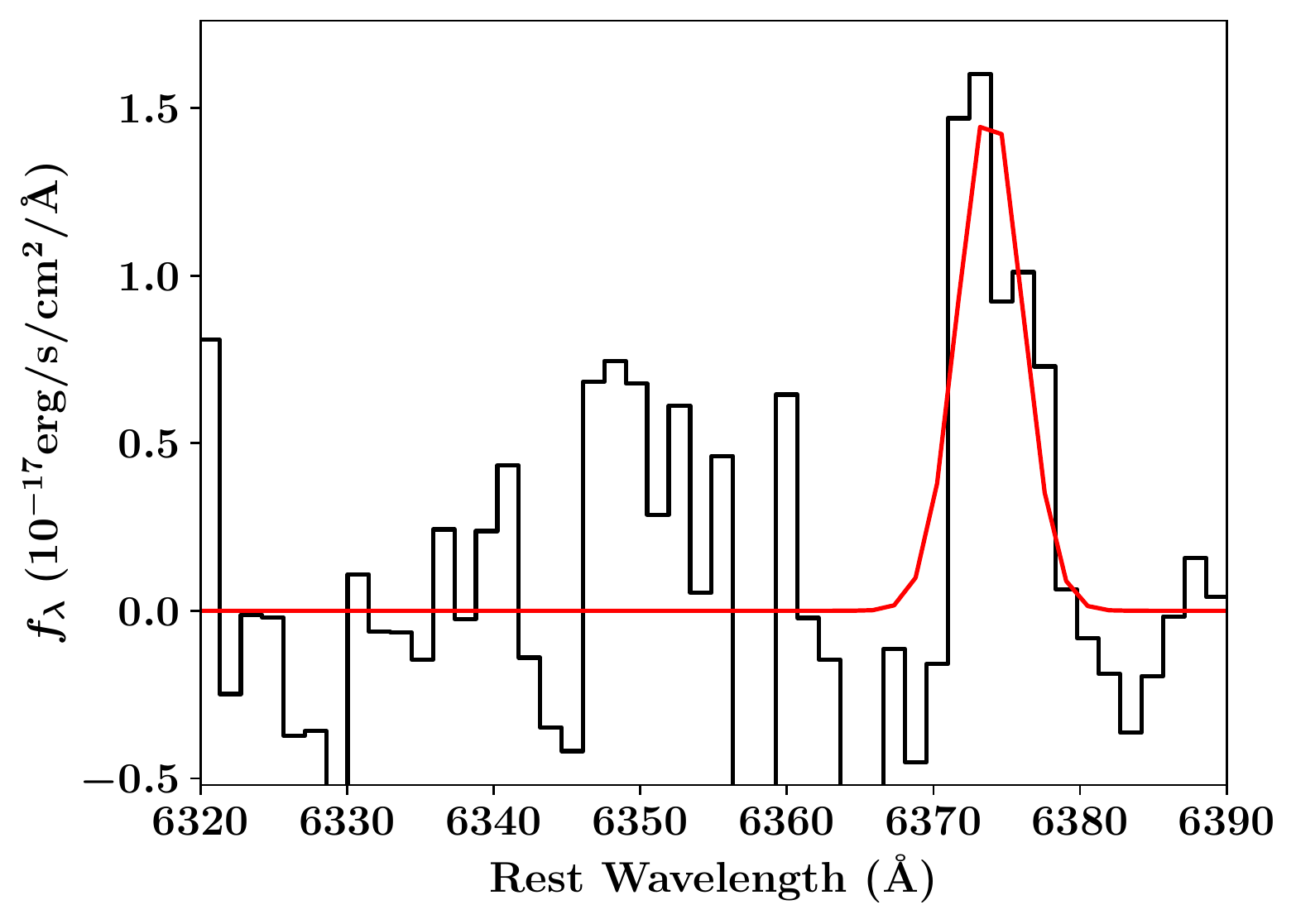}  
        \hspace{1.5mm}
    \includegraphics[width=0.13\textwidth]{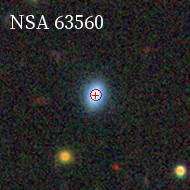}
        \hspace{-3mm}
    \includegraphics[width=0.19\textwidth]{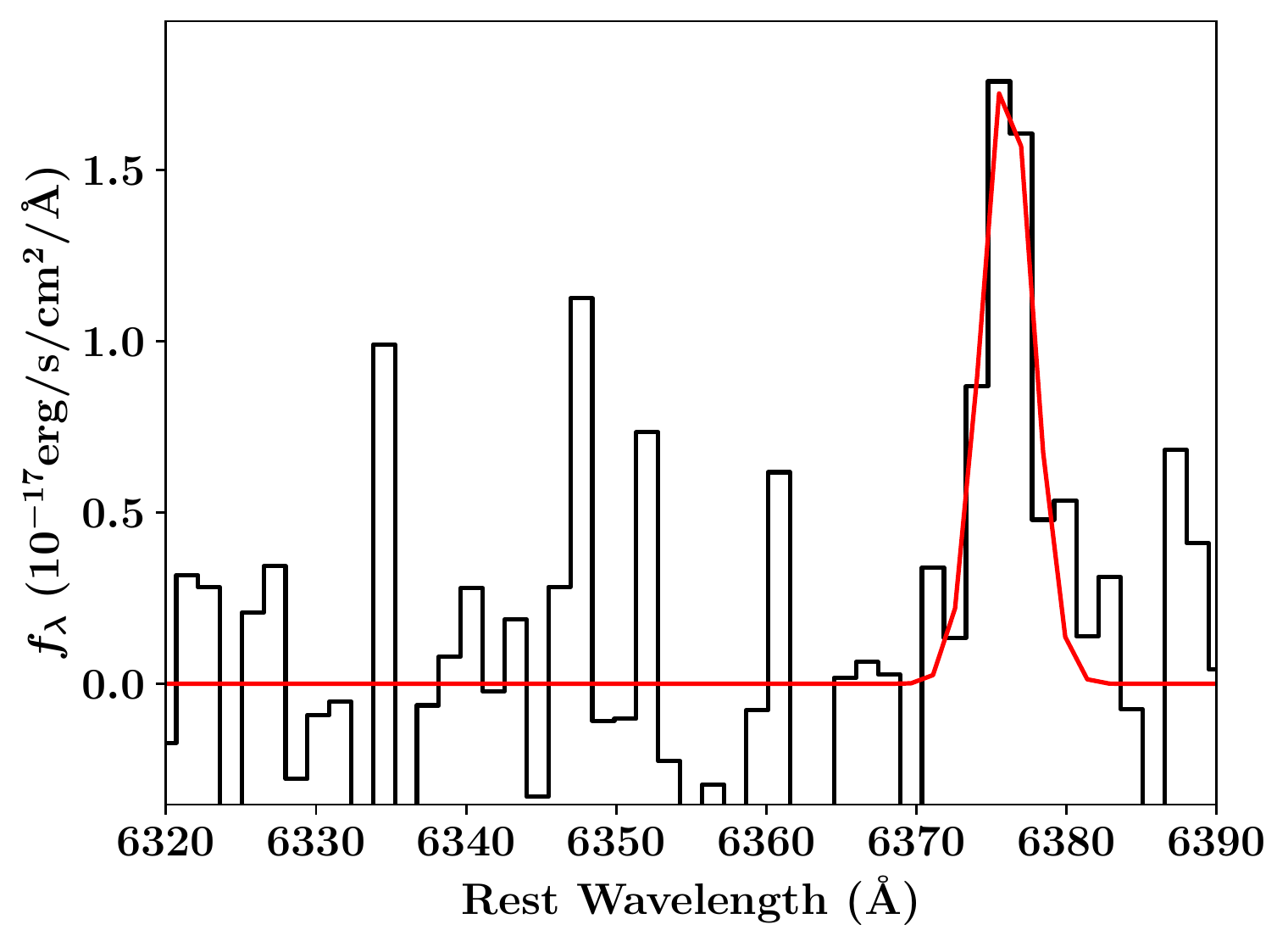} 
    \includegraphics[width=0.13\textwidth]{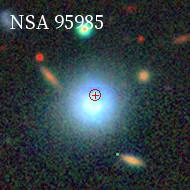}
    \hspace{-3mm}
    \includegraphics[width=0.19\textwidth]{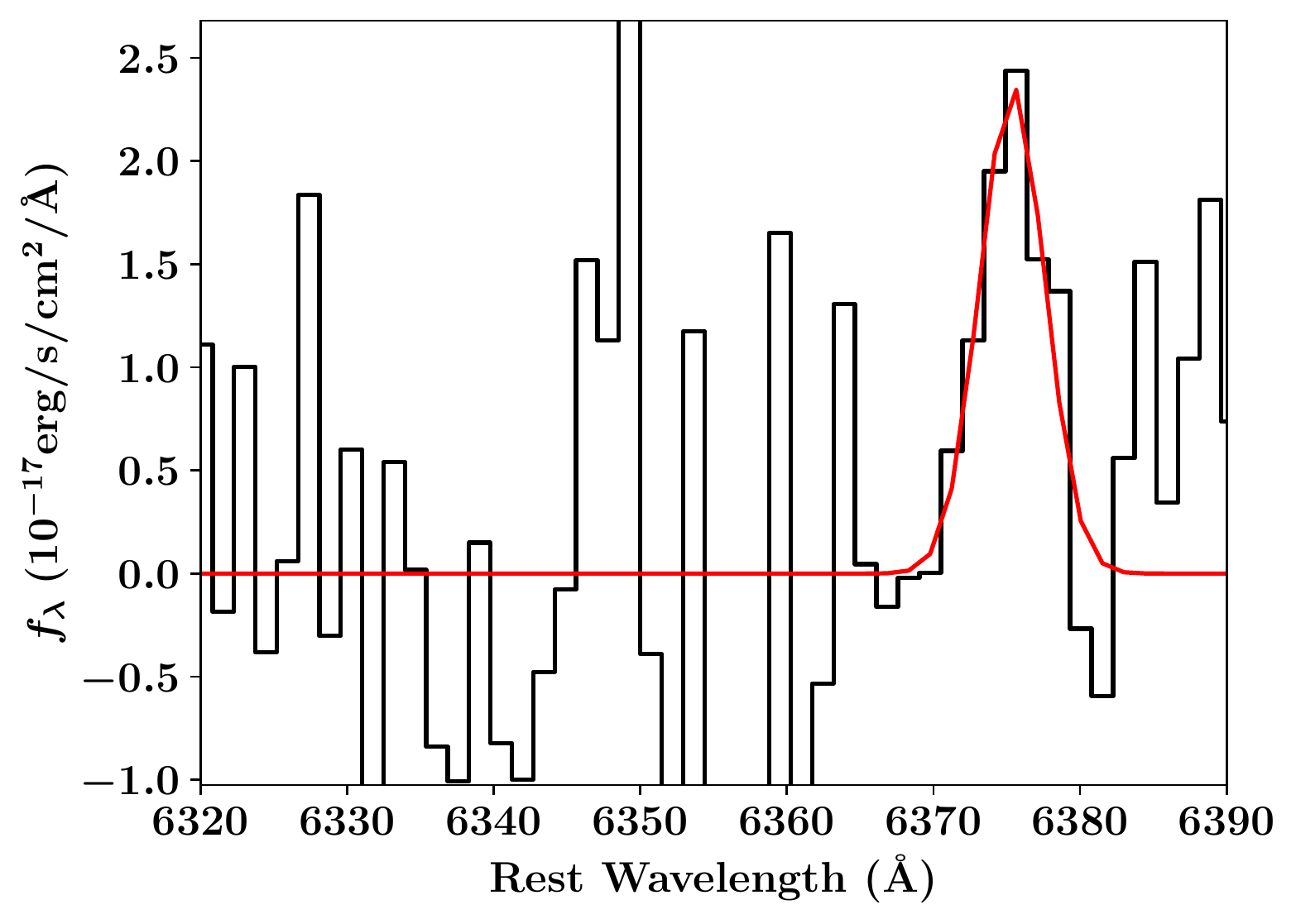}
    \hspace{1.5mm}
    \includegraphics[width=0.13\textwidth]{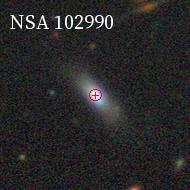}
        \hspace{-3mm}
    \includegraphics[width=0.19\textwidth]{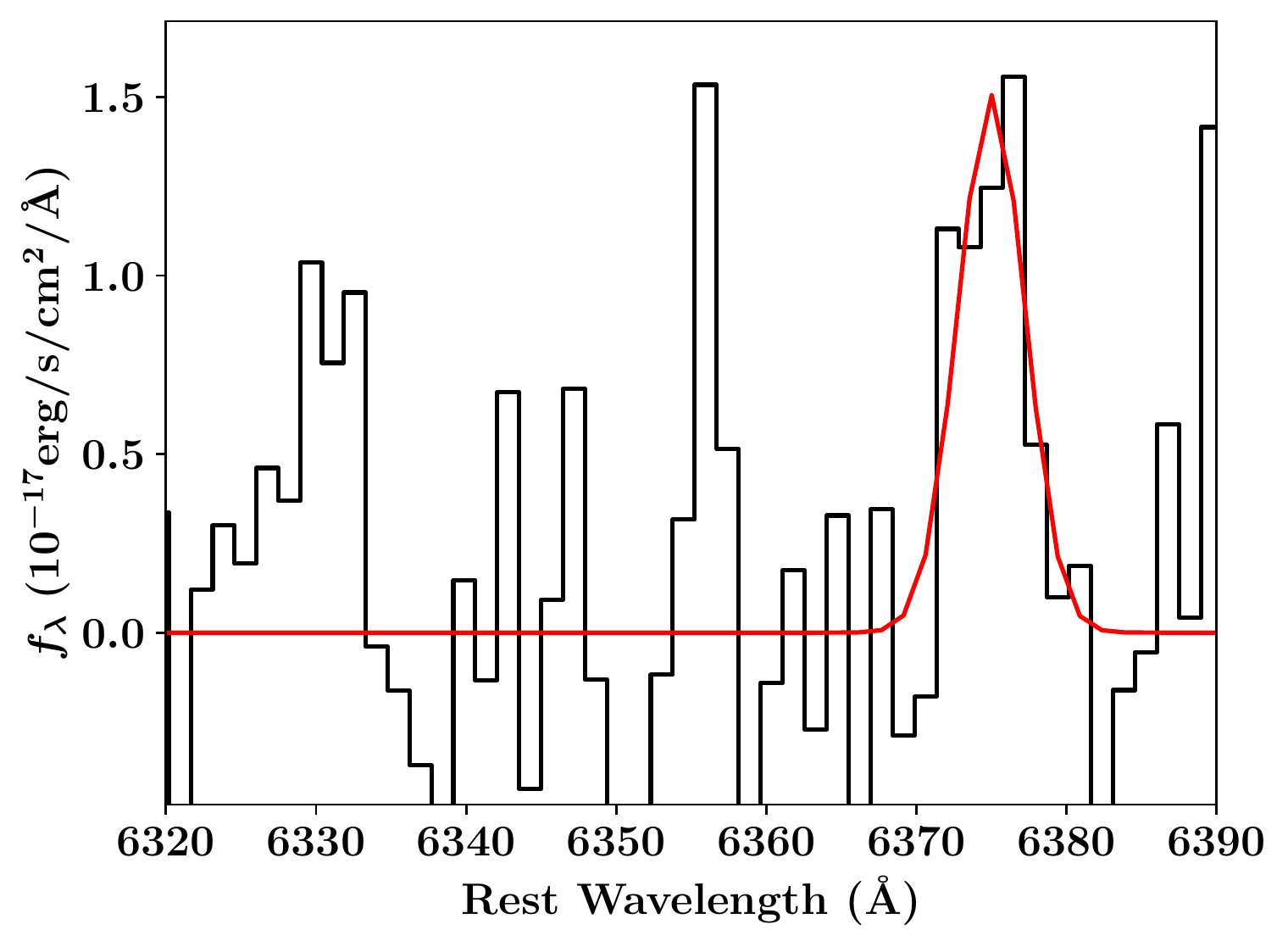}  
        \hspace{1.5mm}
    \includegraphics[width=0.13\textwidth]{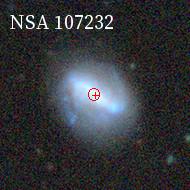}
        \hspace{-3mm}
    \includegraphics[width=0.19\textwidth]{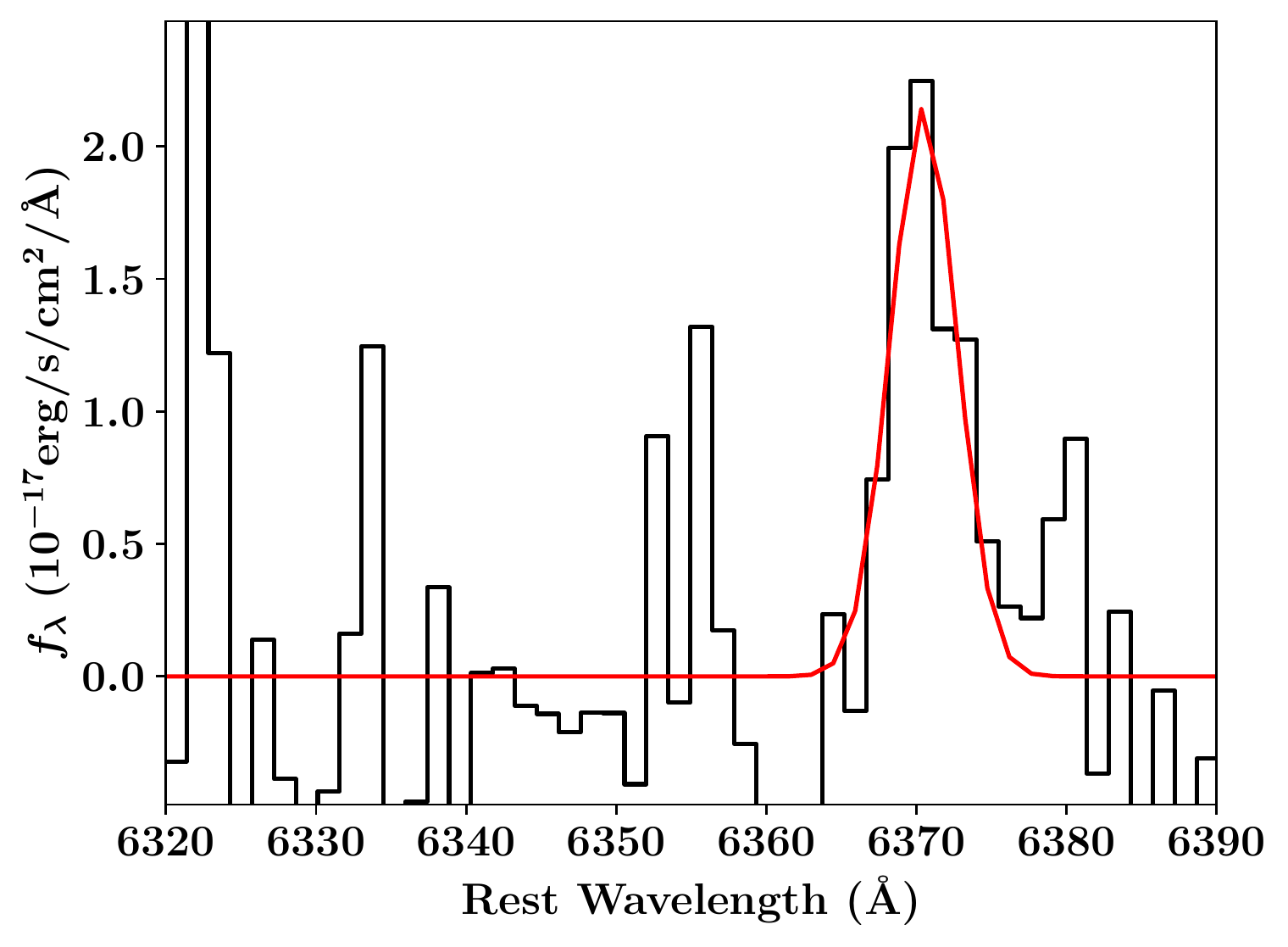} 
    \includegraphics[width=0.13\textwidth]{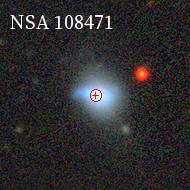}
    \hspace{-3mm}
    \includegraphics[width=0.19\textwidth]{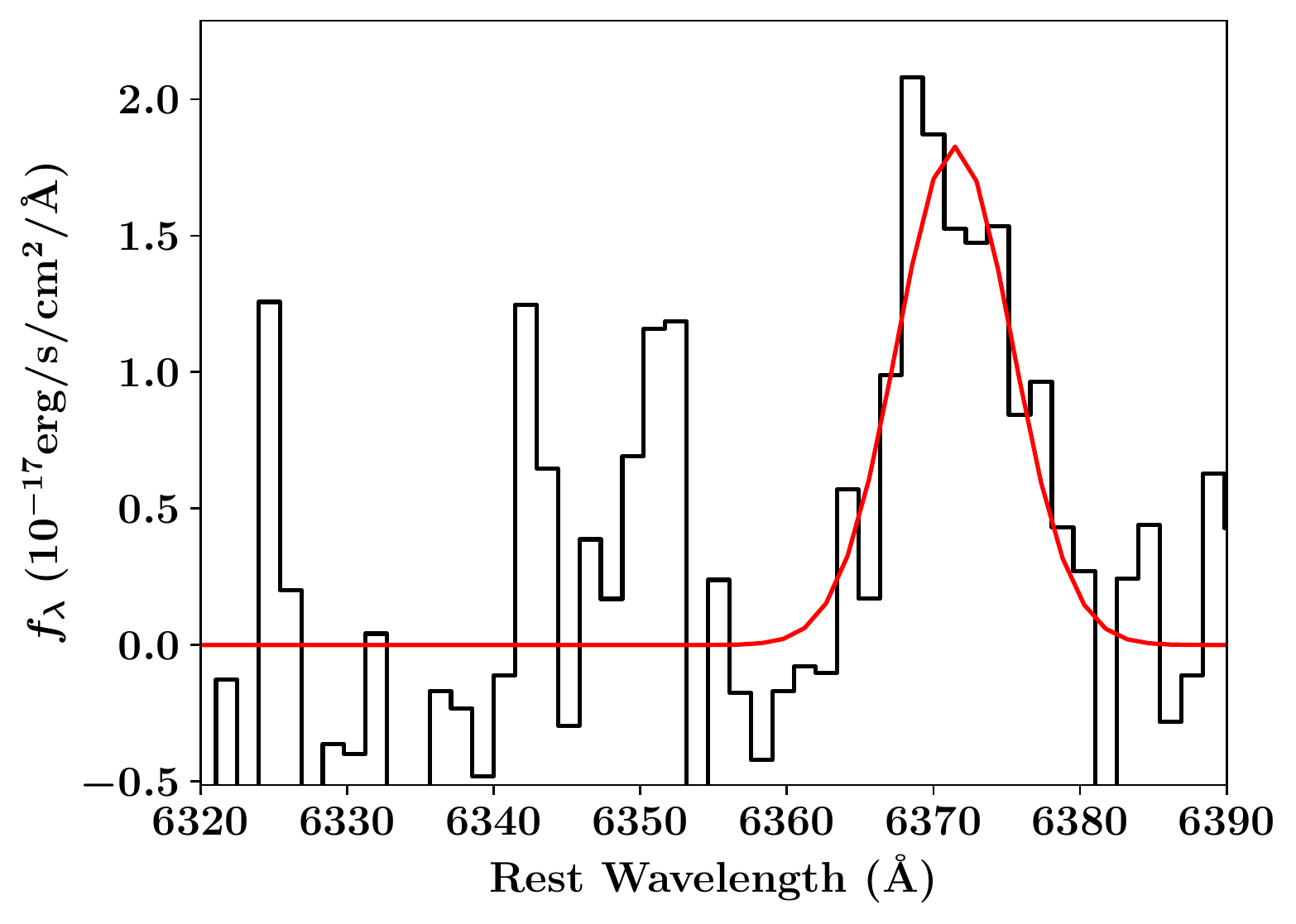}
    \hspace{1.5mm}
    \includegraphics[width=0.13\textwidth]{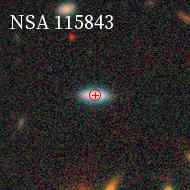}
        \hspace{-3mm}
    \includegraphics[width=0.19\textwidth]{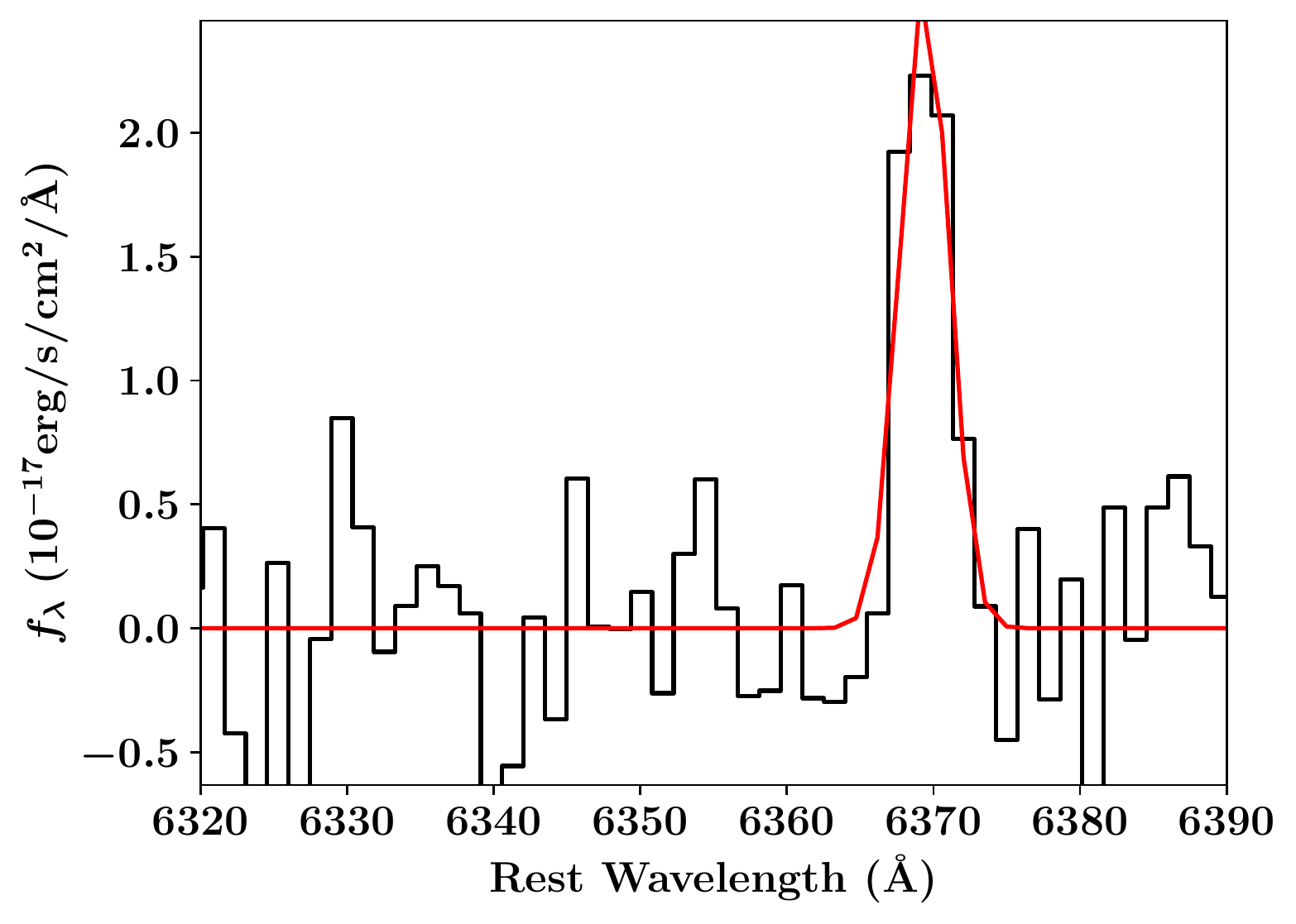}  
        \hspace{1.5mm}
    \includegraphics[width=0.13\textwidth]{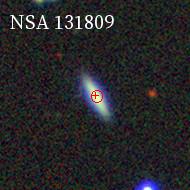}
        \hspace{-3mm}
    \includegraphics[width=0.19\textwidth]{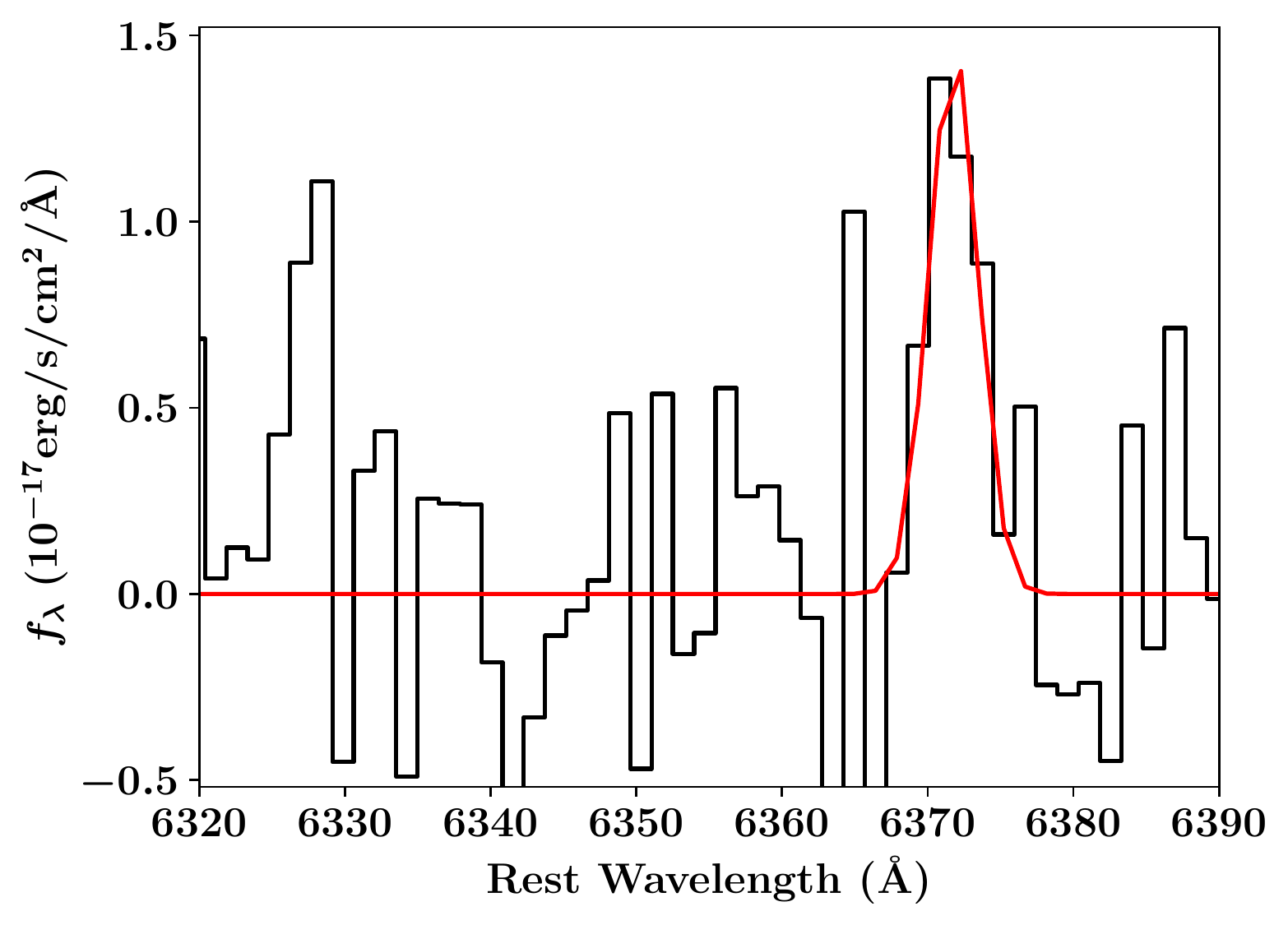} 
    \includegraphics[width=0.13\textwidth]{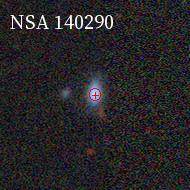}
    \hspace{-3mm}
    \includegraphics[width=0.19\textwidth]{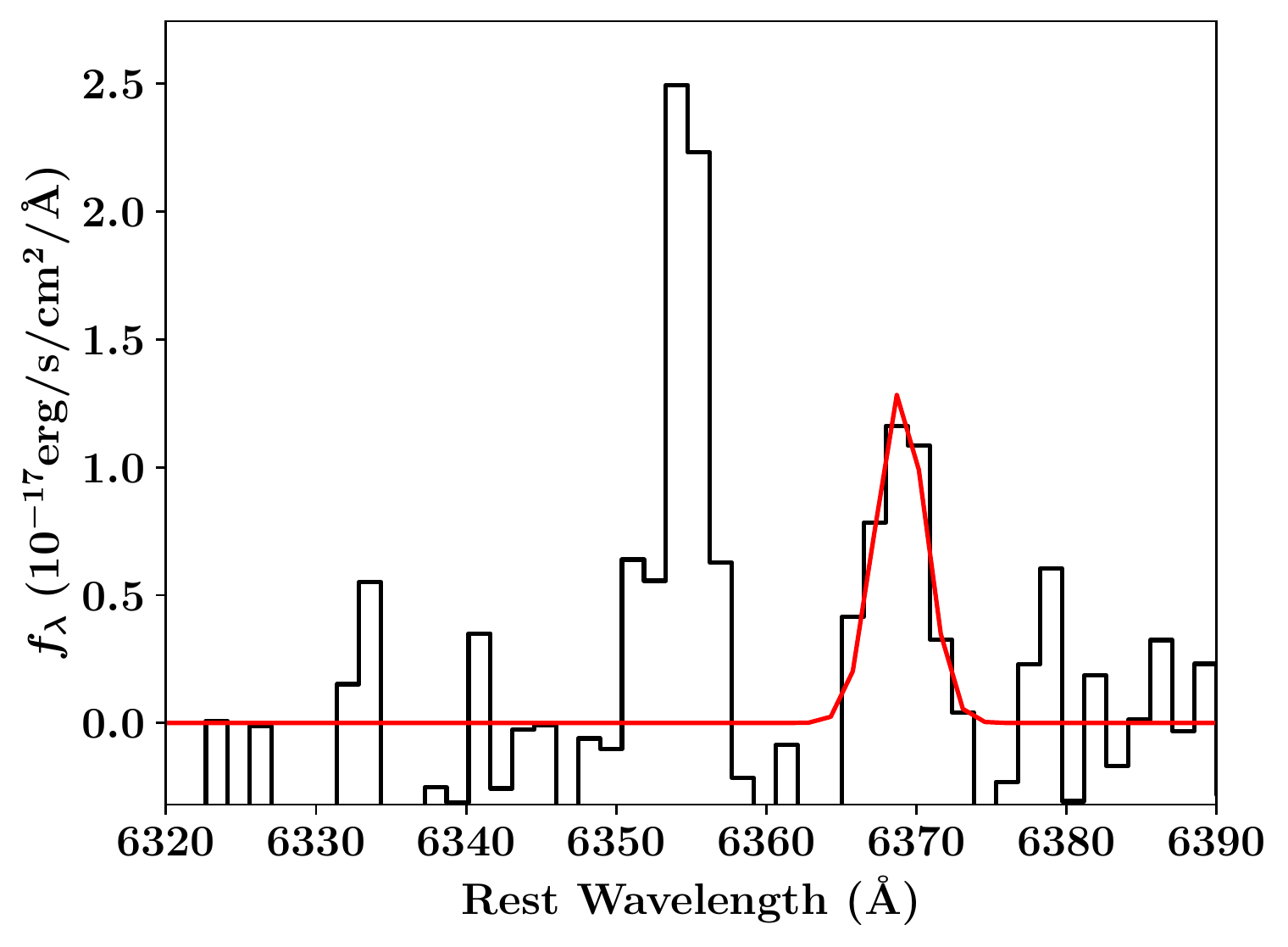}
    \hspace{1.5mm}
    \includegraphics[width=0.13\textwidth]{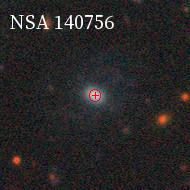}
        \hspace{-3mm}
    \includegraphics[width=0.19\textwidth]{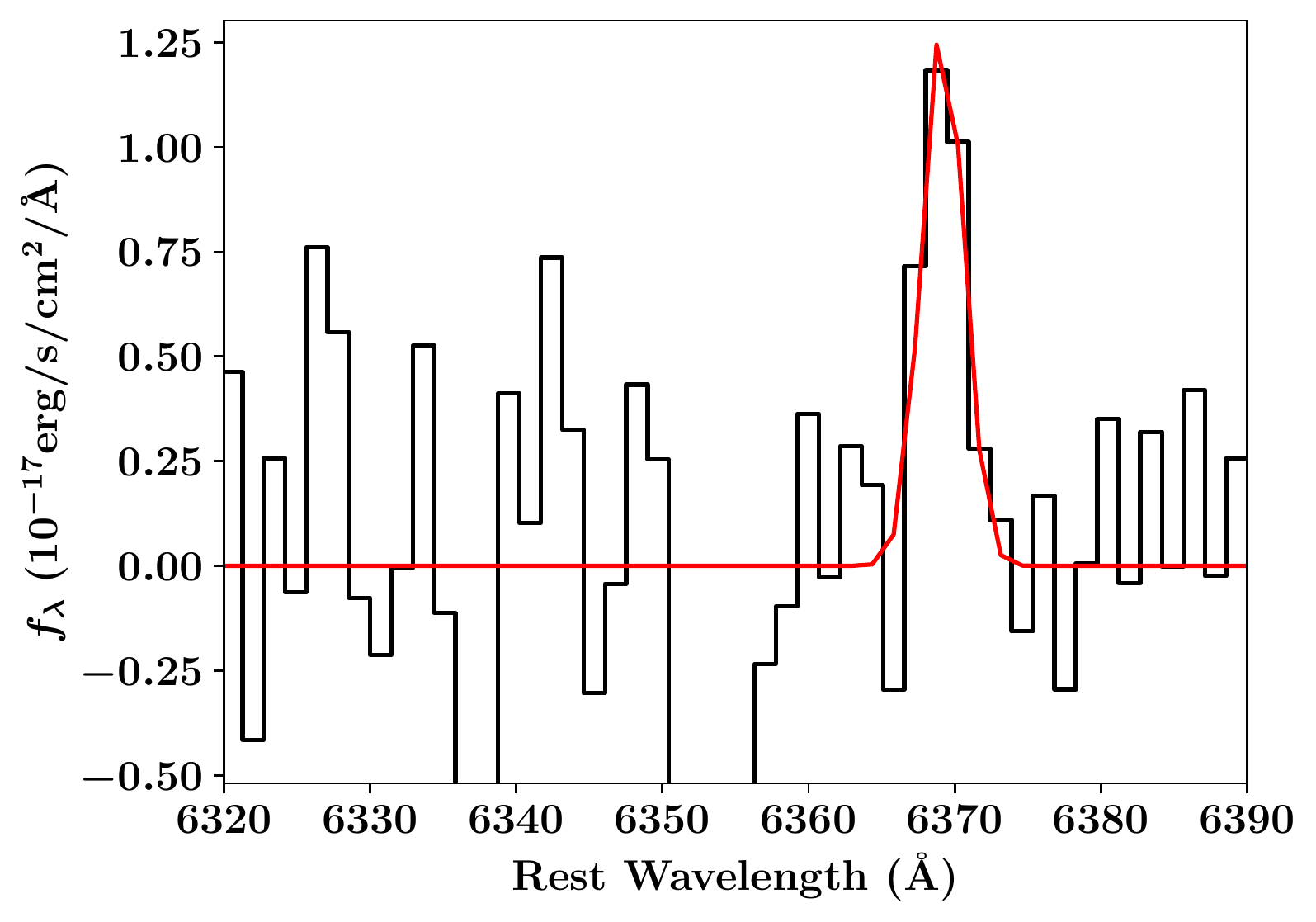}  
        \hspace{1.5mm}
    \includegraphics[width=0.13\textwidth]{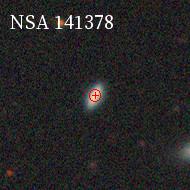}
        \hspace{-3mm}
    \includegraphics[width=0.19\textwidth]{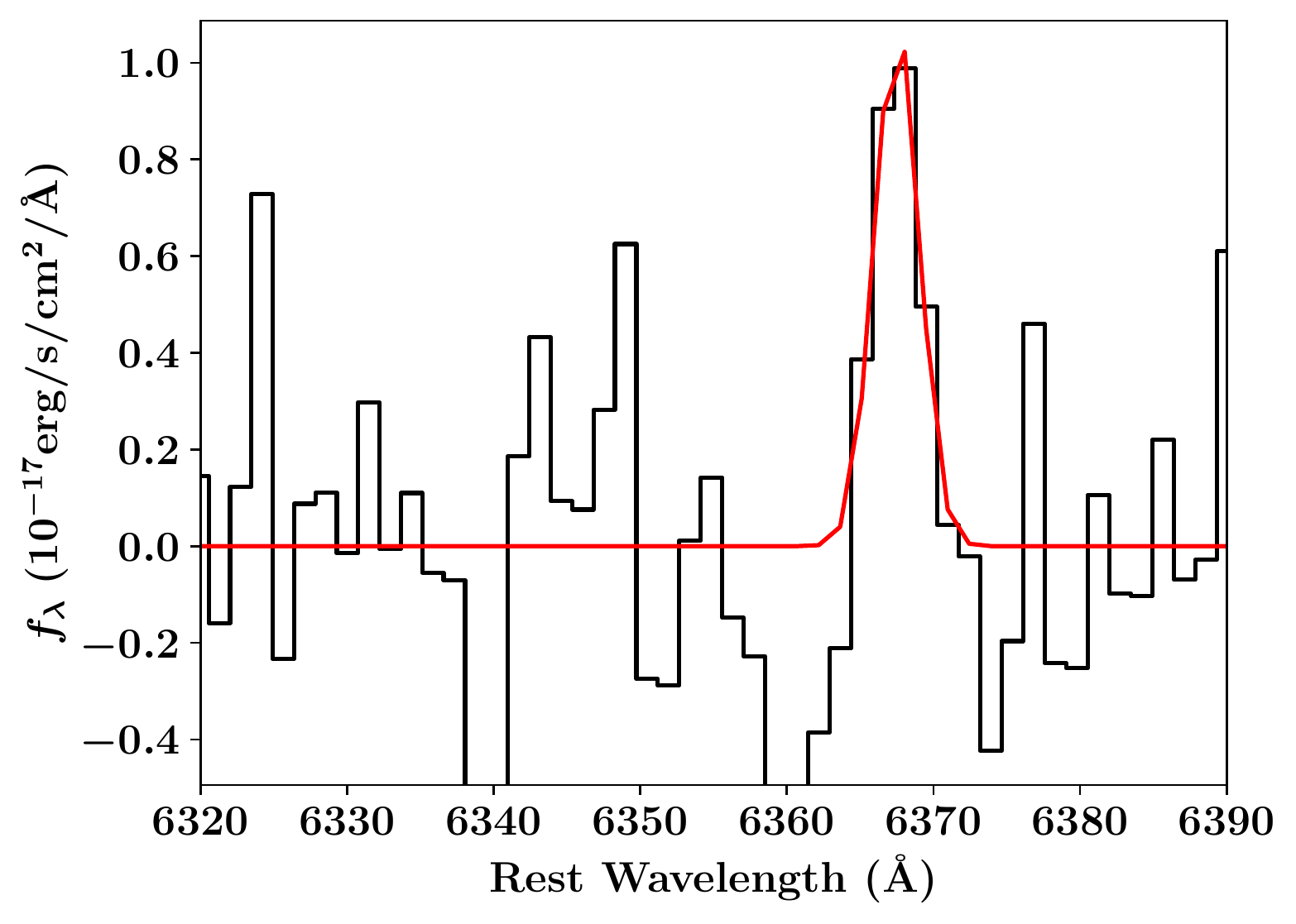} 
        \includegraphics[width=0.13\textwidth]{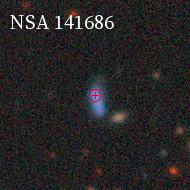}
    \hspace{-3mm}
    \includegraphics[width=0.19\textwidth]{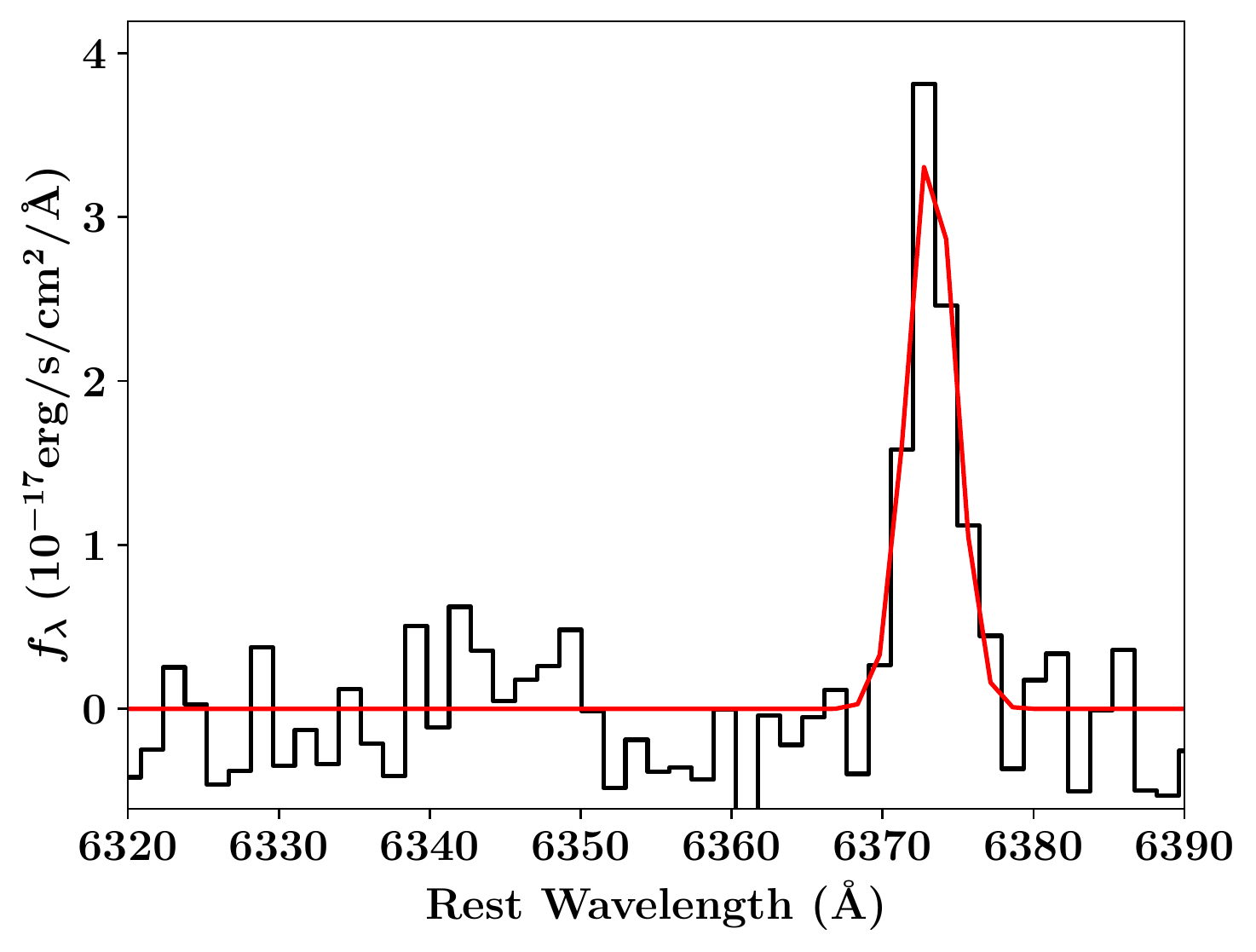}
    \hspace{1.5mm}
    \includegraphics[width=0.13\textwidth]{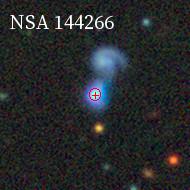}
        \hspace{-3mm}
    \includegraphics[width=0.19\textwidth]{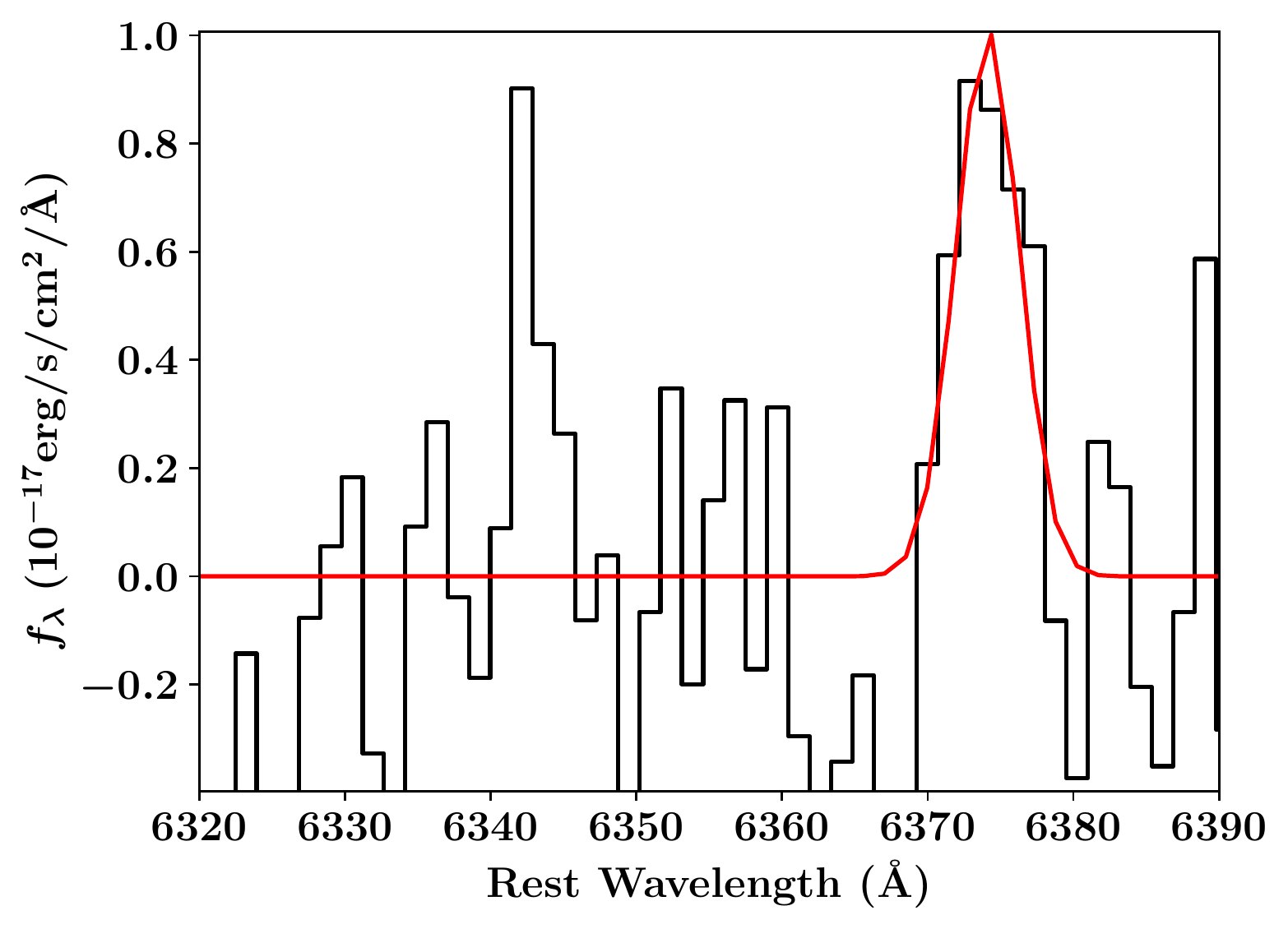}  
        \hspace{1.5mm}
    \includegraphics[width=0.13\textwidth]{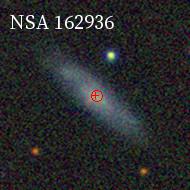}
        \hspace{-3mm}
    \includegraphics[width=0.19\textwidth]{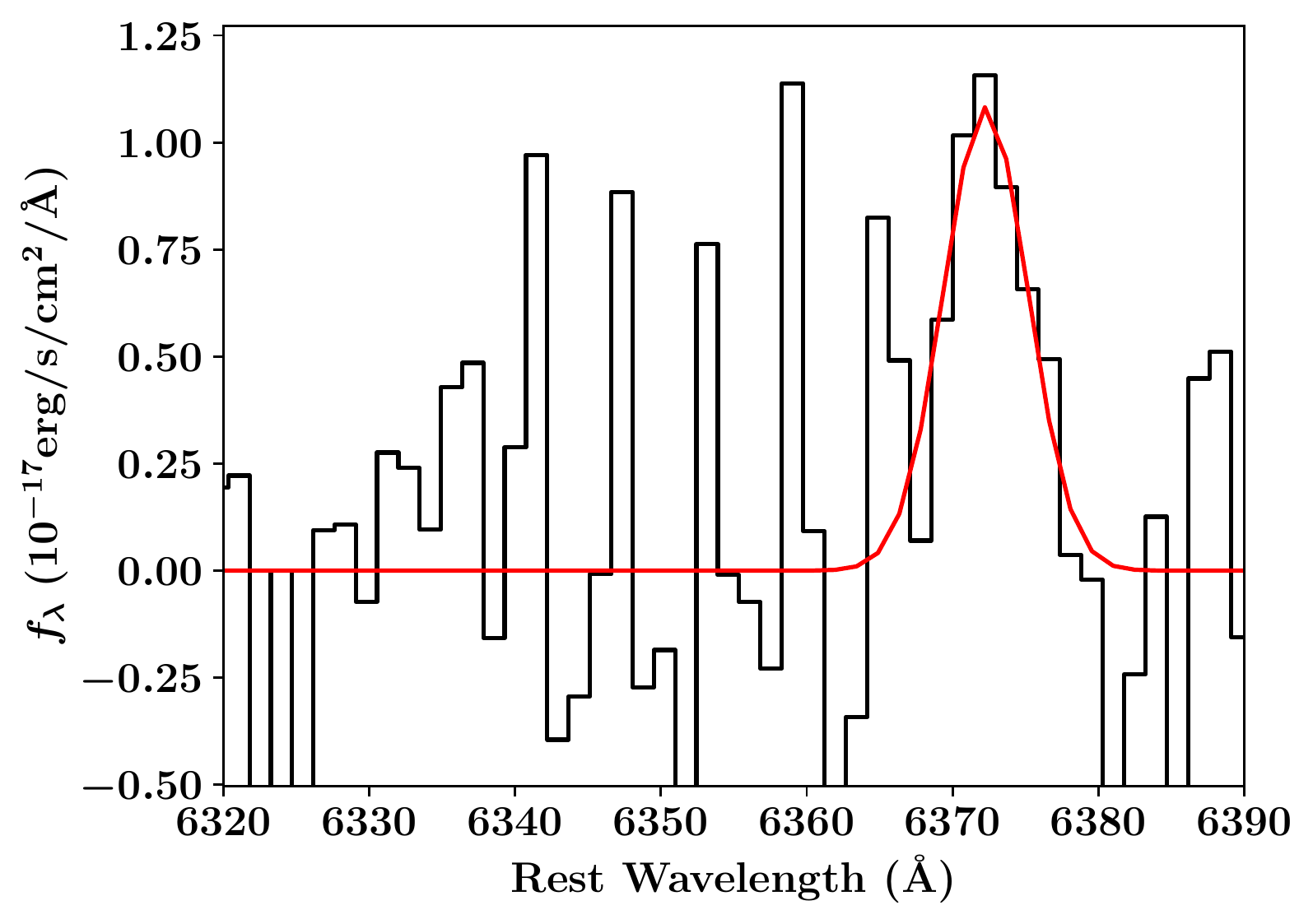} 
     \includegraphics[width=0.13\textwidth]{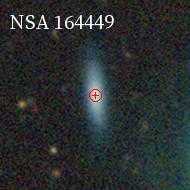}
    \hspace{-3mm}
    \includegraphics[width=0.19\textwidth]{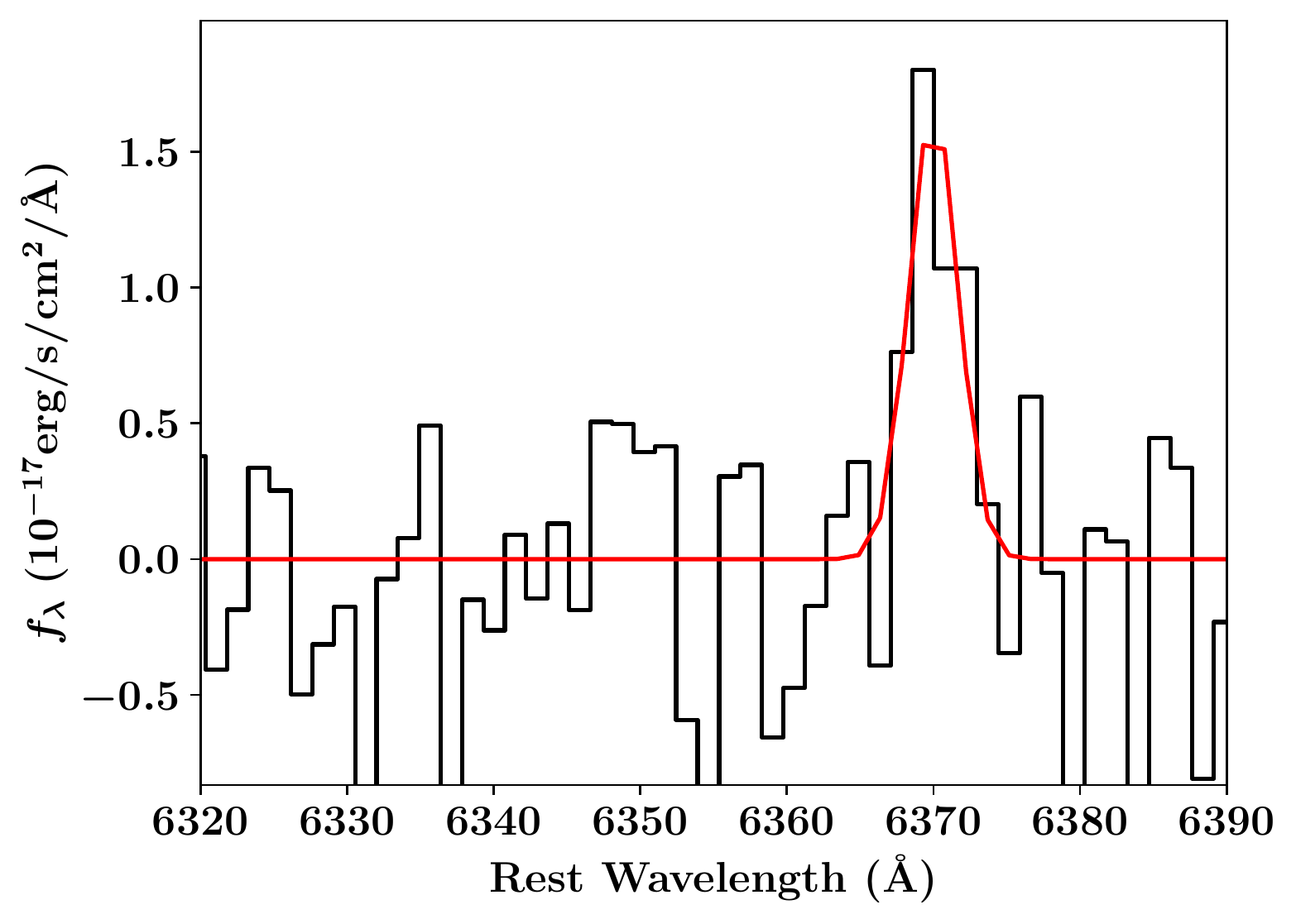}
    \hspace{1.5mm}
    \includegraphics[width=0.13\textwidth]{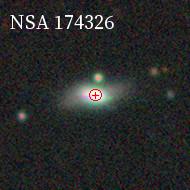}
        \hspace{-3mm}
    \includegraphics[width=0.19\textwidth]{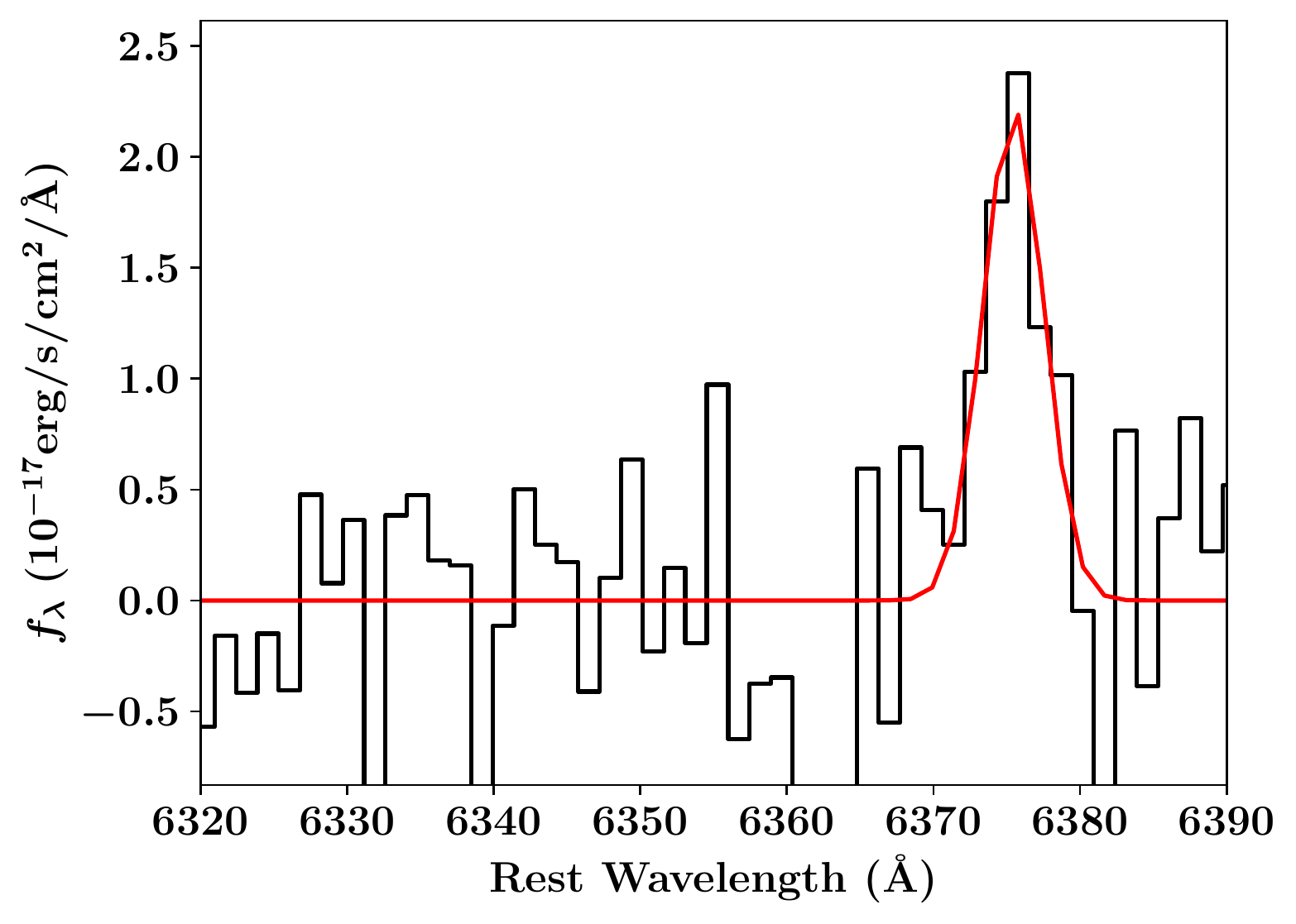}  
        \hspace{1.5mm}
    \includegraphics[width=0.13\textwidth]{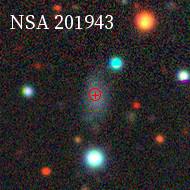}
        \hspace{-3mm}
    \includegraphics[width=0.19\textwidth]{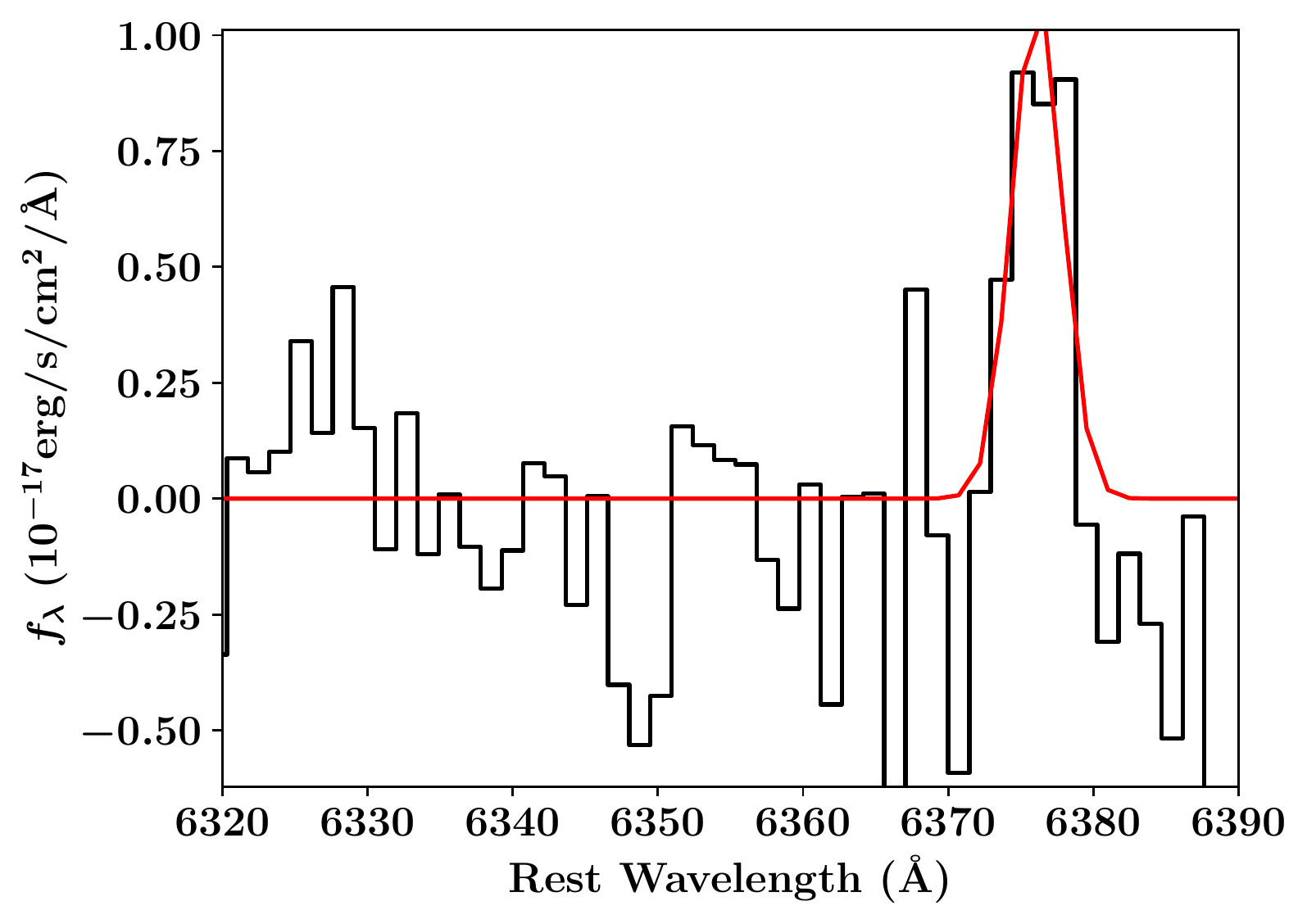}
    \caption{Pairs of the \fexl-emitting dwarf galaxy DESI Legacy Imaging Survey SkyViewer $grz$-band images and the fit to the \fexl\ emission line using the [\ion{O}{1}] and continuum-subtracted spectrum. For each pair, the image is on the left and the spectral fit is on the right. In each of the SkyViewer images, the position of the galaxy nucleus as defined by the NSA is shown by the red cross, and the position of the SDSS 3\arcsec\ spectroscopic fiber is shown by the red circle. For each SDSS spectral fit, the data are shown in black and the \fexl\ emission-line model is overplotted in red.}
    \label{fig:decals}
\end{figure*}

\begin{figure*}[h]
    \centering
    \includegraphics[width=0.13\textwidth]{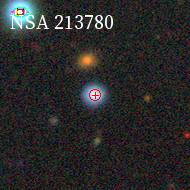}
    \hspace{-3mm}
    \includegraphics[width=0.19\textwidth]{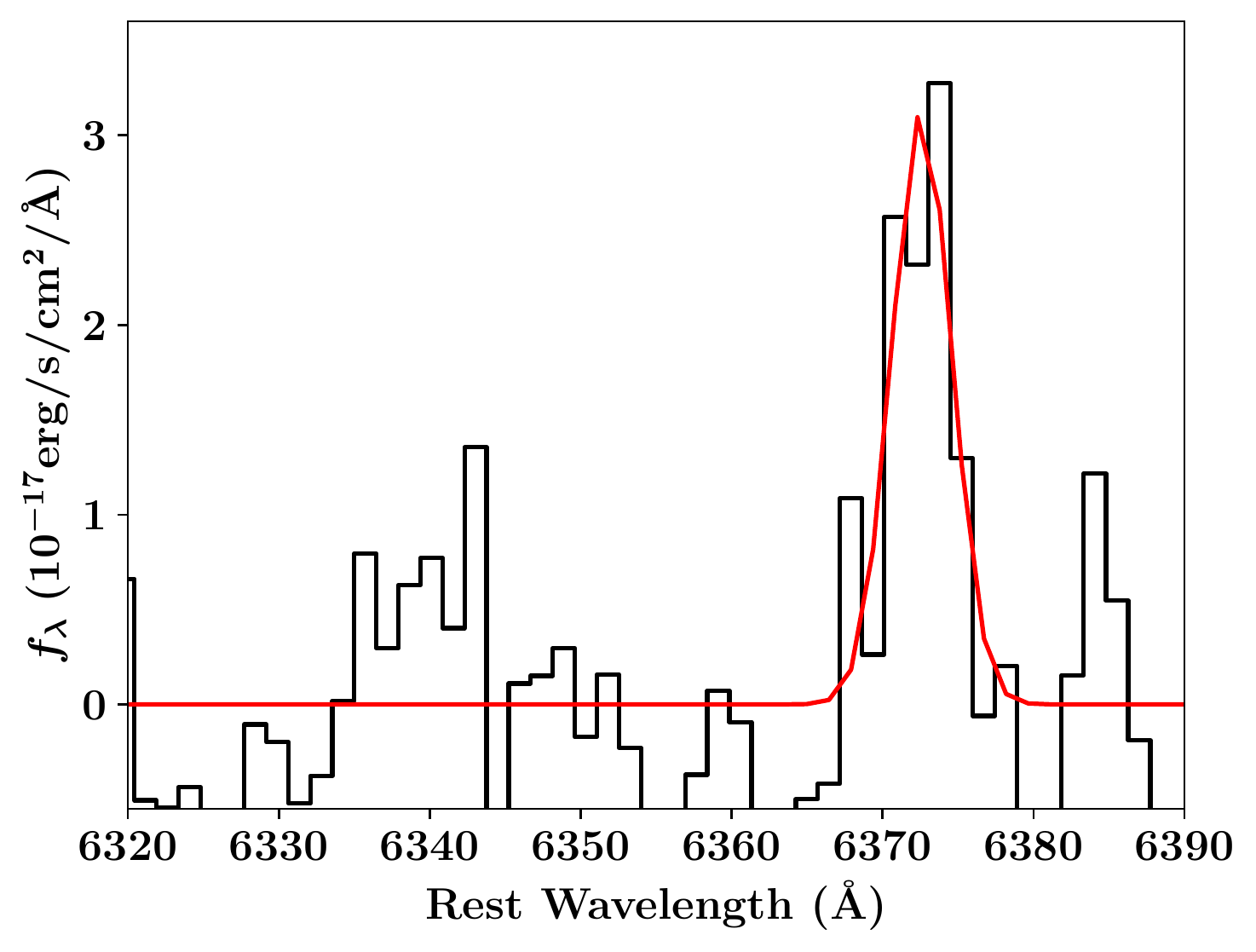}
    \hspace{1.5mm}
    \includegraphics[width=0.13\textwidth]{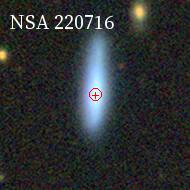}
        \hspace{-3mm}
    \includegraphics[width=0.19\textwidth]{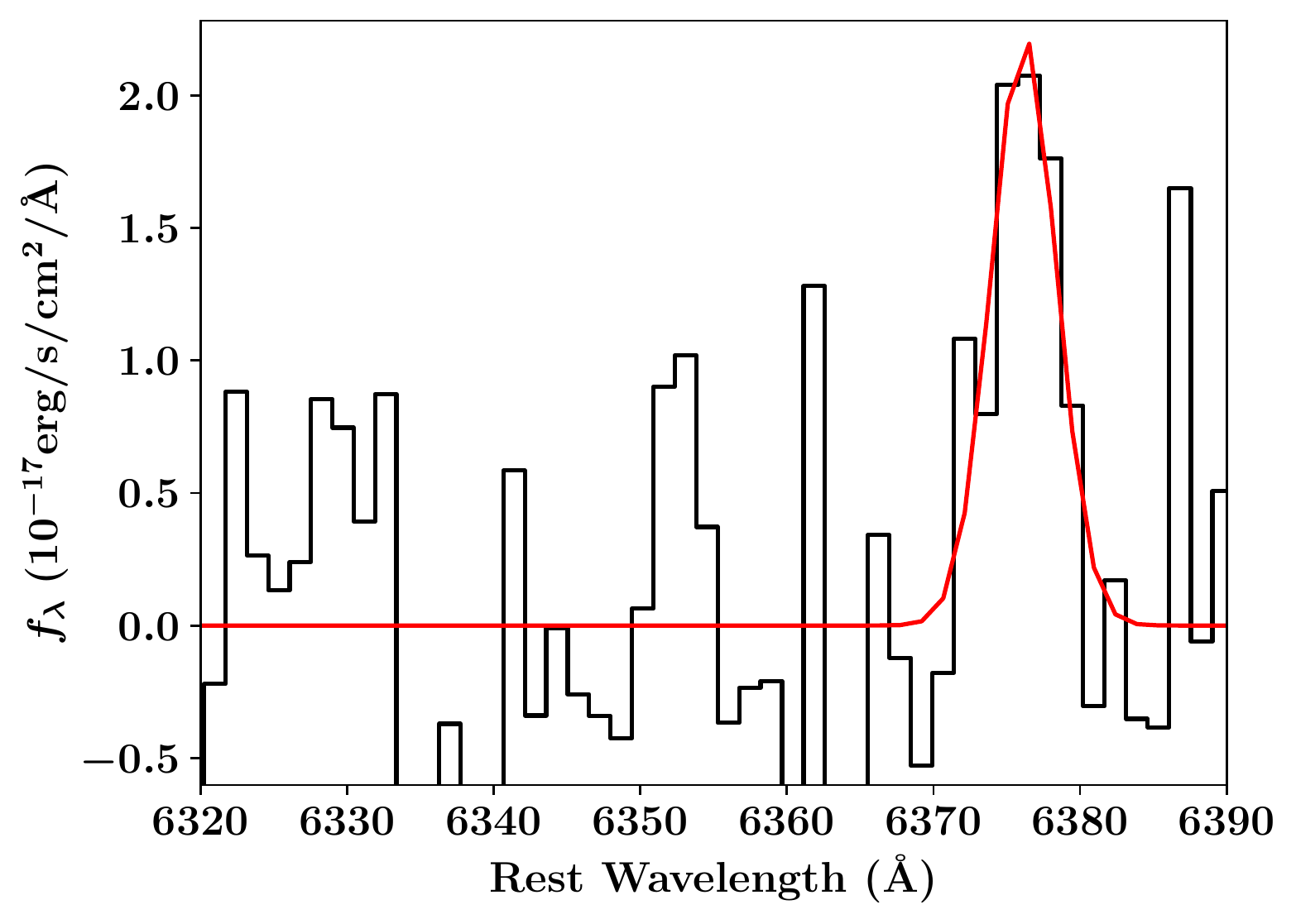}  
        \hspace{1.5mm}
    \includegraphics[width=0.13\textwidth]{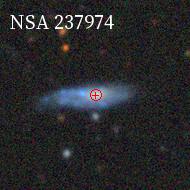}
        \hspace{-3mm}
    \includegraphics[width=0.19\textwidth]{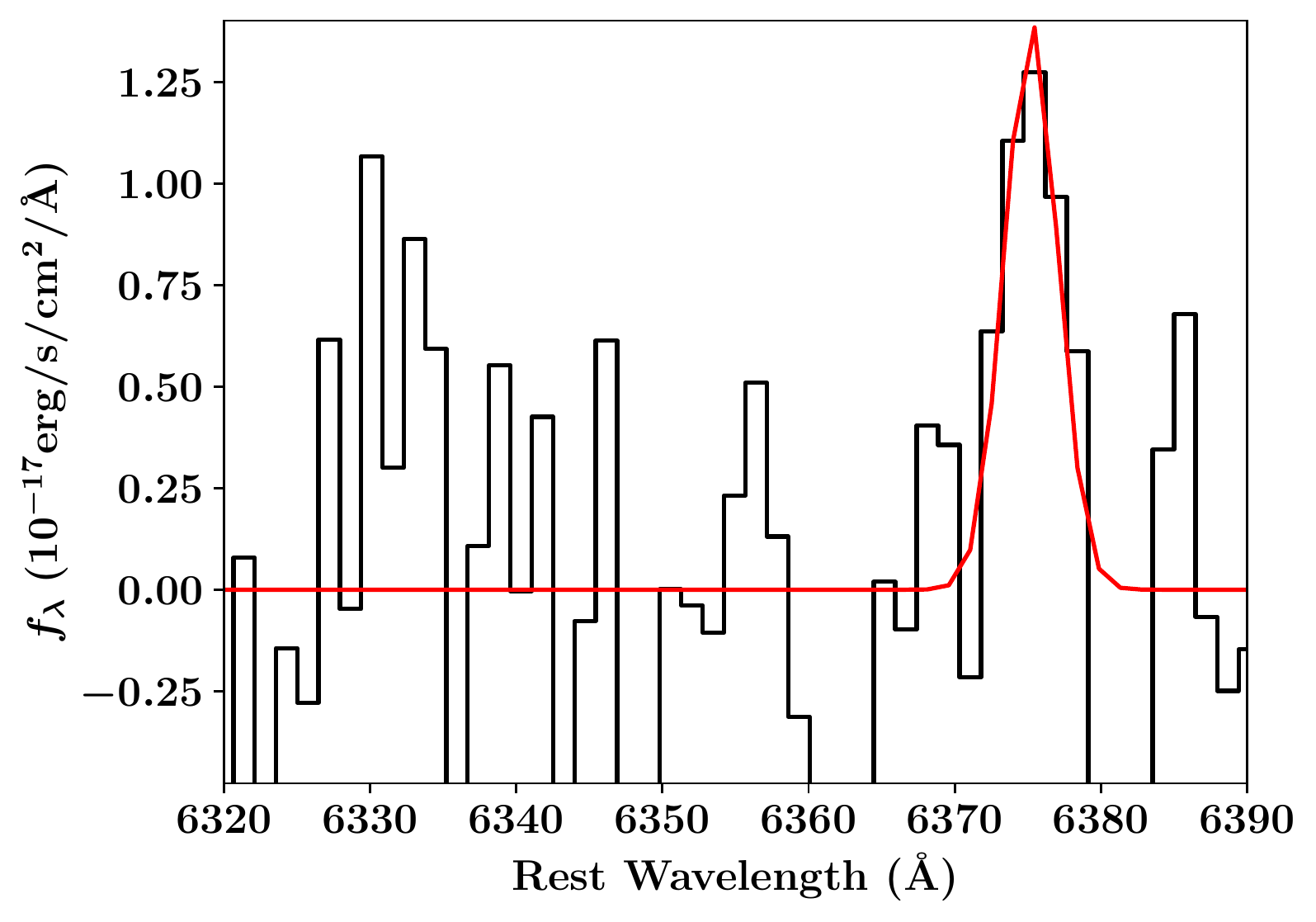}
        \includegraphics[width=0.13\textwidth]{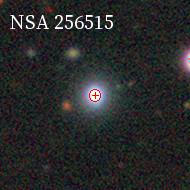}
    \hspace{-3mm}
    \includegraphics[width=0.19\textwidth]{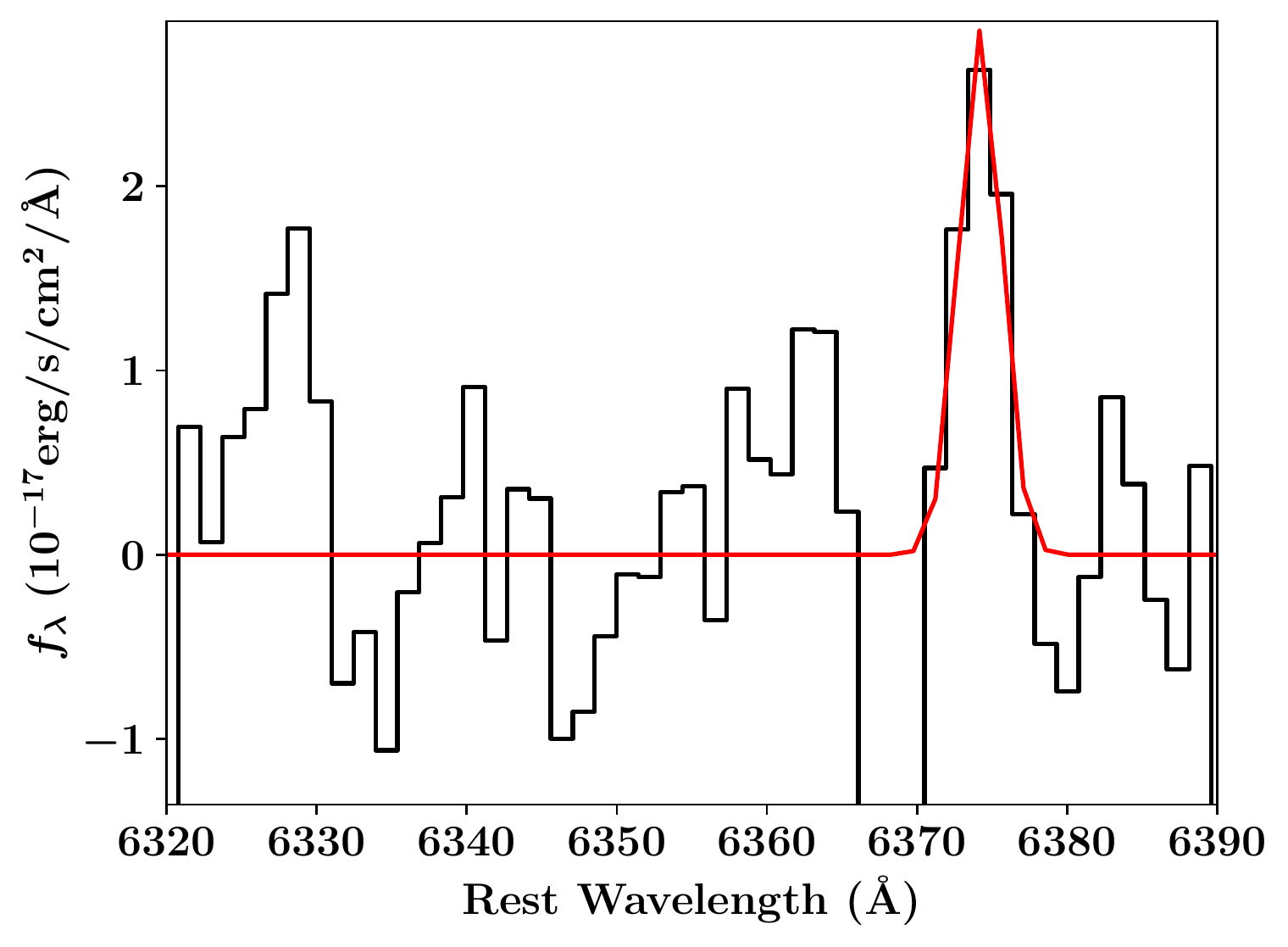}
    \hspace{1.5mm}
    \includegraphics[width=0.13\textwidth]{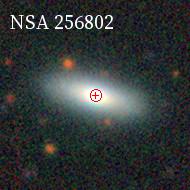}
        \hspace{-3mm}
    \includegraphics[width=0.19\textwidth]{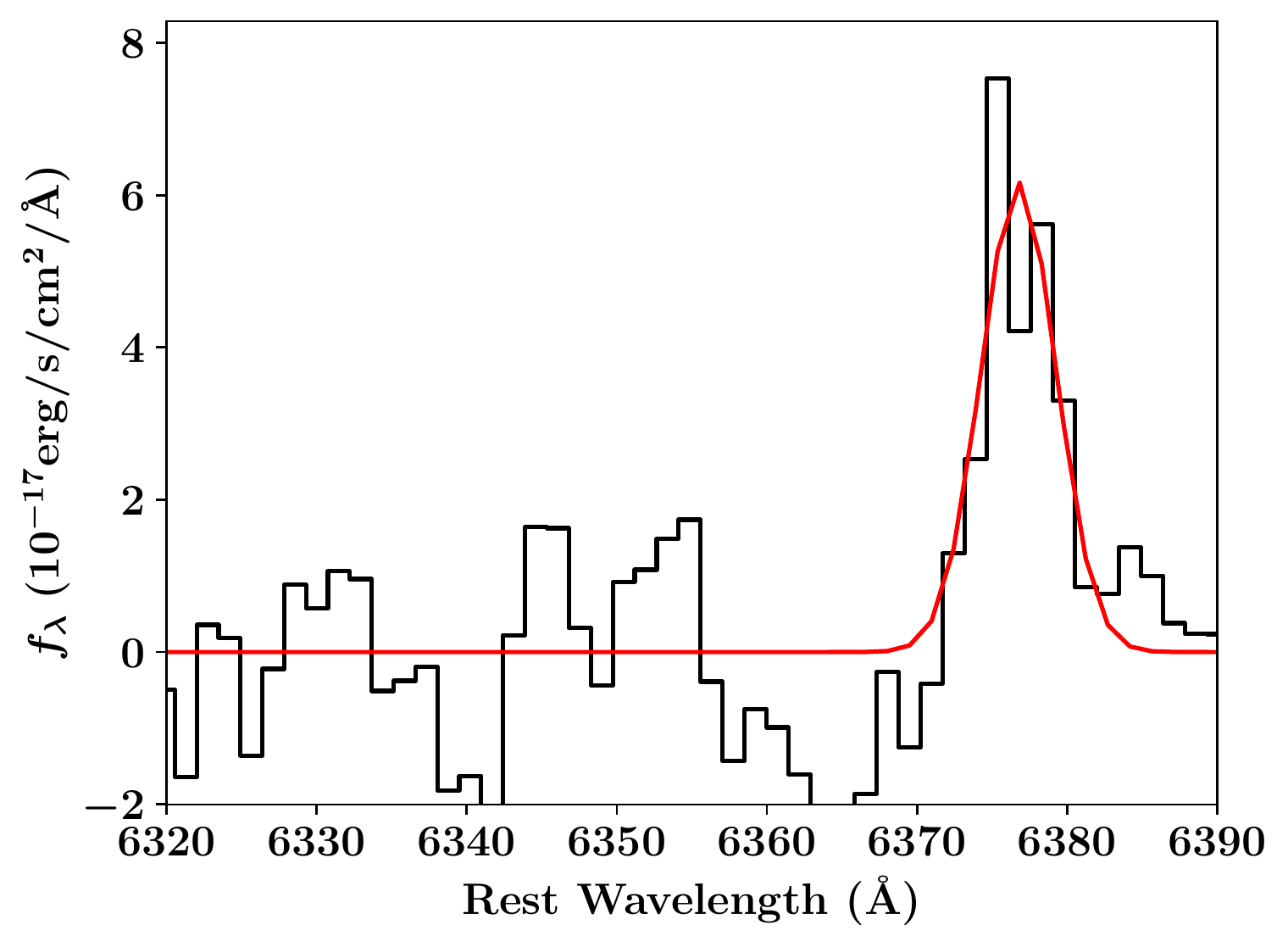}  
        \hspace{1.5mm}
    \includegraphics[width=0.13\textwidth]{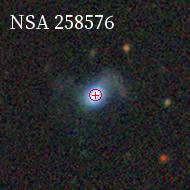}
        \hspace{-3mm}
    \includegraphics[width=0.19\textwidth]{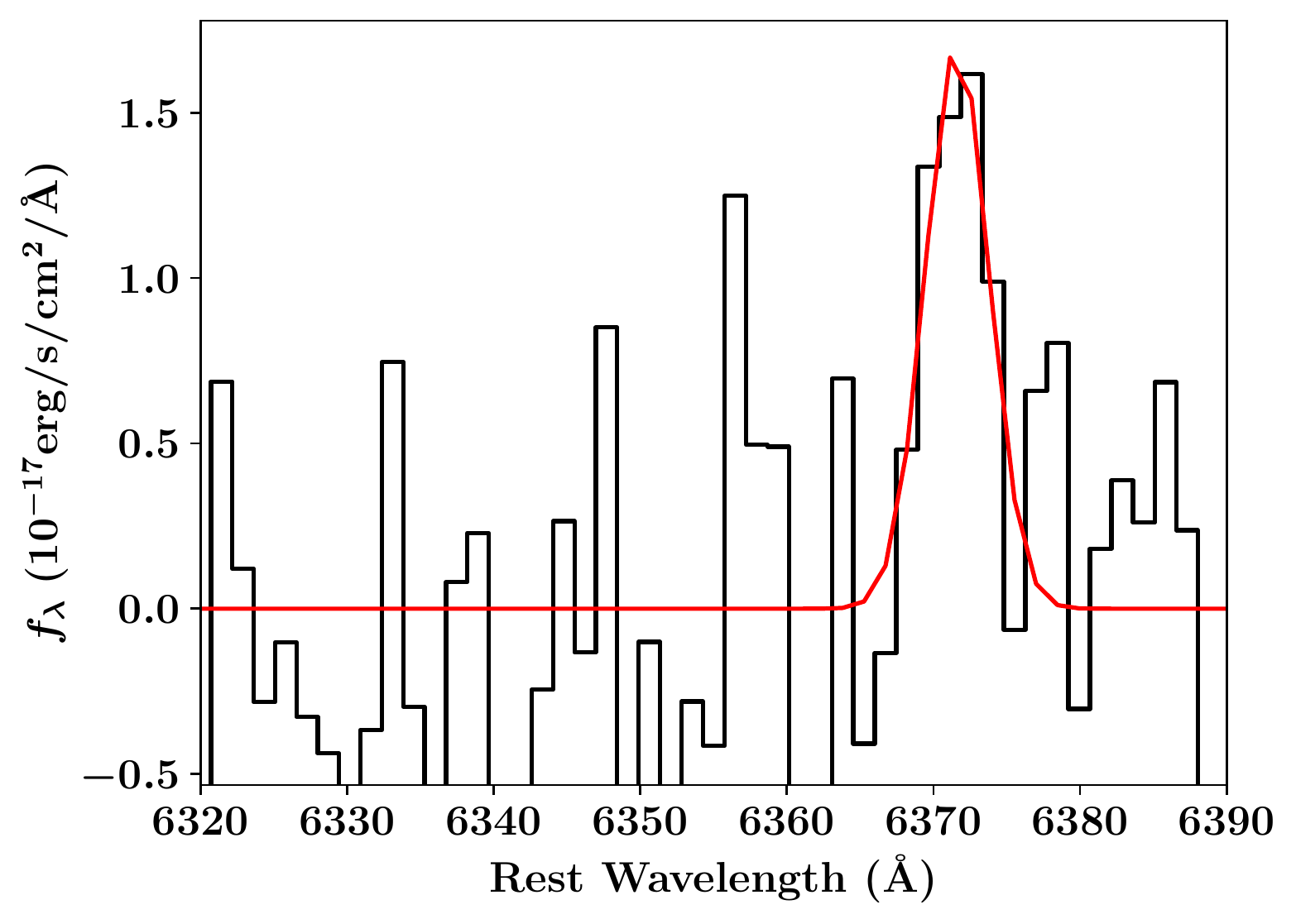}
        \includegraphics[width=0.13\textwidth]{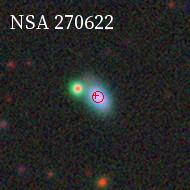}
    \hspace{-3mm}
    \includegraphics[width=0.19\textwidth]{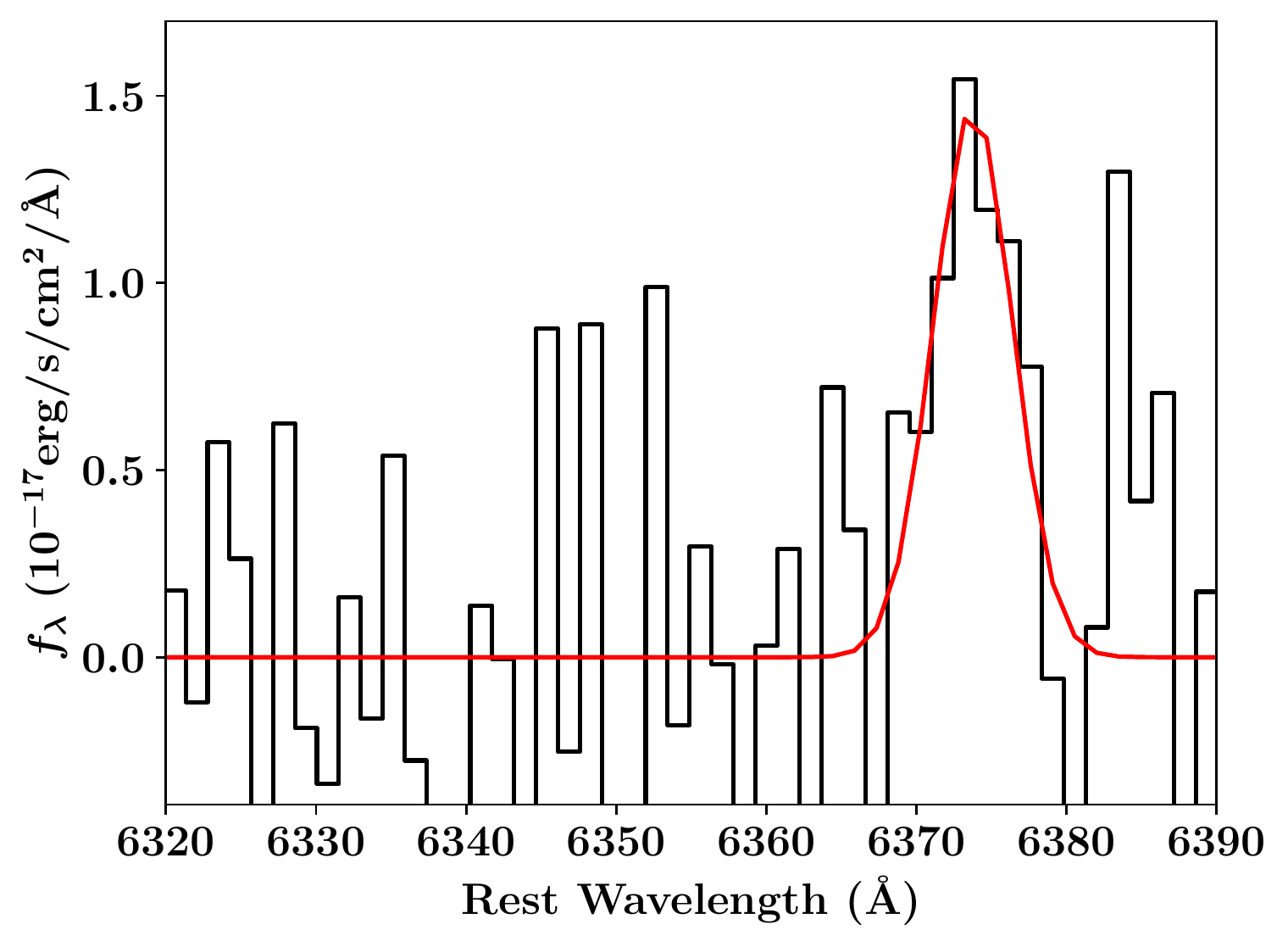}
    \hspace{1.5mm}
    \includegraphics[width=0.13\textwidth]{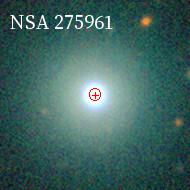}
        \hspace{-3mm}
    \includegraphics[width=0.19\textwidth]{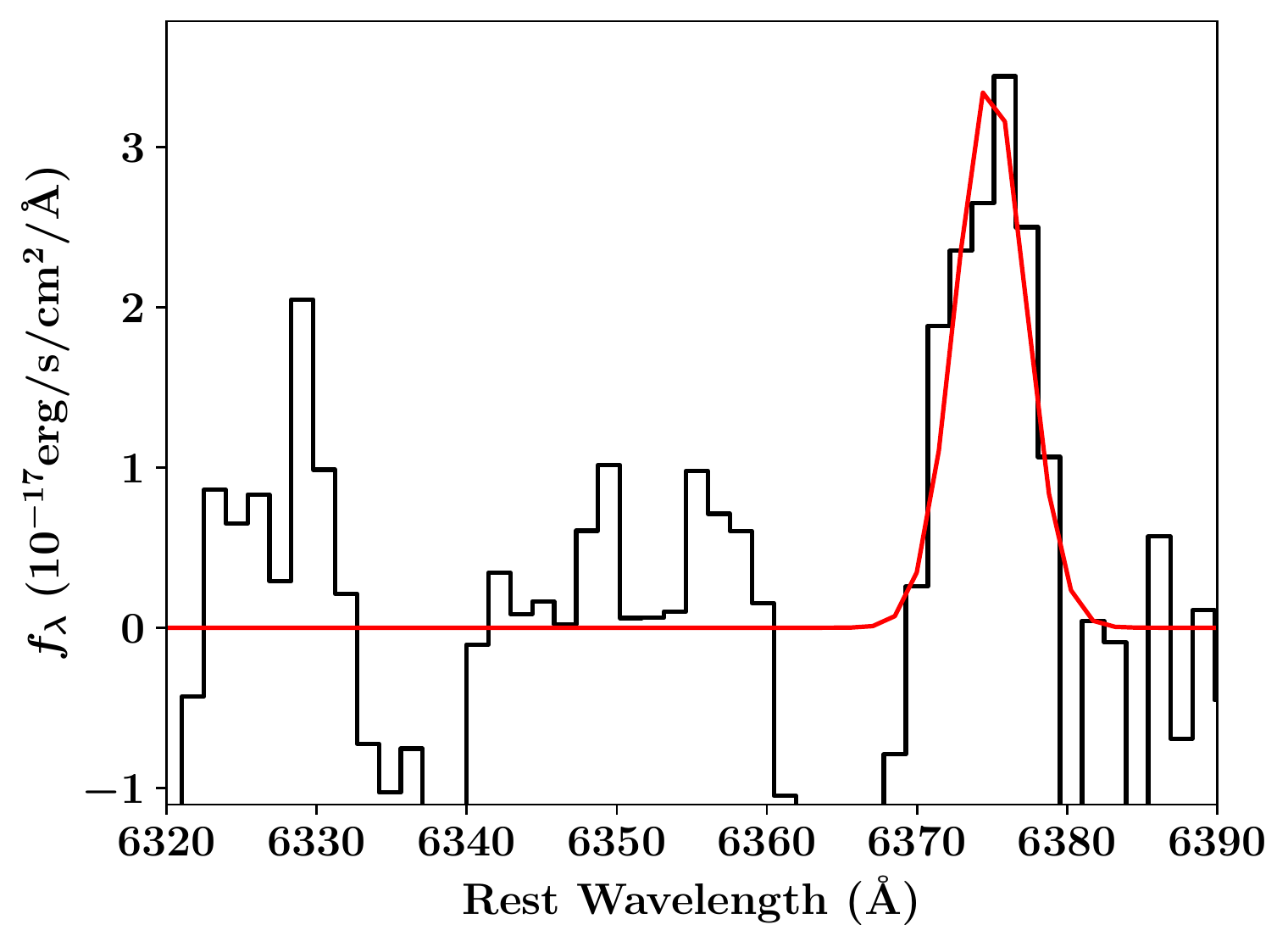}  
        \hspace{1.5mm}
    \includegraphics[width=0.13\textwidth]{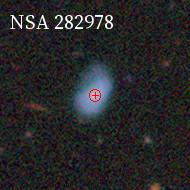}
        \hspace{-3mm}
    \includegraphics[width=0.19\textwidth]{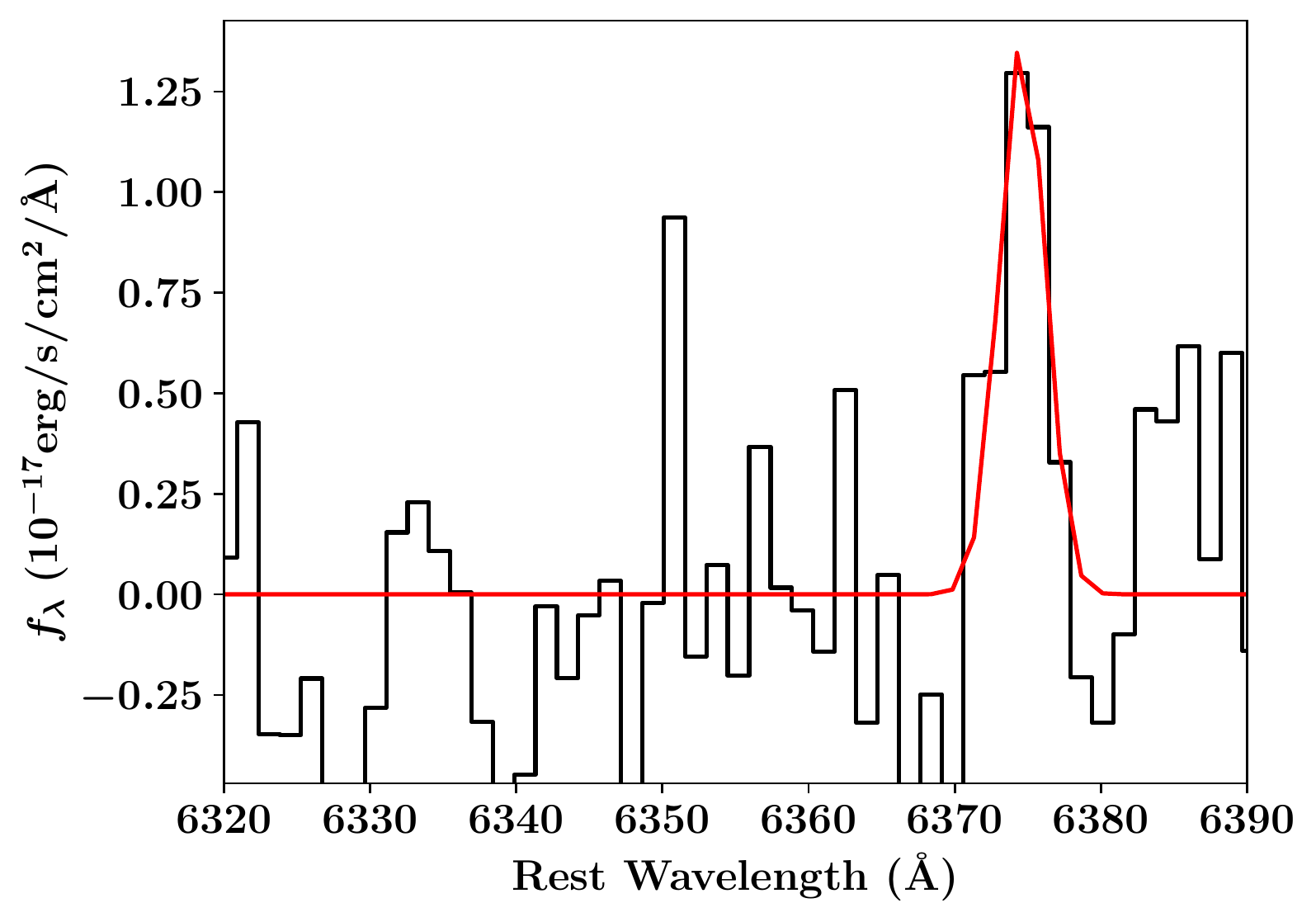}
    \includegraphics[width=0.13\textwidth]{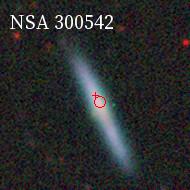}
    \hspace{-3mm}
    \includegraphics[width=0.19\textwidth]{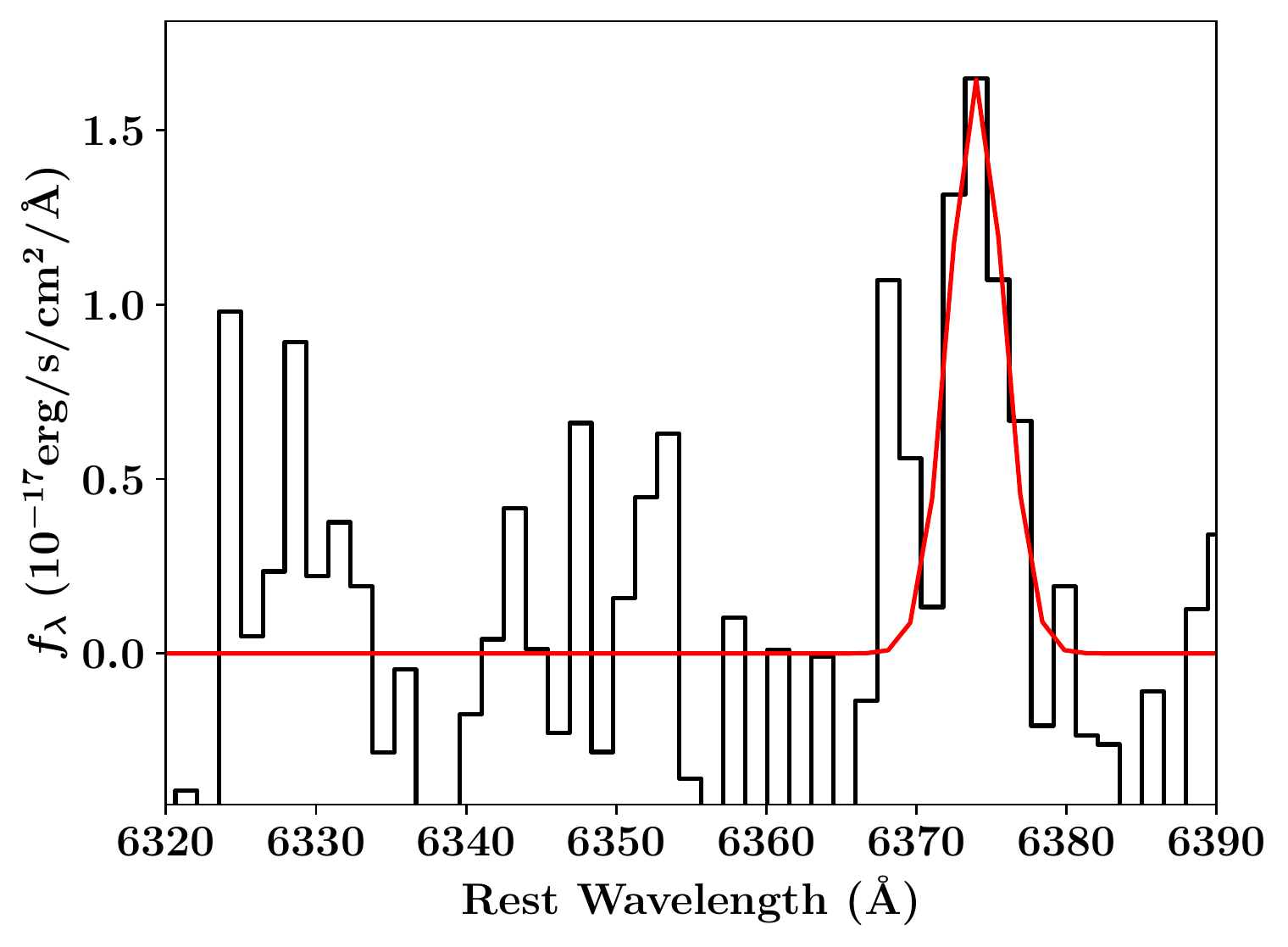}
    \hspace{1.5mm}
    \includegraphics[width=0.13\textwidth]{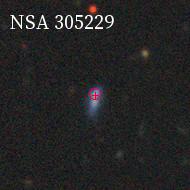}
        \hspace{-3mm}
    \includegraphics[width=0.19\textwidth]{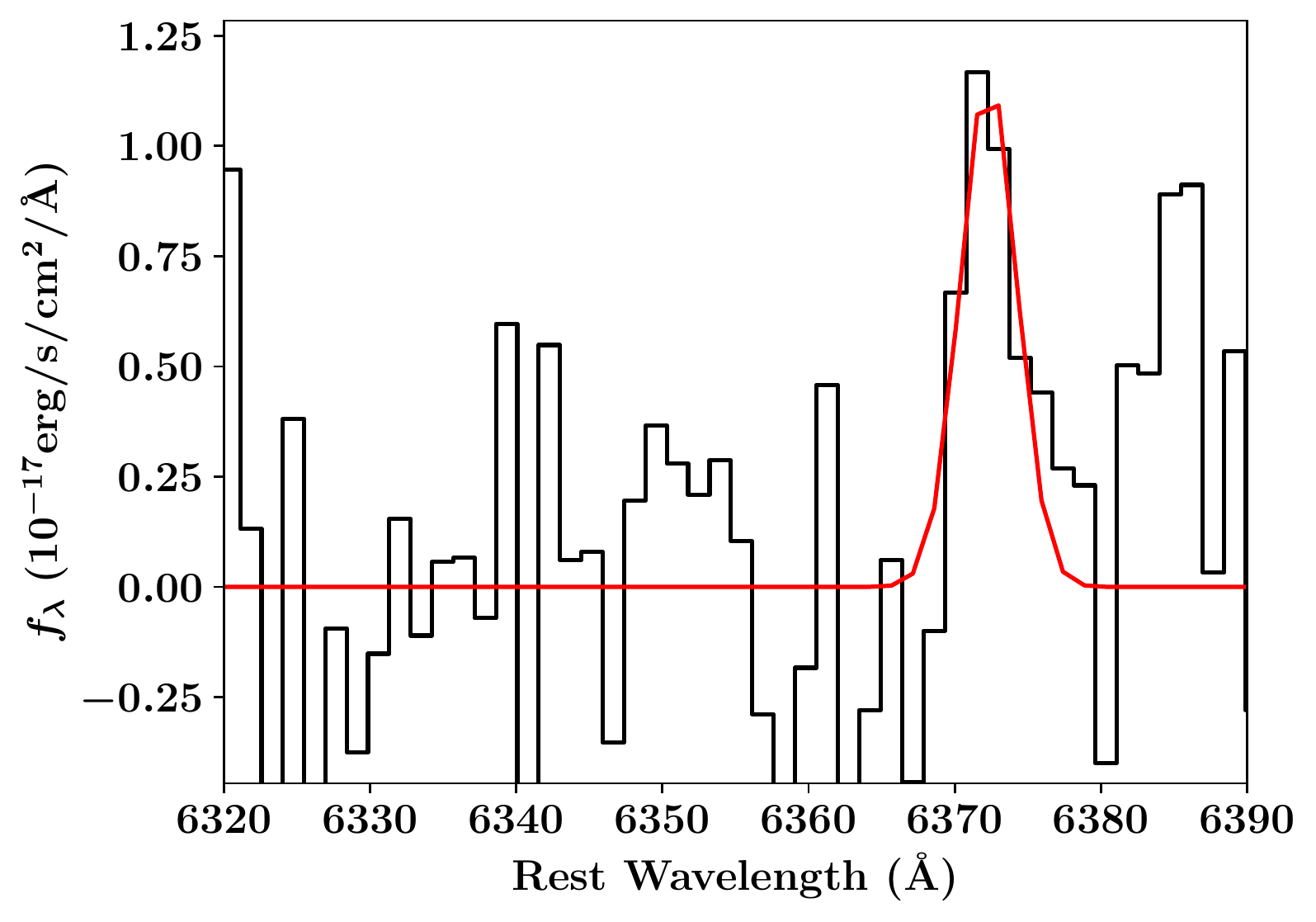}  
        \hspace{1.5mm}
    \includegraphics[width=0.13\textwidth]{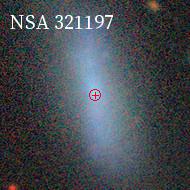}
        \hspace{-3mm}
    \includegraphics[width=0.19\textwidth]{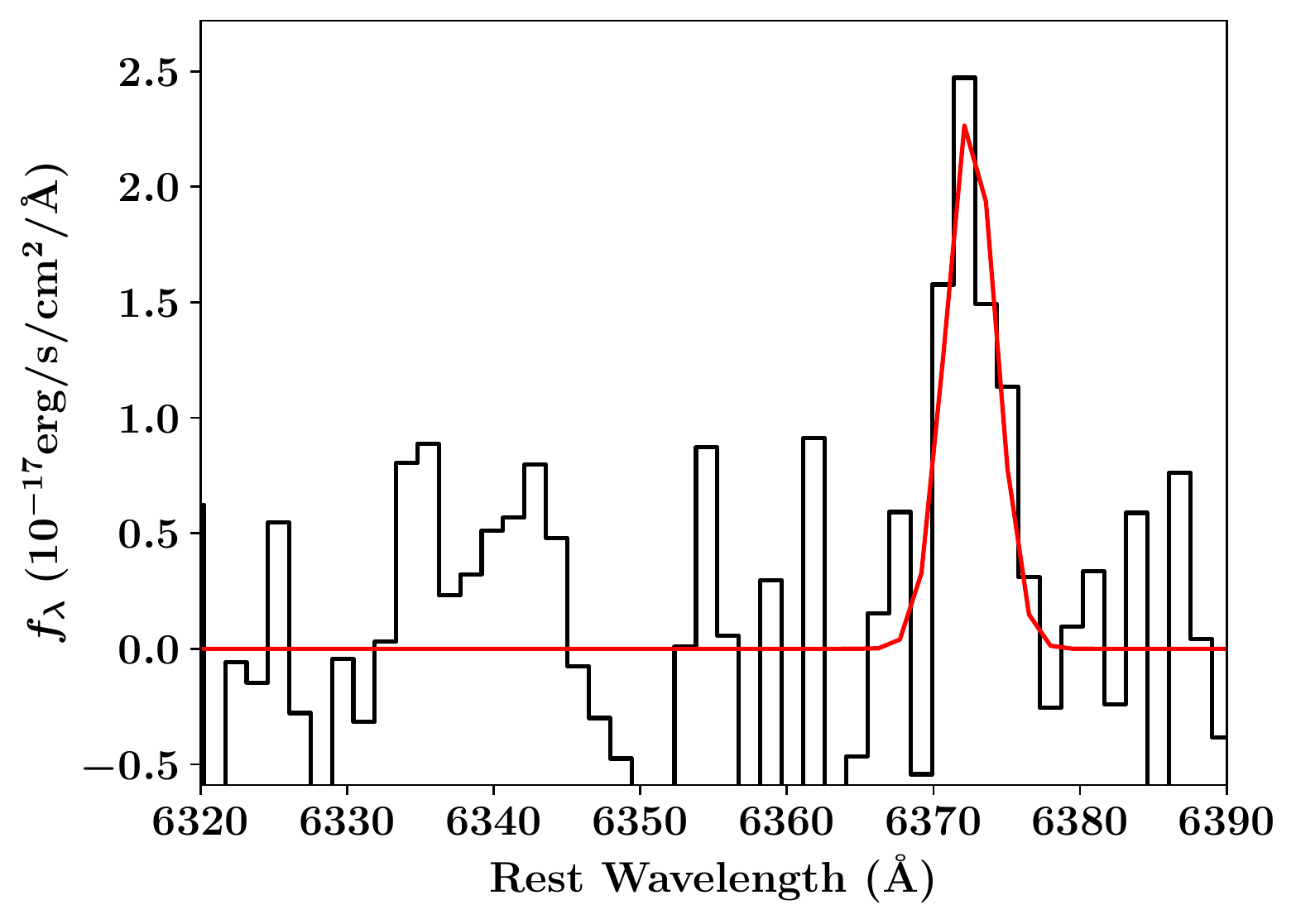}
    \includegraphics[width=0.13\textwidth]{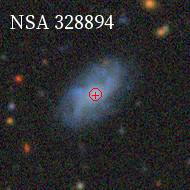}
    \hspace{-3mm}
    \includegraphics[width=0.19\textwidth]{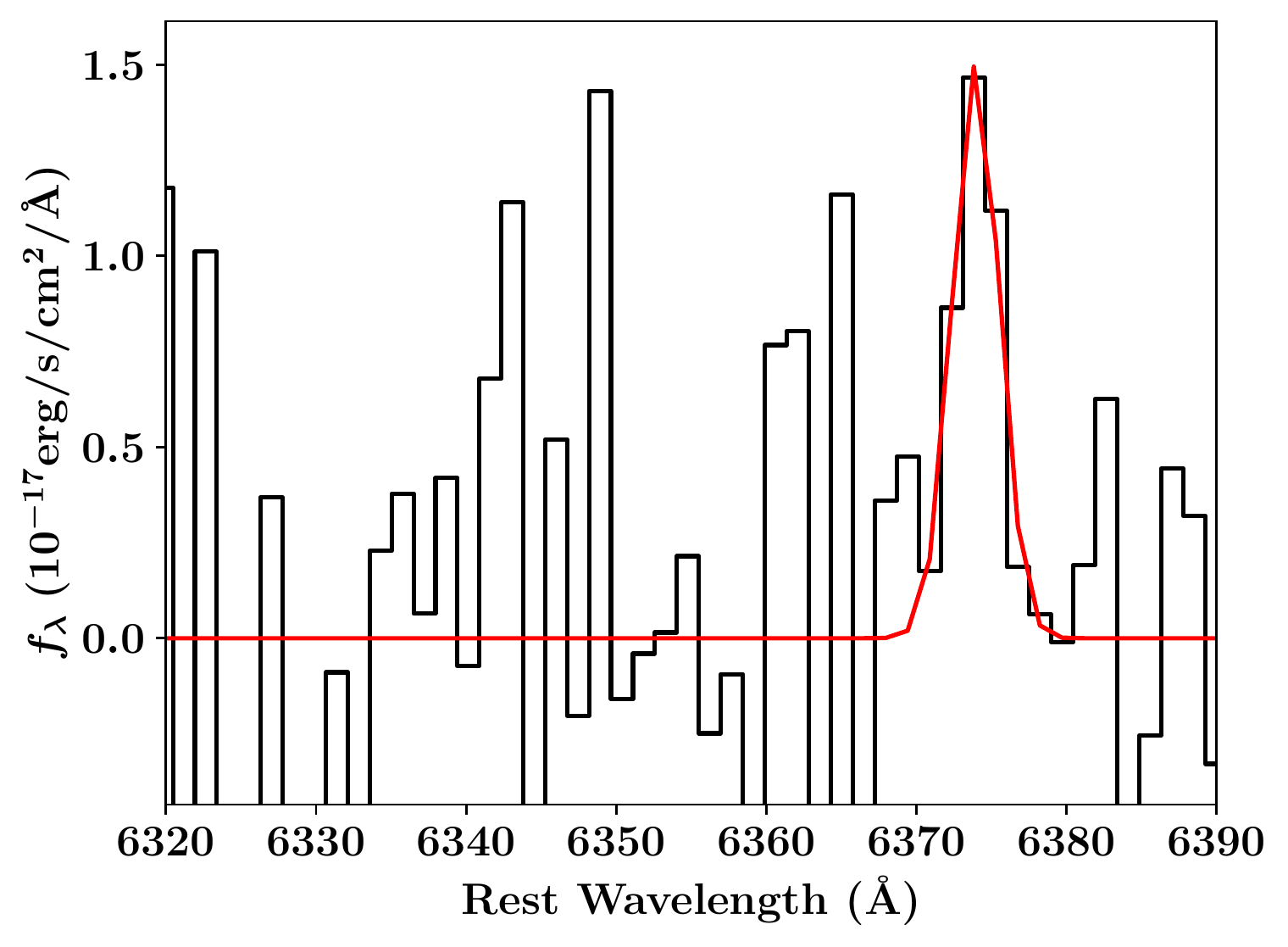}
    \hspace{1.5mm}
    \includegraphics[width=0.13\textwidth]{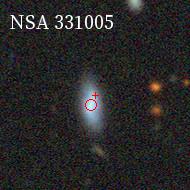}
        \hspace{-3mm}
    \includegraphics[width=0.19\textwidth]{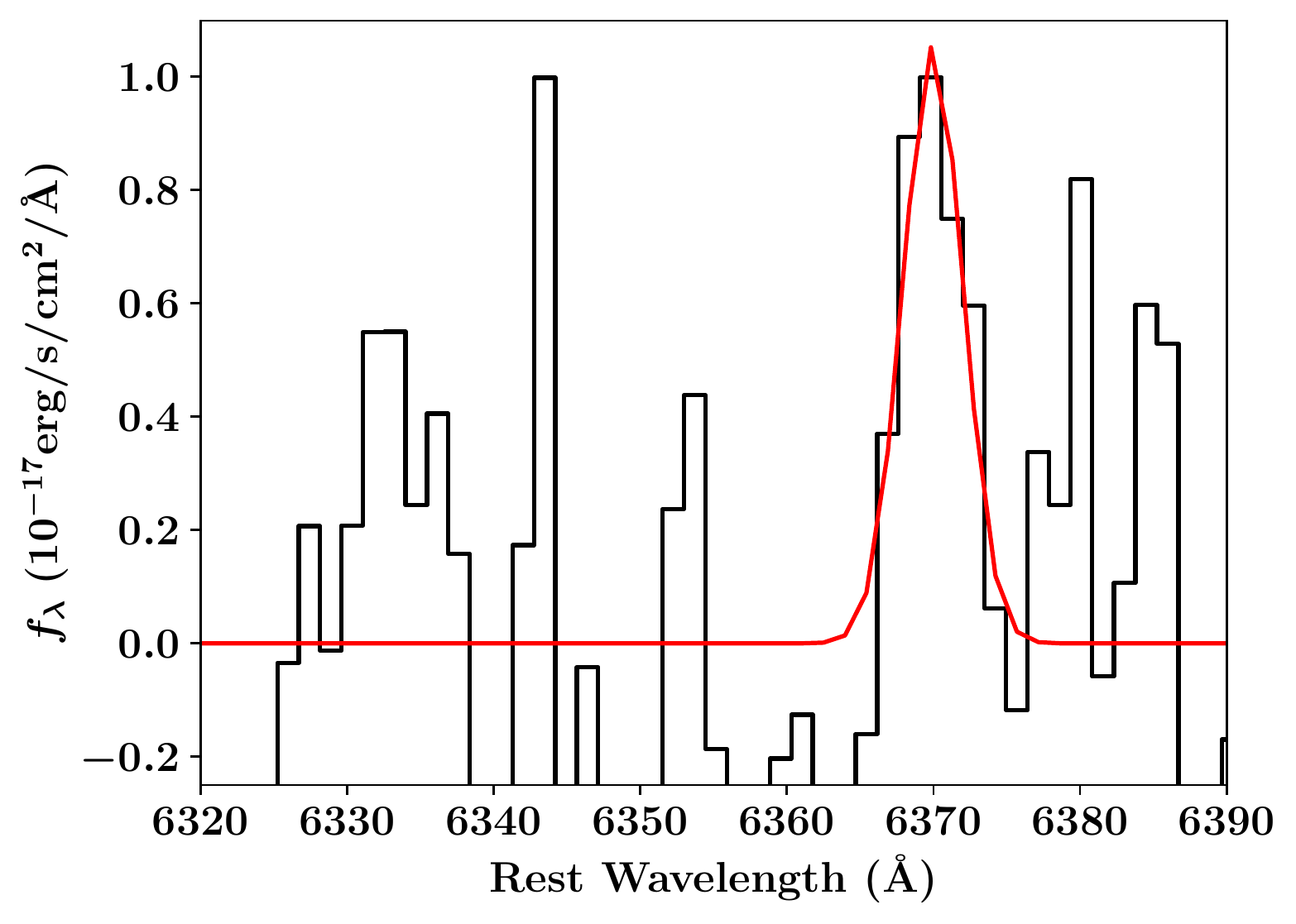}  
        \hspace{1.5mm}
    \includegraphics[width=0.13\textwidth]{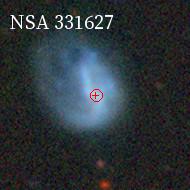}
        \hspace{-3mm}
    \includegraphics[width=0.19\textwidth]{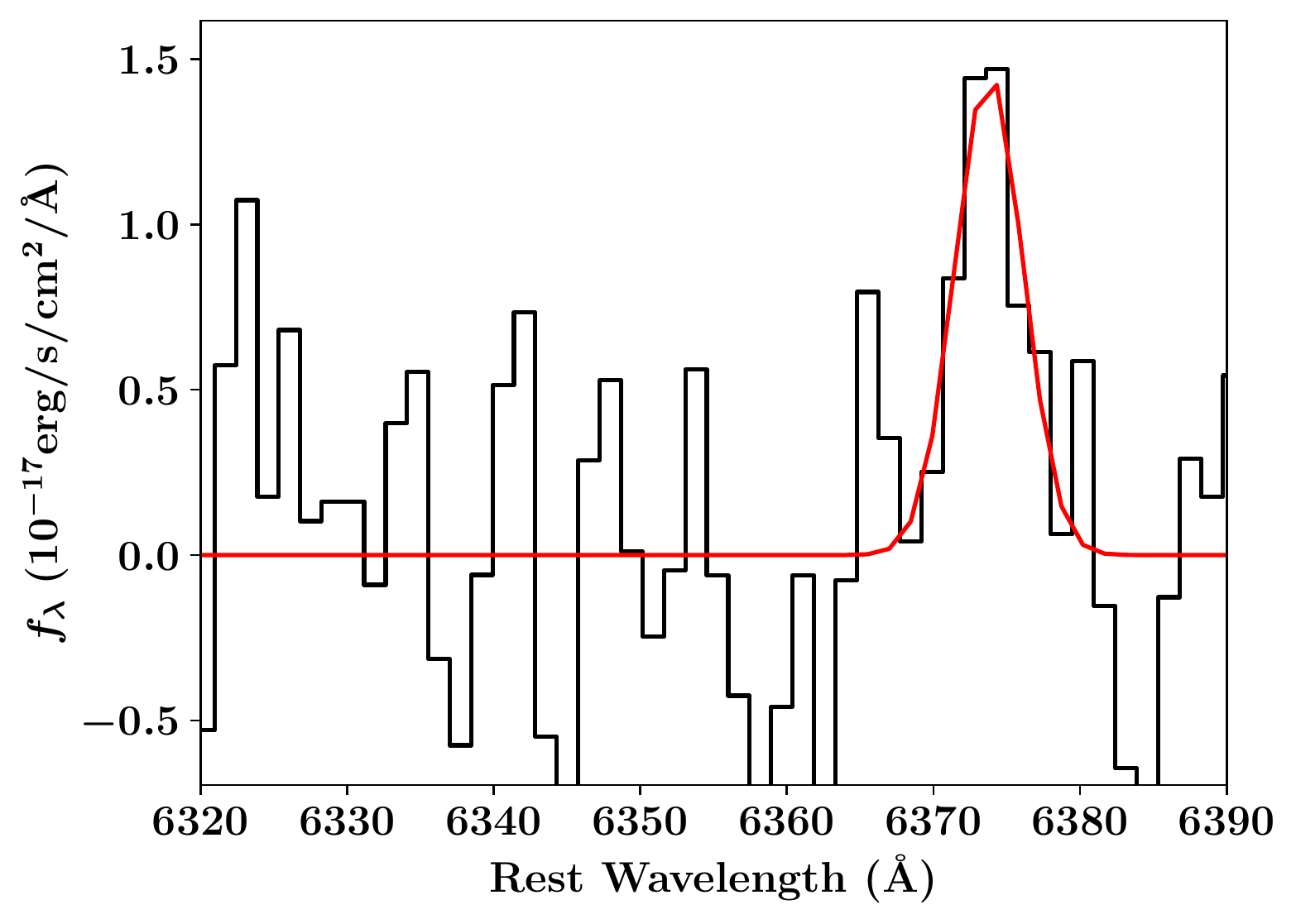}
    \includegraphics[width=0.13\textwidth]{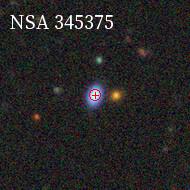}
    \hspace{-3mm}
    \includegraphics[width=0.19\textwidth]{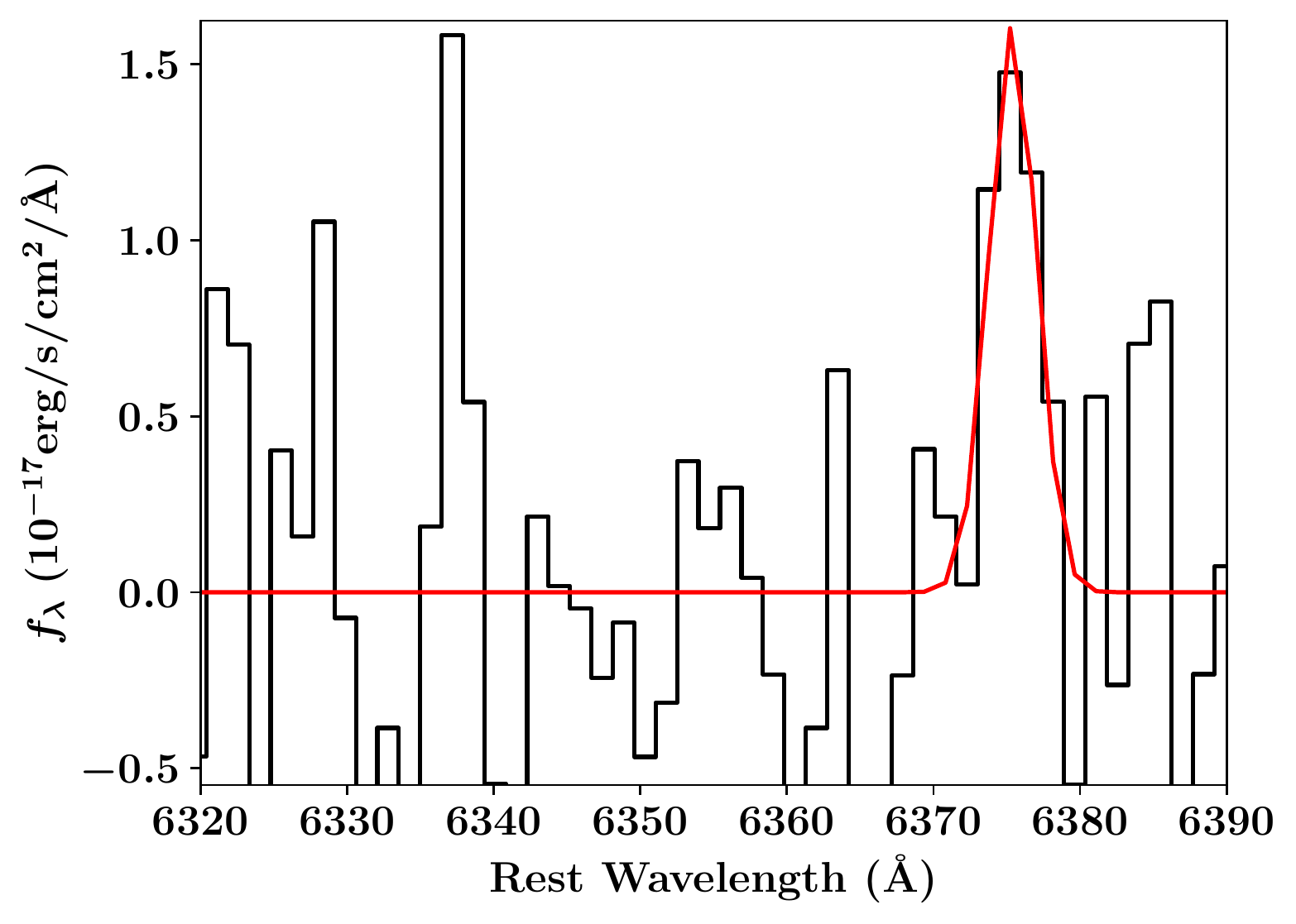}
    \hspace{1.5mm}
    \includegraphics[width=0.13\textwidth]{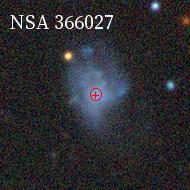}
        \hspace{-3mm}
    \includegraphics[width=0.19\textwidth]{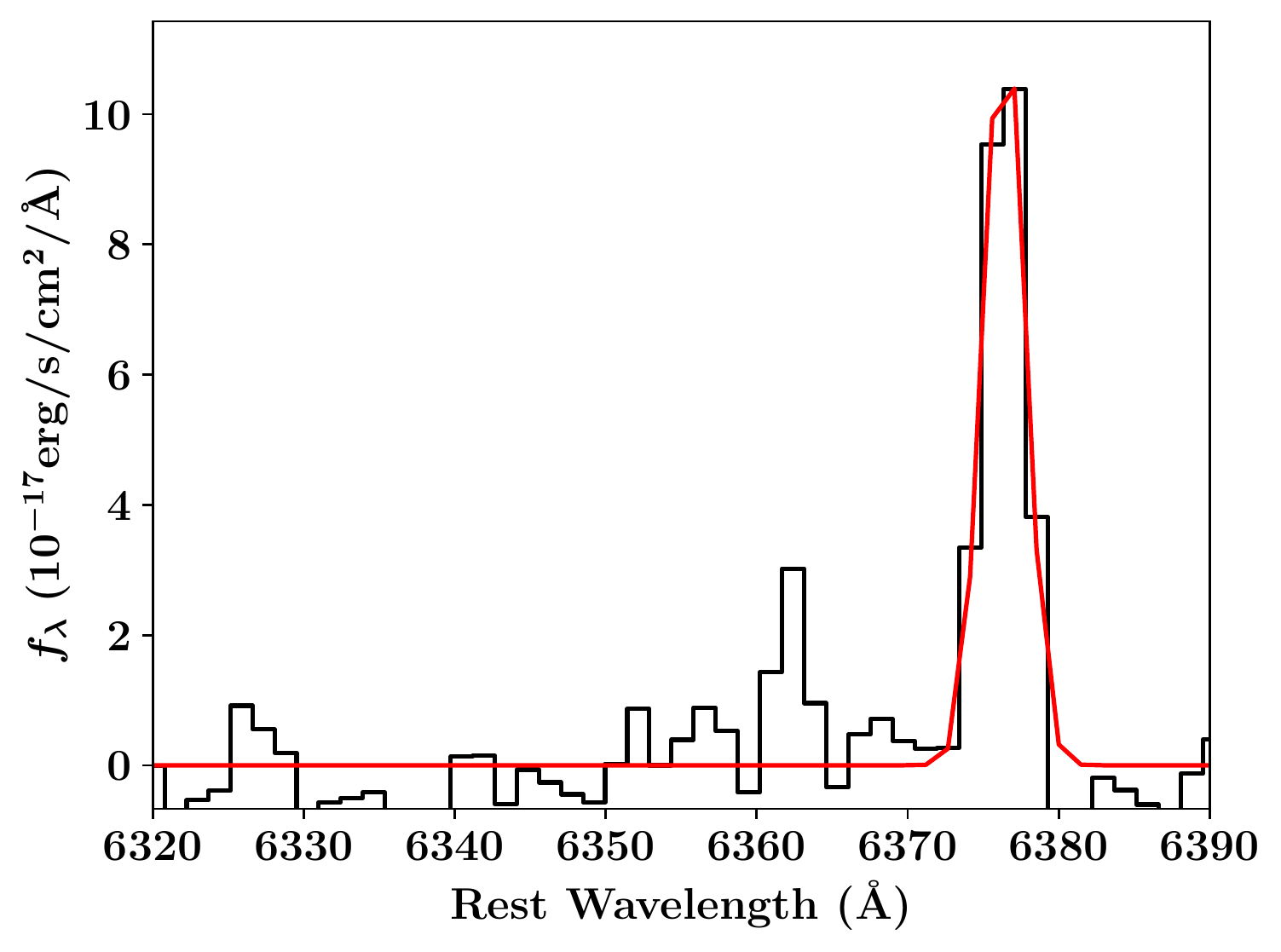}  
        \hspace{1.5mm}
    \includegraphics[width=0.13\textwidth]{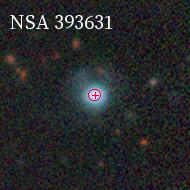}
        \hspace{-3mm}
    \includegraphics[width=0.19\textwidth]{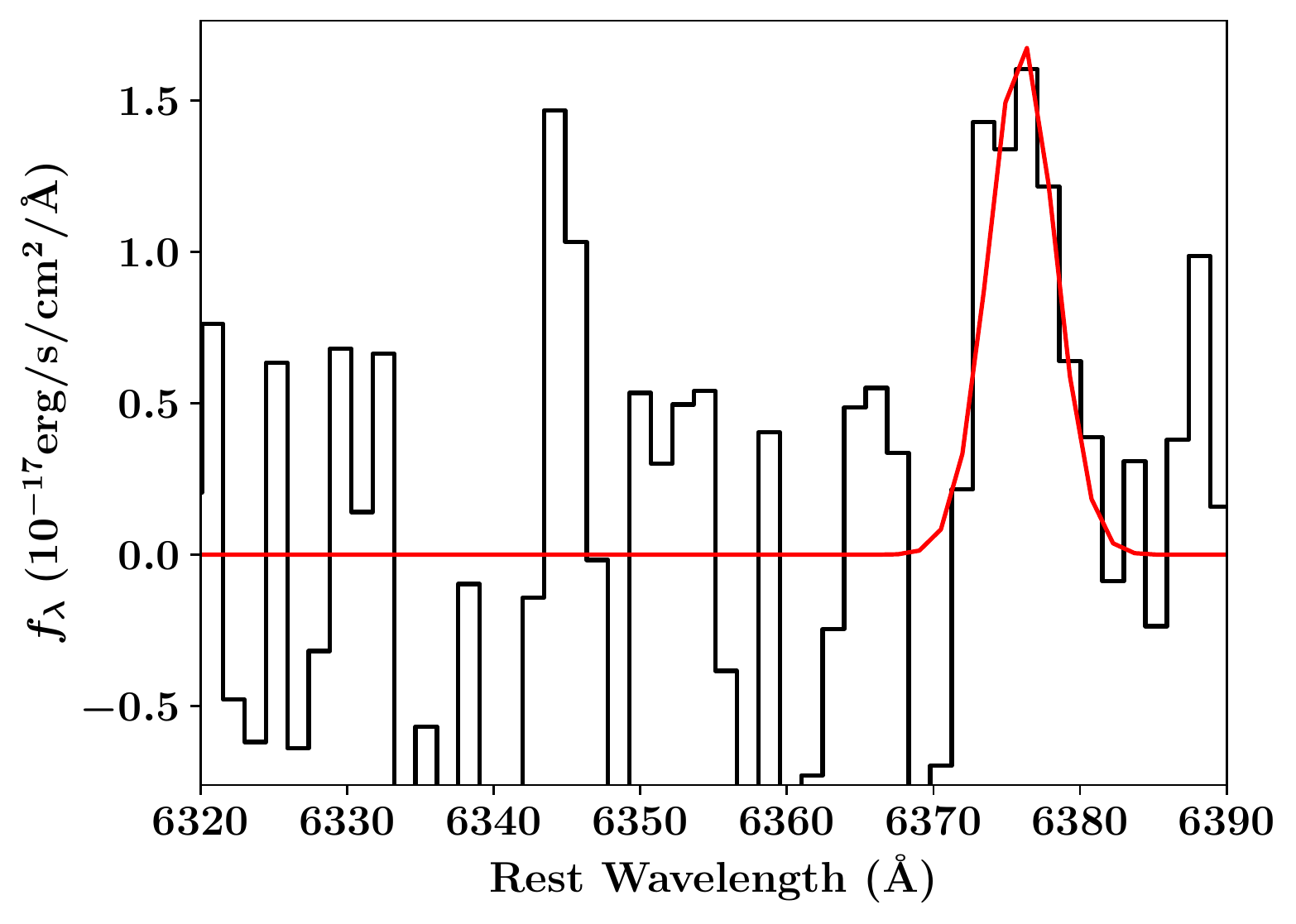}
    \includegraphics[width=0.13\textwidth]{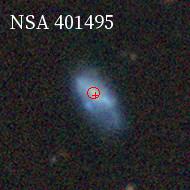}
    \hspace{-3mm}
    \includegraphics[width=0.19\textwidth]{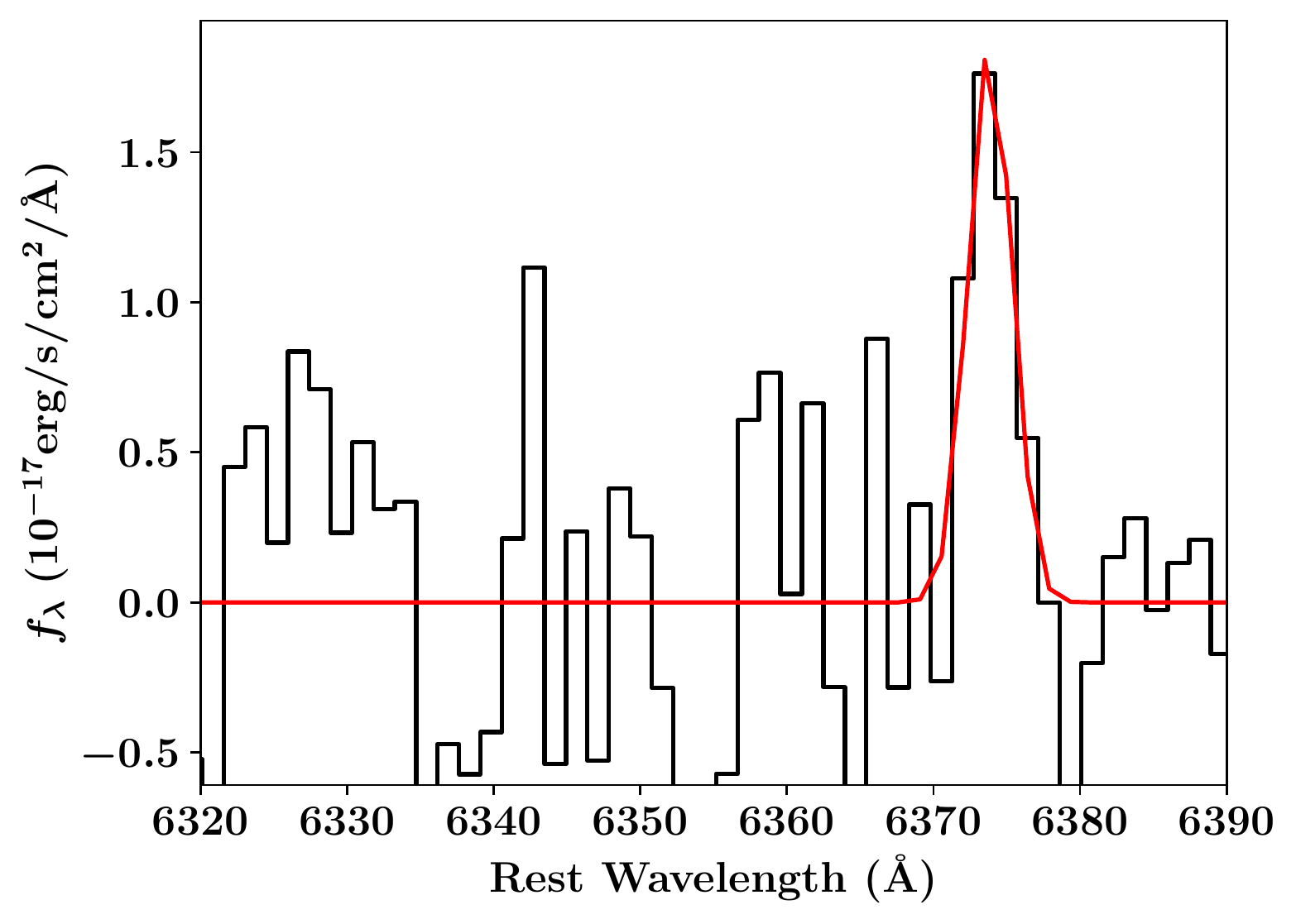}
    \hspace{1.5mm}
    \includegraphics[width=0.13\textwidth]{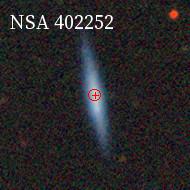}
        \hspace{-3mm}
    \includegraphics[width=0.19\textwidth]{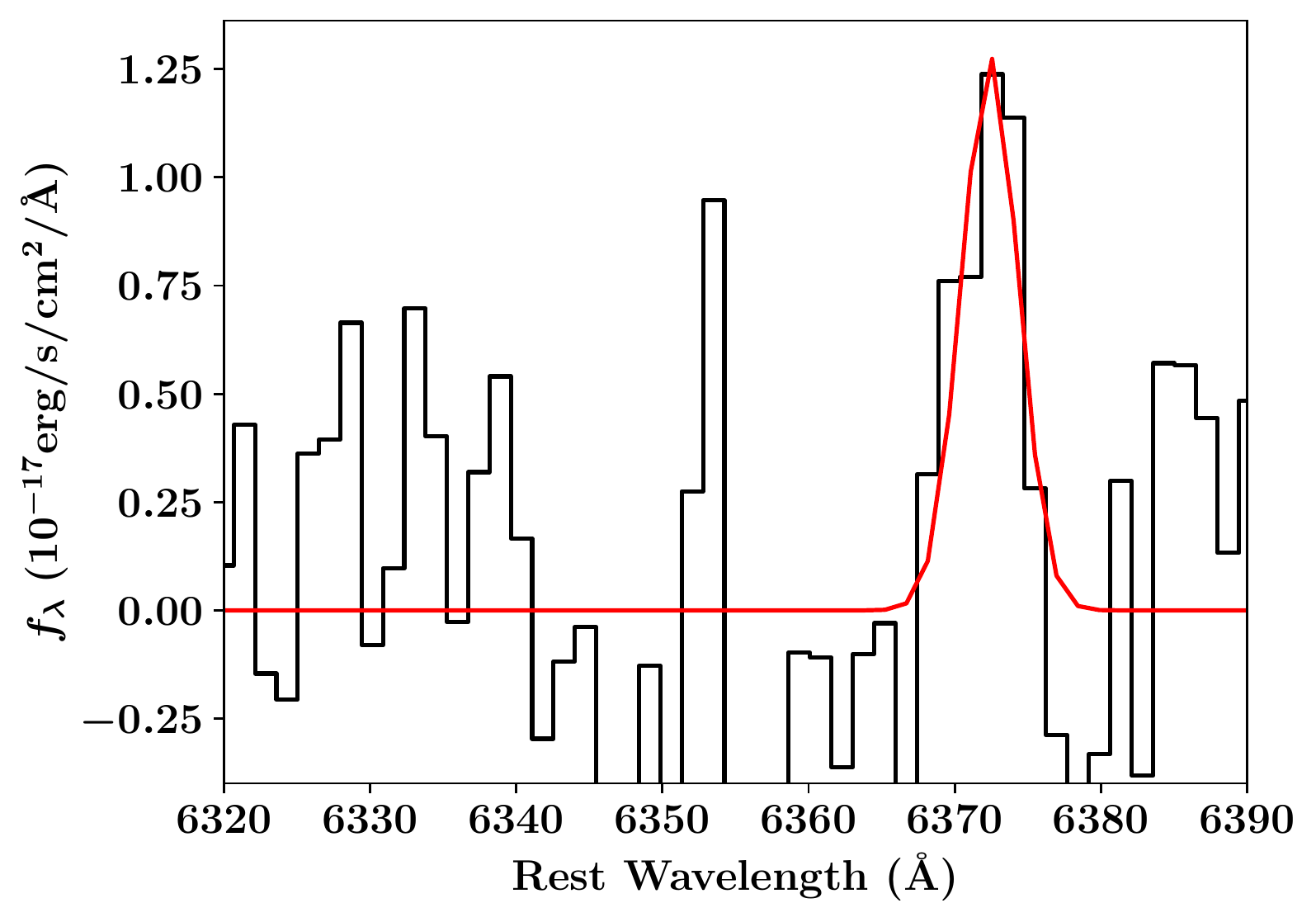}  
        \hspace{1.5mm}
    \includegraphics[width=0.13\textwidth]{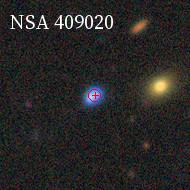}
        \hspace{-3mm}
    \includegraphics[width=0.19\textwidth]{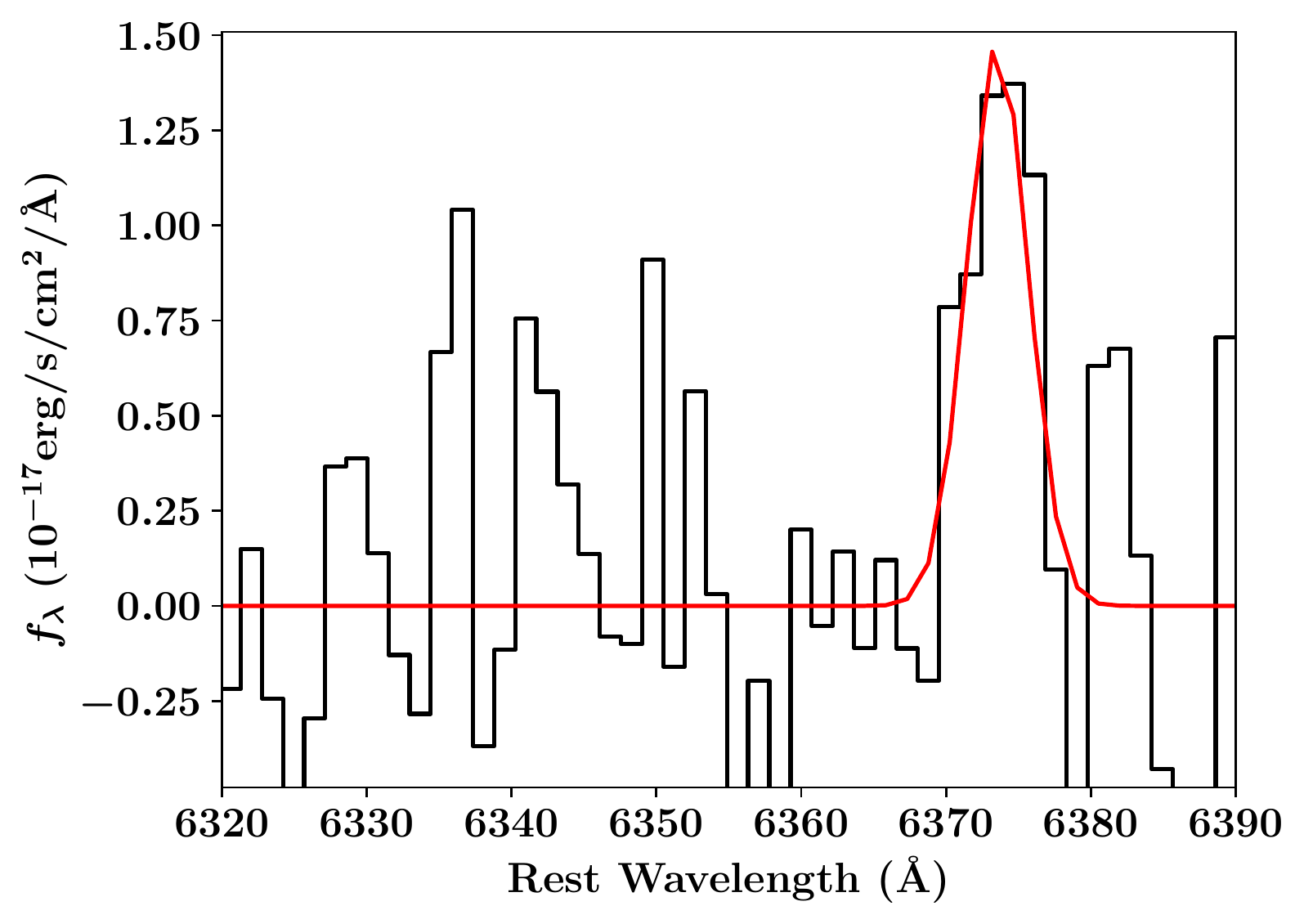}
    \includegraphics[width=0.13\textwidth]{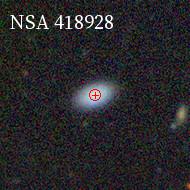}
    \hspace{-3mm}
    \includegraphics[width=0.19\textwidth]{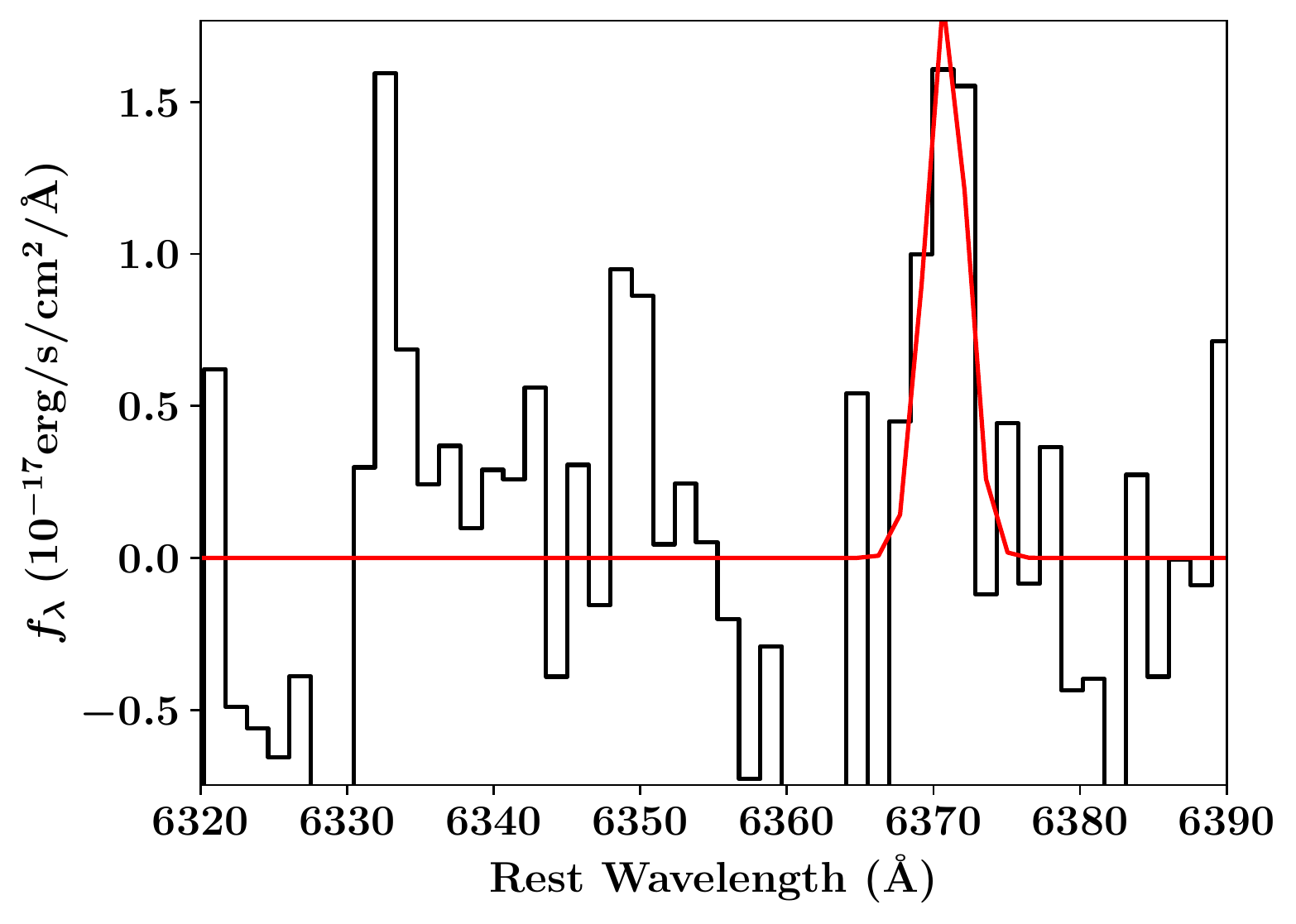}
    \hspace{1.5mm}
    \includegraphics[width=0.13\textwidth]{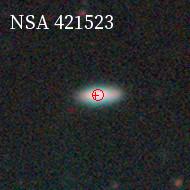}
        \hspace{-3mm}
    \includegraphics[width=0.19\textwidth]{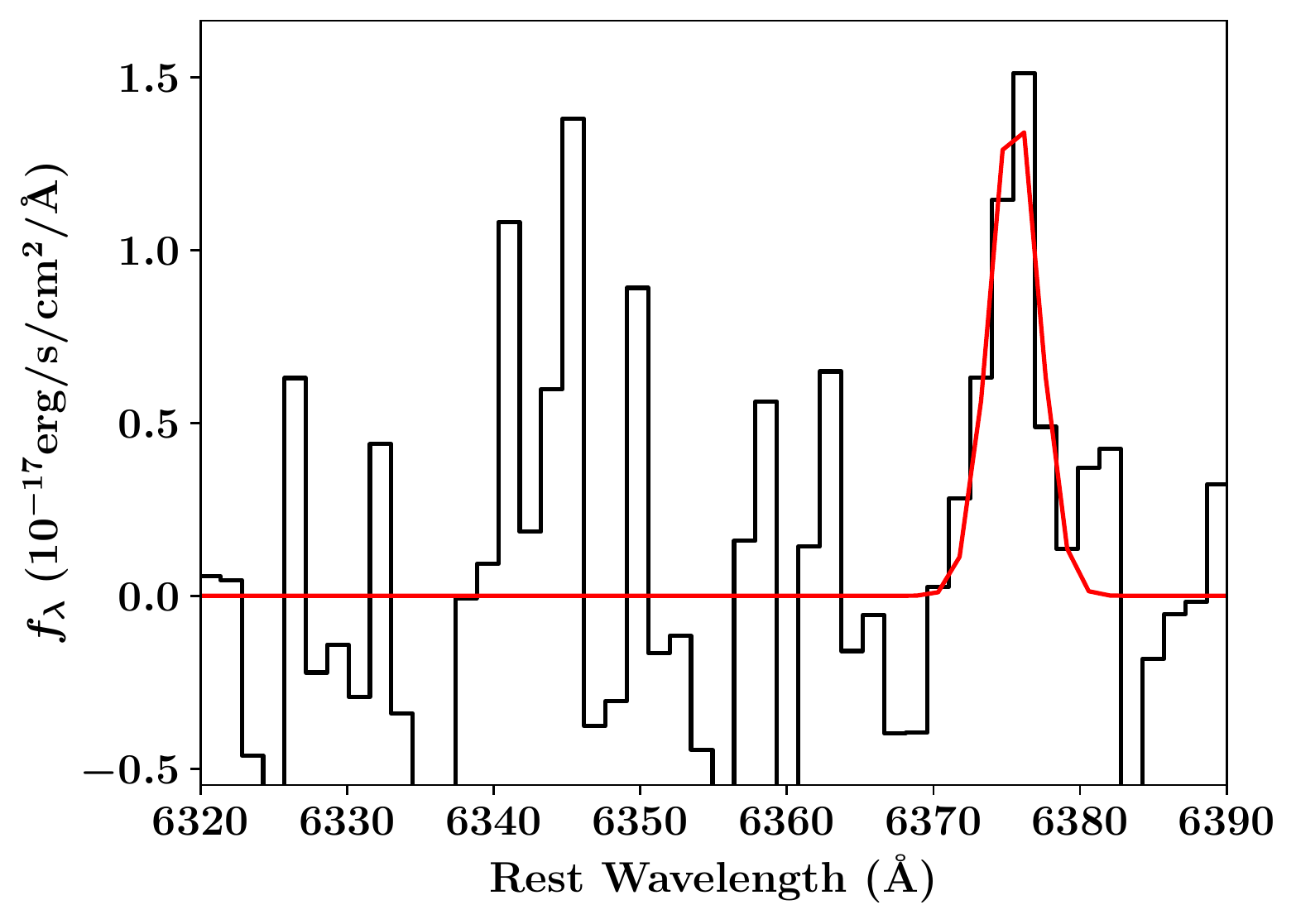}  
        \hspace{1.5mm}
    \includegraphics[width=0.13\textwidth]{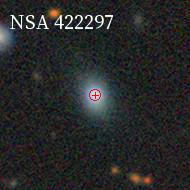}
        \hspace{-3mm}
    \includegraphics[width=0.19\textwidth]{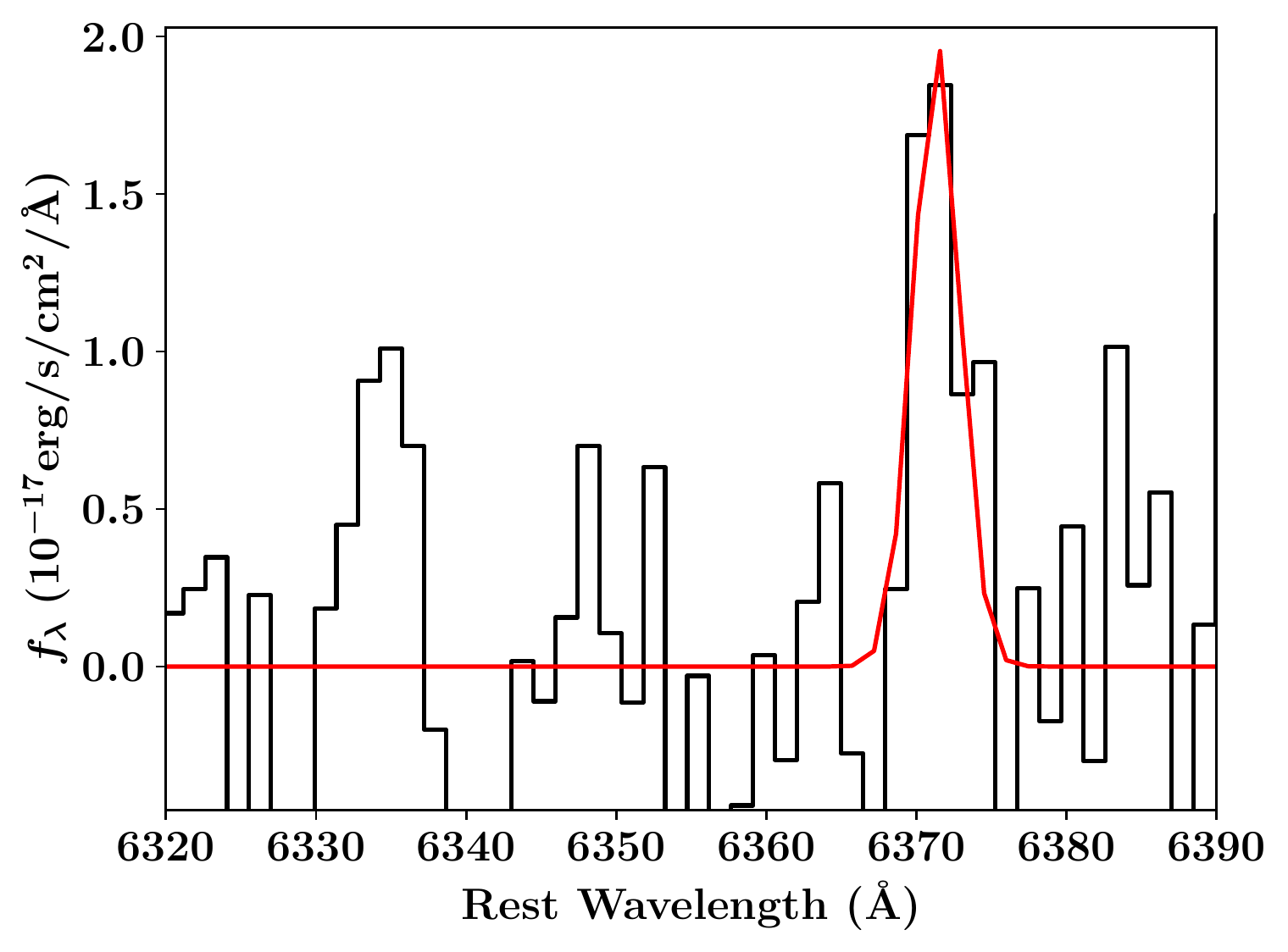}
        \includegraphics[width=0.13\textwidth]{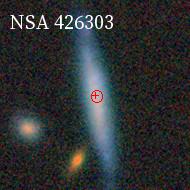}
    \hspace{-3mm}
    \includegraphics[width=0.19\textwidth]{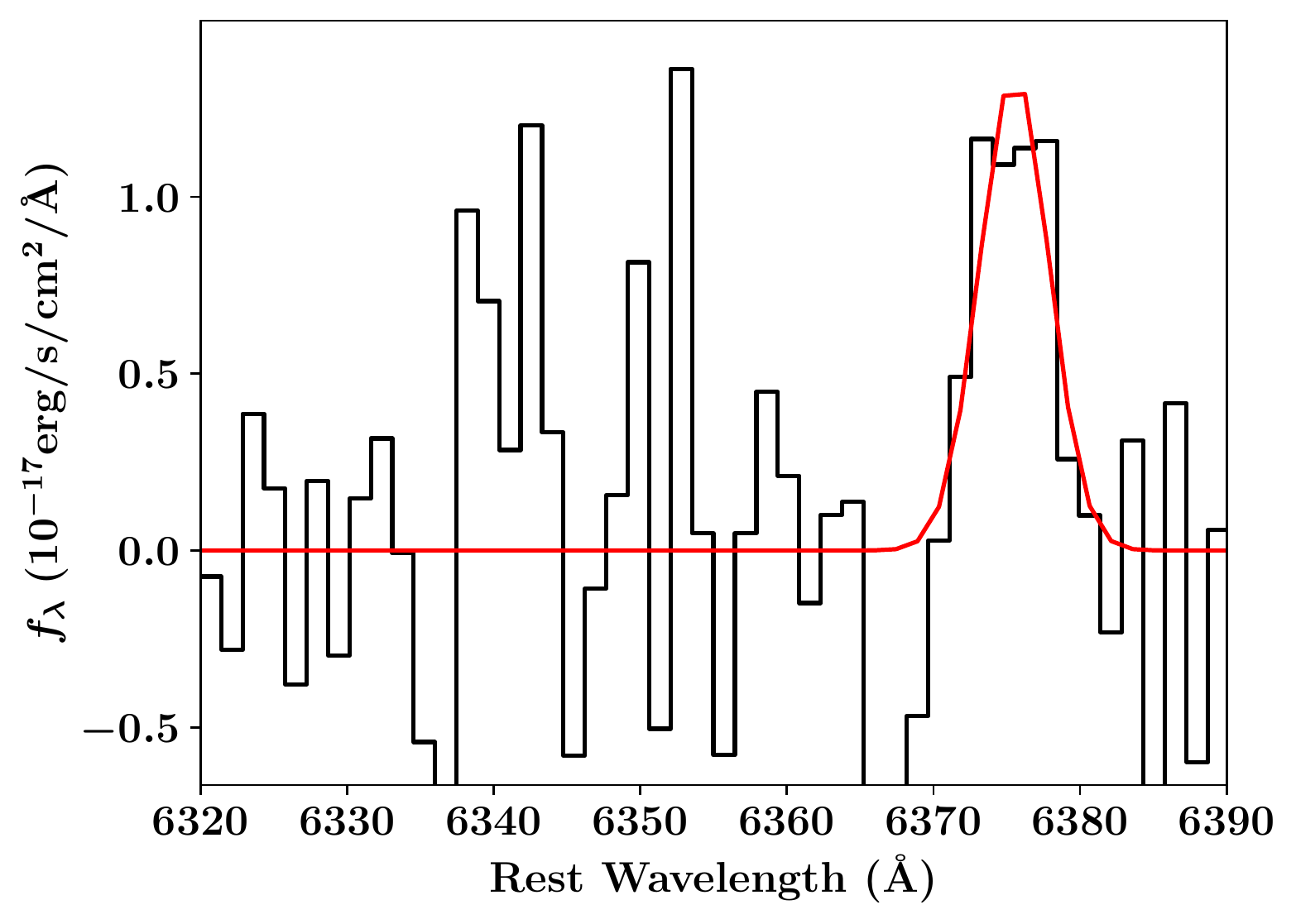}
    \hspace{1.5mm}
    \includegraphics[width=0.13\textwidth]{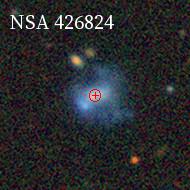}
        \hspace{-3mm}
    \includegraphics[width=0.19\textwidth]{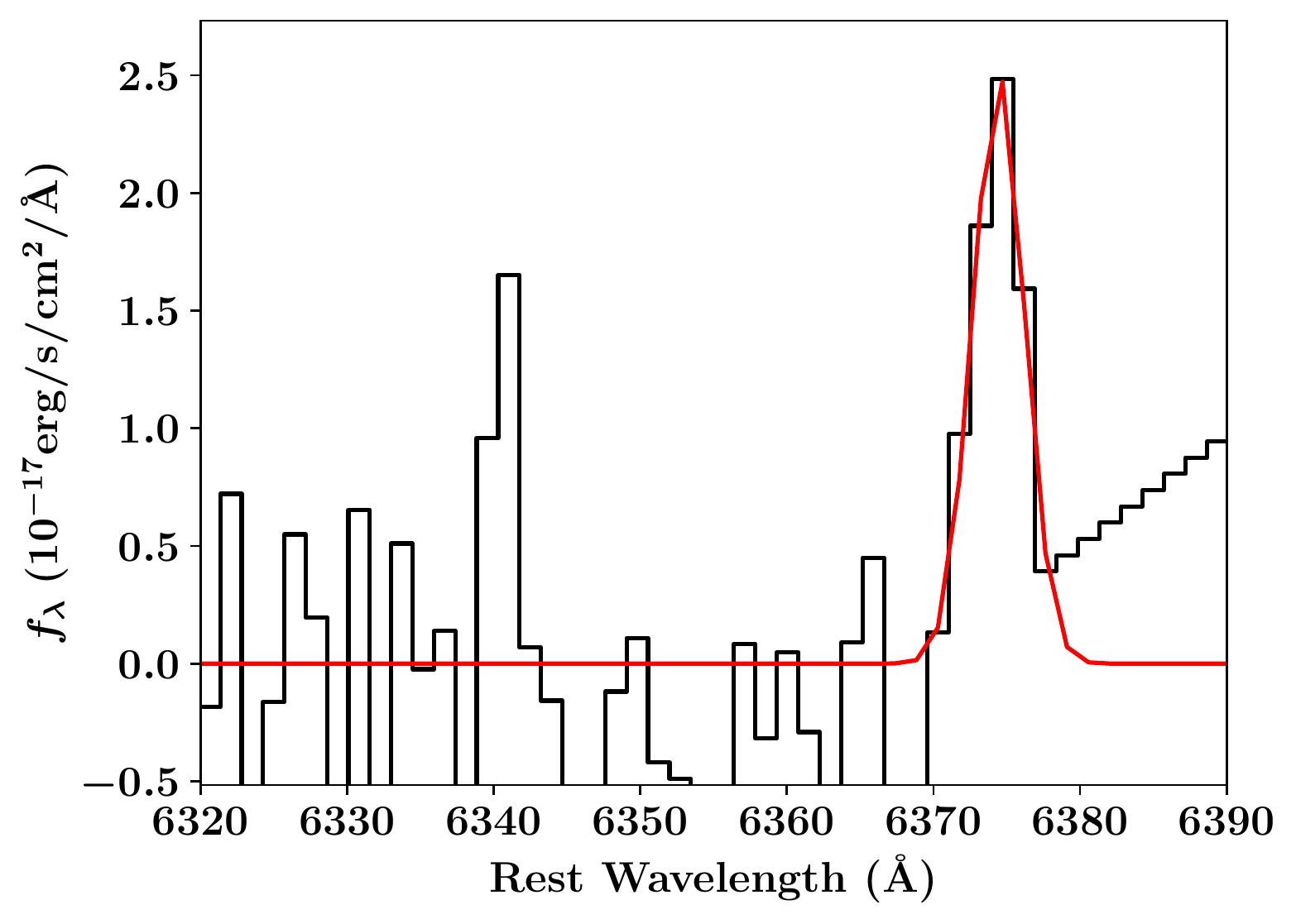}  
        \hspace{1.5mm}
    \includegraphics[width=0.13\textwidth]{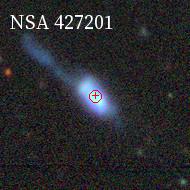}
        \hspace{-3mm}
    \includegraphics[width=0.19\textwidth]{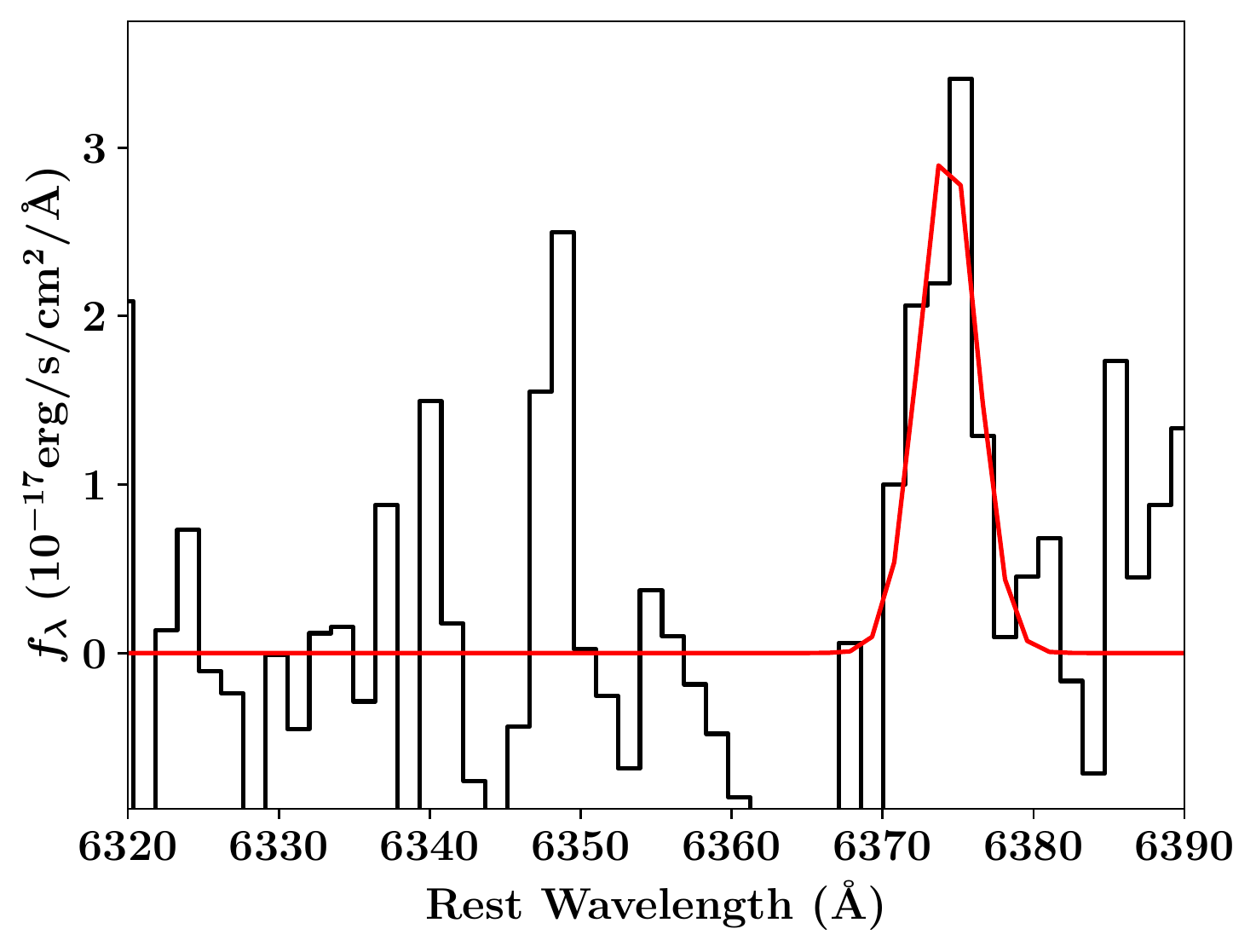}
    \caption{Same as Figure~\ref{fig:decals}.}\label{fig:dec2}
\end{figure*}

\begin{figure*}[h]
    \centering
        \includegraphics[width=0.13\textwidth]{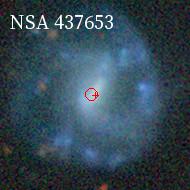}
    \hspace{-3mm}
    \includegraphics[width=0.19\textwidth]{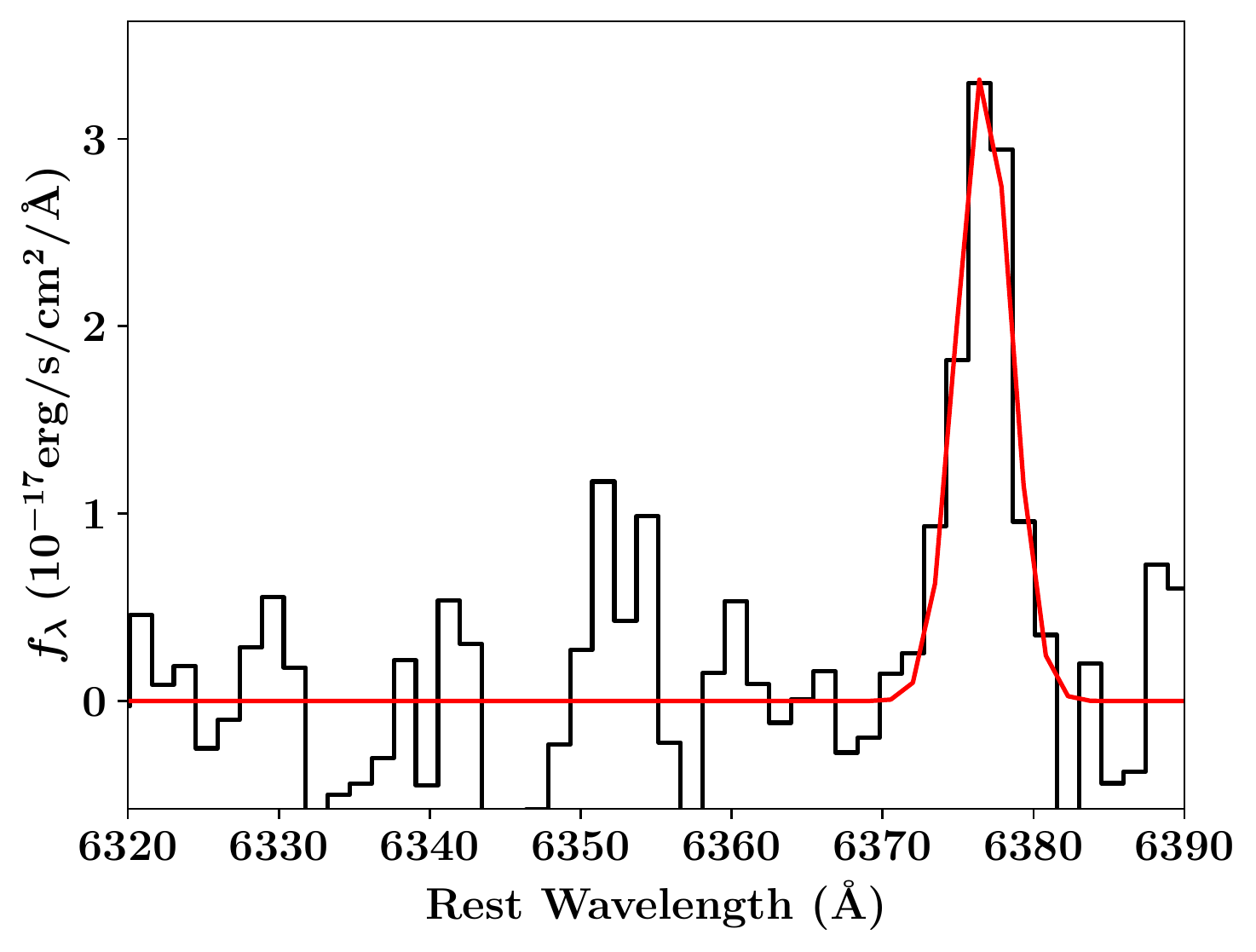}
    \hspace{1.5mm}
    \includegraphics[width=0.13\textwidth]{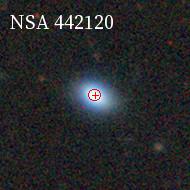}
        \hspace{-3mm}
    \includegraphics[width=0.19\textwidth]{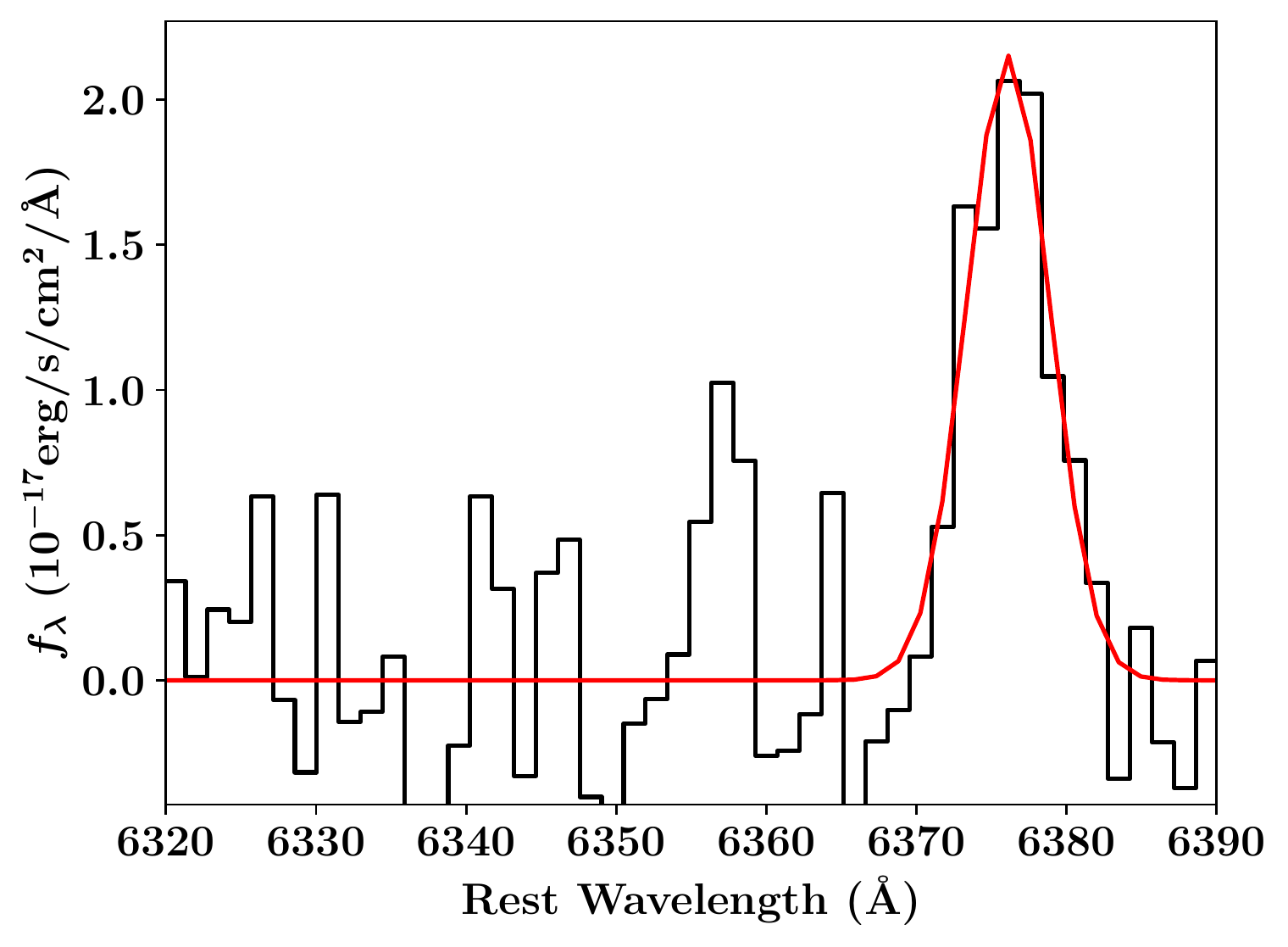}  
        \hspace{1.5mm}
    \includegraphics[width=0.13\textwidth]{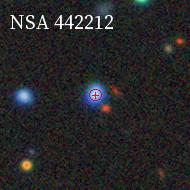}
        \hspace{-3mm}
    \includegraphics[width=0.19\textwidth]{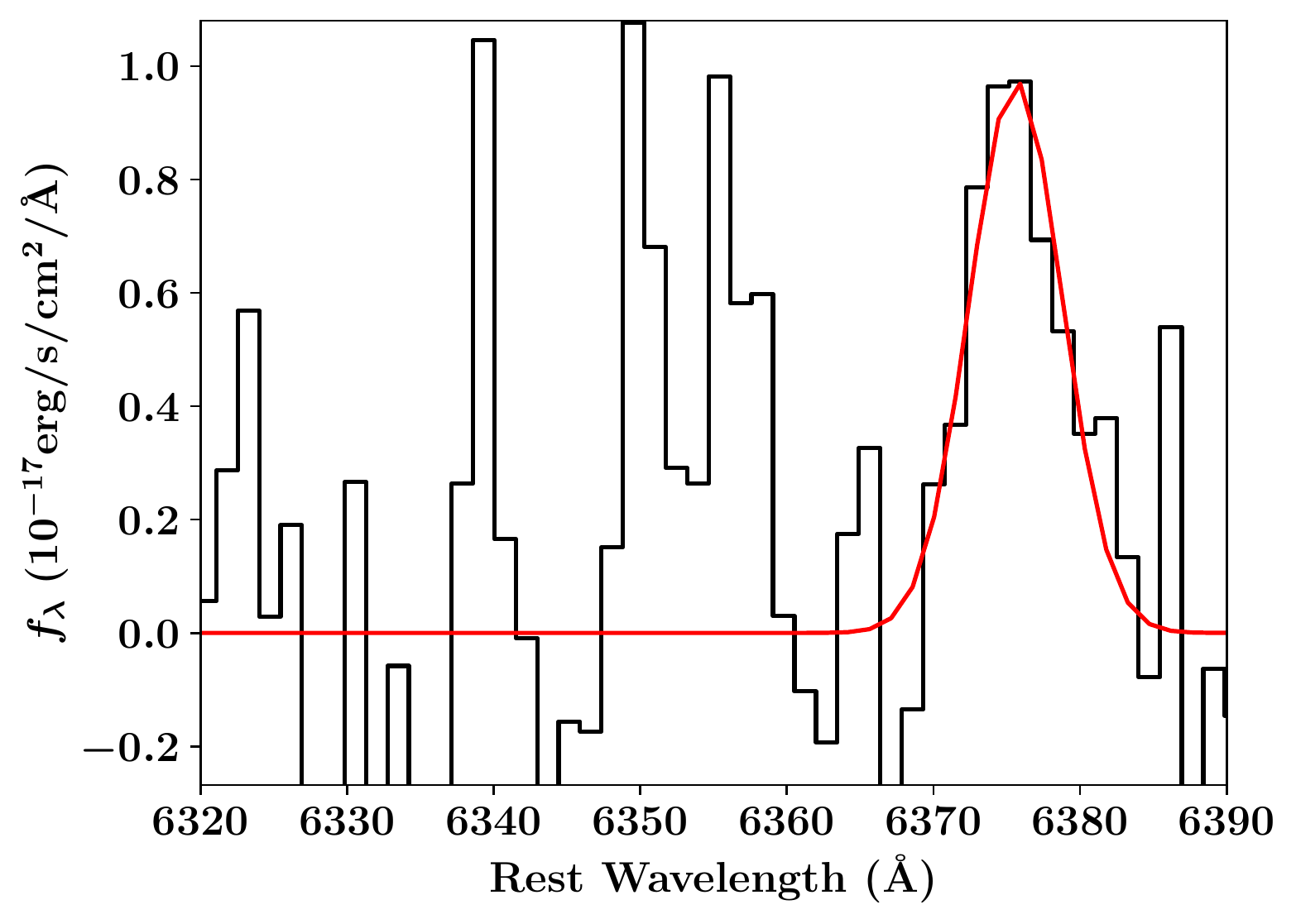}
        \includegraphics[width=0.13\textwidth]{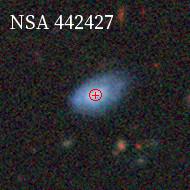}
    \hspace{-3mm}
    \includegraphics[width=0.19\textwidth]{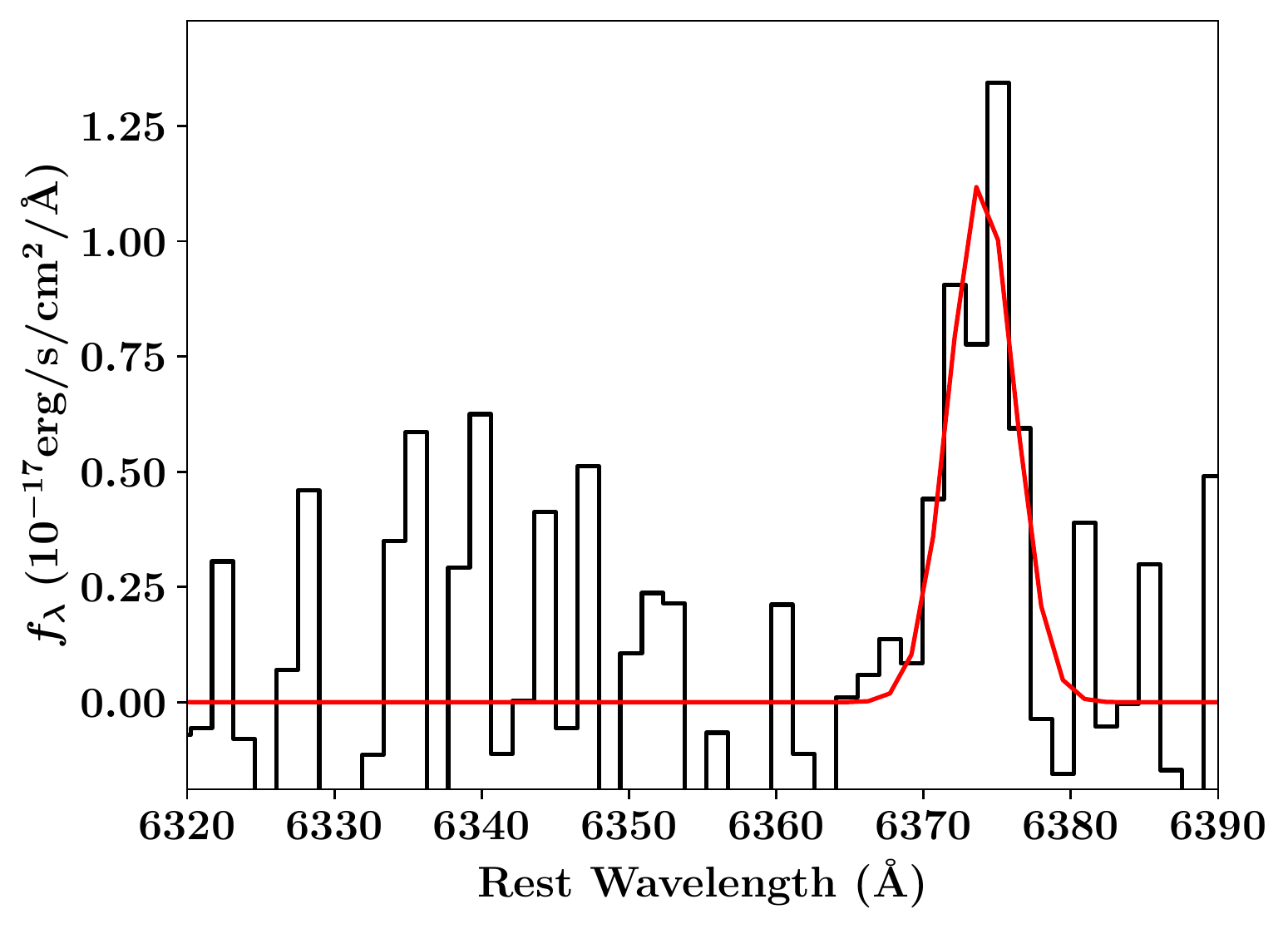}
    \hspace{1.5mm}
    \includegraphics[width=0.13\textwidth]{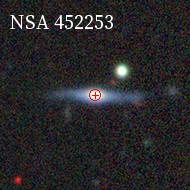}
        \hspace{-3mm}
    \includegraphics[width=0.19\textwidth]{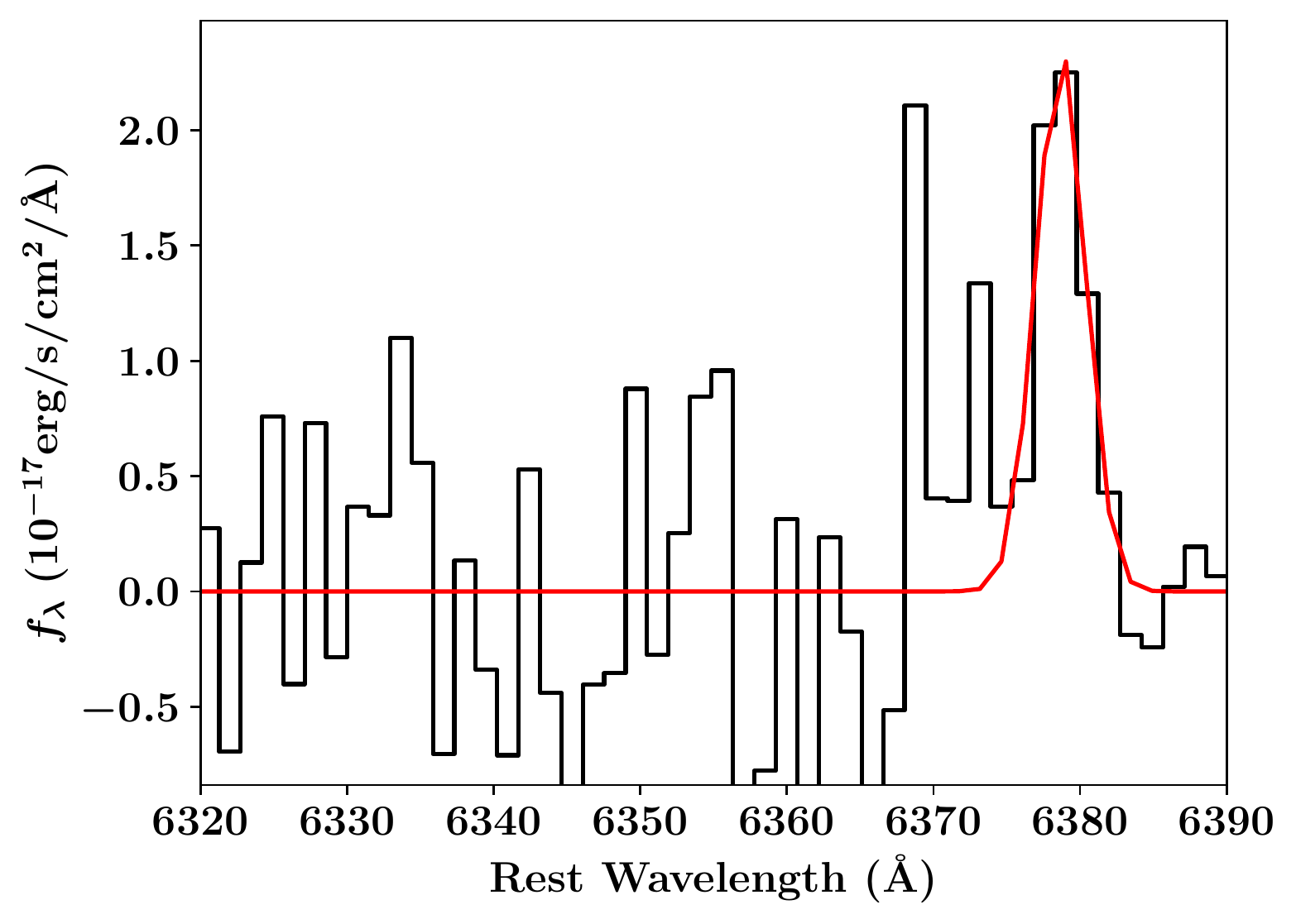}  
        \hspace{1.5mm}
    \includegraphics[width=0.13\textwidth]{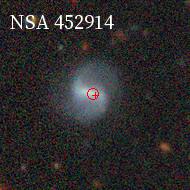}
        \hspace{-3mm}
    \includegraphics[width=0.19\textwidth]{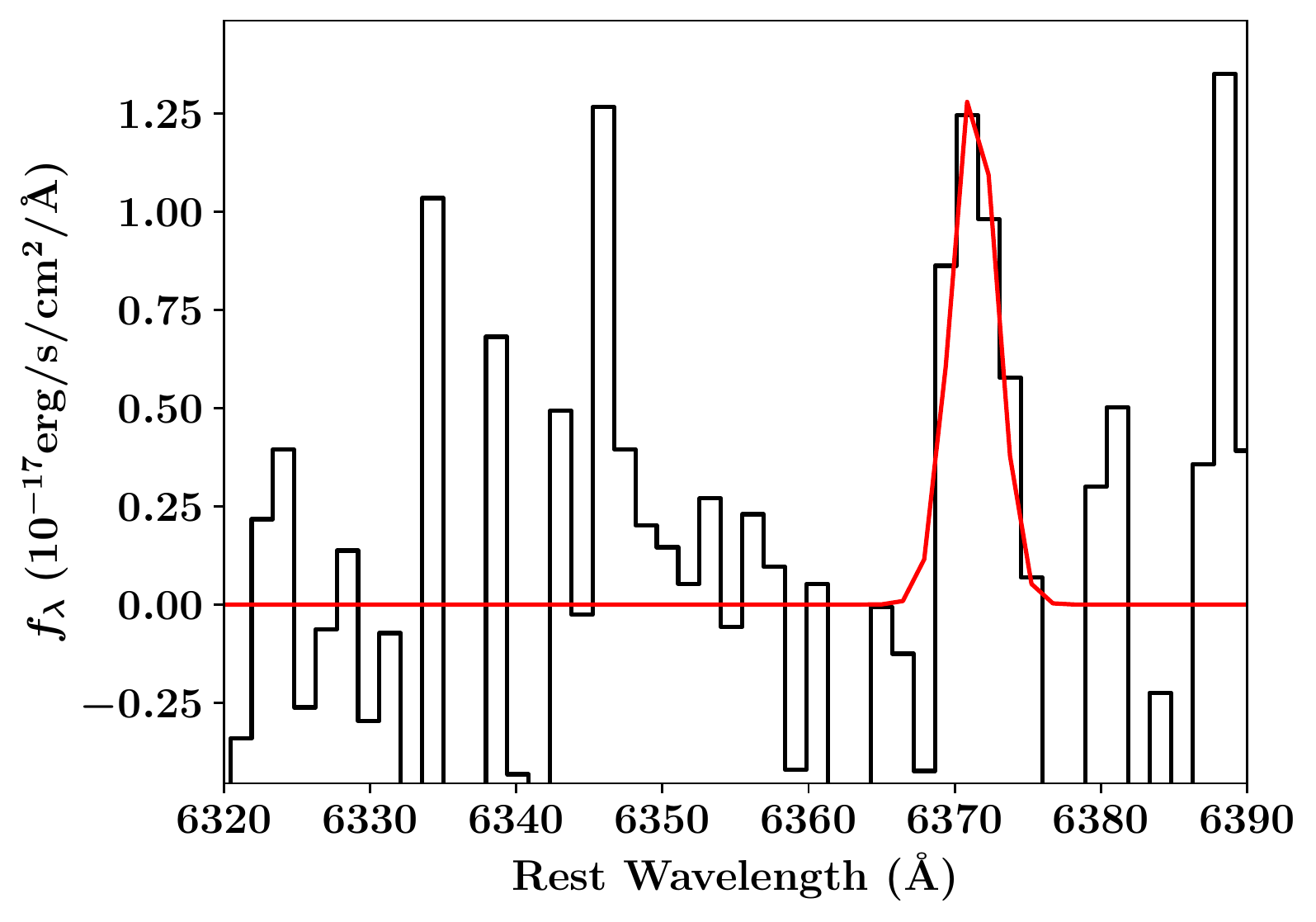}
    \includegraphics[width=0.13\textwidth]{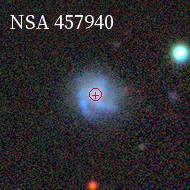}
    \hspace{-3mm}
    \includegraphics[width=0.19\textwidth]{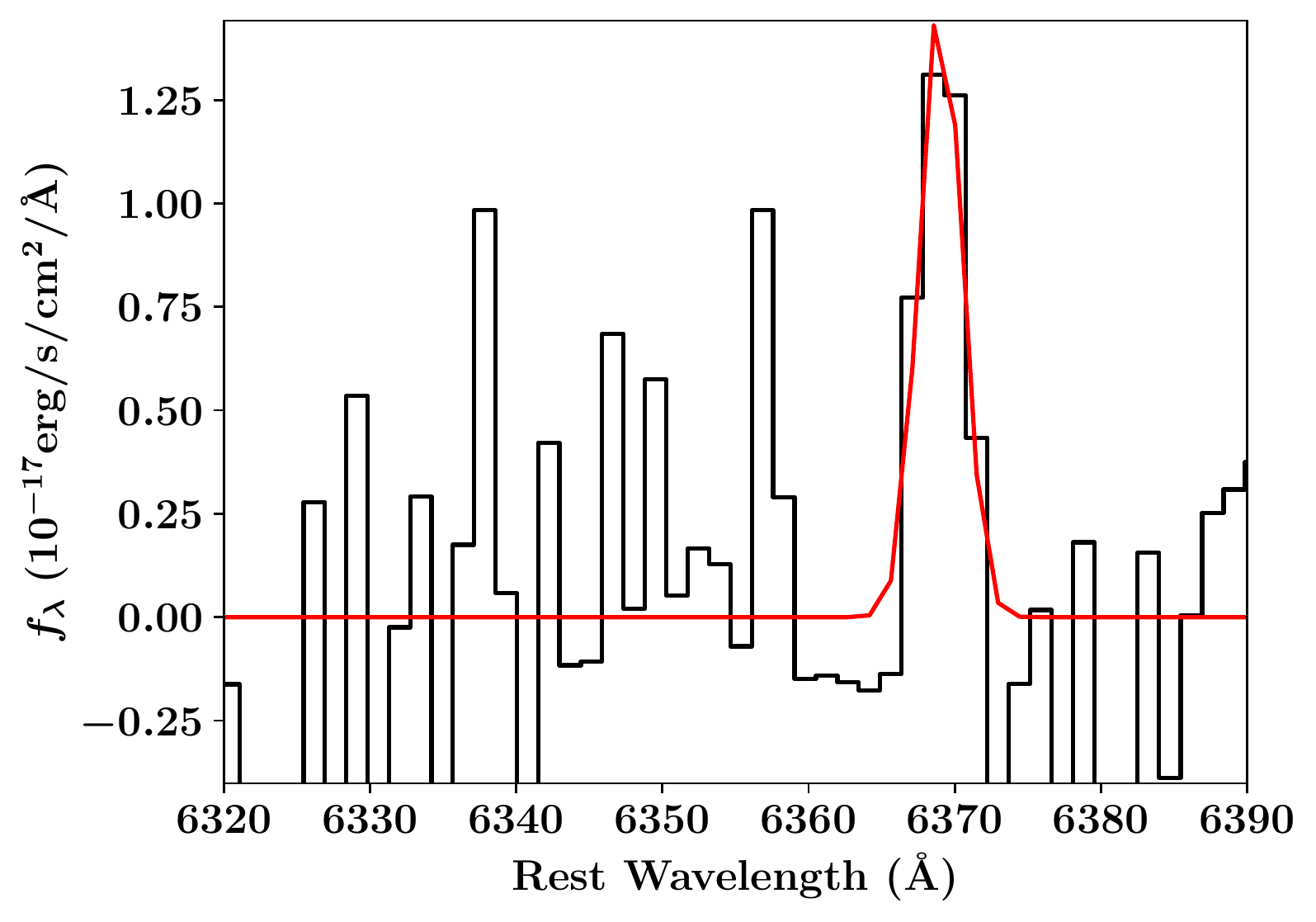}
    \hspace{1.5mm}
    \includegraphics[width=0.13\textwidth]{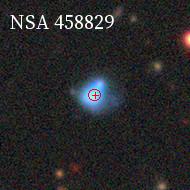}
        \hspace{-3mm}
    \includegraphics[width=0.19\textwidth]{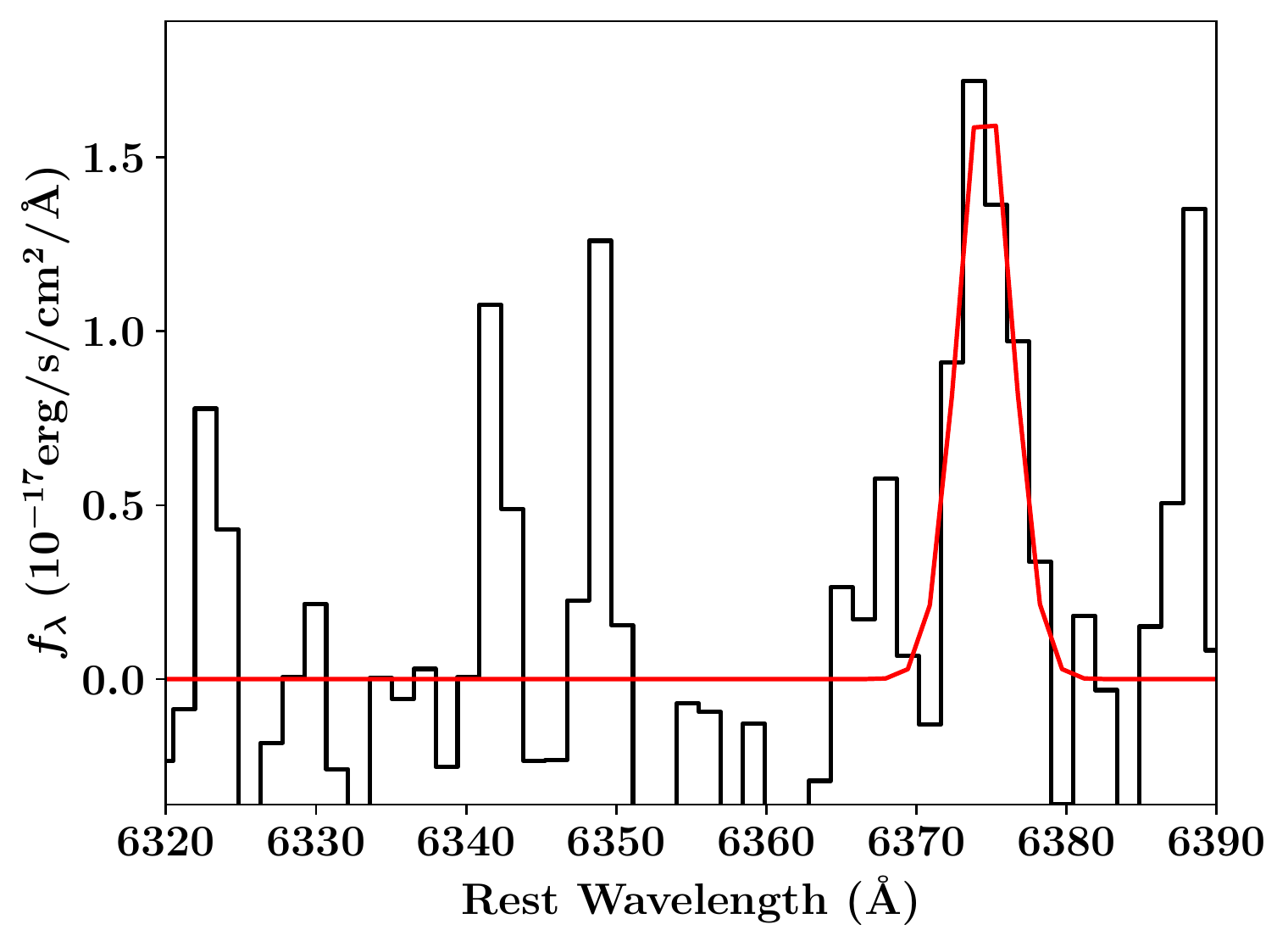}  
        \hspace{1.5mm}
    \includegraphics[width=0.13\textwidth]{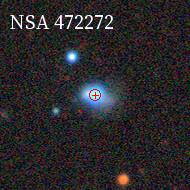}
        \hspace{-3mm}
    \includegraphics[width=0.19\textwidth]{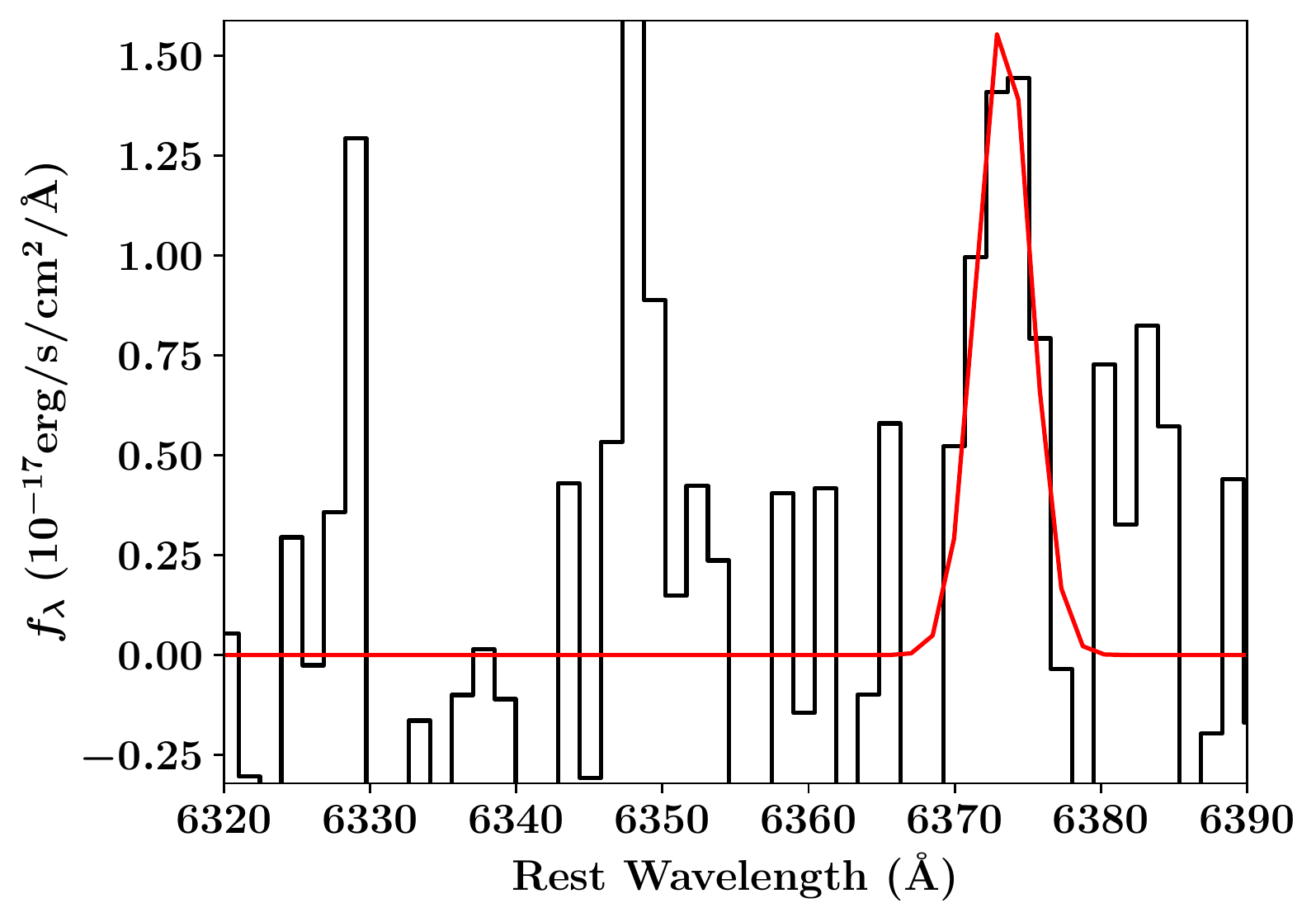}
    \includegraphics[width=0.13\textwidth]{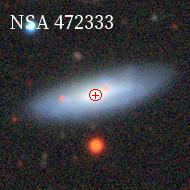}
    \hspace{-3mm}
    \includegraphics[width=0.19\textwidth]{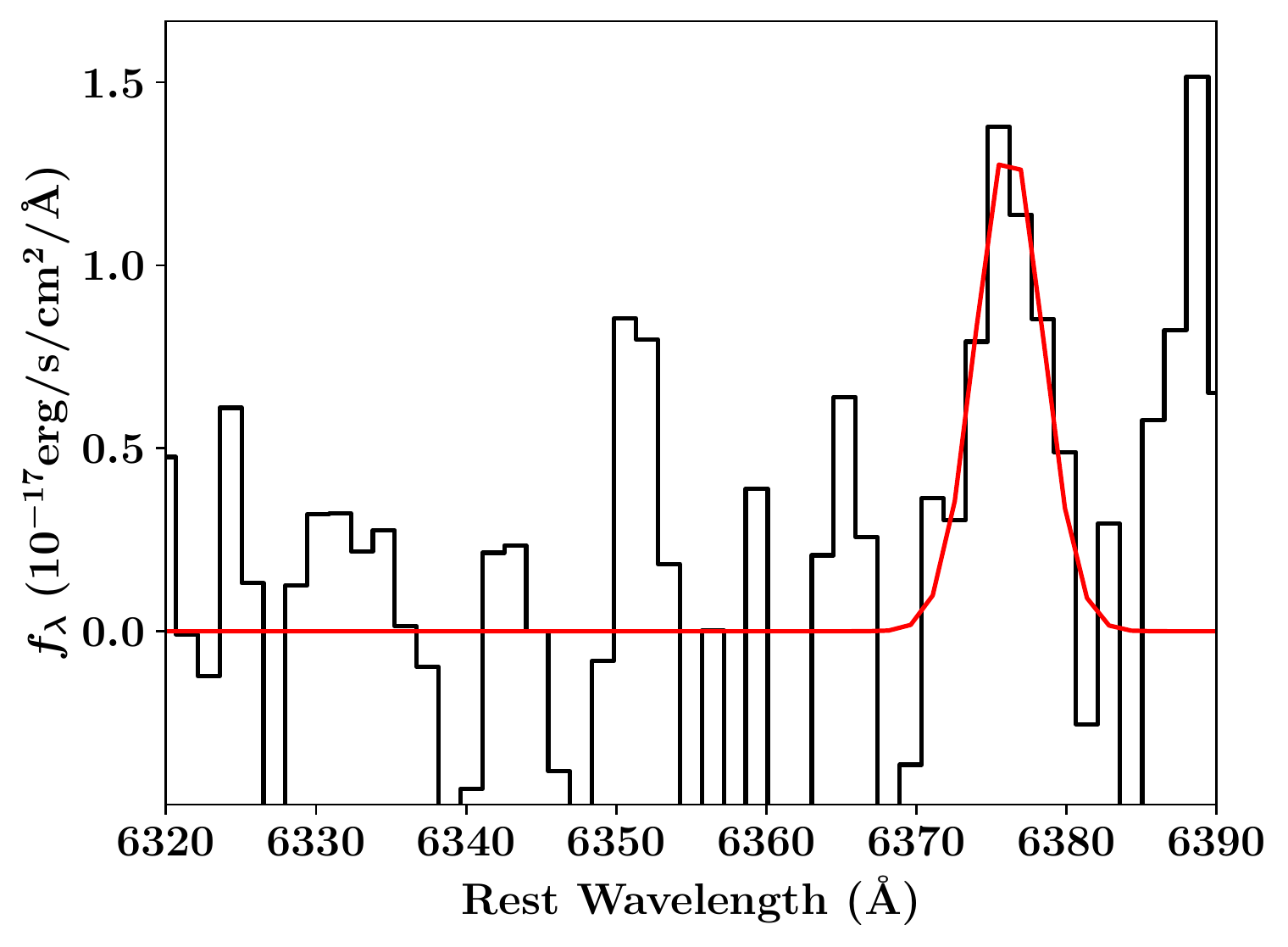}
    \hspace{1.5mm}
    \includegraphics[width=0.13\textwidth]{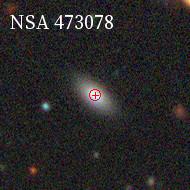}
        \hspace{-3mm}
    \includegraphics[width=0.19\textwidth]{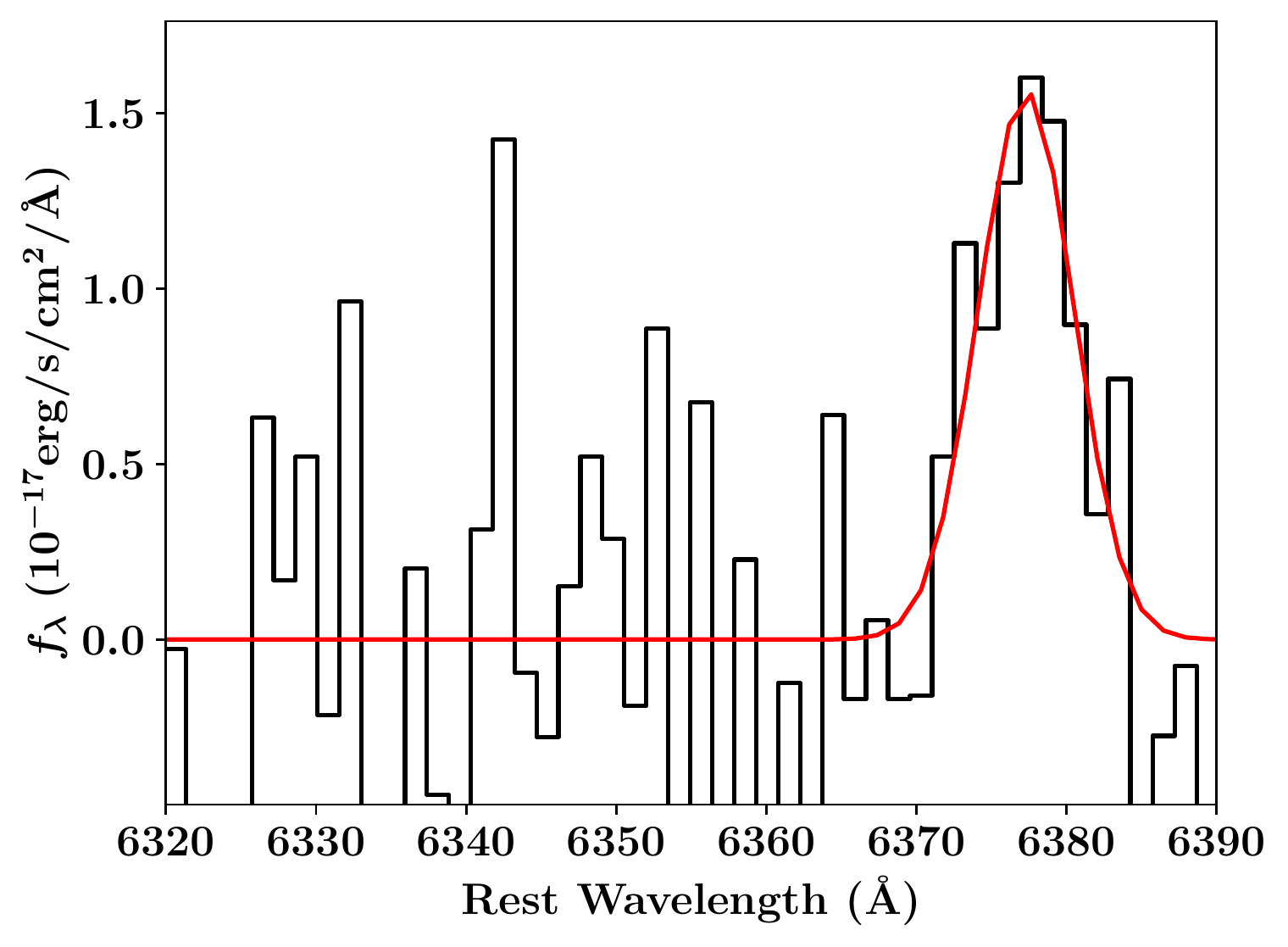}  
        \hspace{1.5mm}
    \includegraphics[width=0.13\textwidth]{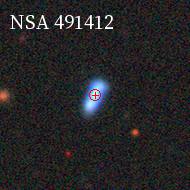}
        \hspace{-3mm}
    \includegraphics[width=0.19\textwidth]{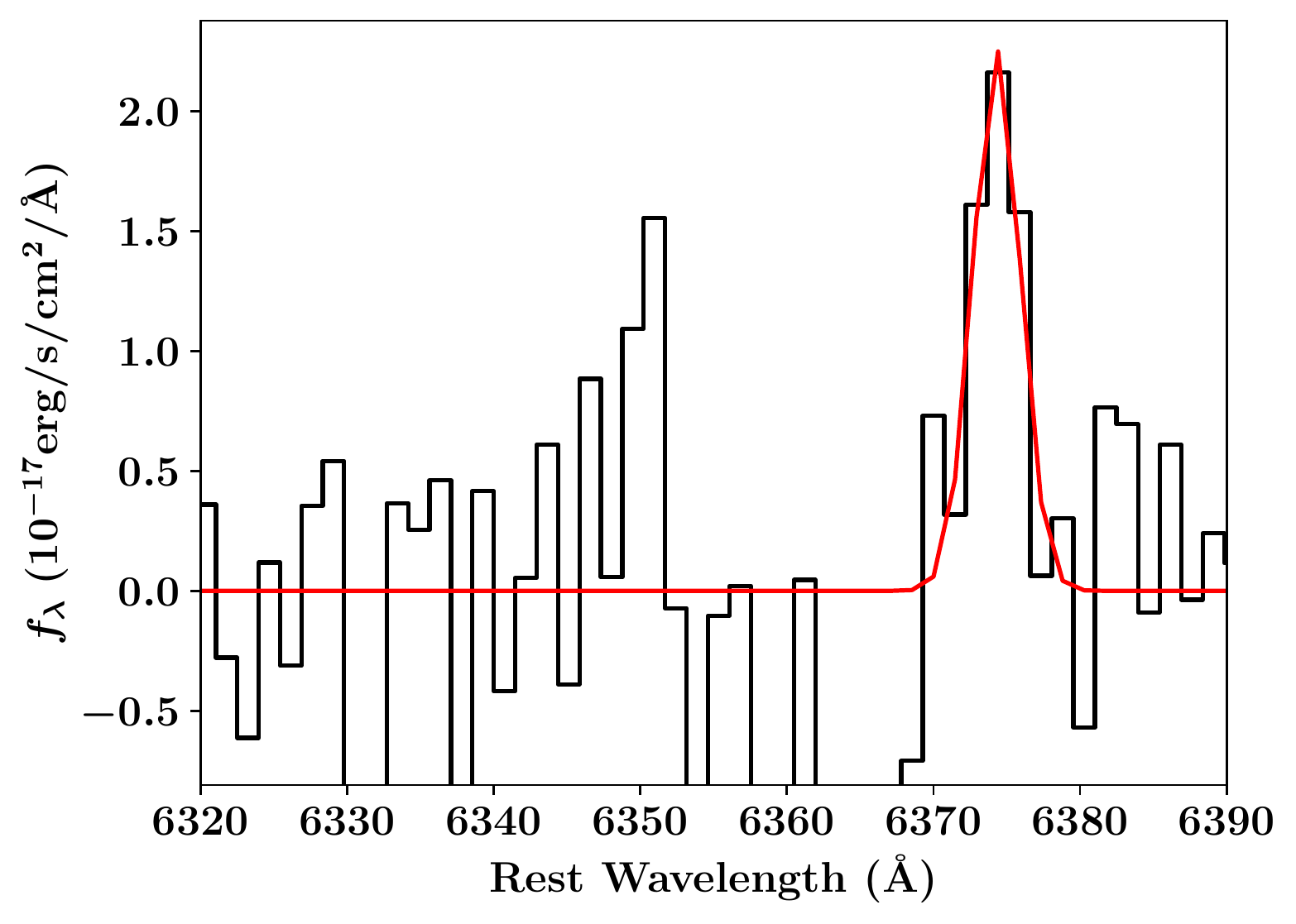}
    \includegraphics[width=0.13\textwidth]{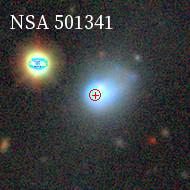}
    \hspace{-3mm}
    \includegraphics[width=0.19\textwidth]{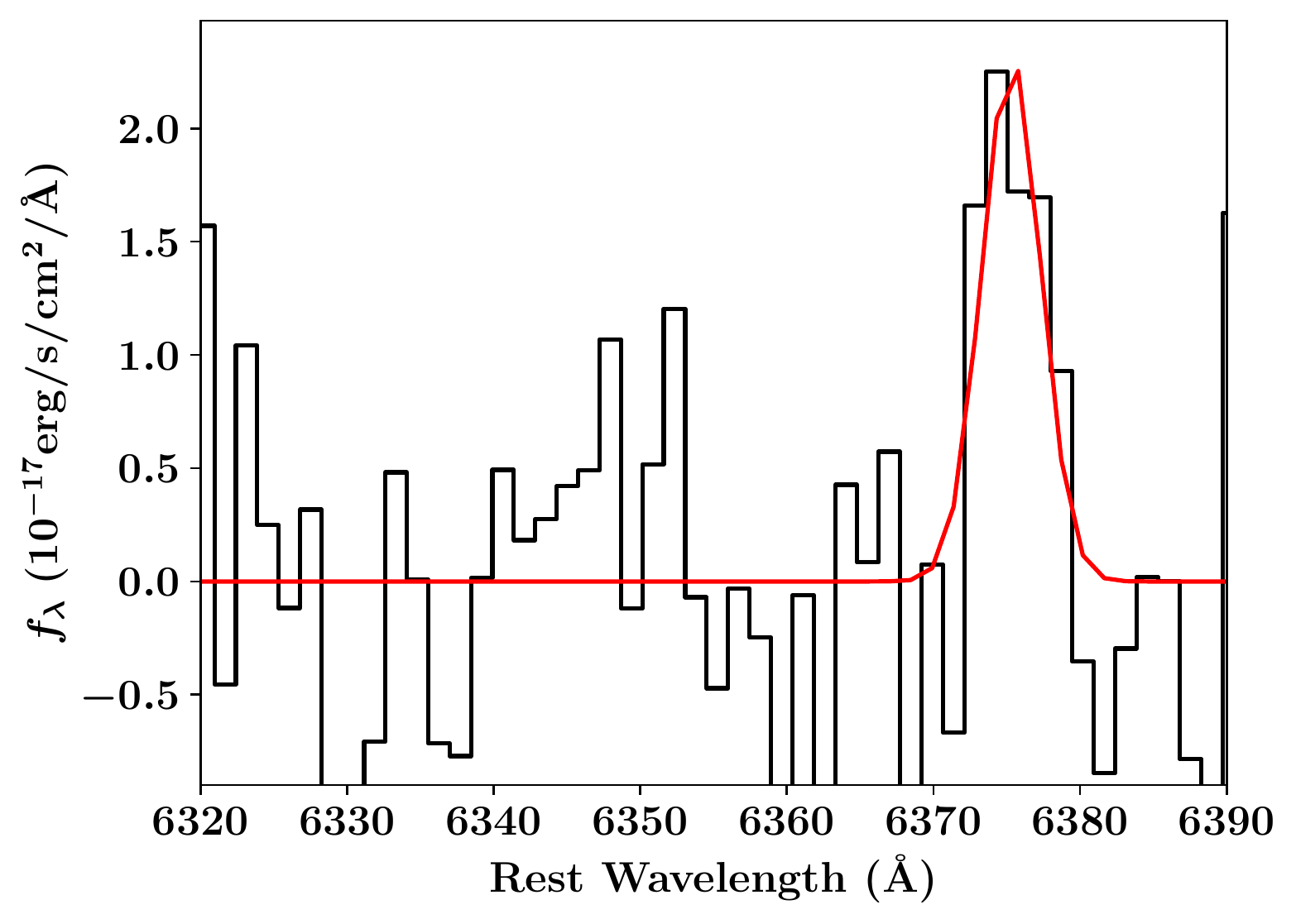}
    \hspace{1.5mm}
    \includegraphics[width=0.13\textwidth]{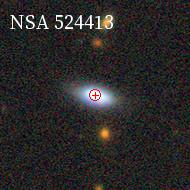}
        \hspace{-3mm}
    \includegraphics[width=0.19\textwidth]{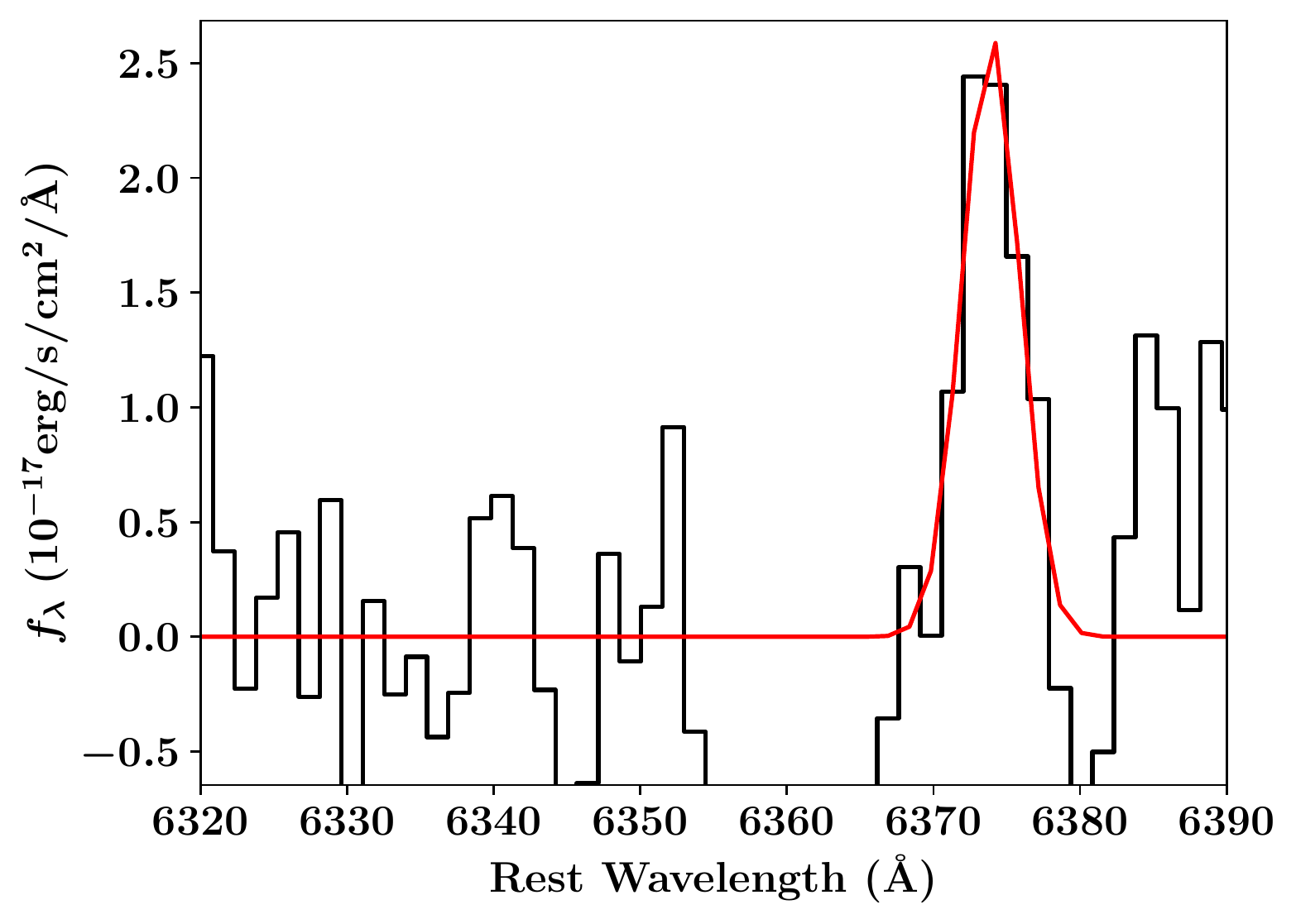}  
        \hspace{1.5mm}
    \includegraphics[width=0.13\textwidth]{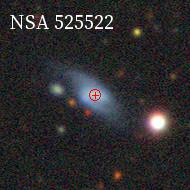}
        \hspace{-3mm}
    \includegraphics[width=0.19\textwidth]{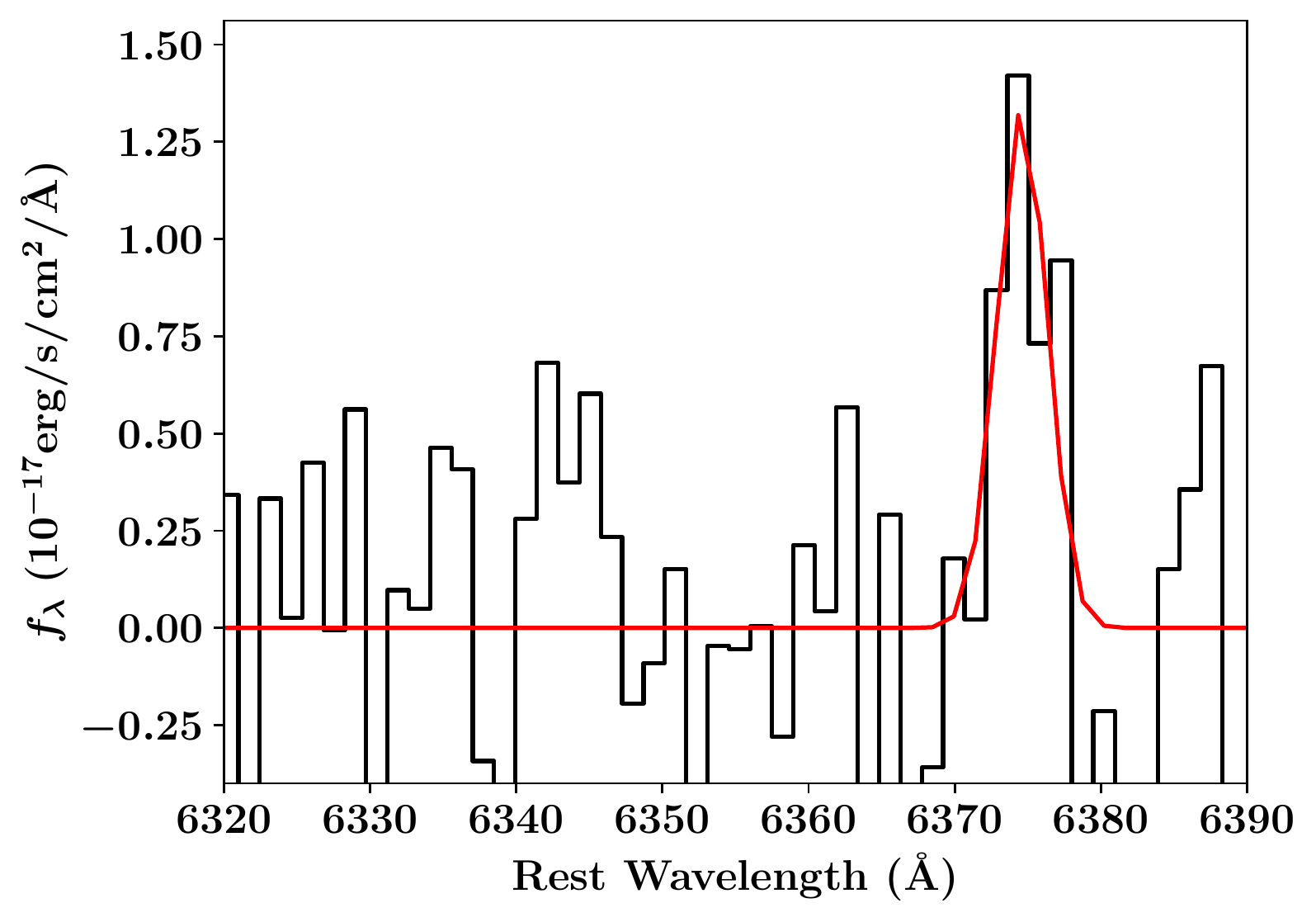}
    \includegraphics[width=0.13\textwidth]{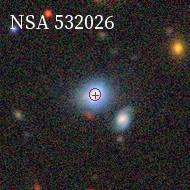}
    \hspace{-3mm}
    \includegraphics[width=0.19\textwidth]{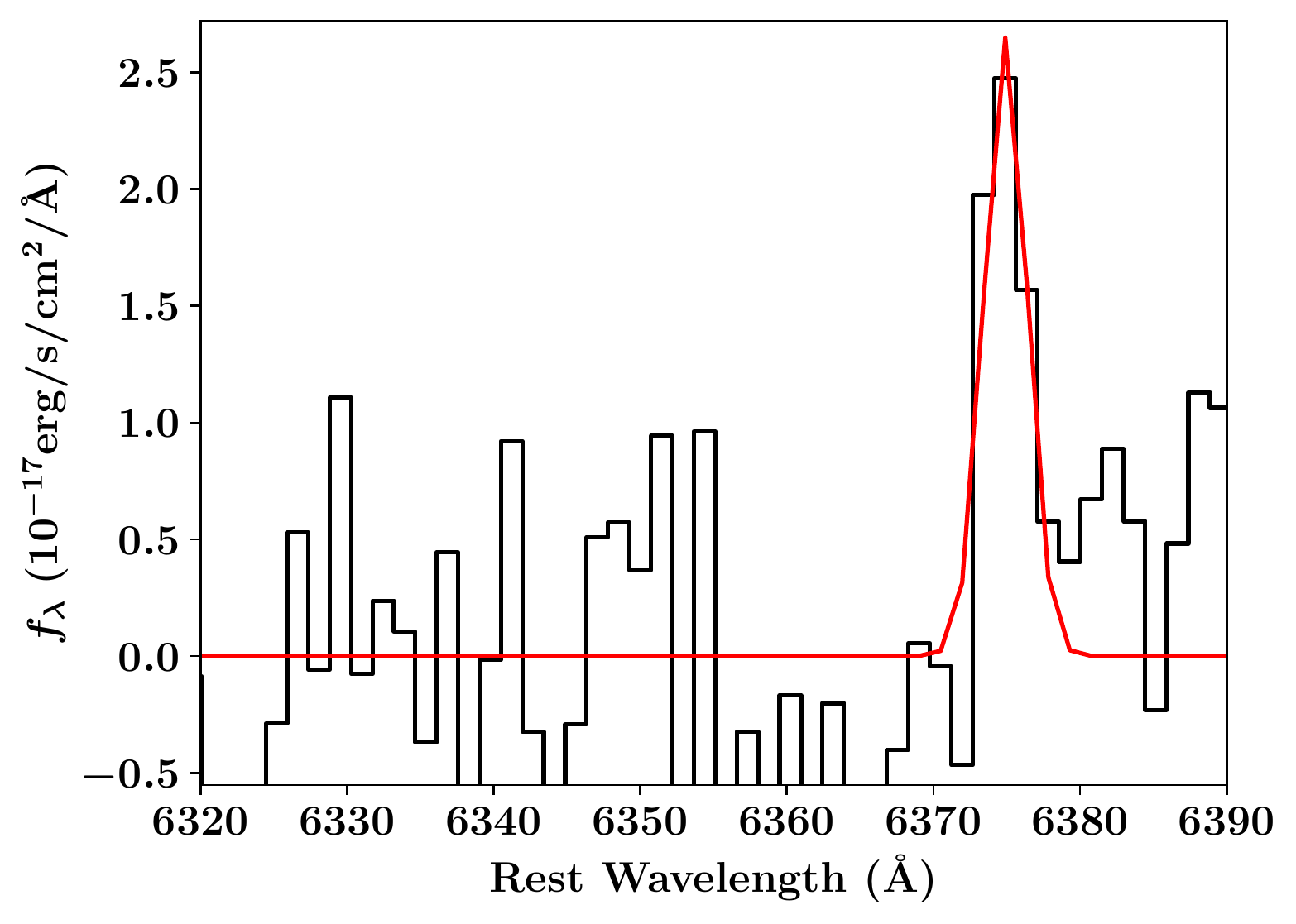}
    \hspace{1.5mm}
    \includegraphics[width=0.13\textwidth]{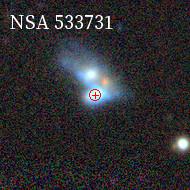}
        \hspace{-3mm}
    \includegraphics[width=0.19\textwidth]{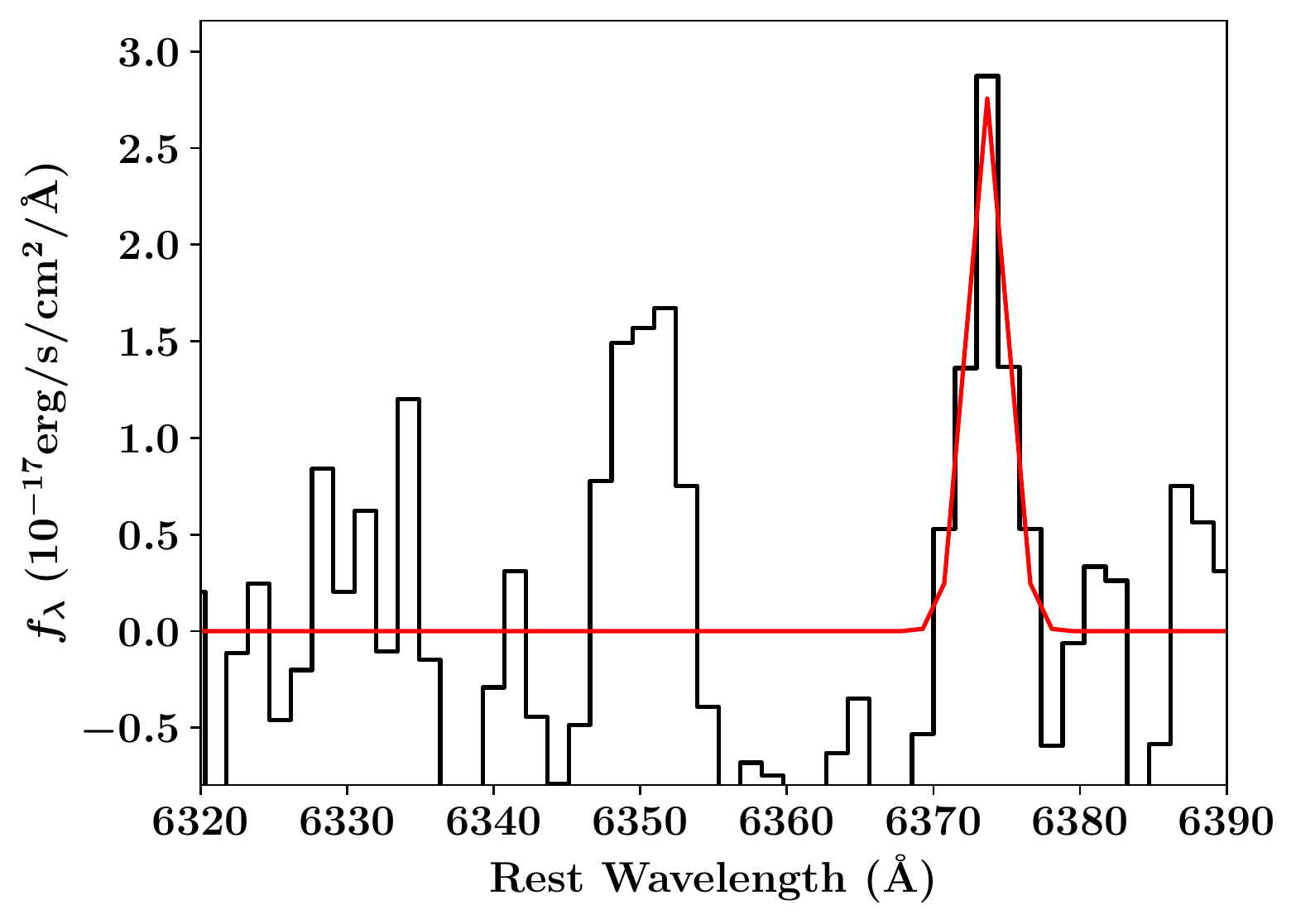}  
        \hspace{1.5mm}
    \includegraphics[width=0.13\textwidth]{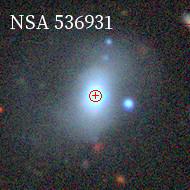}
        \hspace{-3mm}
    \includegraphics[width=0.19\textwidth]{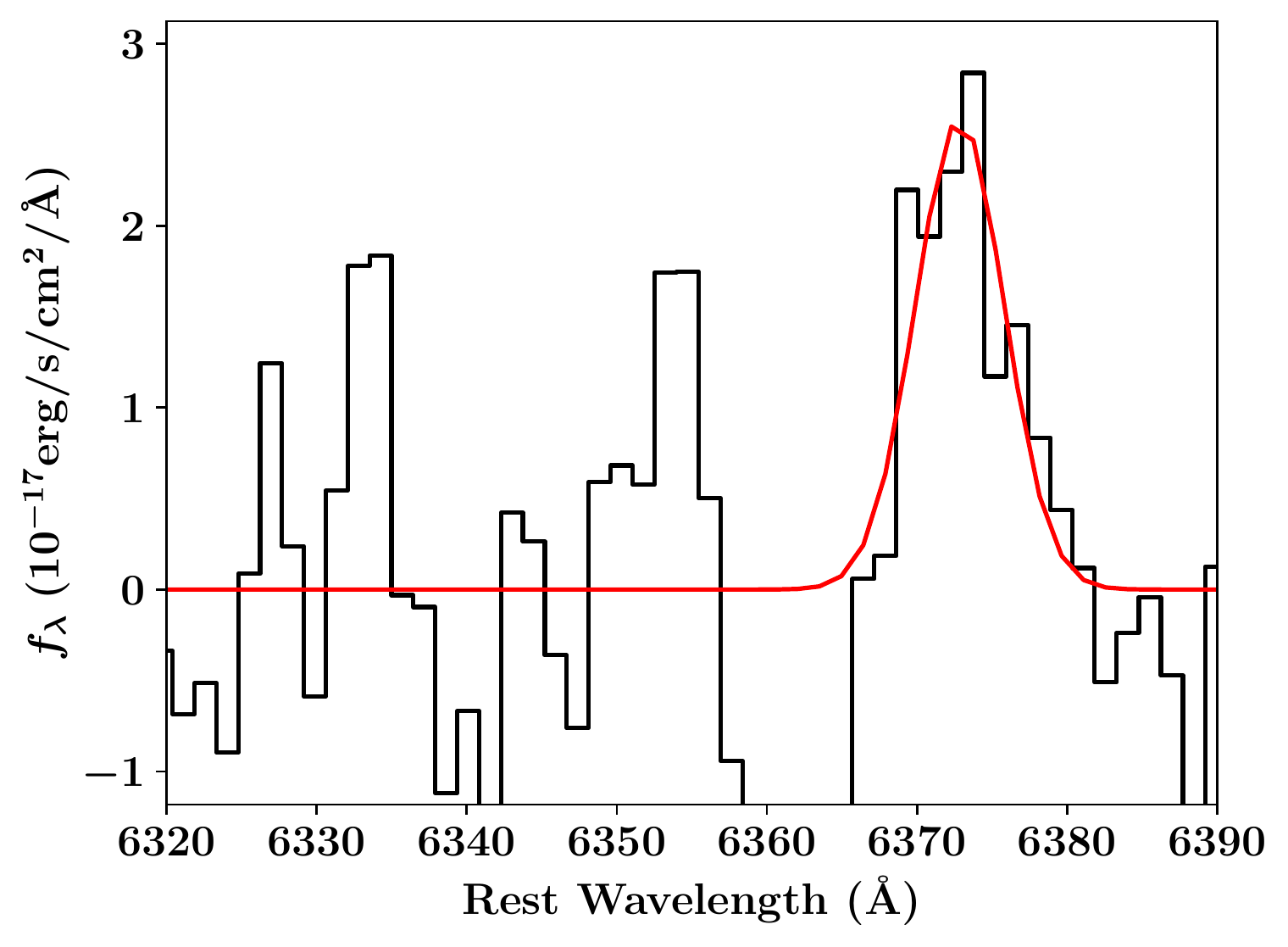}
    \includegraphics[width=0.13\textwidth]{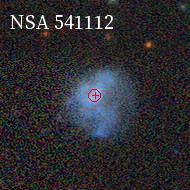}
    \hspace{-3mm}
    \includegraphics[width=0.19\textwidth]{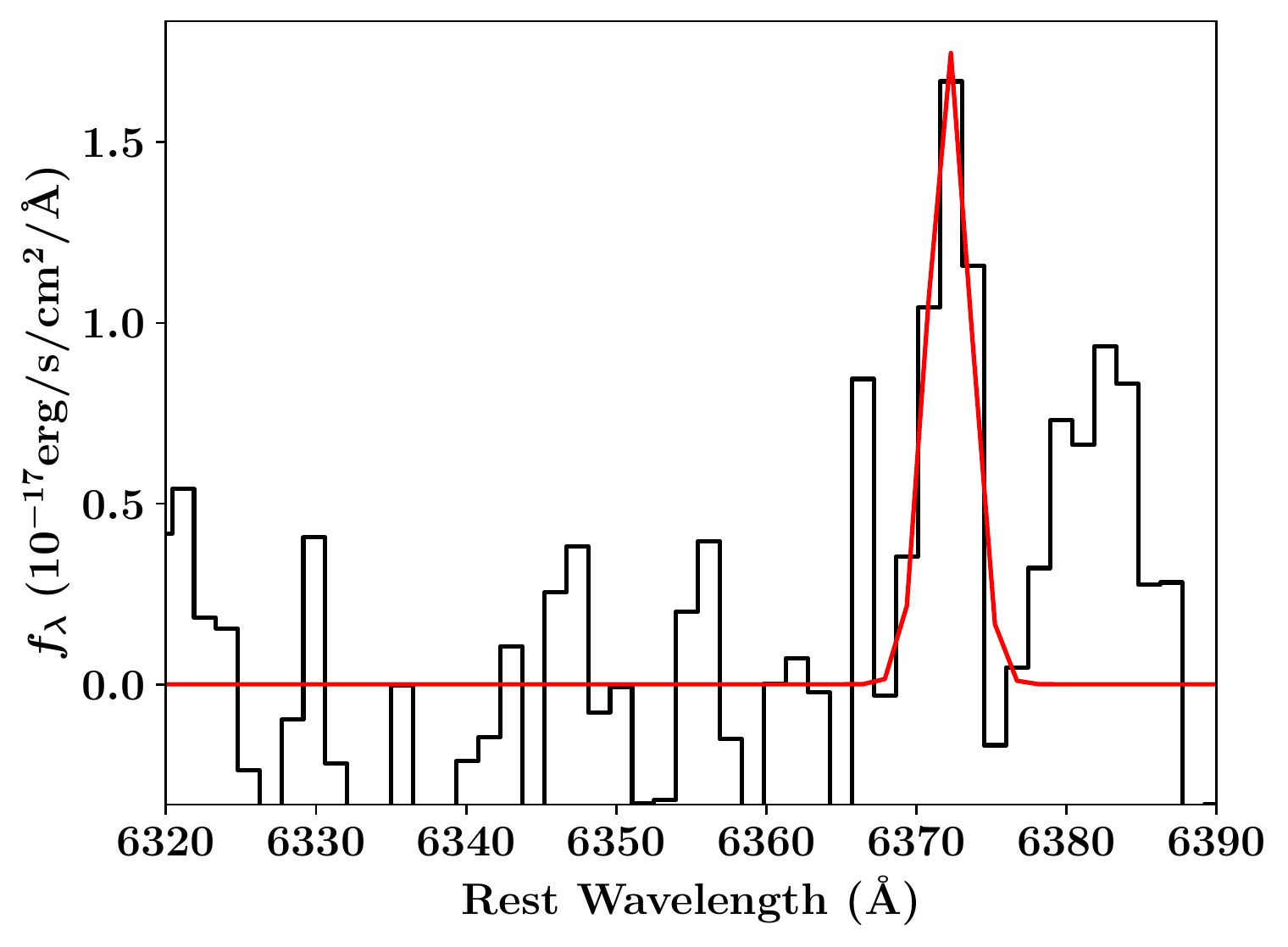}
    \hspace{1.5mm}
    \includegraphics[width=0.13\textwidth]{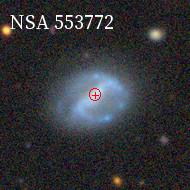}
        \hspace{-3mm}
    \includegraphics[width=0.19\textwidth]{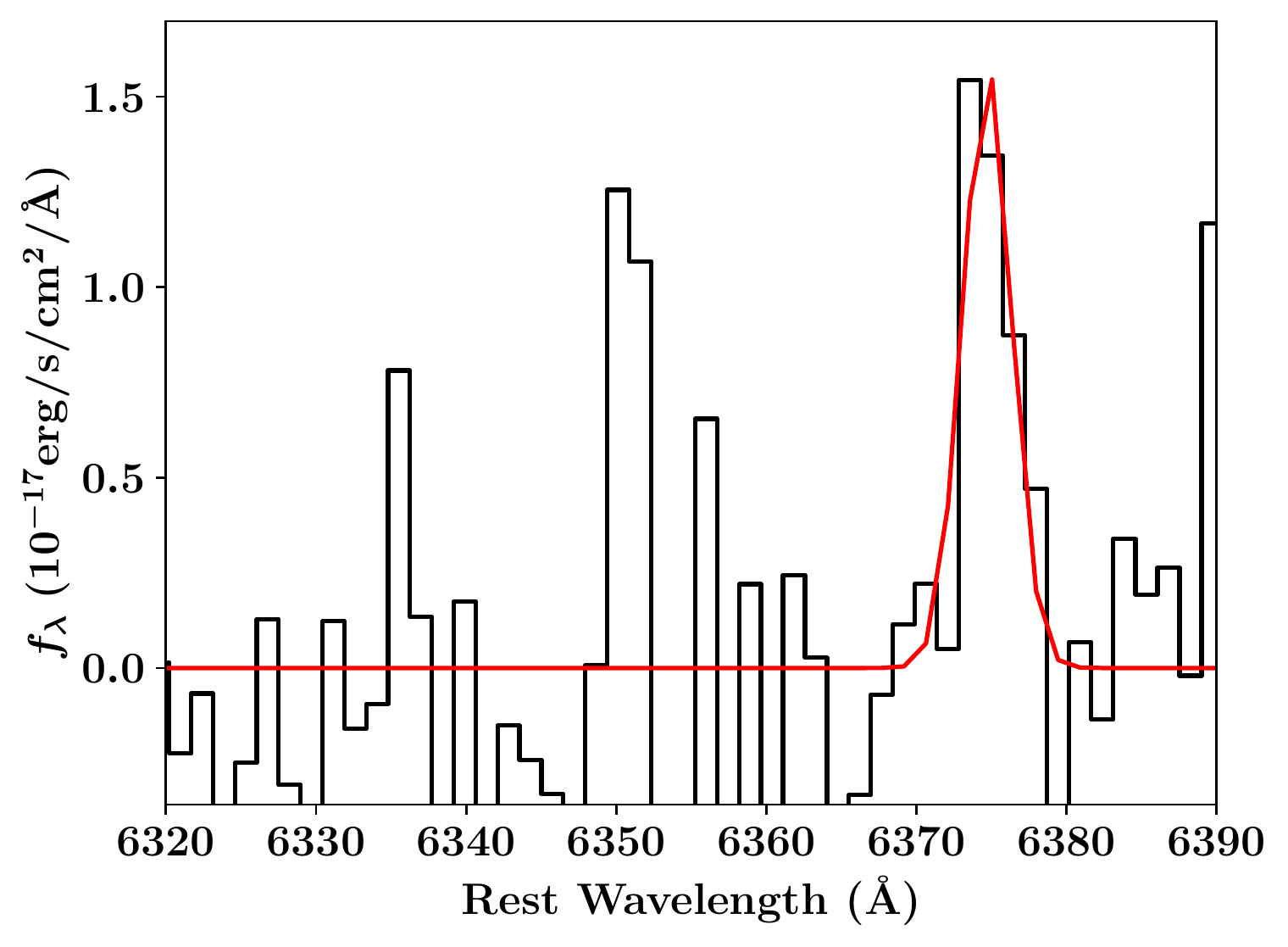}  
        \hspace{1.5mm}
    \includegraphics[width=0.13\textwidth]{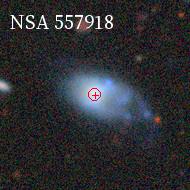}
        \hspace{-3mm}
    \includegraphics[width=0.19\textwidth]{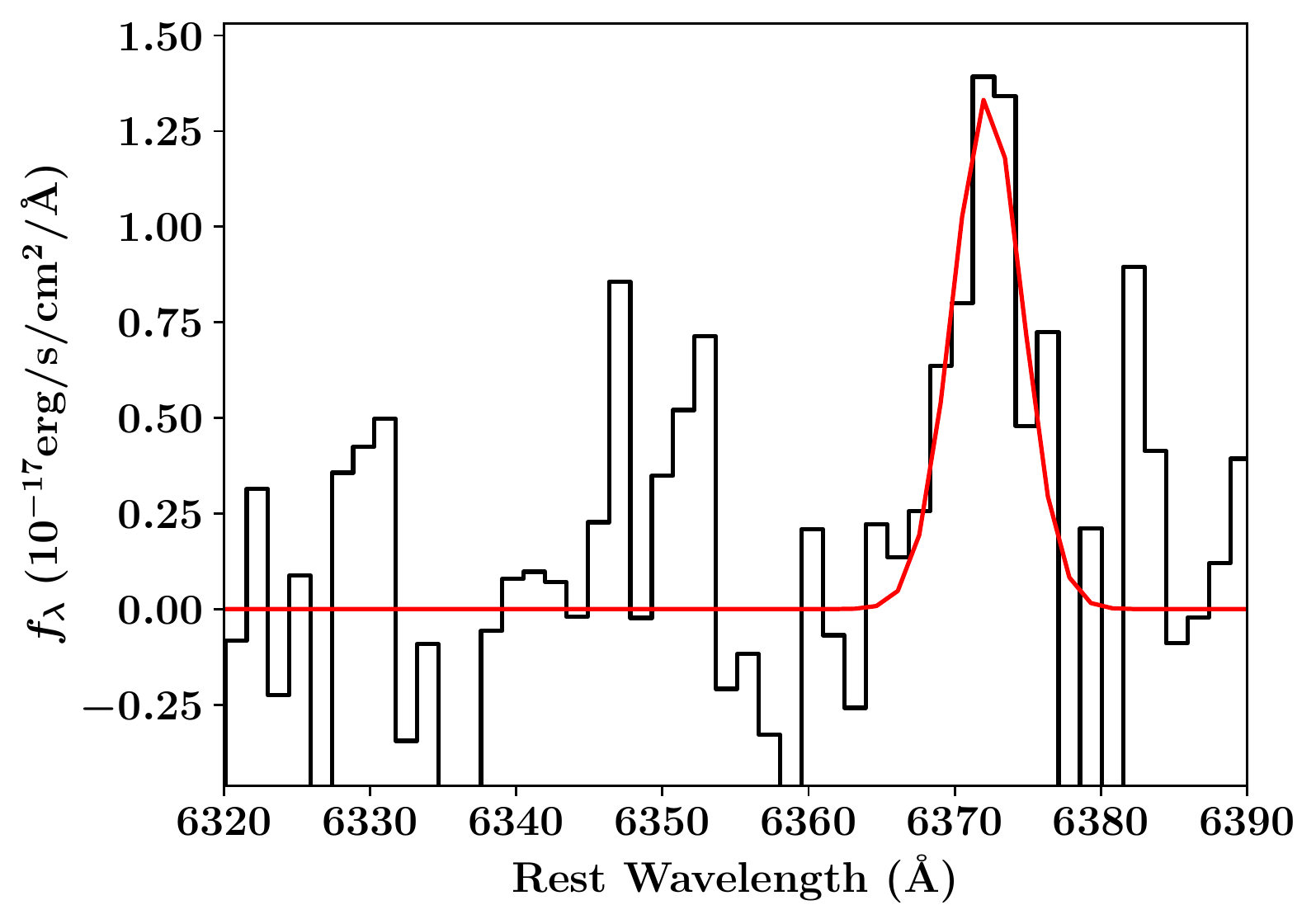}
  \includegraphics[width=0.13\textwidth]{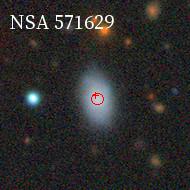}
    \hspace{-3mm}
    \includegraphics[width=0.19\textwidth]{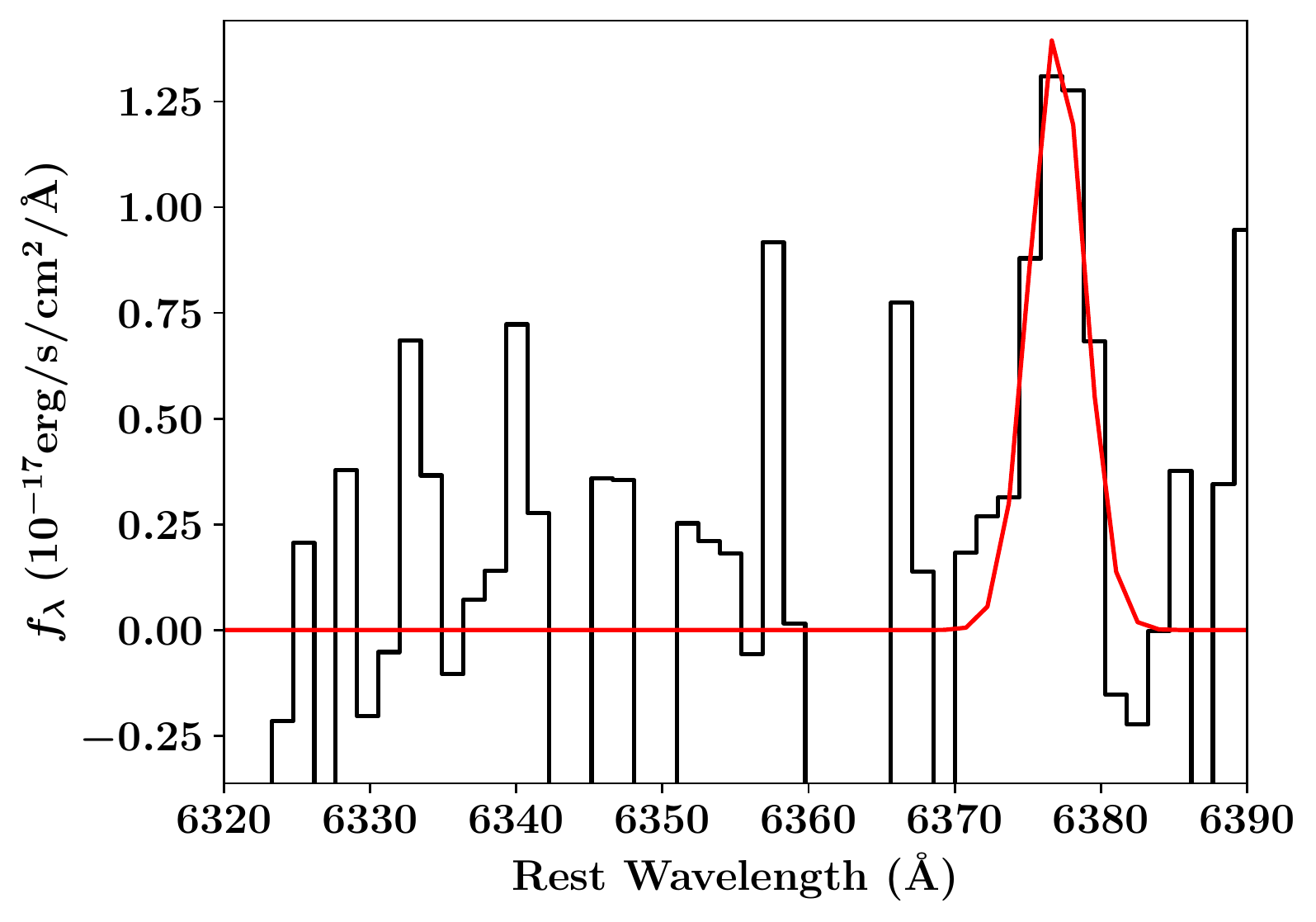}
    \hspace{1.5mm}
    \includegraphics[width=0.13\textwidth]{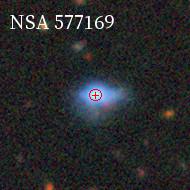}
        \hspace{-3mm}
    \includegraphics[width=0.19\textwidth]{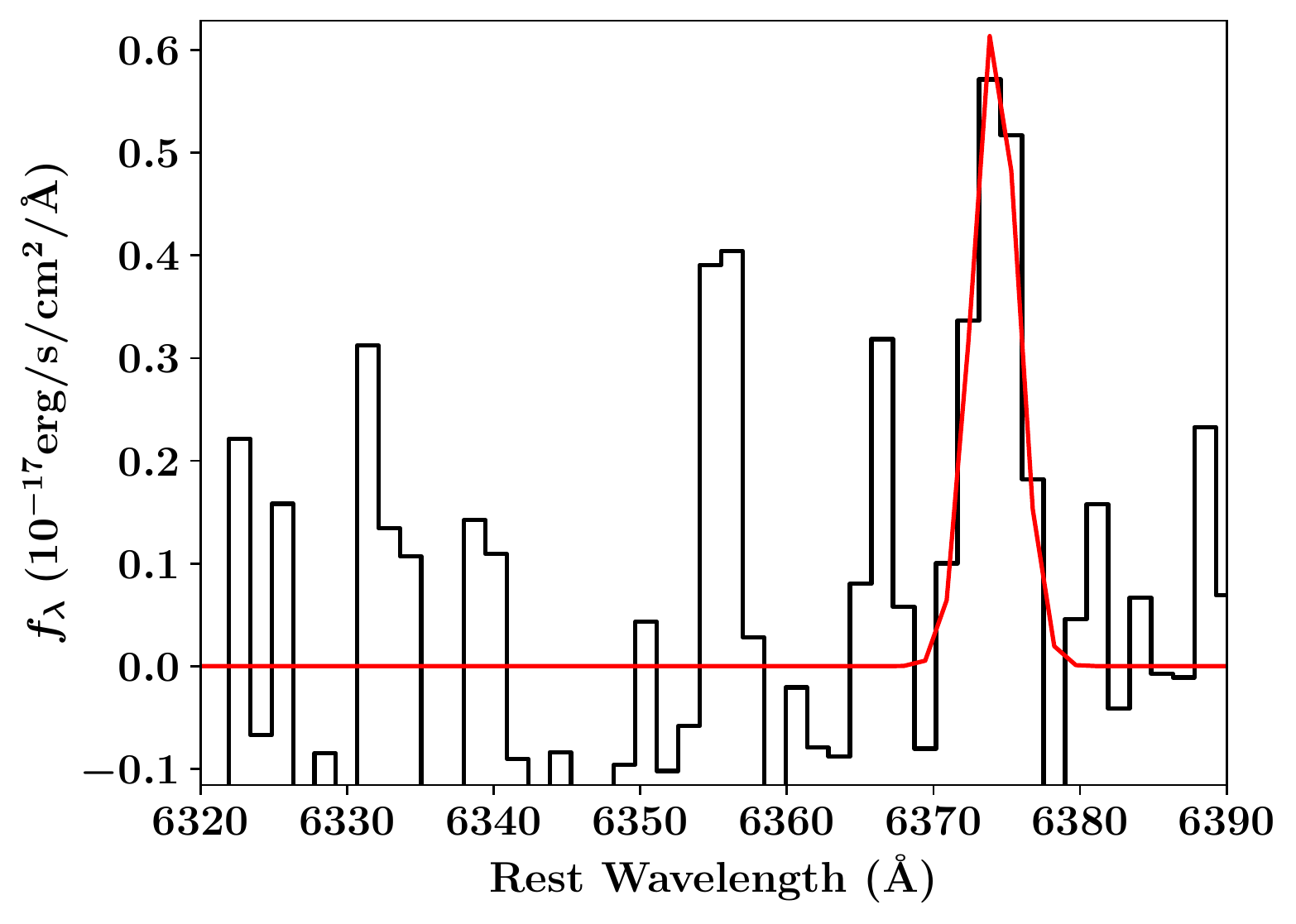}  
        \hspace{1.5mm}
    \includegraphics[width=0.13\textwidth]{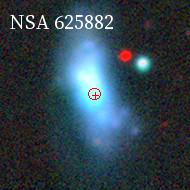}
        \hspace{-3mm}
    \includegraphics[width=0.19\textwidth]{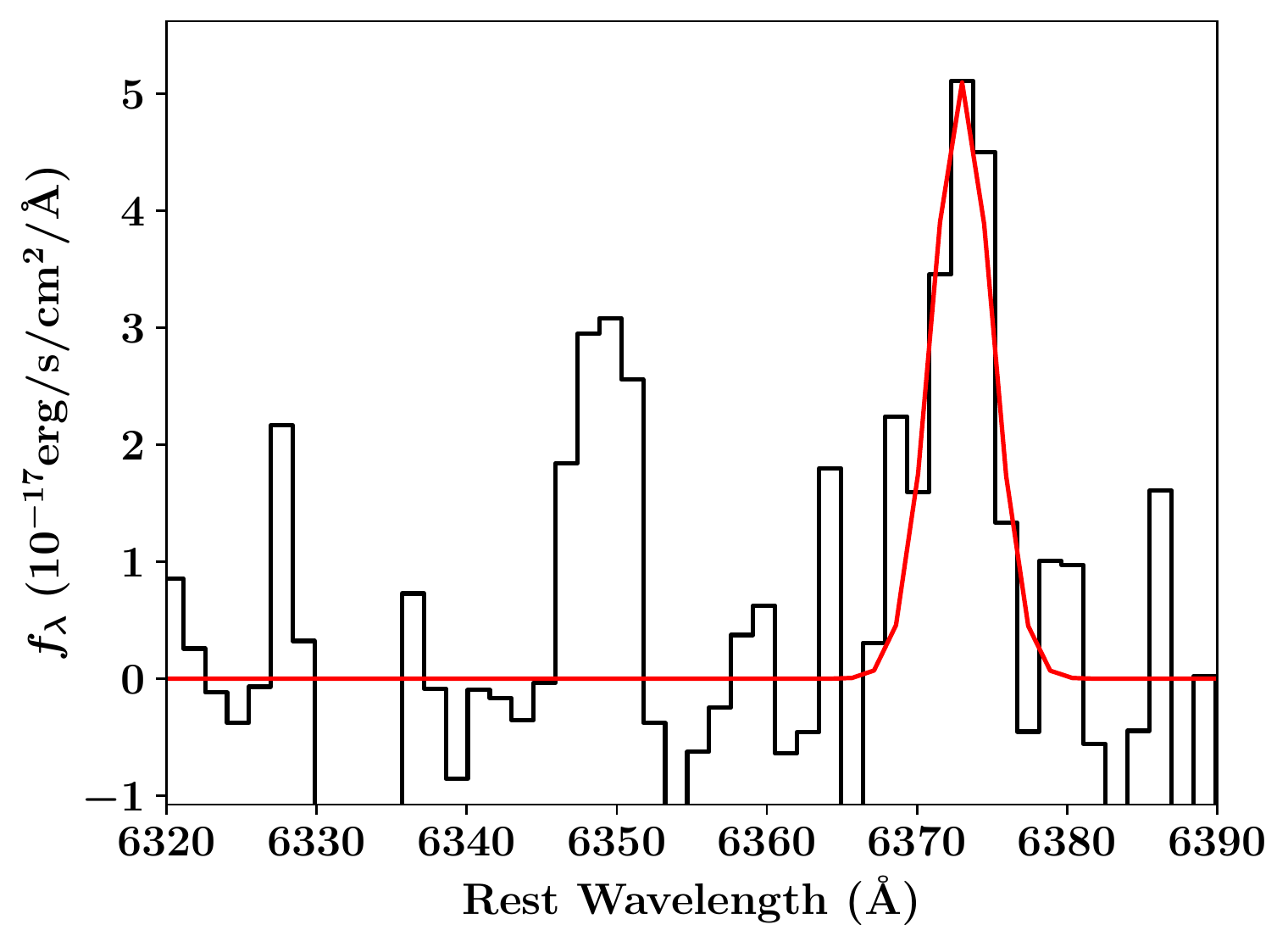}
  \includegraphics[width=0.13\textwidth]{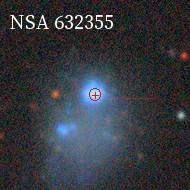}
    \hspace{-3mm}
    \includegraphics[width=0.19\textwidth]{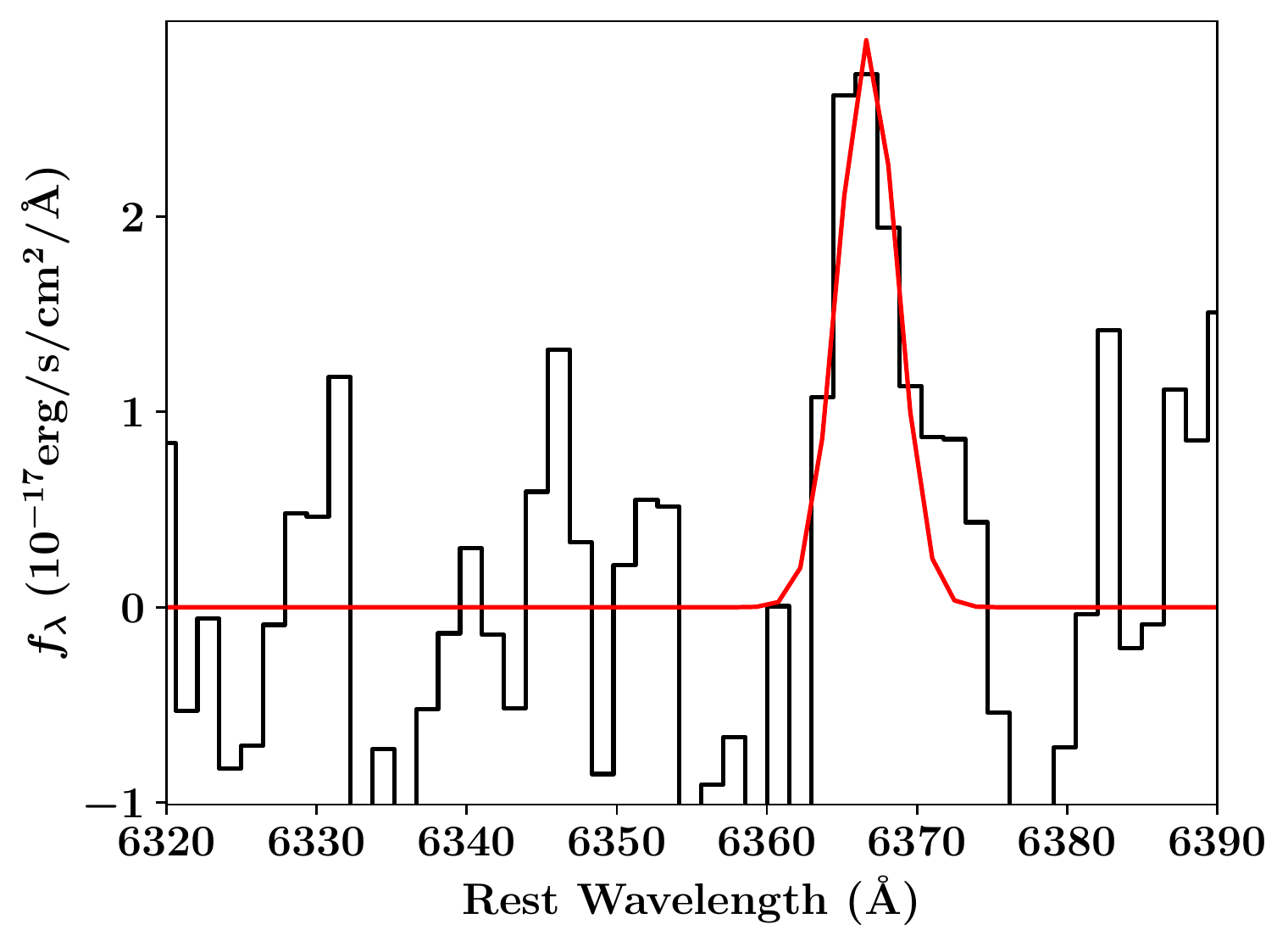}
    \hspace{1.5mm}
    \includegraphics[width=0.13\textwidth]{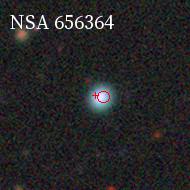}
        \hspace{-3mm}
    \includegraphics[width=0.19\textwidth]{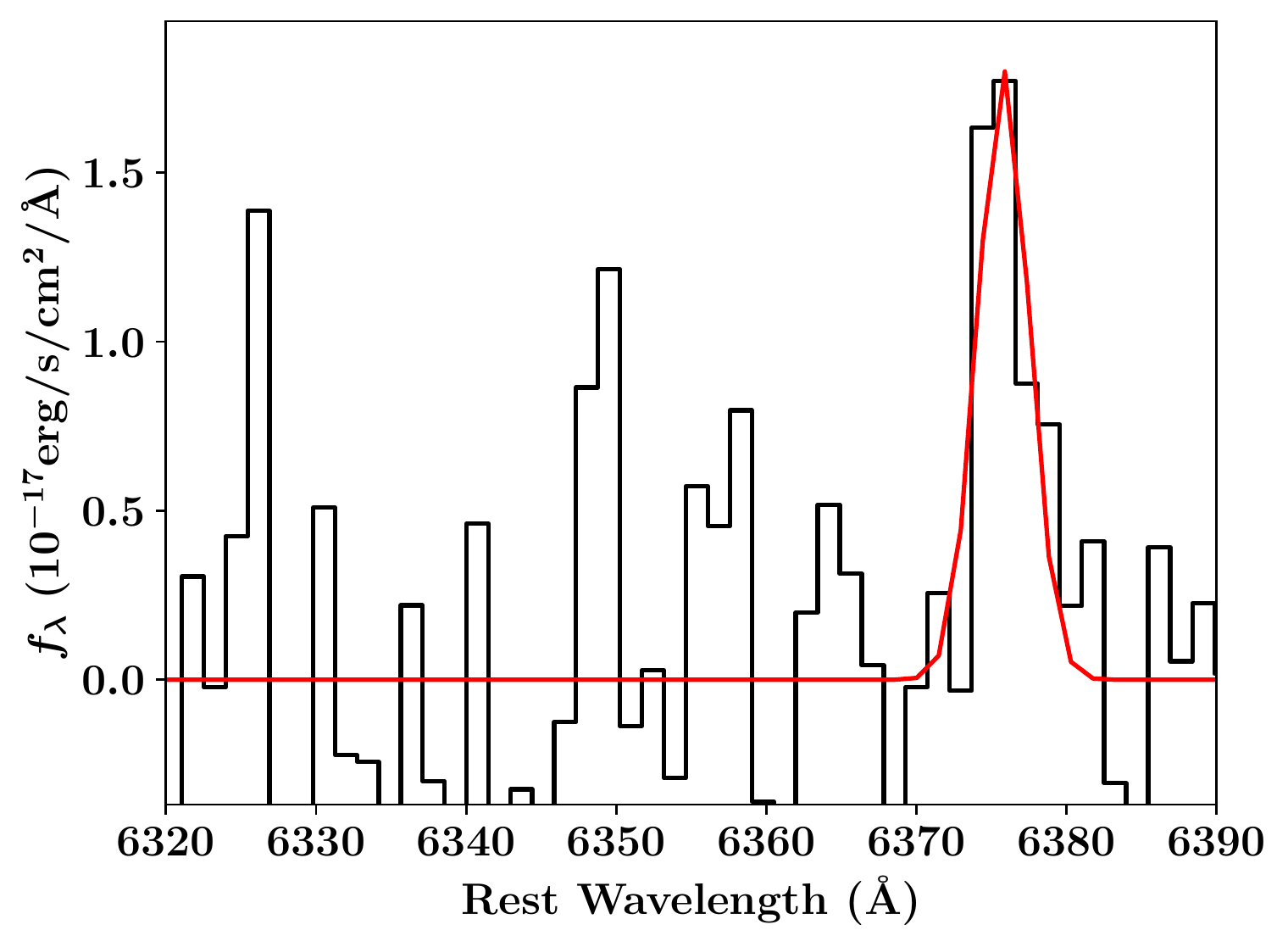}  
        \hspace{1.5mm}
    \includegraphics[width=0.13\textwidth]{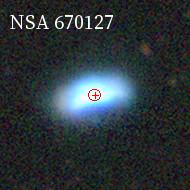}
        \hspace{-3mm}
    \includegraphics[width=0.19\textwidth]{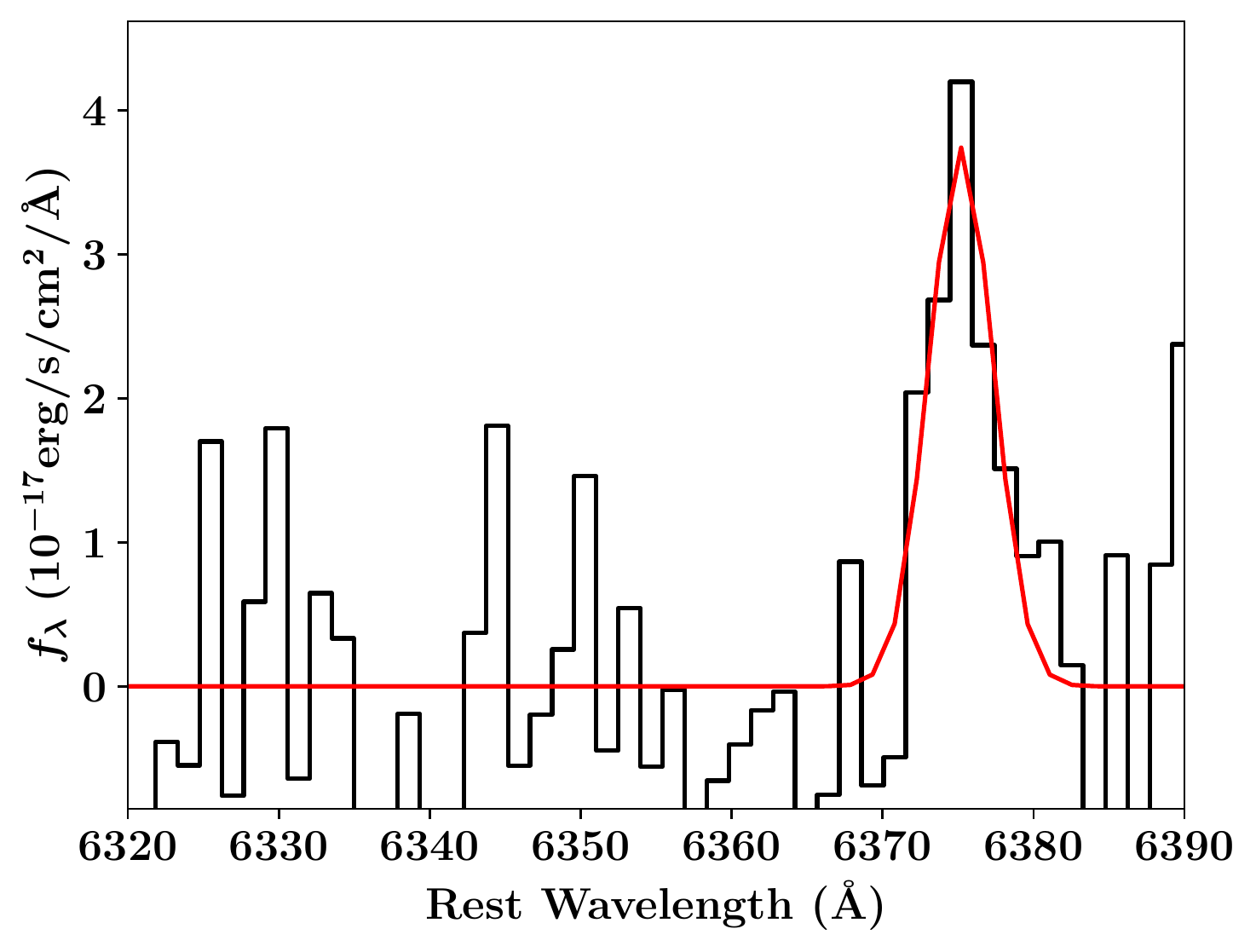}
    \caption{Same as Figure~\ref{fig:decals}.}\label{fig:decals_last}
\end{figure*}

The DESI Legacy Imaging Survey images of each \fexl-emitting galaxy along with the \fex\ emission-line fit are shown in Figures~\ref{fig:decals}--\ref{fig:decals_last}. The \fex\ fitting process is described in Section~\ref{sec:samp}.

\onecolumngrid

\clearpage

\clearpage

\section{Notes on Individual Objects in Previous Studies}

\begin{itemize}
    \item {\bf NSA 50:} This galaxy is a known \ion{He}{2} emitting galaxy \citep{Shirazi2012}, and the high-resolution VLA data taken by \cite{reines2020} could not rule out star formation as the cause of the observed radio emission. We confirm the result from \cite{reines2020} using the FIRST survey radio detections in this work. Additionally, we could not rule out high-mass XRBs as the cause of the observed X-ray emission. Furthermore, the \fex\ emission seen in the SDSS spectrum is consistent with that of SN2005ip. We do note that the \ion{He}{2}/H$\beta$ ratio indicates the presence of an AGN, and this galaxy appeared Seyfert-like in the [\ion{O}{3}]/H$\beta$ vs.~[\ion{S}{2}]/H$\alpha$ and [\ion{O}{1}]/H$\alpha$ diagrams. Given the low \fex\ detection, it is unclear whether the optical emission is driven by AGN activity or a very energetic Type IIn SNe driving strong shocks into the circumstellar medium. 
    \item {\bf NSA 95985:} This galaxy was designated as an X-ray AGN candidate by \cite{Birchall2020}. While the detected \fex\ emission is consistent with that seen in SN2005ip, the object is Seyfert-like in both the [\ion{O}{3}]/H$\beta$ vs.~[\ion{S}{2}]/H$\alpha$ and [\ion{O}{1}]/H$\alpha$ diagrams. 
    \item {\bf NSA 131809:} This galaxy was presented as a variable AGN candidate by \cite{Baldassare2020}. In this work, we find that the observed \fexl\ emission is about 7 times stronger than that seen in the Type IIn supernova SN2005ip, providing additional support for an AGN. 
    \item {\bf NSA 256802:} This galaxy was designated as a broad-line AGN by \cite[][their ID 20]{reines2013}. We recover this designation, and note that this is the only AGN object from \cite{reines2013} that overlaps with our sample. Given the strong X-ray emission shown in Figure~\ref{fig:expxrbalt} as well as the prominent broad H$\alpha$ emission, we conclude the \fex\ emission seen in this object is likely produced by gas photoionized by a radiatively efficient accretion disk. 
    \item {\bf NSA 275961:} This galaxy was designated as a composite object by \citet[][their ID 105]{reines2013}. In this work, we recover the designation of composite using the VO87 diagrams, and find that the object appears AGN-like via the \ion{He}{2}/H$\beta$ ratio and the [\ion{S}{2}]/H$\alpha$ vs.~[\ion{O}{2}]$\lambda\lambda$7320,7330/H$\alpha$ diagram. Additionally, the archival {\it Chandra} data indicate that this object is also AGN-like (see Figure~\ref{fig:expxrbalt}). 
    \item {\bf NSA 300542:} This galaxy was labeled as a SNe candidate by \cite{reines2013} via the detection of a P Cygni profile in H$\alpha$ (NSAv0 61339 in their work). However, the P Cygni profile was very subtle in this object, as shown in Figure 7 in \cite{reines2013}. In this work, we found that NSA 300542 appears Seyfert-like in the [\ion{O}{3}]/H$\beta$ vs.~[\ion{S}{2}]/H$\alpha$ and [\ion{O}{1}]/H$\alpha$ diagrams as well as the [\ion{S}{2}]/H$\alpha$ vs.~[\ion{O}{2}]$\lambda\lambda$7320,7330/H$\alpha$ diagram. Additionally, the \fexl\ emission is about 6 times stronger than SN2005ip. 
    \item {\bf NSA 427201:} This galaxy was labelled as a star-forming galaxy with broad H$\alpha$ by \citet[][their object N]{reines2013}, and not detected in the radio by \cite{reines2020}. We recover the broad component first seen by \cite{reines2013}, and also find that the observed radio emission in both VLASS and FIRST cannot be explained by stellar processes. Additionally, the \fexl\ emission seen in this object is $\sim19$ times higher than that seen in SN2005ip. Given the lack of broad-line emission seen by \cite{Baldassare2016} and radio emission seen by \citet{reines2020}, we conclude this object is a TDE candidate. See Section~\ref{ssec:origin} for a more thorough discussion.
    \item {\bf NSA 533731:} This galaxy is also known as Mrk 709S, is in a pair of interacting galaxies first studied in depth by \cite{Reines2014}, and is one of the two objects that inspired this work. Mrk 709S was confirmed to harbor a LLAGN with a RIAF engine and strong out-flowing winds by \cite{Kimbro2021}. In this work, we find that the observed \fexl\ emission is almost 23 times higher than that seen in SN2005ip, and has an additional marker of AGN activity via strong radio emission seen in both the FIRST and VLASS surveys.
\end{itemize}
\clearpage


\bibliographystyle{apj}
\bibliography{all_papers.bib}

\end{document}